\pdfoutput=1
\documentclass[a4paper,11pt]{article}
\usepackage{amsmath,amssymb,amsfonts}
\usepackage[multiple,stable]{footmisc}
\usepackage[amsmath,amsthm,thmmarks]{ntheorem}
\allowdisplaybreaks[4]
\usepackage[noconfig]{refstyle}
\usepackage[dvipsnames]{xcolor}
\usepackage[hyperfootnotes=false, linktocpage=true, colorlinks, citecolor=blue, linkcolor=blue, urlcolor=Maroon]{hyperref}
\usepackage{array,tabularx,booktabs,geometry,subfigure,graphicx,longtable}
\usepackage[bf]{caption}
\usepackage{tikz}
\usepackage{cite}
\usepackage{multirow}
\usetikzlibrary{shapes,arrows}
\tikzstyle{block} = [rectangle, draw, text width=7em, text centered, rounded corners, minimum height=3em]

\let\eqref=\relax
\newref{eq}{name={eq.~},Name={Eq.~},names={eqs.~},Names={Eqs.~},rngtxt={-},refcmd=(\ref{#1})}
\newref{f}{name={footnote~},Name={Footnote~},names={footnotes~},Names={Footnotes~}}
\newref{app}{name={appendix~},Name={Appendix~},names={appendixes~},Names={Appendixes~}}
\newref{tab}{name={table~},Name={Table~},names={tables~},Names={Tables~}}
\newref{ch}{name={chapter~},Name={Chapter~},names={chapters~},Names={Chapters~}}
\newref{sec}{name={section~},Name={Section~},names={sections~},Names={Sections~}}
\newref{fig}{name={figure~},Name={Figure~},names={figures~},Names={Figures~}}
\numberwithin{equation}{section}

\newcommand{\be}{\begin{equation}}
\newcommand{\ee}{\end{equation}}
\newcommand{\bea}{\begin{equation}\begin{aligned}}	
\newcommand{\eea}{\end{aligned}\end{equation}}		

\newcommand{\iddots}{\mathinner{\mkern2mu\raise1pt\hbox{.}\mkern2mu \raise4pt\hbox{.}\mkern2mu\raise7pt\hbox{.}\mkern1mu}}

\providecommand{\id}{\leavevmode\hbox{\small$\mathrm{1}$\kern-3.8pt\normalsize$\mathrm{1}$}}
\def\fnote#1#2{\begingroup\def\thefootnote{#1}\footnote{#2}
     \addtocounter{footnote}{-1}\endgroup}

\begin{document}

\vspace{1cm}

\title{       {\Large \bf Jumping Spectra and Vanishing Couplings \\in Heterotic Line Bundle Standard Models}}

\vspace{2cm}

\author{
James~Gray${}{}$ and
Juntao~Wang${}{}$
}
\date{}
\maketitle
\begin{center} {\small ${}${\it Department of Physics, 
Robeson Hall, Virginia Tech,\\ Blacksburg, VA 24061, U.S.A.}}\\
\vspace{0.1cm}
{\small ${}${\it Simons Center for Geometry and Physics,\\Stony Brook,~NY 11794,~U.S.A.}}

\fnote{}{jamesgray@vt.edu}
\fnote{}{wjunt88@vt.edu}

\end{center}

\begin{abstract}
\noindent
We study two aspects of the physics of heterotic Line Bundle Standard Models on smooth Calabi-Yau threefolds. First, we investigate to what degree modern moduli stabilization scenarios can affect the standard model spectrum in such compactifications. Specifically, we look at the case where some of the complex structure moduli are fixed by a choice of hidden sector bundle. In this context, we study the frequency with which the system tends to be forced to a point in moduli space where the cohomology groups determining the spectrum in the standard model sector jump in dimension. Second, we investigate to what degree couplings, that are permitted by all of the obvious symmetries of the theory, actually vanish due to certain topological constraints associated to their higher dimensional origins. We find that both effects are prevalent within the data set of heterotic Line Bundle Standard Models studied.
\end{abstract}

\thispagestyle{empty}
\setcounter{page}{0}
\newpage

\tableofcontents

\section{Introduction}

In the last ten to fifteen years a lot of progress has been made in understanding supersymmetric four dimensional effective theories, descending from smooth Calabi-Yau compactifications of heterotic M-theory. In terms of model building, solutions to the theory which give rise to a charged matter spectrum identical to that of the Minimal Supersymmetric Standard Model (MSSM) have been obtained \cite{Candelas:1985en,gsw,Greene:1986bm,Greene:1986jb,Braun:2011ni,Braun:2005ux,Braun:2005bw,Braun:2005zv,Distler:1987ee,Kachru:1995em,Bouchard:2005ag,Braun:2005nv,Bouchard:2006dn,Blumenhagen:2006ux,Blumenhagen:2006wj,Anderson:2007nc,Anderson:2008uw,Anderson:2009mh,Anderson:2011ns,Anderson:2012yf,Anderson:2013xka,He:2013ofa,Constantin:2015bea,Braun:2017feb,Constantin:2018xkj}. These were first constructed in small numbers in the context of irreducible higher rank bundles with non-abelian structure groups \cite{Braun:2005ux,Bouchard:2005ag,Braun:2005zv,Anderson:2009mh,Braun:2011ni}. Later, the concept of Line Bundle Standard Models was introduced: it was realized that simple sums of line bundles could be phenomenological viable in this context \cite{Anderson:2011ns,Anderson:2012yf}. This work is of course complemented by extensive model building efforts in other heterotic constructions, see for example \cite{Buchmuller:2005jr,Buchmuller:2006ik,Lebedev:2006kn,Kim:2007mt,Lebedev:2007hv,Lebedev:2008un,Nibbelink:2009sp,Blaszczyk:2009in,Blaszczyk:2010db,Kappl:2010yu,Assel:2009xa,Christodoulides:2011zs,Cleaver:2011ir,Mutter:2018sra,Faraggi:2017cnh,Maio:2011qn,GatoRivera:2009yt,GatoRivera:2010xn}. This lead to very large numbers of heterotic models being produced with exactly the standard models charged matter content. In another advance, that will be directly relevant to this paper, good progress has been made in understanding Yukawa couplings in this context \cite{Strominger:1985ks,Candelas:1987se,Candelas:1990pi,Greene:1986bm,Greene:1986jb,Distler:1987gg,Distler:1987ee,Greene:1987xh,Distler:1995bc,Braun:2006me,Bouchard:2006dn,Anderson:2009ge,Anderson:2010tc,Buchbinder:2014sya,Blesneag:2015pvz,Blesneag:2016yag}. Algebraic methods for computing tree-level superpotential trilinear couplings have long been understood \cite{Strominger:1985ks,Candelas:1987se,Candelas:1990pi,Greene:1986bm,Greene:1986jb,Distler:1987gg,Distler:1987ee,Greene:1987xh,Distler:1995bc,Braun:2006me,Bouchard:2006dn,Anderson:2009ge,Anderson:2010tc,Buchbinder:2014sya}. Recently, however, techniques based upon differential geometry have been developed \cite{Blesneag:2015pvz,Blesneag:2016yag} which, perhaps surprisingly, can be more powerful in many situations. In particular, this work provides a very strong vanishing theorem on these tree-level Yukawa couplings and also makes the computation of the moduli dependence of these quantities more tractable in many contexts.

Although heterotic compactifications have traditionally proven to be extremely promising from the point of view of particle physics model building, they have struggled more in the context of moduli stabilization. Nevertheless there have been a number of recent advances in understanding the ${\cal N}=1$ effective theories associated to these compactifications which have lead to new moduli stabilization mechanisms in this context. Of particular note for the current paper, it has been realized that the holomorphic poly-stable slope zero vector bundles that appear in this context can stabilize the complex structure moduli of the base Calabi-Yau manifold \cite{Donagi:2009ra,Anderson:2010mh,Anderson:2011ty,Anderson:2011cza,Anderson:2013qca}. It is important to note in this context that concrete examples of this effect have been provided. While it is still difficult to fix one final over-all modulus in a controlled manner in heterotic compactifications (see \cite{Anderson:2011cza} for example), it is clear that progress is being made. In addition, there is much that is still not understood about the effective theories' potential - particularly at higher order in curvature expansions.

Given this progress in model building and moduli stabilization it is natural to take the analysis of these models to a finer level of detail. In this paper we wish to achieve this in two particular regards. First, we wish to begin a study of how modern moduli-stabilization mechanisms in Calabi-Yau compactifications of heterotic M-theory interact with model building concerns. More specifically, we will examine the interplay of the moduli stabilization of \cite{Anderson:2010mh,Anderson:2011ty} with Line Bundle Standard Model building \cite{Anderson:2011ns,Anderson:2012yf}. Using hidden sector vector bundles to stabilize complex structure moduli, as was proposed in \cite{Anderson:2010mh,Anderson:2011ty}, forces the base Calabi-Yau threefold to a computable sub-locus of its moduli space. Given this concrete knowledge as to where in complex structure moduli space the system is forced, one can investigate how this stabilization mechanism affects model building considerations. In particular, the bundle valued cohomologies that determine particle spectra in heterotic theories are only quasi-topological in nature. They can jump in dimension at higher co-dimensional loci in complex structure moduli space causing the matter spectrum of the associated four dimensional effective theory to jump in an index preserving manner \cite{Donagi:2004qk,Donagi:2004ia}. If the moduli stabilization mechanism of \cite{Anderson:2010mh,Anderson:2011ty} happens to force the system to a locus where the bundle cohomologies associated to standard model degrees of freedom jump, then that mechanism and model building considerations can not be divorced. 

This effect can be either good or bad. If the jump causes the addition of an extra standard model family degree of freedom and its partner from a mirror family, then the moduli stabilization mechanism will have forced the addition of standard model exotics - a phenomenologically undesirable result. In contrast to this, one could envisage a situation where a model which had no Higgs, Higgs conjugate pair, was forced to a locus where the cohomologies of such degrees of freedom where forced to jump. This would render previously unviable models phenomenologically interesting. 

One might think that such effects would be extremely rare in heterotic models, given the relatively uncoupled nature of the visible and hidden sector vector bundles. Nevertheless, we will show that, in the class of models we study, this interaction of moduli stabilization and model building considerations occurs rather frequently. More precisely, we find that, in cases where the particle spectrum of the standard model bundle is capable of jumping, such phenomena are common in the known examples of Line Bundle Standard Models. This indicates that one should be aware, in pursuing studies that divorce model building from moduli stabilization, that including the latter concern may be relevant to many of the models obtained. 

It should be noted that this effect, where the system is driven to a locus in moduli space where extra degrees of freedom occur, might be naively thought to be rather similar in nature to the work presented in \cite{Brandle:2002fa,Jarv:2003qy,Kofman:2004yc,Mohaupt:2004pr,Lukas:2004du,Abel:2005jx,Greene:2007sa}. In fact the phenomena being considered here are completely distinct to that work, being rather different in nature and not as ubiquitous in effect.

The second issue we will consider in this paper concerns vanishing of Yukawa couplings. As was mentioned above, in \cite{Blesneag:2015pvz,Blesneag:2016yag} a vanishing theorem was presented wherein tree-level trilinear couplings that are consistent with all of the obvious gauge symmetries of the four dimensional effective theory are nevertheless zero due to seemingly topological restrictions. We will investigate to what degree this vanishing theorem comes in to effect in the known set of Line Bundle Standard Models \cite{Anderson:2011ns,Anderson:2012yf}. By the simple method of direct computation in every model in this data set, we discern how many of the couplings that are consistent with the symmetries of these theories, as presented in \cite{Anderson:2011ns,Anderson:2012yf}, are actually vanishing due to this theorem. In total $17.9\%$ of the potentially allowed couplings are actually zero, with some forms of interaction vanishing at the $35.4\%$ level. This is therefore, once again, a significant effect which should be borne in mind when constructing heterotic standard models with an eye toward phenomenological viability. That this effect is common was anticipated in \cite{Blesneag:2015pvz,Blesneag:2016yag} - here we compute exact numbers in a standard model building context. In addition to this straight forward computation we briefly suggest, based on the work of \cite{Anderson:2010tc,Buchbinder:2014sya}, a gauge-theoretic mechanism which may underly these severe restrictions on the Yukawa-Couplings of these heterotic effective theories. It will be important to understand whether this conjecture is correct going forwards as, if it is indeed responsible for these vanishings, then one  could expect many higher order couplings to suffer a similar fate.

The structure of the rest of this paper is as follows. In Section \ref{lsmreview} we briefly review Line Bundle Standard Models in Calabi-Yau threefold compactifications of heterotic theories. We then review, in Section \ref{stabintro}, the mechanism by which hidden sector bundles can stabilize complex structure moduli in this context. In Section \ref{specjump} we present our work combining moduli stabilization and model building considerations in heterotic Line Bundle Standard Models. Section \ref{yuksec} of the paper contains our analysis of topological vanishing of Yukawa couplings in Line Bundle Standard Models. Finally, in Section \ref{conc} we present our conclusions. Two appendices contain details of the results from our two lines of investigation which complement the summary data given in the main text.


\section{Heterotic Line Bundle Standard Models} \label{lsmreview}

Traditionally, in constructing a heterotic Calabi-Yau compactifications designed to give rise to physics close to the MSSM, one chooses a gauge bundle $V_{\textnormal{SM}}$ with a non-abelian structure group, for example $SU(3)$, $SU(4)$ or $SU(5)$. The low energy gauge group in the visible sector is then simply the commutant of this structure group inside $E_8$, that is $E_6$, $SO(10)$ or $SU(5)$ respectively for the examples mentioned in the previous sentence. These precursor `GUT' groups are then broken down to the standard model gauge group by Wilson lines associated with the fundamental group of the Calabi-Yau threefold.

Line Bundle Standard Models are constructed somewhat differently. Instead of focussing on a non-abelian structure group, the gauge bundle $V_{\textnormal{SM}}$ is chosen to be a simple sum of line bundles. Taking a sum of five such objects as an example, we have the following.
\begin{eqnarray} \label{firstsum}
V_{\textnormal{SM}}= \bigoplus_i^5 {\cal L}_i
\end{eqnarray}
 The structure group of such a bundle is $S(U(1)^5) \cong U(1)^4$. The commutant of this group inside $E_8$ is $SU(5)\times U(1)^4$ which is therefore, naively, the low energy gauge group. However, the four $U(1)$ factors are all typically Green-Schwarz massive, at least in examples with a K\"ahler moduli space of high enough dimension, and thus at low energies this approach can also give us viable GUT groups that can then be broken to $SU(3)\times SU(2) \times U(1)$ by an appropriate Wilson line.
 
 The advantage of working with Line Bundle Standard Models over more conventional approaches to heterotic model building largely center around proving that the gauge fields in the compactification preserve supersymmetry. Showing that an irreducible, higher rank bundle is slope-stable can be a time consuming and complicated affair, involving the consideration of an infinite number of possible sub-sheafs of $V_{\textnormal{SM}}$. In the case of a simple sum of line bundles such as (\ref{firstsum}) proving that supersymmetry is preserved is much simpler. The equivalent condition in this case is slope poly-stability and for such a sum we need only check that the slope of each line bundle is the same (and in fact vanishes in physical examples). This simplification leads to a huge increase in the number of models that can be constructed with thousands of Line Bundle Standard Models being known \cite{Anderson:2011ns,Anderson:2012yf} while only a few irreducible higher rank gauge bundles have ever been constructed which give rise to the exact charged spectrum of the MSSM \cite{Bouchard:2005ag,Braun:2005nv,Anderson:2009mh,Braun:2011ni}.

The spectrum of a Line Bundle Standard Model is determined in a two step process. Firstly, an exercise in group theory tells us what matter can possibly appear in the four dimensional effective theory. Secondly, what matter actually does appear is computed in terms of bundle valued cohomology groups.

In terms of group theory, the representations of the four dimensional gauge group that can appear are simply determined by branching rules and the fact that all of the charged matter in ten dimensions is valued in the adjoint representation. Thus, in the $SU(5)$ case for example we find the following decomposition of representations under a maximal subgroup.
\begin{eqnarray} \label{gt}
E_8 &=&SU(5) \times SU(5)\\\nonumber
{\bf 248} &=& ({\bf 24},{\bf 1}) \oplus ({\bf 1},{\bf 24}) \oplus ({\bf 5},{\bf 10}) \oplus (\overline{\bf 5},\overline{\bf 10}) \oplus ({\bf 10},\overline{\bf 5}) \oplus(\overline{\bf 10},{\bf 5})
\end{eqnarray}
If we take the first $SU(5)$ factor to be the low energy GUT group and the second $SU(5)$ factor to be that in which the structure group of the bundle resides we can then read off what representations we can possibly obtain in four dimensions. Here, for example, we could potentially obtain the ${\bf 24}, {\bf 1}, {\bf 5}, \overline{\bf 5}, {\bf 10}$ and $\overline{\bf 10}$ representations of $SU(5)$. In the case of Line Bundle Standard Models, we can also, of course associate a series of $U(1)$ charges to the matter multiplets which we have omitted in (\ref{gt}) in the interests of keeping the expressions uncluttered. We follow the convention of including all five $U(1)$ charges associated to $S(U(1)^5)$ despite the fact that only four of these gauge factors are independent as this simplifies many of the resulting equations. 

In order to see how many copies of each representation we obtain in the low energy spectrum (if any) we must compute the appropriate bundle cohomology groups. In fact, we wish to incorporate a Wilson line and work out the spectrum at the level of the four dimensional theory with standard model gauge group. Since most Calabi-Yau that we know how to construct are simply connected, this we typically obtain a compactification manifold with non-trivial fundamental group that can support a Wilson line by quotienting some `upstairs' space $X$ by an appropriate freely acting discrete symmetry $\Gamma$. The bundle must be chosen to be equivariant with respect to this symmetry in order that it too is compatible with the quotient. Indeed, following \cite{Anderson:2011ns,Anderson:2012yf} we will consider the case where each line bundle ${\cal L}_i$ in $V_{\textnormal{SM}}$ is equivariant individually. The spectrum on the `downstairs' quotient manifold $\hat{X}=X/\Gamma$ can then be given in terms of just a few pieces of data.

 As described in \cite{Anderson:2011ns,Anderson:2012yf}, if the discrete group $\Gamma$ is a product of abelian factors of the form $\Gamma=\bigotimes_r \mathbb Z_{m_r}$ (as will be considered here), then the definition of the Wilson line proceeds via the choice of two sets of integers $p_r$ and $\tilde{p}_r$. These integers must satisfy the conditions
\begin{eqnarray} 
 3 p_r + 2 \tilde{p}_r = 0 \; \textnormal{mod} \; m_r \; \forall r \;\; \textnormal{such that} \;\; p_r \neq \tilde{p}_r \;\; \textnormal{for at least one $r$}
 \end{eqnarray}
 We can then define some representations $W(g)=\bigotimes_r e^{ p_r g 2 \pi i/m_r}$ and $\tilde{W}(g)=\bigotimes_r e^{ \tilde{p}_r g 2 \pi i/m_r}$. These representations encode all of the information we require about the Wilson line in order to complete a spectrum computation. Indeed, if we combine this information with the characters of $\Gamma$, $\chi_i^*$, which define the equivariant structure associated to the line bundle ${\cal L}_i$, we can write down the spectrum of the Line Bundle Standard Model associated to these choices, as given in Table \ref{spectab}.
\begin{table}
\begin{center}
\begin{tabular}{|l|l|l|l|}\hline
 $SU(5)$ repr.&$G_{\rm SM}$ repr.&name&cohomology\\\hline\hline
 ${\bf 10}_{{\bf e}_i}$&$({\bf 3},{\bf 2})_1$&$Q_i$&$h^1(X,{\cal L}_i,\chi_i\otimes W^*\otimes\tilde{W}^*)$\\
 &$(\bar{\bf 3},{\bf 1})_{-4}$&$u_i$&$h^1(X,{\cal L}_i,\chi_i\otimes W^*\otimes W^*)$\\
 &$({\bf 1},{\bf 1})_6$&$e_i$&$h^1(X,{\cal L}_i,\chi_i\otimes\tilde{W}^*\otimes\tilde{W}^*)$\\\hline
 $\bar{\bf 5}_{{\bf e}_i+{\bf e}_j}$&$(\bar{\bf 3},{\bf 1})_2$&$d_{i,j},\, T_{i,j}$&$h^1({\cal L}_i\otimes {\cal L}_j,\chi_i\otimes \chi_j\otimes W)$\\
 &$({\bf 1},{\bf 2})_{-3}$&$L_{i,j},\, H_{i,j}$&$h^1({\cal L}_i\otimes {\cal L}_j,\chi_i\otimes\chi_j\otimes\tilde{W})$\\\hline
 ${\bf 5}_{-{\bf e}_i-{\bf e}_j}$&$({\bf 3},{\bf 1})_{-2}$&$\bar{T}_{i,j}$&$h^2({\cal L}_i\otimes {\cal L}_j,\chi_i\otimes \chi_j\otimes W)$\\
 &$({\bf 1},{\bf 2})_3$&$\bar{H}_{i,j}$&$h^2({\cal L}_i\otimes {\cal L}_j,\chi_i\otimes\chi_j\otimes\tilde{W})$\\\hline
  ${\bf 1}_{{\bf e}_i-{\bf e}_j}$&$({\bf 1},{\bf 1})_0$&$S_{i,j}$&$h^1({\cal L}_i\otimes {\cal L}_j^{\vee},\chi_i\otimes \chi_j^*)$\\\hline
\end{tabular}
\caption{\it\small Cohomologies which determine the downstairs spectrum of Line Bundle Standard Models. The cohomological notation including a representation after a comma simply denotes that only the piece of the cohomology forming that representation under the discrete group $\Gamma$ should be considered. The representations $W$, $\tilde{W}$ and $\chi_i$ are described in the text. The number of mirror particles is determined by the second cohomology valued in the same bundles and representations.}\label{spectab}
\end{center}
\end{table}
Note that in this table we use the same notation for the (potentially anomalous) $U(1)$ charges as given in \cite{Anderson:2011ns,Anderson:2012yf}. That is, the ${\bf e_i}$ are unit vectors such that, for example $10_{\bf e_1}$ has a single unit of positive charge under the first abelian factor. Note that, because these five $U(1)$ factors are related in $S(U(1)^5)$, a combination of fields that has a single unit of charge under each factor is a gauge invariant. We will frequently specify the spectrum of a Line Bundle Standard Model by giving a set of GUT multiplets with $U(1)$ charges. Such a notation is consistent because, despite the fact that the different standard model degrees of freedom that would form a single irreducible $SU(5)$ multiplet all descend from different ten-dimensional antecedents and thus no such symmetry is present in the four-dimensional theory, the standard model multiplets do arise in complete GUT multiplets.

A key point for the latter sections of the current paper is that the cohomologies appearing in Table \ref{spectab} are complex structure dependent. At higher codimension loci in complex structure moduli space, the dimensions of these cohomology groups, and thus the matter spectrum of the resulting four dimensional theory can jump in an index preserving manner.

We will consider a particular existent Line Bundle Standard Model data set \cite{Anderson:2011ns,Anderson:2012yf} built over Calabi-Yau manifolds which can be described as quotients of complete intersections in products of projective spaces (CICYs) \cite{Yau:1986gu,Hubsch:1986ny,Candelas:1987kf,Candelas:1987du,Green:1986ck,Gray:2013mja,Gray:2014fla}. Note that analogous constructions could be pursued over different base spaces, such as quotients of gCICYs \cite{Anderson:2015iia} or toric hypersurfaces \cite{Batyrev,Berglund:1991pp,Kreuzer:2000qv,Kreuzer:2000xy,Altman:2014bfa}. It would be interesting to see if such constructions mirror the structure that we will describe in this paper.

A family of CICYs can be represented by a configuration matrix of the following form.
 \begin{eqnarray} \label{config}
X=\left [ \begin{array}{c|cccc}
\mathbb{P}^{n_1} & q^1_1 & q^1_2 & \dots & q^1_k  \\
\mathbb{P}^{n_2} & q^2_1 & q^2_2 & \dots & q^2_k\\
\vdots & \vdots & \vdots & \ddots & \vdots  \\
\mathbb{P}^{n_m} & q^{m}_1 & q^{m}_2 & \dots & q^{m}_k
\end{array}
\right ],
\end{eqnarray}
Here, the first column specifies the ambient space $A$ in which the Calabi-Yau manifold $X$ will be defined, $A=\mathbb{P}^{n_1}\times\dots\times\mathbb{P}^{n_k}$. The manifold $X$ is defined within this ambient space as the common zero locus of a set of $k$ defining polynomials. The remaining columns each determine the multi-degree of one of these defining polynomials. In a given column each row specifies the degree of that defining relation with respect to the homogeneous coordinates of the corresponding ambient space factor. Throughout this paper we will denote by $x_{r,a}$ the $a^{\textnormal{th}}$ homogeneous coordinate on the $r^{\textnormal{th}}$ ambient space projective factor.

The dimension of a complete intersection described by a configuration matrix of the form (\ref{config}) is simply $\sum_{r} n_r -k$. That is, the dimension is simply given as the dimension of the ambient space minus the number of constraints being imposed. The condition for a vanishing first Chern class for $X$, meaning that the manifold is indeed Calabi-Yau, can be achieved if the following condition is met.
\begin{eqnarray}
n_r+1=\Sigma_{l=1}^{k}q^r_l \;\; \forall \; r
\end{eqnarray}

The CICYs are all simply connected and therefore, in order to accommodate Wilson line breaking, quotients of these manifolds by appropriate freely acting discrete symmetries are considered. Braun has classified all such actions, allowing for a set of defining relations which respect the symmetry while remaining transverse, that descend from a linear action on the ambient spaces $A$ that appear in the original classifying list of such constructions \cite{Braun:2010vc}.

Having specified the Calabi-Yau manifolds to be utilized $\hat{X}$ in the above way, in \cite{Anderson:2011ns,Anderson:2012yf} the authors then produce Heterotic Line Bundle Standard Models by specifying appropriate sums of line bundles on $X$. These are chosen to be equivariant under the symmetries by which the manifolds are quotiented and to give rise to spectra on $\hat{X}$ which precisely match that of the standard model in the sector charged under $SU(3) \times SU(2) \times U(1)$. If one works with favorable CICYs, where all of $H^{1,1}(X)$ descends from forms dual to divisors on the ambient space, general line bundles on $X$ can be specified by the following notation.
\begin{eqnarray}
{\cal L} = \mathcal{O}_{\bold{X}}(p_1, p_2,\hdots, p_{k})  \Leftrightarrow c_1({\cal L}) = \sum_r p_r J_r
\end{eqnarray}
Here $J_r$ is the K\"ahler forms of the $r^\textnormal{th}$ ambient space factor, restricted to the Calabi-Yau threefold. These are the restriction of the analogous line bundles on the ambient space $\mathcal{O}_{\mathbb{A}}(p_1, p_2,\hdots, p_{k}))$ to $X$. The bundle $V_\textnormal{SM}$ is then taken to be an equivariant sum of five such objects. In fact, in the data set of \cite{Anderson:2011ns,Anderson:2012yf}, each line bundle in $V_\textnormal{SM}$ is taken to be equivariant individually. As we will see in concrete examples in later sections, the cohomology of various products of these line bundles can be computed using a combination of a theorem due to Bott, Borel and Weil and the Koszul sequence \cite{Hubsch:1992nu,Anderson:2008ex}. For a discussion of equivariance in this setting, and induced symmetry actions on cohomology see for example the appendices of \cite{Anderson:2012yf}. Once the cohomology, and its representation content, of the line bundles is known, the spectrum of the associated heterotic theory can be read off from Table \ref{spectab}.

Using such a construction, in \cite{Anderson:2011ns,Anderson:2012yf}, a data set of 2012 Line Bundle Standard Models was produced. It is properties of this data set that will be examined in the rest of this paper. It would certainly be interesting to apply a similar analysis to larger data sets of this type which could be obtained by extending the work of \cite{Anderson:2013xka}, for example.

\section{Moduli Stabilization, Potentials and Couplings}\label{stabintro}

The moduli stabilization mechanism we will consider in this paper concerns the complex structure degrees of freedom and was presented in \cite{Anderson:2010mh,Anderson:2011ty} (a similar description of moduli stabilization in Type II was considered in \cite{Gray:2018kss}). We will be particularly interested in the mechanism for fixing these particular moduli in the current work as the cohomology groups determining the spectrum of a model, as presented in the previous section, depend upon these degrees of freedom. The basic mechanism is as follows.

An ${\cal N}=1$ compactification of heterotic string theory on a Calabi-Yau threefold $X$ includes a gauge connection on a gauge bundle $V$ which satisfies the Hermitian Yang-Mills equations.
\begin{eqnarray}
g^{a\overline{b}} F_{a \overline{b}}=0 \;,\; F_{ab}= F_{\overline{a} \overline{b}} =0 
\end{eqnarray}

Starting with a good solution to these equations, one can consider a perturbation of all of the degrees of freedom of the problem about that vacuum. In particular, focusing on the holomorphy condition $F_{\overline{a}\overline{b}}=0$, we can perturb the complex structure and gauge connection and ask what constraints maintaining supersymmetry places on those variations. The following condition is obtained \cite{Anderson:2010mh,Anderson:2011ty}.
\begin{eqnarray} \label{ftatiyah}
\delta {\cal J}_{[\overline{a}}^{\;d} F_{\overline{b}]d}^{(0)} + 2 i  D_{[\overline{a}}^{(0)} \delta A_{\overline{b}]} =0
\end{eqnarray}

Here $\delta {\cal J} \in  H^1(TX)$ is a variation of the complex structure tensor, $\delta A$ is the perturbation in the gauge connection and objects with a superscript $(0)$ are constructed from unperturbed quantities. What (\ref{ftatiyah}) states is that a complex structure fluctuation is a true low energy degree of freedom only if there exists a gauge field fluctuation which solves this constraint. Otherwise, such a variation of complex structure will necessarily cause the bundle to become non-holomorphic, breaking supersymmetry.

Equation (\ref{ftatiyah}) can be interpreted cohomologically as saying that the complex structure moduli of the base Calabi-Yau threefold that are true massless degrees of freedom of the four dimensional effective theory are given as the following kernel.
\begin{eqnarray} \label{mrker}
\ker \left( H^1(TX) \stackrel{F^{(0)}}{\longrightarrow} H^2(\textnormal{End}_0(V))\right)
\end{eqnarray}
The allowed deformations of the connection are much easier to understand. A gauge field fluctuation living in the usual cohomology describing the bundle moduli, $\delta A \in H^1(\textnormal{End}_0(V))$, satisfies (\ref{ftatiyah}) for a vanishing $\delta {\cal J}$ and is therefore always consistent with holomorphy as one would expect.

The permitted combined deformations of the base complex structure and bundle moduli of holomorphic vector bundles is in fact very well studied in the mathematics literature. Indeed, the above discussion is simply a field theory manifestation of Atiyah's discussion of the tangent to the moduli space of holomorphic bundles \cite{atiyah}. Atiyah states that the allowed deformations are given by $H^1({\cal Q})$ where the bundle ${\cal Q}$ is defined by the following short exact extension sequence.
\begin{eqnarray} \label{qdef}
0 \to \textnormal{End}_0(V) \to {\cal Q} \to TX \to 0
\end{eqnarray}
Analyzing the long exact sequence in cohomology associated to (\ref{qdef}) one then finds the following,
\begin{eqnarray}
H^1({\cal Q}) = H^1(\textnormal{End}_0(V)) \oplus \textnormal{ker}\left( H^1(TX) \longrightarrow H^2(\textnormal{End}_0(V))\right)\;,
\end{eqnarray}
which agrees with the field theoretic analysis given above.

The above discussion shows in general terms a choice of bundle can restrict complex structure moduli via the requirement of holomorphy of that object. However, it will be crucial for the purposes of this paper to construct explicit examples of such bundles and compute to exactly which locus in complex structure moduli space the system is constrained.

Fortunately such examples have indeed been provided in the literature \cite{Anderson:2010mh,Anderson:2011ty,Anderson:2013qca}. Perhaps the simplest such examples take the form of bundles of $SU(2)$ structure group which are constructed as extensions of a line bundle and its dual. To see how this works it is simplest to look at an explicit case. The example that follows was first presented in \cite{Anderson:2013qca}.

As a base manifold, let us consider a freely acting quotient $\hat{X}$ of the the following CICY,
\begin{eqnarray}
X= \left[ \begin{array}{c|c} \mathbb{P}^1 & 2 \\\mathbb{P}^1 & 2 \\\mathbb{P}^1 & 2 \\\mathbb{P}^1 & 2 \\\end{array}\right]\;,
\end{eqnarray}
by the following $\mathbb{Z}_2 \times \mathbb{Z}_4$ symmetry action.
\begin{eqnarray}\label{firstsym}
\gamma_1 &:& x_{r,a} \to (-1)^{a+r+1}x_{r,a} \\ \nonumber
\gamma_2 &:& x_{r,a} \to x_{\sigma(r),a+r+1} \; \textnormal{where} \; \sigma=(12)(34)
\end{eqnarray}
It will useful going forward to know the most general form of the polynomial defining relation for $X$ that is consistent with the symmetry (\ref{firstsym}). This is explicitly given by the following expression.
\begin{eqnarray} 
p&=& \text{c}_1 x_{1,0} x_{1,1} x_{2,0} x_{2,1} x_{3,0} x_{3,1} x_{4,0}
   x_{4,1}+   \text{c}_9 \left(x_{1,0}^2 x_{3,0} x_{3,1} x_{4,0} x_{4,1}
   x_{2,0}^2+x_{1,1}^2 x_{3,0} x_{3,1} x_{4,0} x_{4,1} x_{2,0}^2  \right. \nonumber \\
&&  \left.+x_{1,0}^2 x_{2,1}^2
   x_{3,0} x_{3,1} x_{4,0} x_{4,1}+x_{1,1}^2 x_{2,1}^2 x_{3,0} x_{3,1} x_{4,0}
   x_{4,1}\right)+   \text{c}_3 \left(x_{1,1}^2 x_{2,0} x_{2,1} x_{4,0} x_{4,1}
   x_{3,0}^2  \right. \nonumber \\
&&  \left. +x_{1,0} x_{1,1} x_{2,1}^2 x_{3,1} x_{4,0}^2 x_{3,0}+x_{1,0} x_{1,1}
   x_{2,0}^2 x_{3,1} x_{4,1}^2 x_{3,0}+x_{1,0}^2 x_{2,0} x_{2,1} x_{3,1}^2 x_{4,0}
   x_{4,1}\right)+ \nonumber \\
 & &  \text{c}_4 \left(x_{1,0} x_{1,1} x_{2,0}^2 x_{4,0} x_{4,1}
   x_{3,0}^2+x_{1,1}^2 x_{2,0} x_{2,1} x_{3,1} x_{4,0}^2 x_{3,0}+x_{1,0}^2 x_{2,0}
   x_{2,1} x_{3,1} x_{4,1}^2 x_{3,0}  \right. \nonumber \\
&&  \left. +x_{1,0} x_{1,1} x_{2,1}^2 x_{3,1}^2 x_{4,0}
   x_{4,1}\right)+    \text{c}_5 \left(x_{1,0} x_{1,1} x_{2,1}^2 x_{4,0} x_{4,1}
   x_{3,0}^2+x_{1,0}^2 x_{2,0} x_{2,1} x_{3,1} x_{4,0}^2 x_{3,0}  \right. \nonumber \\
&&  \left. +x_{1,1}^2 x_{2,0}
   x_{2,1} x_{3,1} x_{4,1}^2 x_{3,0}+x_{1,0} x_{1,1} x_{2,0}^2 x_{3,1}^2 x_{4,0}
   x_{4,1}\right)+ \text{c}_6 \left(x_{1,0}^2 x_{2,0} x_{2,1} x_{4,0} x_{4,1}
   x_{3,0}^2  \right. \nonumber \\
&&  \left. +x_{1,0} x_{1,1} x_{2,0}^2 x_{3,1} x_{4,0}^2 x_{3,0}+x_{1,0} x_{1,1}
   x_{2,1}^2 x_{3,1} x_{4,1}^2 x_{3,0}+x_{1,1}^2 x_{2,0} x_{2,1} x_{3,1}^2 x_{4,0}
   x_{4,1}\right)+ \nonumber \\
&   &  \text{c}_7 \left(x_{1,1}^2 x_{2,1}^2 x_{3,0}^2 x_{4,0}^2+x_{1,0}^2
   x_{2,1}^2 x_{3,1}^2 x_{4,0}^2+x_{1,1}^2 x_{2,0}^2 x_{3,0}^2 x_{4,1}^2+x_{1,0}^2
   x_{2,0}^2 x_{3,1}^2 x_{4,1}^2\right)+ \nonumber \\
& &  \text{c}_8 \left(x_{1,0}^2 x_{2,1}^2
   x_{3,0}^2 x_{4,0}^2+x_{1,0}^2 x_{2,0}^2 x_{3,1}^2 x_{4,0}^2+x_{1,1}^2 x_{2,1}^2
   x_{3,0}^2 x_{4,1}^2+x_{1,1}^2 x_{2,0}^2 x_{3,1}^2 x_{4,1}^2\right)+ \nonumber \\
  & & \text{c}_2
   \left(x_{1,0} x_{1,1} x_{2,0} x_{2,1} x_{3,0}^2 x_{4,0}^2+x_{1,0} x_{1,1} x_{2,0}
   x_{2,1} x_{3,1}^2 x_{4,0}^2+x_{1,0} x_{1,1} x_{2,0} x_{2,1} x_{3,0}^2
   x_{4,1}^2  \right. \nonumber \\
&&  \left. +x_{1,0} x_{1,1} x_{2,0} x_{2,1} x_{3,1}^2 x_{4,1}^2\right)+   \text{c}_{10}
   \left(x_{1,1}^2 x_{2,0}^2 x_{3,0}^2 x_{4,0}^2+x_{1,1}^2 x_{2,1}^2 x_{3,1}^2
   x_{4,0}^2+x_{1,0}^2 x_{2,0}^2 x_{3,0}^2 x_{4,1}^2  \right. \nonumber \\ \nonumber
&&  \left. +x_{1,0}^2 x_{2,1}^2 x_{3,1}^2
   x_{4,1}^2\right)+
 \text{c}_{11} \left(x_{1,0}^2 x_{2,0}^2 x_{3,0}^2
   x_{4,0}^2+x_{1,1}^2 x_{2,0}^2 x_{3,1}^2 x_{4,0}^2+x_{1,0}^2 x_{2,1}^2 x_{3,0}^2
   x_{4,1}^2 \right. \\  && \left.  +x_{1,1}^2 x_{2,1}^2 x_{3,1}^2 x_{4,1}^2\right)   \label{p}
\end{eqnarray}
Here the coefficients $c$ are general constants which form a redundant description of the complex structure moduli of the manifold.

Over this base we construct the extension,
\begin{eqnarray} \label{mrext}
0 \to {\cal L} \to V \to {\cal L}^{\vee} \to 0 \;,
\end{eqnarray}
where ${\cal L}$ is the line bundle that descends from the object ${\cal O}_X(-2,-2,1,1)$ on the covering space in the language outlined in the previous subsection. This line bundle is equivariant with respect to the $\mathbb{Z}_2 \times \mathbb{Z}_4$ symmetry and thus the construction does indeed respect the symmetry being quotiented by. The non-trivial nature of the bundle (\ref{mrext}) is controlled by extension group $\textnormal{Ext}^1({\cal L}^{\vee}, {\cal L}) = H^1(X,{\cal L}^2)$, or rather by the appropriately transforming piece of this that survives in the downstairs theory. For the line bundle specified here, this cohomology vanishes for a generic enough choice of complex structure of $X$. As such, generically, the only extension of the form (\ref{mrext}) is the split bundle which has structure group $S(U(1) \times U(1))$. However, at higher codimension loci in complex structure moduli space the cohomology $H^1(X,{\cal L}^2)$ jumps in dimension to non-zero values. At such loci, one can define a non-split $SU(2)$ bundle of the form (\ref{mrext}). 

The essential idea, then is to start with a background wherein the complex structure is fixed to a jumping locus of $H^1(X,{\cal L}^2)$ and the vector bundle $V$ is taken to be an irreducible rank 2 object of the form (\ref{mrext}). One would expect that complex structure fluctuations that took the system off of this loci would not lie in the kernel (\ref{mrker}) as there would then be no appropriate $SU(2)$ bundle to perturb to and going to the split bundle would be more than an infinitesimal perturbation of the gauge connection. It was shown in \cite{Anderson:2011ty} that this is indeed the case. In such a situation, the requirement of bundle holomorphy stabilizes the system to the jumping locus of the extension group.
 
In fact, the computations that one performs to explicitly find the stabilization locus associated to such a bundle reveal an extremely rich structure. To perform such calculations one examines the Koszul sequence which, in the current example, takes the following form.
\begin{eqnarray} \label{koszulcodim1}
0 \to {\cal N}^{\vee} \otimes {\cal L}_A^2 \to {\cal L}_A^2 \to {\cal L}^2_X \to 0
\end{eqnarray}
Performing sequence chasing on the long exact sequence in cohomology associated to (\ref{koszulcodim1}) and using some facts associated to the specific example we have described above one can find that the cohomology group describing the extension classes of (\ref{mrext}) is given by the following expression.
\begin{eqnarray} \label{mrkery}
H^1(X, {\cal L}^2) = \textnormal{ker} \left( H^2(A,{\cal N}^{\vee} \otimes {\cal L}_A^2) \to H^2(A,{\cal L}^2_A) \right)
\end{eqnarray}
In the case at hand, we can denote a general element of the cohomology $H^2(A,{\cal N}^{\vee} \otimes {\cal L}_A^2)=H^2(A,{\cal O}(-6,-6,0,0))$, in polynomial language via the Bott-Borel-Weil theorem, as follows.
\begin{eqnarray}  \nonumber
&& s_1 \left( \frac{1}{x_{1,0}^2 x_{1,1}^2 x_{2,0}^2 x_{2,1}^2} \right) + s_3 \left( \frac{1}{x_{1,0}^4 x_{2,0}^2 x_{2,1}^2 } + \frac{1}{x_{1,1}^4 x_{2,0}^2 x_{2,1}^2} + \frac{1}{x_{1,0}^2 x_{1,1}^2 x_{2,0}^4} + \frac{1}{x_{1,0}^2 x_{1,1}^2 x_{2,1}^4}\right) \\ \label{thischap}  &&+ s_2 \left(\frac{1}{x_{1,0}^3 x_{1,1} x_{2,0}^3 x_{2,1}} +\frac{1}{x_{1,0} x_{1,1}^3 x_{2,0}^3 x_{2,1}} + \frac{1}{x_{1,0}^3 x_{1,1} x_{2,0} x_{2,1}^3} +\frac{1}{x_{1,0} x_{1,1}^3 x_{2,0} x_{2,1}^3}\right) 
\\ \nonumber &&+s_4 \left( \frac{1}{x_{1,0}^4 x_{2,0}^4} +\frac{1}{x_{1,1}^4 x_{2,0}^4} + \frac{1}{x_{1,0}^4 x_{2,1}^4}+\frac{1}{x_{1,1}^4 x_{2,1}^4} \right)\;.
\end{eqnarray}
Here the $s_k$ are arbitrary coefficients. Given this, via (\ref{mrkery}) any potential extension class can be represented by an object of the form (\ref{thischap}).

By performing an explicit computation, examples of which can be found in \cite{Anderson:2010mh,Anderson:2011ty,Anderson:2013qca} or later sections of this paper, one can obtain a set of loci, that is a reducible algebraic variety, in a combined space of complex structure moduli and potential extension classes of (\ref{mrext}). In the case at hand, the generators of this reducible variety are as follows.
\begin{eqnarray} \nonumber
&& c_7 s_1+c_2 s_2+c_8 s_3+c_{10} s_3+c_{11} s_4,c_8 s_1+c_2 s_2+c_7 s_3+c_{11} s_3+c_{10} s_4,c_9 s_1+c_1 s_2+2 c_9 s_3 \\ \nonumber &&+c_9 s_4,c_{10} s_1+c_2 s_2+c_7 s_3+c_{11} s_3+c_8 s_4,c_{11} s_1+c_2 s_2+c_8 s_3+c_{10} s_3+c_7 s_4,c_3 s_1+c_4 s_2+c_5 s_2 \\ \nonumber &&+c_6 s_3,c_4 s_1+c_3 s_2+c_6 s_2+c_5 s_3,c_5 s_1+c_3 s_2+c_6 s_2+c_4 s_3,c_6 s_1+c_4 s_2+c_5 s_2+c_3 s_3, \\ \nonumber && c_{10} s_1+c_2 s_2+c_7 s_3+c_{11} s_3+c_8 s_4,c_7 s_1+c_2 s_2+c_8 s_3+c_{10} s_3+c_{11} s_4,c_9 s_1+c_1 s_2+2 c_9 s_3 \\ \nonumber &&+c_9 s_4,c_{11} s_1+c_2 s_2+c_8 s_3+c_{10} s_3+c_7 s_4,c_8 s_1+c_2 s_2+c_7 s_3+c_{11} s_3+c_{10} s_4,c_5 s_1+c_3 s_2+c_6 s_2 \\ \nonumber &&+c_4 s_3,c_3 s_1+c_4 s_2+c_5 s_2+c_6 s_3,c_6 s_1+c_4 s_2+c_5 s_2+c_3 s_3,c_4 s_1+c_3 s_2+c_6 s_2+c_5 s_3,c_2 s_1 \\ \nonumber &&+c_7 s_2+c_8 s_2+c_{10} s_2+c_{11} s_2,c_2 s_1+c_7 s_2+c_8 s_2+c_{10} s_2+c_{11} s_2,c_1 s_1+4 c_9 s_2,c_2 s_1+c_7 s_2 \\ \nonumber &&+c_8 s_2+c_{10} s_2+c_{11} s_2,c_2 s_1+c_7 s_2+c_8 s_2+c_{10} s_2+c_{11} s_2,c_4 s_1+c_3 s_2+c_6 s_2+c_5 s_3,c_6 s_1 \\ \nonumber &&+c_4 s_2+c_5 s_2+c_3 s_3,c_3 s_1+c_4 s_2+c_5 s_2+c_6 s_3,c_5 s_1+c_3 s_2+c_6 s_2+c_4 s_3,c_8 s_1+c_2 s_2+c_7 s_3 \\ \nonumber &&+c_{11} s_3+c_{10} s_4,c_{11} s_1+c_2 s_2+c_8 s_3+c_{10} s_3+c_7 s_4,c_9 s_1+c_1 s_2+2 c_9 s_3+c_9 s_4,c_7 s_1 \\ \nonumber && +c_2 s_2+c_8 s_3+c_{10} s_3+c_{11} s_4,c_{10} s_1+c_2 s_2+c_7 s_3+c_{11} s_3+c_8 s_4,c_6 s_1+c_4 s_2+c_5 s_2+c_3 s_3, \\ \nonumber && c_5 s_1+c_3 s_2+c_6 s_2+c_4 s_3,c_4 s_1+c_3 s_2+c_6 s_2+c_5 s_3,c_3 s_1+c_4 s_2+c_5 s_2+c_6 s_3,c_{11} s_1+c_2 s_2 \\ \nonumber &&+c_8 s_3+c_{10} s_3+c_7 s_4,c_{10} s_1+c_2 s_2+c_7 s_3+c_{11} s_3+c_8 s_4,c_9 s_1+c_1 s_2+2 c_9 s_3+c_9 s_4, \\ \nonumber &&c_8 s_1+c_2 s_2+c_7 s_3+c_{11} s_3+c_{10} s_4,c_7 s_1+c_2 s_2+c_8 s_3+c_{10} s_3+c_{11} s_4 
\end{eqnarray}
Essentially, if one substitutes in a specific complex structure into these equations then the possible solutions for the $s_k$ specify all of the possible extensions classes at that point in moduli space in terms of the description given in (\ref{thischap}).

Next, this reducible algebraic variety can be primary decomposed to find its irreducible pieces. A discussion of the methods that we use for computations such as this can be found in, for example, \cite{Gray:2006gn,Gray:2009fy}. We utilized the specific implementations found in \cite{Gray:2008zs,cicypackage} in this work. Each of these pieces can then be processed further by an elimination of the variables $s_k$ describing the possible extension classes. This provides a set of irreducible varieties in complex structure moduli space which are the loci to which the associated choices of extension classes stabilize the system. There can be a great many such loci. For example, in \cite{Anderson:2013qca}, it was shown for the example described above that there are $25$ such loci in complex structure moduli space to which one could be stabilized, varying from points to $7$ dimensional surfaces. The specific loci that were found in that work are reproduced in Table \ref{introtab}
\begin{table}[!h]
\begin{eqnarray} \nonumber
\begin{array}{|c|c|c|}
\hline
\textnormal{\bf Equations} & \textnormal{\bf Dimension} & \textnormal{\bf Singular} \\
\hline
c_3 - c_4 - c_5 + c_6=c_2 - c_7 - c_8 - c_{10} - c_{11}= c_1 - 4 c_9=0 & 7 & \textnormal{singular} \\ \hline 
c_3+c_4+c_5+c_6=c_2+c_7+c_8+c_{10}+c_{11}=c_1+4 c_9=0 & 7 & \textnormal{singular} \\ \hline
c_9=c_2=c_1=c_7+c_8+c_{10}+c_{11}=c_4+c_5=c_3+c_6=0 & 4 & \textnormal{singular} \\ \hline
c_7-c_8-c_{10}+c_{11}=c_4-c_5=c_3-c_6=c_2=c_1=0 & 5 & \textnormal{singular}\\  \hline
c_7-c_8-c_{10}+c_{11}=c_6=c_5=c_4=c_3=c_1 c_8-2 c_2 c_9+c_1c_{10}=0&4&\textnormal{singular}\\ \hline
c_{11}=c_{10}=c_{9}=c_8=c_7=0 &5 & \textnormal{singular} \\ \hline
c_9=c_6=c_5=c_4=c_3=c_2=c_1=c_8+c_{10}=c_7+c_{11}=0 &1&\textnormal{singular}\\ \hline
c_9=c_2=c_1=c_8+c_{10}=c_7+c_{11}=c_5+c_6=c_4+c_6=c_3-c_6=0&2&\textnormal{singular}\\\hline
c_9=c_2=c_1=c_8+c_{10}=c_7+c_{11}=c_5-c_6=c_4-c_6=c_3-c_6=0&2&\textnormal{singular}\\ \hline
c_{11}=c_{10}=c_9=c_8=c_7=c_2=c_1=c_3-c_4-c_5+c_6=0 &2&\textnormal{singular}\\ \hline
c_{11}=c_{10}=c_9=c_8=c_7=c_2=c_1=c_3+c_4+c_5+c_6=0 &2&\textnormal{singular}\\ \hline
c_{11}=c_{10}=c_9=c_8=c_7=c_2=c_1=c_4+c_5=c_3+c_6=0 &1&\textnormal{singular} \\ \hline
c_{11}=c_{10}=c_9=c_8=c_7=c_2=c_1=c_4-c_5=c_3-c_6=0 &1&\textnormal{singular} \\ \hline
c_{11}=c_{10}=c_9=c_8=c_7=c_2=c_1=c_5+c_6=c_4+c_6=c_3-c_6=0 & 0 & \textnormal{singular}\\ \hline
c_{11}=c_{10}=c_9=c_8=c_7=c_2=c_1=c_5-c_6=c_4-c_6=c_3-c_6=0 & 0 & \textnormal{singular}\\ \hline
c_{10}-c_{11}=c_8-c_{11}=c_7-c_{11}=c_6=c_5=c_4=c_3=0 &3&\textnormal{singular} \\ \hline
c_{10}-c_{11}=c_{8}-c_{11}=c_7-c_{11}=c_6=c_5=c_4=c_3=c_2 c_9 - c_1 c_{11}=0 &2&\textnormal{singular} \\ \hline
c_{10}-c_{11}=c_8-c_{11}=c_7-c_{11}=c_4+c_5=c_3+c_6=c_2 c_9-c_1 c_{11}=0 &4&\textnormal{smooth} \\ \hline
c_{10}-c_{11}=c_8-c_{11}=c_7-c_{11}=c_5+c_6=c_4+c_6=c_3-c_6=c_2 c_9-c_1 c_{11}=0 & 3 & \textnormal{singular} \\ \hline
c_{10}-c_{11}=c_8-c_{11}=c_7-c_{11}=c_5-c_6=c_4-c_6=c_3-c_6=c_2 c_9-c_1 c_{11}=0 & 3 & \textnormal{singular} \\ \hline
c_8-c_{10}=c_7-c_{11}=c_6=c_5=c_4=c_3=c_2 c_9+50 c_1 c_{10} + 50 c_1 c_{11}=0 & 3 & \textnormal{singular}\\ \hline
c_{10}+c_{11}=c_9=c_6=c_5=c_4=c_3=c_2=c_1=c_8+c_{11}=c_7-c_{11}=0 & 0 & \textnormal{singular} \\ \hline
c_{10}+c_{11}=c_9=c_2=c_1=c_8+c_{11}=c_7-c_{11}=c_4-c_5=c_3-c_6=0 &2& \textnormal{singular} \\ \hline
c_{10}+c_{11}=c_9=c_2=c_1=c_8+c_{11}=c_7-c_{11}=c_5+c_6=c_4+c_6=c_3-c_6=0 & 1 & \textnormal{singular} \\ \hline
c_{10}+c_{11}=c_9=c_2=c_1=c_8+c_{11}=c_7-c_{11}=c_5-c_6=c_4-c_6=c_3-c_6=0 & 1 & \textnormal{singular} \\ 
\hline
\end{array}
\end{eqnarray}
\caption{A table of results taken from \cite{Anderson:2013qca} showing the loci in complex structure moduli space to which the Calabi-Yau three-fold $\tilde{X}/(\mathbb{Z}_2 \times \mathbb{Z}_4)$ can be stabilized by the bundle $V$ in equation (\ref{mrext}). The column ``Dimension'' denotes the complex dimension of the given locus. The column ``Singular'' specifies whether the Calabi-Yau three-fold associated with a generic complex structure in each locus is singular or smooth.}
\label{introtab}
\end{table}

Finally one should check, for each locus that the system could be stabilized to, that for a generic enough choice of complex structure moduli in that set the Calabi-Yau manifold is smooth. This can be quite constraining, especially for quotients of CICYs, and in fact only one of the loci for the example being discussed here, of dimension 4, turns out to correspond to a smooth threefold (see Table \ref{introtab}).

\section{Particle Spectrum Jumping due to Moduli Stabilization} \label{specjump}

In this section we will consider the interplay between bundles constructed in the visible sector in order to engineer a standard model like spectrum in the low energy theory, and bundles inserted into the hidden $E_8$ in order to stabilize complex structure moduli. In particular, we will be investigating to what degree hidden sector bundles can force the complex structure of the Calabi-Yau threefold to a locus in moduli space where the visible sector is forced to jump. Such an effect could be either beneficial (in introducing a Higgs doublet pair into a model which previously had none for example) or undesired (for example in causing additional generations and anti-generations to appear).

\begin{figure}[h]\centering
\includegraphics[width=0.85\textwidth]{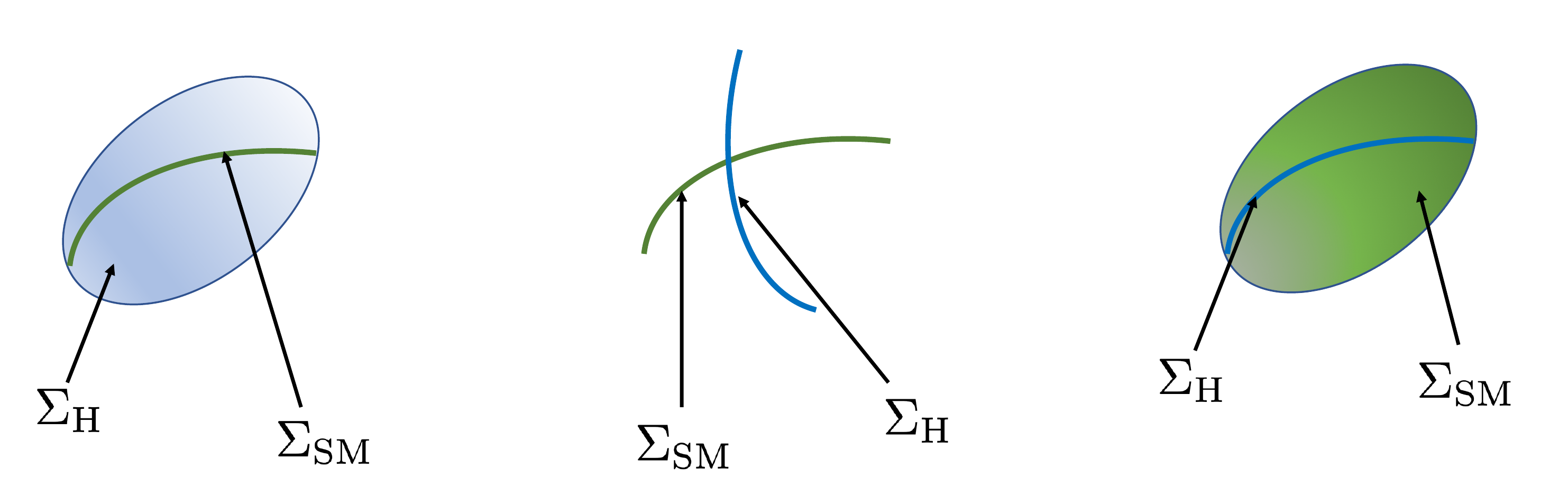}
\caption{{\it Three possible situations involving intersection between the locus in moduli space $\Sigma_H$ where the hidden sector bundle stabilizes the complex structure and the locus $\Sigma_{\textnormal{SM}}$ where the spectrum of the standard model bundle jumps. The only case in which the moduli stabilization mechanism forces the standard model spectrum to change is the last, as discussed further in the text.}}
\label{locuspic}
\end{figure}

There are several possibilities for intersection of the jumping locus of the standard model sector and hidden sector bundles in complex structure moduli space. These are depicted in Figure \ref{locuspic}.
In the first situation depicted in the figure, the locus of jumping of the standard model bundle lies inside the locus of jumping of the hidden sector one. In the second situation the two loci intersect at a higher codimension locus in complex structure moduli space. In both of these cases, the moduli stabilization mechanism does not force the standard model matter sector to change from that found at an arbitrary point in complex structure moduli space, where most such models are analyzed during their construction. In both situations, if the stabilization mechanism forces the complex structure to a generic enough point in $\Sigma_{\textnormal{H}}$, then $\Sigma_{\textnormal{SM}}$ will miss this point.

On the other hand, the third situation in Figure \ref{locuspic}, or its extreme limit where $\Sigma_{\textnormal{H}}=\Sigma_{\textnormal{SM}}$ is of interest to us here. In this case, wherever we are on the hidden sector locus the standard model spectrum jumps from that which is observed at a generic point in complex structure moduli space. As such, if model building  was carried out without thinking about the moduli stabilization mechanism, then incorrect conclusions would be reached about the particle content of the four dimensional effective theory.

Naively, one might think that such a phenomenon would be extremely rare. After all, the visible and hidden $E_8$'s of heterotic string or M-theory are rather separate in nature and are only coupled to each other gravitationally. Given this, why should the locus of jumping of a bundle in one sector lie exactly inside that of another ($\Sigma_{\textnormal{H}} \subset \Sigma_{\textnormal{SM}}$)? There are some conditions linking the two bundles, however, and we will find that these are strong enough to make the phenomenon we are talking about surprisingly common. 

The first condition we will consider is the standard one following from requiring integrability of the Bianchi Identity. 
\begin{eqnarray} \label{c2cond}
\textnormal{ch}_2(V_{\textnormal{H}})_a + \textnormal{ch}_2(V_{\textnormal{SM}})_a - \textnormal{ch}_2(TX)_a + [W]_a = 0 \;\; \forall a
\end{eqnarray}
Here the indices $a=1,\ldots, h^{(1,1)}(X)$ label the harmonic $(2,2)$ forms on the Calabi-Yau threefold, $V_{\textnormal{SM}}$ and $V_{\textnormal{H}}$ are the visible and hidden sector bundles respectively, $X$ is the Calabi-Yau manifold and $[W]$ is a form proportional to the dual of the class of the (in general reducible) curve wrapped by NS five-branes/M5 branes in the vacuum configuration being considered (in the heterotic string/heterotic M-theory respectively). Allowing for M5 branes that preserve supersymmetry, and thus lead to a class $[W]$ that is effective, (\ref{c2cond}) leads us to the following inequality.
\begin{eqnarray} \label{ineq}
\textnormal{ch}_2(V_{\textnormal{H}})_a+ \textnormal{ch}_2(V_{\textnormal{SM}})_a \leq \textnormal{ch}_2(TX)_a \;\; \forall \;\;a
\end{eqnarray} 

\vspace{0.1cm}

In addition to the second Chern character constraint (\ref{ineq}), there is the constraint that both the visible and hidden sector bundles must be slope poly-stable and slope zero for the same choice of four dimensional K\"ahler moduli. Due to the warping of heterotic M-theory, there is a slight difference between the polarizations experienced between the two bundles, but nevertheless this is easy to account for. As with (\ref{ineq}), providing that both bundles are indeed stable in reasonably large chambers of the K\"ahler cone, this constraint is not seemingly too difficult to satisfy.

 Although the inequality (\ref{ineq}) and the requirement for simultaneous stability of the hidden and visible sector gauge bundles may not seem like a very strong set of constraints, in some cases it can become so once one considers quotienting the Calabi-Yau manifold in order to introduce Wilson lines\footnote{Which has been shown to be essentially the only way to break the GUT group in such compactifications \cite{Anderson:2014hia}}. The issue is that equivariance constraints, ensuring that the gauge bundles are consistent with the symmetry by which the Calabi-Yau threefold is being quotiented, can mean that quite a few bundles are not available in building models and hidden sectors. The resulting combination of equivariance, stability and second Chern class constraints can be quite restrictive.

\subsection{A systematic investigation of a class of bundle constructions}

To illustrate the issues discussed above, and to obtain an idea of how commonly moduli stabilization affects the visible sector spectrum, at least in a class of examples, we will look at specific types of construction of visible and hidden sector bundles. The visible sector will be taken to be a sum of line bundles (more specifically a line bundle standard model). The hidden sector will be taken to be a simple extension of two line bundles of the following form.
\begin{eqnarray} \label{typeconsidered}
0 \to {\cal L} \to {V}_{\textnormal{H}} \to {\cal L}^\vee \to 0
\end{eqnarray}

Constructions of the type (\ref{typeconsidered}) are perhaps the simplest types of bundles that can lead to complex structure stabilization of the type described in Section \ref{stabintro}. They have structure group $SU(2)$ and the only simpler possibility, that of an abelian structure group, is ruled out by the fact that sums of line bundles do not exhibit this phenomenon.

In considering examples of such hidden and visible sector bundles, one immediately sees one compatibility constraint that arises. Consider a line bundle sum $V_{\textnormal{SM}}$ in the visible sector containing a line bundle ${\cal L}_1$. That same line bundle can not be used in creating an extension of the form (\ref{typeconsidered}) for the hidden sector bundle ${V}_{\textnormal{H}}$. The issue is simply one of stability. As can be seen from the defining sequence (\ref{typeconsidered}), if ${\cal L}={\cal L}_1$ then that line bundle injects into ${V}_{\textnormal{H}}$. It must therefore be of negative slope if ${V}_{\textnormal{H}}$ is to be stable. However, on the locus in K\"ahler moduli space where $V_{\textnormal{SM}}$ is poly-stable $\mu({\cal L}_1)=0$ and thus the visible and hidden sectors can't simultaneously preserve supersymmetry in such a situation. Similarly one can not set ${\cal L}={\cal L}_1^\vee$. In such a situation we find an exactly analogous situation when we consider the stability of ${V}_{\textnormal{H}}^{\vee}$. Since ${V}_{\textnormal{H}}$ is stable iff ${V}_{\textnormal{H}}^{\vee}$ is, this leads us to the same conclusion.

Overall, the observation of the previous paragraph can be quite a big constraint on the possible hidden sectors that can be included to complete a line bundle standard model and stabilize complex structure moduli. As stated earlier, there are frequently not many choices of equivariant line bundles that can be included in an extension such as (\ref{typeconsidered}) in the hidden sector without violating the bound on $\textnormal{ch}_2({V}_{\textnormal{H}})$ imposed by integrability of the Bianchi Identity and supersymmetry. Given that all of the line bundles that appear in the line bundle standard model (and their duals) are ruled out on grounds of stability, in some cases one can be left with very few, or even no, possibilities.

Assuming simultaneously stable hidden and visible sector bundles can be found we must then study the relevant jumping loci in complex structure moduli space and compare them. Ideally the procedure would be as follows, using the discussion of Section \ref{stabintro}.
\begin{enumerate}
\item Find the locus in complex structure/potential extension space to which the hidden sector bundle forces the system.
\item Primary decompose that locus to find its irreducible components. For each individual locus, eliminate the degrees of freedom corresponding to potential extensions to obtain a variety living purely in complex structure moduli space $\Sigma_I^H$. We will denote the reducible variety composed of all of these irreducible components $\Sigma_H = \bigcup_I \Sigma_I^H$.
\item In a similar manner, find the locus in complex structure moduli space, $\Sigma_{\textnormal{SM}}$ on which the visible sector matter spectrum jumps.
\item For each irreducible piece of $\Sigma_{\textnormal{H}}$ ask if that locus is contained in $\Sigma_{\textnormal{SM}}$. I.e. check whether there exists an $\Sigma_I^H$ such that $\Sigma_I^H \subseteq \Sigma_{\textnormal{SM}}$.
\end{enumerate}

Unfortunately, in practice the above procedure is often prohibitively computationally intensive. The problem is that the jumping cohomology of relevance for $\Sigma_{\textnormal{H}}$ is $H^1({\cal L}^2)$ where ${\cal L}$ is an equivariant line bundle. This cohomology very often involves large numbers in the first Chern class of ${\cal L}^2$ and this leads to a primary decomposition which is extremely costly in the second step in the list just given.

If at all computationally feasible, we use the above procedure when analyzing examples. However, if this computation can not be completed in practice, then we carry out the following analysis instead.
\begin{enumerate}
\item Find a set of example points in complex structure moduli space lying on $\Sigma_{\textnormal{SM}}$.
\item Determine if these points also lie on $\Sigma_{\textnormal{H}}$.
\item For those that do, if any, perform a linear perturbation analysis around that point in complex structure moduli space to determine if the hidden sector locus $\Sigma_I^H$ on which it lies is localized within $\Sigma_{\textnormal{SM}}$.
\end{enumerate}

We will give more details as to how this is achieved in the examples we will present going forward. In this manner, we can check whether any of the random points in $\Sigma_{\textnormal{SM}}$ that were picked lie on a component of the hidden sector locus such that $\Sigma_I^H \subset \Sigma_{\textnormal{SM}}$. When we do find such points it is most likely that we have found a case where the equidimensional hulls of the two reducible varieties in complex structure moduli space coincide. We should mention in addition that, throughout the work presented in this section, we check the smoothness of the Calabi-Yau threefolds involved at both the specific points we chose and the loci we consider in complex structure moduli space.

Applying the procedure described above provides us with a, presumably rather weak, lower bound on the frequency with which the hidden sector bundle can cause the standard model matter content to jump. We will see later that this is already good enough to illustrate one must be cautious in combining moduli stabilization and model building. We note that we start by finding points on $\Sigma_{\textnormal{SM}}$ rather than $\Sigma_{\textnormal{H}}$ here because the line bundles involved then tend to have smaller entries in their first Chern class. This leads to a more tractable computation.

\subsection{An example} \label{egsec}

Let us illustrate the above general discussion with a concrete example. We will work on a freely acting $\mathbb{Z}_2$ quotient of CICY number 6777 which is described by the following configuration matrix.
\begin{eqnarray} \label{conf}
X=\left [ \begin{array}{c|cccc}
\mathbb{P}^1 & 1 & 1 & 0 & 0  \\
\mathbb{P}^1 & 0 & 0 & 0 & 2 \\
\mathbb{P}^1 & 0 & 0 & 2 & 0  \\
\mathbb{P}^1 & 2 & 0 & 0 & 0  \\
\mathbb{P}^3 & 1 & 1 & 1 & 1
\end{array}
\right ].
\end{eqnarray}

We label the homogeneous coordinates of the four ambient space $\mathbb{P}^1$ factors as $x_{r,a}$ where $r=1,\ldots, 4$ runs over the projective spaces and $a=0,1$ labels the homogeneous coordinates on each factor. The homogeneous coordinates of the $\mathbb{P}^3$ factor are labeled as $x_{5,\alpha}$ where $\alpha=0,\ldots,3$. Given this notation, we can write the ambient space coordinate action of the $\mathbb{Z}_2$ symmetry by which we will quotient as follows.
\begin{eqnarray} \label{sym1}
\Gamma_{\mathbb{Z}_2}^x: ( x_{r,a} : x_{5,\alpha} )\to ((-1)^{a+1} x_{r,a} : (-1)^{\textnormal{max}(2 \alpha,3)} x_{5,\alpha})
\end{eqnarray}
In addition, the symmetry has a non-trivial normal bundle action, or equivalently action on the defining polynomials. Labeling the four defining relations corresponding to the columns of (\ref{conf}) as $p_A$ where $A=1,\ldots,4$, we have the following.
\begin{eqnarray} \label{sym2}
\Gamma_{\mathbb{Z}_2}^p : (p_1,p_2,p_3,p_4) \to (-p_1,p_2,p_3,-p_4)
\end{eqnarray}

Given the action (\ref{sym1}) and (\ref{sym2}), the most general defining relations for the configuration matrix (\ref{conf}) that are compatible with the symmetry are as follows.
\begin{eqnarray} \label{defrel}
p_1 &=&c_{1,7} x_{1,1} x_{5,0} x_{4,0}^2+c_{1,8} x_{1,1} x_{5,1} x_{4,0}^2+c_{1,1} x_{1,0} x_{5,2} x_{4,0}^2+c_{1,2} x_{1,0} x_{5,3} x_{4,0}^2\\ \nonumber
   &&+c_{1,3}
   x_{1,0} x_{4,1} x_{5,0} x_{4,0} +c_{1,4} x_{1,0} x_{4,1} x_{5,1} x_{4,0}+c_{1,9} x_{1,1} x_{4,1} x_{5,2} x_{4,0}+c_{1,10} x_{1,1} x_{4,1} x_{5,3}
   x_{4,0}\\ \nonumber
   &&+c_{1,11} x_{1,1} x_{4,1}^2 x_{5,0}+c_{1,12} x_{1,1} x_{4,1}^2 x_{5,1}+c_{1,5} x_{1,0} x_{4,1}^2 x_{5,2}+c_{1,6} x_{1,0} x_{4,1}^2 x_{5,3} \\ \nonumber
p_2 &=&c_{2,1} x_{1,0} x_{5,0}+c_{2,2} x_{1,0} x_{5,1}+c_{2,3} x_{1,1} x_{5,2}+c_{2,4} x_{1,1} x_{5,3} \\ \nonumber
p_3 &=&c_{3,1} x_{5,2} x_{3,0}^2+c_{3,2} x_{5,3} x_{3,0}^2+c_{3,3} x_{3,1} x_{5,0} x_{3,0}+c_{3,4} x_{3,1} x_{5,1} x_{3,0}+c_{3,5} x_{3,1}^2 x_{5,2}+c_{3,6}
   x_{3,1}^2 x_{5,3} \\ \nonumber
p_4 &=&c_{4,1} x_{5,0} x_{2,0}^2+c_{4,2} x_{5,1} x_{2,0}^2+c_{4,3} x_{2,1} x_{5,2} x_{2,0}+c_{4,4} x_{2,1} x_{5,3} x_{2,0}+c_{4,5} x_{2,1}^2 x_{5,0}+c_{4,6}
   x_{2,1}^2 x_{5,1}
   \end{eqnarray}
  In these expressions, the $c$'s are arbitrary coefficients associated to the complex structure moduli space. We call the manifold obtained by quotienting $X$ by the symmetry action (\ref{sym1}) and (\ref{sym2}) $\hat{X}$.

\vspace{0.2cm}

On the quotient manifold described above we now define the visible sector bundle (first constructed in \cite{Anderson:2011ns,Anderson:2012yf}). On $X$ we define,
\begin{eqnarray}  \label{sm1}
V_{\textnormal{SM}} = \bigoplus_{i=1}^5 {\cal L}_i \;,
\end{eqnarray}
where
 \begin{eqnarray} \label{sm2}
 &\mathcal{L}_1=\mathcal{O}(1, -1, 1, -1, 0)\;,\;\mathcal{L}_2= \mathcal{O}(0, 1, 1, 1, -1)\;,\; \mathcal{L}_3=\mathcal{O}(0, 1, -2, 1, 0) \;, \\ \nonumber
&\mathcal{L}_4=\mathcal{O}(0, -1, 0, -2,1)\;,\; \mathcal{L}_5=\mathcal{O}(-1, 0, 0, 1, 0) \;.
\end{eqnarray}
Each line bundle in $V_{\textnormal{SM}}$ is individually equivariant, and thus this does indeed define a line bundle standard model, with bundle $\hat{V}_{\textnormal{SM}}$, on the quotient $\hat{X}$. We take the parameters defining the Wilson line and equivariant structure on the sum of line bundles, as described in Section \ref{lsmreview} to be $W=1$, $\tilde{W}=-1$, $\chi_i=1$ for $i=1,3 \ldots, 5$ and $\chi_2=-1$. With these choices, the downstairs standard model charged matter spectrum, expressed concisely in terms of GUT multiplets as described earlier, is as follows at a general point in complex structure moduli space.
  \begin{eqnarray}\label{spec}
  \left\{2 \, \bold{10}_{{\bf e}_3},\bold{10}_{{\bf e}_4},2\, \overline{\bold{5}}_{{{\bf e}_1},{{\bf e}_4}},\overline{\bold{5}}_{{{\bf e}_2},{{\bf e}_3}},\bold{5}_{-{{\bf e}_1},-{{\bf e}_2}},\overline{\bold{5}}_{{{\bf e}_1},{{\bf e}_2}}\right\}
  \end{eqnarray}
    
 The multiplicity of $\overline{\bold{5}}_{{{\bf e}_2},{{\bf e}_3}}$ representations in this example has the potential to jump (along with the multiplicity of $\bold{5}_{-{{\bf e}_2},-{{\bf e}_3}}$ multiplets in an index preserving manner) at higher dimensional loci in complex structure moduli space. To see this, we must consider the cohomology $H^1(\hat{X}, \hat{{\cal L}}_2 \otimes \hat{{\cal L}}_3)=H^1(\hat{X},\hat{{\cal O}}(0,2,-1,2,-1))$, the dimension of which counts the multiplicity of these degrees of freedom (here hatted bundles correspond to projections of the associated upstairs objects). To compute this jumping, we work on the covering space $X$ with the cohomology of the associated equivariant bundles and then pick out the relevant subspace (which descends to the cohomology on $\hat{X}$) by comparing representation content of that space with the Wilson line and equivariant structure.
  
 To calculate the cohomology $H^1(X, {\cal L}_2 \otimes {\cal L}_3)=H^1(X,{\cal O}(0,2,-1,2,-1))$ we make use of the Koszul sequence
 \begin{eqnarray} \label{koszul1}
 0\rightarrow \wedge^4\mathcal{N}^{\vee}\otimes{\cal L}_2 \otimes {\cal L}_3\rightarrow\wedge^3\mathcal{N}^{\vee}\otimes{\cal L}_2 \otimes {\cal L}_3 \rightarrow\dots\rightarrow {\cal L}_2 \otimes {\cal L}_3|_A\rightarrow {\cal L}_2 \otimes {\cal L}_3|_X\rightarrow 0
 \end{eqnarray}
This long exact sequence can be broken into several short exact sequences as follows.
\begin{eqnarray} \label{koszul2}
 0\rightarrow \wedge^4 \mathcal{N}^{\vee}\otimes{\cal L}_2 \otimes {\cal L}_3\rightarrow \wedge^3\mathcal{N}^{\vee}\otimes {\cal L}_2 \otimes {\cal L}_3\rightarrow \mathcal{K}_1\rightarrow 0
\\\nonumber
0\rightarrow \mathcal{K}_1\rightarrow\wedge^2 \mathcal{N}^{\vee}\otimes{\cal L}_2 \otimes {\cal L}_3\rightarrow \mathcal{K}_2\rightarrow 0
\\ \nonumber
\vdots
\\ \nonumber
0\rightarrow \mathcal{K}_3\rightarrow {\cal L}_2 \otimes {\cal L}_3\rightarrow {\cal L}_2 \otimes {\cal L}_3|_X\rightarrow 0
\end{eqnarray}
Here the $\mathcal{K}_i$ where $i=1,\ldots 3$ are kernels and cokernels of the relevant maps. The ambient space cohomologies of all of the line bundles appearing in the sequences (\ref{koszul2}) (excluding those of the ${\cal K}$'s) are vanishing with two exceptions: $h^5(A, \wedge^4\mathcal{N}^{\vee}\otimes{\cal L}_2 \otimes {\cal L}_3)=8$ and $h^5(A,\wedge^3\mathcal{N}^{\vee}\otimes{\cal L}_2 \otimes {\cal L}_3)=6$. Chasing the associated long exact sequences in cohomology we find the following.
\begin{eqnarray} \label{themap}
H^1(X, {\cal L}_2 \otimes {\cal L}_3) \cong \ker \left( H^5(A, \wedge^4\mathcal{N}^{\vee}\otimes{\cal L}_2 \otimes {\cal L}_3) \to H^5(A,\wedge^3\mathcal{N}^{\vee}\otimes{\cal L}_2 \otimes {\cal L}_3) \right)
\end{eqnarray}
Thus, at a generic enough point in complex structure moduli space, we find that $h^1(X, {\cal L}_2 \otimes {\cal L}_3) =2$, leading to the single $\overline{\bold{5}}_{{{\bf e}_2},{{\bf e}_3}}$ representation in (\ref{spec}), after quotienting by the $\mathbb{Z}_2$ symmetry, by applying the correspondence of Table \ref{spectab}. 

In order to present concrete formula which are concise, we will focus on calculating the locus in complex structure moduli space where the subspace $H^1(X,\hat{ {\cal L}}_2 \otimes \hat{{\cal L}}_3 ,\chi_2 \otimes \chi_3 \otimes \tilde{W}) \in H^1(X,\hat{ {\cal L}}_2 \otimes \hat{{\cal L}}_3)$ jumps in dimension (corresponding to a jump in the number of left handed $SU(2)$ doublets in the four dimensional effective theory). To do this, we need to study the map in (\ref{themap}) in more detail. We now form an explicit description of the cohomologies in (\ref{themap}) as polynomials in ambient space coordinates, take the relevant subset of such objects that are picked out in the downstairs cohomology of interest by the choice of equivariant structure and Wilson line, and study the map in more detail. 

A general element of the relevant subspace of the source cohomology group in (\ref{themap}) can be written as follows.
\begin{eqnarray} \label{source}
S=\frac{s_1}{x_{3,0} x_{5,0}}+\frac{s_2}{x_{3,0} x_{5,1}}+\frac{s_3}{x_{3,1} x_{5,2}}+\frac{s_4}{x_{3,1} x_{5,3}}.
\end{eqnarray}
Here the $s_k$ are arbitrary coefficients. Note that $h^6(A, \wedge^4\mathcal{N}^{\vee}\otimes  {\cal L}_{2}\otimes {\cal L}_3)=8$ and we are dividing by a $\mathbb{Z}_2$ symmetry, so the four dimensional space obtained in (\ref{source}) is as expected.

Next we consider the target space in (\ref{themap}). We have that
\begin{eqnarray}
\wedge^3 \mathcal{N}^{\vee}\otimes {\cal L}_{2}\otimes {\cal L}_3 &=& {\cal O}(-2,0,-1,0,-4) \oplus {\cal O}(-2,2,-3,0,-4) \\ \nonumber && \oplus\; {\cal O}(-1,0,-3,2,-4) \oplus {\cal O}(-1,0,-3,0,-4) \;.
\end{eqnarray}
Given this, the only contribution to $h^6(A,\wedge^3\mathcal{N}^{\vee}\otimes {\cal L}_{2}\otimes {\cal L}_3)$ comes from $h^6(A, {\cal O}(-2,2,-3,0,-4))$, with the other three cohomologies vanishing. The map from (\ref{source}) to this cohomology is given by multiplication by the fourth defining relation, followed by deleting terms in the resulting polynomial that are not of the correct degree to appear in $h^6(A, {\cal O}(-2,2,-3,0,-4))$. Performing this computation we obtain the following image of the general source element (\ref{source}).
 \begin{eqnarray} \label{image}
\textnormal{Im}(S)= \frac{s_1 c_{4,1} x_{2,0}^2}{x_{3,0}}+\frac{s_2 c_{4,2} x_{2,0}^2}{x_{3,0}}+\frac{s_3 c_{4,3} x_{2,1} x_{2,0}}{x_{3,1}}+\frac{s_4 c_{4,4} x_{2,1}
   x_{2,0}}{x_{3,1}}+\frac{s_1 c_{4,5} x_{2,1}^2}{x_{3,0}}+\frac{s_2 c_{4,6} x_{2,1}^2}{x_{3,0}}
 \end{eqnarray}
In order to find the kernel of the map, we then simply require that the coefficients of each of the rationomes in (\ref{image}) vanishes. Doing so we obtain the following constraints on the $s_k$, in terms of the complex structure choice $c_{A,\gamma}$ in (\ref{defrel}), in order for an element of the source of the form in (\ref{source}) to be in the kernel.
\begin{eqnarray} \label{kercond}
s_1 c_{4,5}+s_2 c_{4,6}=0 \;,\; s_3 c_{4,3}+s_4 c_{4,4}=0 \;,\; s_1 c_{4,1}+s_2 c_{4,2}=0 
 \end{eqnarray}
 Writing these conditions in matrix form we obtain the following.
 \begin{eqnarray} \label{matkercond}
 \left(
\begin{array}{cccc}
 c_{4,5} & c_{4,6} & 0 & 0 \\
 0 & 0 & c_{4,3} & c_{4,4} \\
 c_{4,1} & c_{4,2} & 0 & 0 \\
\end{array}
\right) \cdot  \left(
\begin{array}{c}
 s_1 \\
 s_2 \\
 s_3 \\
 s_4 \\
\end{array}
\right)=0
 \end{eqnarray} 

Given (\ref{matkercond}), it is easy to see that for a generic choice of complex structure the kernel will be one dimensional as stated earlier. However, on the locus\;,
\begin{eqnarray}
c_{4,2}c_{4,5}-c_{4,1}c_{4,6}=0\;,
\end{eqnarray}
the rank of the matrix in (\ref{matkercond}) changes from 3 to 2, and thus the dimension of the kernel  will change from one to two. Therefore, on this special locus the number of $SU(2)$ doublets descending from the $\bold{\overline{5}_{{\bf e}_2,{\bf e}_3}}$ representation increases. Naturally, in order for the index to be preserved, there is also an increase in the number of associated anti-doublets on the same locus in complex structure moduli space.

\vspace{0.2cm}

Next we turn our attention to the hidden sector and the bundle which is added to constraint the complex structure of the compactification. In searching for bundles of the form (\ref{typeconsidered}), we find that the following two possibilities

\begin{eqnarray} \label{thels}
{\cal L}={\cal O}(-2,-1,1,1,0) \text{ }\text{and}\text{ } {\cal L}= {\cal O}(1,-1,1,-2,0).
\end{eqnarray}
are equivariant and satisfy all of the constraints given earlier in this Section. 

To examine this in more detail we first note that the second Chern class of $X$ can be presented as a two index quantity, where we expand the $(2,2)$ form in a redundant basis given by products of $(1,1)$ forms spanning $H^{1,1}(X)$. We can then contract this description of the Chern class with the intersection form to get a description of  $c_2(X)$ as a vector of length $h^{1,1}(X)=h^{2,2}(X)$. When we do this we obtain the following.
\begin{eqnarray}
c_2(X) = (24,24,24,24,56)
\end{eqnarray}
The second Chern class of the standard model we are examining here, expressed in the same manner is:
\begin{eqnarray}
c_2(V_\textnormal{SM}) = (12,12,12,12,32)\;.
\end{eqnarray}
Finally, the second Chern classes of the extensions (\ref{typeconsidered}) of the ${\cal L}$'s given in (\ref{thels}) are respectively the following.
\begin{eqnarray}
c_2({V}_{\textnormal{H}}) = (4, 12, 4, 4, 20) \;\; \textnormal{or} \;\; c_2({V}_{\textnormal{H}}) =  (4, 12, 4, 4, 20).
\end{eqnarray}
It is easy to see that the $SU(2)$ bundles we are choosing satisfy the second Chern character condition (\ref{ineq}).

Since all line bundles are equivariant with respect to the $\mathbb{Z}_2$ symmetry we are considering, the only constraint that we have left to consider is that of stability. It is straight forward to show \cite{Anderson:2013qca} that an extension of the form (\ref{typeconsidered}) is stable iff $\mu({\cal {\cal L}}) <0$, that is if the slope of ${\cal L}$ is strictly negative.

We recall the expression for the slope of a line bundle,
\begin{eqnarray}
\mu({\cal L}_i)=\sum_{r,s,t=1}^{h^{1,1}(X)}d_{rst}c^r_1({\cal L}_i)t^st^t=0 \;,
\end{eqnarray}
and give a definition of a set of variables ${\sigma}_r$:
\begin{eqnarray}
\sigma_r=\sum_{s,t=1}^{h^{1,1}(X)}d_{rst}t^st^t \;.
\end{eqnarray}
Then, examining the standard model bundle given in (\ref{sm1}) and (\ref{sm2}), we obtain the following conditions for the slopes of the line bundles involved to vanish (a necessary and sufficient condition for its poly-stability).
\begin{eqnarray} 
 \sigma_1-\sigma_2+\sigma_3-\sigma_4=0\;,\;\sigma_2+\sigma_3+\sigma_4-\sigma_5=0\;,\;\sigma_2-2\sigma_3+\sigma_4=0\;, \\ \nonumber -\sigma_2-2\sigma_4+\sigma_5=0\;,\;-\sigma_1+\sigma_4=0
 \end{eqnarray}
The general solution to these equations is given by the following.
\begin{eqnarray}
 \sigma_1=\sigma_2=\sigma_3=\sigma_4\;,\;\sigma_5=3\sigma_1
 \end{eqnarray}
 
We can now ask about the slope of the possible ${\cal L}$'s given in (\ref{thels}) on this locus, and thus about the stability of the hidden sector bundles. We find that,
 \begin{eqnarray}
 \mu({\cal O}(-2,-1,1,1,0))=-\sigma_1<0 \;\;,\;\; \mu({\cal O}(1,-1,1,-2,0))=-\sigma_1<0 \;.
 \end{eqnarray}
 Thus, the proposed hidden sector extensions are indeed stable on the same locus in K\"ahler moduli space as the $V_{\textnormal{SM}}$ and our last constraint is satisfied.
 
 \vspace{0.3cm}
 
 To proceed further we will focus on ${\cal L}={\cal O}(1,-1,1,-2,0)$, although a similar analysis can be followed for the other possibility in (\ref{thels}). The next step is to study the jumping locus of the extension group defining (\ref{typeconsidered}) and compare this jumping locus to that of $V_{\textnormal{SM}}$. The extension class of (\ref{typeconsidered}) lies in $H^1(X,{\cal L}^2)$. Performing an analogous chasing of Koszul sequence to the one we performed for the visible sector bundle, we arrive at the following description of this cohomology.
 \begin{eqnarray} \label{hidmap}
H^1(X,\mathcal{L}^2) \cong \text{ker}(H^5(A,\wedge^4 {\cal N}^{\vee} \otimes \mathcal{L}^2)\rightarrow H^2(A,\mathcal{L}^2))
\end{eqnarray}
In fact, we will be interested in the associated cohomology on the quotiented manifold $\hat{X}$. For simplicity in this example we choose the trivial equivariant structure on ${\cal L}^2$, and thus this will correspond to simply considering the invariant elements in the cohomology groups concerned under the naive transformation induced from the coordinate action of the symmetry. This choice is consistent with non-trivial equivariant structures on the normal bundle such as (\ref{sym2}) in this example. More generally in this work we consider all possible choices of equivariant structure.

A general element of the down-stairs cohomology describing the source space of the map corresponding to (\ref{hidmap}) is found to be the following.
\begin{eqnarray} \label{hidsource}
 S&=&\frac{s_1}{x_{2,0}^2 x_{4,0}^4}+\frac{s_4}{x_{2,0} x_{2,1} x_{4,0}^3 x_{4,1}}+\frac{s_2}{x_{2,0}^2 x_{4,0}^2 x_{4,1}^2}+\frac{s_7}{x_{2,1}^2 x_{4,0}^2
   x_{4,1}^2}+\frac{s_5}{x_{2,0} x_{2,1} x_{4,0} x_{4,1}^3}\\ \nonumber
   &&+\frac{s_6}{x_{2,1}^2 x_{4,0}^4}+\frac{s_3}{x_{2,0}^2 x_{4,1}^4}+\frac{s_8}{x_{2,1}^2
   x_{4,1}^4}
\end{eqnarray}
Where, as in previous examples, the $s_k$ are a set of arbitrary constants.  The map itself, from an examination of (\ref{hidmap}), should be built out of a combination of four defining relations. This map is in fact constructed in a somewhat non-trivial fashion as follows:
\begin{eqnarray} \label{hidmap}
f= \epsilon^{\alpha\beta\gamma\delta}p_{1\alpha}p_{2\beta}p_{3\gamma}p_{4\delta},
\end{eqnarray}
 Here $\epsilon^{\alpha\beta\gamma\delta}$ is the totally antisymmetric tensor and $p_{A\alpha}$ denotes the partial differentiation of $p_A$ with respect to the variable $x_{5,\alpha}$ where $\alpha$ runs from $0$ to $3$. It is easy to see that $f$ then has multi-degree $(2,2,2,2,0)$ which is precisely what is needed to match the source and target degrees in (\ref{hidmap}).
 
 As in the computation of the kernel in (\ref{themap}), we can now multiply the general source polynomial (\ref{hidsource}) by the map polynomial (\ref{hidmap}) and demand that all of the coefficients of terms appearing in the target space vanish. When we do so we obtain a very long expression depending upon the $s_k$ in (\ref{hidsource}) and the coefficients in the defining relations $c_{A,\gamma}$. While there are only $14$ constraints obtained in this manner, which we will denote by ${\cal I}_{\alpha}$ where $\alpha=1,\ldots 14$, they are over two pages in length and so we do not reproduce them here.
 
 For general defining relations, the kernel of this map is found to be trivial. For special loci in complex structure moduli space, however, a non-trivial kernel is obtained, and thus the question arises how best to find this locus. As discussed earlier in this subsection, ideally we would like to primary decompose the ideal associated to these constraints and analyze each irreducible component of the associated variety separately. However, in the case at hand, this method is too computationally intensive, especially as part of a large scan over cases.

Given this situation, this is an example where we follow the methodology outlined earlier for cases where primary decomposition is too slow. We begin by finding a set of points in complex structure moduli space lying on the jumping locus $\Sigma_{SM}$ of the Standard Model sector bundle $V_{\textnormal{SM}}$. In other words, denoting the generators of the ideal that define the locus in complex structure moduli space where the cohomology of $V_{\textnormal{SM}}$ jumps as $S_\kappa(c_{A,\gamma})$, we find sets of $c_{A,\gamma}=c_{A,\gamma}^0$ such that $S_\kappa(c_{A,\gamma}^0)=0 \;\; \forall \; \kappa$. This is achievable in almost all cases we encounter, as the standard model bundle ideal is somewhat less complicated than its hidden sector cousin. This is simply due to the fact that the extension classes of (\ref{typeconsidered}) is the first cohomology of ${\cal L}^2$, and the square that appears tends to make the associated ideals larger.

 Next we ask whether any of the solutions $c_{A,\gamma}^0$ also lie on the variety describing the kernel of the map (\ref{hidmap}) for some non-vanishing value of the $s_k$. That is we plug each solution $c_{A,\gamma}=c^0_{A,\gamma}$ into $\mathcal{I}(c_{A,\gamma},s_k)$ and get a new ideal $\mathcal{I}^{\prime}(s_k)$:
 \begin{eqnarray}
 \mathcal{I}(c_{A,\gamma},s_k)\xrightarrow{c_{A,\gamma}=c^0_{A,\gamma}} \mathcal{I}(c^0_{A,\gamma},s_k)\equiv \mathcal{I}^{\prime}(s_k) \;.
 \end{eqnarray}
 We then find sets of points $s_k=s^0_k$ which lie on the locus described by $\mathcal{I}^{\prime}(s_k)$, that is, we find a series of associated possible kernel elements of (\ref{hidmap}), if any non-trivial solutions exist. Assuming all of this can be achieved, which it can in the example at hand, we end up with a set of solutions, each comprised of a set of values $c_{A,\gamma}=c^0_{A,\gamma}$, which lie on the jumping locus of both $V_{\textnormal{SM}}$ and ${V}_{\textnormal{H}}$, along with some associated non-trivial examples of kernel elements for (\ref{hidmap}) given by the $s_k=s^0_k$.

  Given these sets of points common to $\Sigma_{SM}$ and $\Sigma_{H}$, we must now decide which of the cases given in Figure \ref{locuspic} these points lie on. We are most interested in the third possibility depicted in that figure where the component of the hidden sector jumping locus that the starting point we have isolated lies on  is a subset of the standard model jumping locus: $\Sigma_{I}^{H}\subset \Sigma_{\textnormal{SM}}$. It is in this case that the moduli stabilization mechanism will cause the standard model spectrum to jump.
  
  To ascertain if the situation described in the last paragraph is indeed the one we have, we perform a linearized perturbation analysis of the equations given by setting the generators of the relevant ideal to zero. To do this, we substitute $c_{A,\gamma}=c^0_{A,\gamma}+\delta c_{A,\gamma}$ and $s_k=s^0_k+\delta s_k$ into $\mathcal{I}(c_{A,\gamma},s_k)$ and keep only the linear terms in $\delta c_{A,\gamma}$ and $\delta s_k$ to obtain a new set of generators for an ideal $\mathcal{I}^{\prime\prime}(\delta c_{A,\gamma},\delta s_k)$. For this ideal, the generators are nothing but a set of linears in the variables $\delta c_{A,\gamma}$ and $\delta s_k$, and thus it is very easy to perform an elimination on the variables $\delta s_k$ and obtain a set of constraints, $\mathcal{S}^{\prime}(\delta c_{A,\gamma})$, purely in terms of the $\delta c_{A,\gamma}$. Now our task is to compare the two ideals $\mathcal{S}(c_{A,\gamma})$ and $\mathcal{S}^{\prime}(\delta c_{A,\gamma})$, If all the solutions of  $\mathcal{S}^{\prime}(\delta c_{A,\gamma})=0$ solve $\mathcal{S}(c^0_{A,\gamma}+\delta c_{A,\gamma})=0$ up to linear terms in $\delta c_{A,\gamma}$, then we can conclude that, at least under infinitesimal perturbation, some irreducible component $\Sigma_I^H$ of $\Sigma_{H}$ lies on $\Sigma_{\textnormal{SM}}$. 

 In the case at hand, the locus on the standard model side is as follows.
 \begin{eqnarray} \label{Seg}
 \mathcal{S}=c_{4,2}c_{4,5}-c_{4,1}c_{4,6}=0
 \end{eqnarray}
 Assuming that $c_{4,1}^0 \neq0$ we can then use the following solution for $\mathcal{S}$:
 \begin{eqnarray} \label{cseg}
 c_{4,1}^0= a,c_{4,2}^0= b,c_{4,5}^0= c,c_{4,6}^0= \frac{b c}{a}
 \end{eqnarray}
The other complex structure coefficients $c_{A,\gamma}$ in the problem can be taken to be any number since no constraint arises on them.
  
Substituting these solutions into the equations $\mathcal{I}$ on the extension side, we obtain an ${\cal I}'$ which can easily be seen to have the following solutions for $s_k^0$:
 \begin{eqnarray} \label{freedom}
s_4^0= 0,s_5^0= 0,s_6^0= -\frac{a}{c} s_1,s_7^0= -\frac{a}{c} s_2, s_8^0=-\frac{a}{c} s_3.
 \end{eqnarray}
Here we can take $s_1$,$s_2$ and $s_3$ to be any value. 

Now that we have some points in moduli space common to both jumping loci, the next step is the linear perturbation analysis. Substituting $c_{A,\gamma} \to c_{A,\gamma}^0 + \delta c_{A,\gamma}$ and $s_k \to s_k^0 +\delta s_k$ into ${\cal I}$, keeping up to linear terms in the perturbations and eliminating the $\delta s_k$ we arrive at the following single constraint on the $\delta c_{A,\gamma}$.
\begin{eqnarray} \label{lin1}
-a^2\delta c_{4,6}+ab\delta c_{4,5}+a c \delta c_{4,2}-bc \delta c_{4,1} =0
\end{eqnarray}

In principle we now should solve this constraint for, for example $\delta c_{4,6}$ and substitute the result into ${\cal S}$ to see if that set of equations is also solved by these fluctuations to linear order. In fact this is not necessary in this case, as it can easily be observed that (\ref{lin1}) is precisely the linearization of (\ref{Seg}) around the starting points we have chosen. In this case this hidden sector locus does not merely lie inside $\Sigma_{\textnormal{SM}}$, it is identical to it.

Thus, even though we don't know the full information about the primary decomposition and elimination of $\mathcal{I}$, it is still possible to show that some of its components lie on the standard model jump locus.  In fact, in this example, we find a locus on the extension side which is precisely $\Sigma_{\textnormal{SM}}$. As a final check, one can verify that for a generic enough choice of complex structure of the form given in (\ref{cseg}) the cohomology on the extension side does indeed jump, from $h^{*}(X,{\cal O}(2,-2,2,-4,0))=(0,0,12,0)$ to $h^{*}(X,{\cal O}(2,-2,2,-4,0))=(0,5,17,0)$. An examination of the representation content of the larger first cohomology group which is obtained shows that three of these five elements survive to the quotient. This is in agreement with the freedom found in (\ref{freedom}) above.
  
\subsection{Results of a systematic scan over a class of Heterotic Line Bundle Standard Models}
As we have seen in the last subsection, moduli stabilization can indeed influence the standard model physics we observe in heterotic compactifications. The question we wish to answer is how common are such phenomena in known examples of heterotic standard model compactifications. That is, how often is it the case that the hidden sector bundle can effect the visible sector spectrum in this manner. To investigate this we have run a scan over the known data set of Heterotic Line Bundle Standard Models \cite{Anderson:2011ns,Anderson:2012yf}. To summarize, for each Line Bundle Standard Model in the data set, we do the following.

\begin{itemize}
\item First, we scan over all of the standard model multiplets to find those which have the potential to jump by using an analagous calculation to that found in Section \ref{egsec}.
\item Second, for all the standard models which are found in the first step to have spectra which can jump, we find all the extension bundles of the form (\ref{typeconsidered}) which satisfy the relevant consistency conditions, such as equivariance under the symmetry being considered and (\ref{ineq}).
\item Third, we calculate the jumping locus for the cohomology on the standard model side, and if it is possible, find the jumping locus on the extension side by using primary decomposition and elimination. If this is not practical in a given case, then we employ the linear perturbation analysis described in Section \ref{egsec}. 
\end{itemize}

By using this process, we scanned over all of the 2012 cases in the data set of \cite{Anderson:2011ns,Anderson:2012yf}. The resulting data detailing which standard model constructions have spectra which can be influenced by moduli stabilization are given in Table \ref{resultstab}.

\begin{table}
\footnotesize
 \begin{center}
 \resizebox{0.8\textwidth}{!}{%
\begin{tabular}{ |c|c|c|c|c| } 
 \hline
 CICY. Num. & Mod. Num & Sym. Num. & Multiplet & Extension Line  \\ 
  \hline
 6784 & 6 & 3,4,5,6 & $\overline{\bold 5}_{{\bf e}_1,{\bf e}_5} , \; {\cal O}(-1,2,2,-1)$ & ${\cal O}(-1,1,1,-1)$ \\ 
 \hline
 6784 & 51 & 3,4,5,6 & $\overline{\bold 5}_{{\bf e}_4,{\bf e}_5} , \; {\cal O}(-3,2,2,-1)$ & ${\cal O}(-1,1,1,-1)$ \\ 
  \hline
 6784 & 54 & 3,4,5,6 & $\overline{\bold 5}_{{\bf e}_1,{\bf e}_5} , \; \overline{\bold 5}_{{\bf e}_2,{\bf e}_5} , \; {\cal O}(-1,2,2,-1)$ & ${\cal O}(-1,1,1,-1)$ \\ 
 \hline
 6828 & 1 & 2 & $\overline{\bold 5}_{{\bf e}_1,{\bf e}_2} , \; {\cal O}(2,-3,2,-1)$ & ${\cal O}(1,-1,1,-1)$ \\ 
 \hline
 6828 & 2-5 & 2 & $\overline{\bold 5}_{{\bf e}_1,{\bf e}_2} , \; {\cal O}(2,-1,2,-1)$ & ${\cal O}(1,-1,1,-1)$ \\ 
 \hline
 6828 & 7 & 2 & $\overline{\bold 5}_{{\bf e}_1,{\bf e}_2} , \; {\cal O}(2,2,-1,-1)$ & ${\cal O}(1,1,-1,-1)$ \\ 
 \hline
 \multirow{4}*{7435} & \multirow{4}*{1} & \multirow{4}*{2} & ${\bold{10}}_{{\bf e}_5} , \; {\cal O}(-2,-2,0,1)$ & ${\cal O}(0,1,1,-1)$\\
\cline{4-5}
&&& $\overline{\bold 5}_{{\bf e}_1,{\bf e}_2} , \; {\cal O}(4,2,-2,-1)$ & ${\cal O}(0,1,1,-1)$\\
\cline{4-5}
&&& $\overline{\bold 5}_{{\bf e}_3,{\bf e}_5} , \; {\cal O}(-3,-2,1,1)$ & ${\cal O}(1,1,0,-1)$\\
\cline{4-5}
&&&$\overline{\bold 5}_{{\bf e}_4,{\bf e}_5} , \; {\cal O}(-3,-2,1,1)$ & ${\cal O}(1,1,0,-1)$\\
\hline
 \multirow{4}*{7435} & \multirow{4}*{2} & \multirow{4}*{2} & ${\bold{10}}_{{\bf e}_5} , \; {\cal O}(-2,0,-2,1)$ & ${\cal O}(0,1,1,-1)$\\
\cline{4-5}
&&& $\overline{\bold 5}_{{\bf e}_1,{\bf e}_2} , \; {\cal O}(4,-2,2,-1)$ & ${\cal O}(0,1,1,-1)$\\
\cline{4-5}
&&&$\overline{\bold 5}_{{\bf e}_3,{\bf e}_5} , \; {\cal O}(-3,1,-2,1)$ & ${\cal O}(1,0,1,-1)$\\
\cline{4-5}
&&& $\overline{\bold 5}_{{\bf e}_4,{\bf e}_5} , \; {\cal O}(-3,1,-2,1)$ & ${\cal O}(1,0,1,-1)$\\
\hline
 \multirow{3}*{7435} & \multirow{3}*{3} & \multirow{3}*{2} & $\overline{\bold 5}_{{\bf e}_1,{\bf e}_5} , \; {\cal O}(-2,4,2,-1)$ & ${\cal O}(1,0,1,-1)$\\
\cline{4-5}
&&&$\overline{\bold 5}_{{\bf e}_2,{\bf e}_4} , \; {\cal O}(1,-3,-2,1)$ & ${\cal O}(0,1,1,-1)$\\
\cline{4-5}
&&&$\overline{\bold 5}_{{\bf e}_3,{\bf e}_4} , \; {\cal O}(1,-3,-2,1)$ & ${\cal O}(0,1,1,-1)$\\
\hline
 \multirow{3}*{7435} & \multirow{3}*{4} & \multirow{3}*{2} &  $\overline{\bold 5}_{{\bf e}_1,{\bf e}_5} , \; {\cal O}(-2,2,4,-1)$ & ${\cal O}(1,1,0,-1)$\\
\cline{4-5}
&&& $\overline{\bold 5}_{{\bf e}_2,{\bf e}_4} , \; {\cal O}(1,-2,-3,1)$ & ${\cal O}(0,1,1,-1)$\\
\cline{4-5}
&&& $\overline{\bold 5}_{{\bf e}_3,{\bf e}_4} , \; {\cal O}(1,-2,-3,1)$ & ${\cal O}(0,1,1,-1)$\\
\hline
 \multirow{4}*{7435} & \multirow{4}*{5} & \multirow{4}*{2} &${\bold{10}}_{{\bf e}_5} , \; {\cal O}(-2,-2,0,1)$ & ${\cal O}(0,1,1,-1)$\\
\cline{4-5}
&&&$\overline{\bold 5}_{{\bf e}_1,{\bf e}_2} , \; {\cal O}(2,4,-2,-1)$ & ${\cal O}(1,0,1,-1)$ \\
\cline{4-5}
&&& $\overline{\bold 5}_{{\bf e}_3,{\bf e}_5} , \; {\cal O}(-2,-3,1,1)$ & ${\cal O}(1,1,0,-1)$\\
\cline{4-5}
&&& $\overline{\bold 5}_{{\bf e}_4,{\bf e}_5} , \; {\cal O}(-2,-3,1,1)$ & ${\cal O}(1,1,0,-1)$\\
\hline
 \multirow{4}*{7435} & \multirow{4}*{6} & \multirow{4}*{2} &${\bold{10}}_{{\bf e}_5} , \; {\cal O}(-2,0,-2,1)$ & ${\cal O}(0,1,1,-1)$\\
\cline{4-5}
&&& $\overline{\bold 5}_{{\bf e}_1,{\bf e}_2} , \; {\cal O}(2,-2,4,-1)$ & ${\cal O}(1,1,0,-1)$\\
\cline{4-5}
&&& $\overline{\bold 5}_{{\bf e}_3,{\bf e}_5} , \; {\cal O}(-2,1,-3,1)$ & ${\cal O}(1,0,1,-1)$\\
\cline{4-5}
&&& $\overline{\bold 5}_{{\bf e}_4,{\bf e}_5} , \; {\cal O}(-2,1,-3,1)$ & ${\cal O}(1,0,1,-1)$\\
\hline
 6732 & 1-2,34-35 & 1,2 & $\overline{\bold 5}_{{\bf e}_1,{\bf e}_4} , \; {\cal O}(0,2,2,-1,-1)$ & ${\cal O}(0,1,1,0,-1)$\\ 
  \hline
 6732 & 3-4 & 1,2 & $\overline{\bold 5}_{{\bf e}_1,{\bf e}_2} , \; {\cal O}(2,0,2,-1,-1)$ & ${\cal O}(1,0,1,0,-1)$\\ 
  \hline
 6732 & 15-17 & 1,2 & $\overline{\bold 5}_{{\bf e}_3,{\bf e}_5} , \; {\cal O}(2,0,2,-1,-1)$ & ${\cal O}(1,0,1,0,-1)$ \\ 
   \hline
  \multirow{2}*{6732} &  \multirow{2}*{19} &  \multirow{2}*{1,2} & $\overline{\bold 5}_{{\bf e}_1,{\bf e}_5} , \; {\cal O}(-1,2,0,2,-1)$ & ${\cal O}(0,1,0,1,-1)$ \\ 
  \cline{4-5}
  &&&$\overline{\bold 5}_{{\bf e}_2,{\bf e}_4} , \; {\cal O}(1,-2,1,-3,1)$ & ${\cal O}(0,1,0,1,-1)$\\
   \hline
 6732 & 26-28 & 1,2 & $\overline{\bold 5}_{{\bf e}_2,{\bf e}_5} , \; {\cal O}(0,-2,-2,1,1)$ & ${\cal O}(0,1,1,0,-1)$ \\ 
   \hline
 6732 & 30-31 & 1,2 & $\overline{\bold 5}_{{\bf e}_1,{\bf e}_2} , \; {\cal O}(2,0,2,-1,-1)$ & ${\cal O}(1,0,1,0,-1)$ \\ 
   \hline
6732 & 32 & 1,2 &$\overline{\bold 5}_{{\bf e}_4,{\bf e}_5} , \; {\cal O}(-2,0,-2,2,1)$ & ${\cal O}(1,0,1,0,-1)$\\ 
   \hline
\multirow{2}*{ 6732} & \multirow{2}*{33} & \multirow{2}*{1,2} & $\overline{\bold 5}_{{\bf e}_1,{\bf e}_2} , \; {\cal O}(2,-1,0,2,-1)$ & ${\cal O}(1,0,0,1,-1)$ \\ 
\cline{4-5}
&&&$\overline{\bold 5}_{{\bf e}_4,{\bf e}_5} , \; {\cal O}(-2,1,1,-3,1)$ & ${\cal O}(1,0,0,1,-1)$\\
   \hline
6732 & 36 & 1,2 & $\overline{\bold 5}_{{\bf e}_3,{\bf e}_4} , \; {\cal O}(0,-2,-2,2,1)$ & ${\cal O}(0,1,1,0,-1)$ \\ 
   \hline
 6770 & 13 & 1,2 & $\overline{\bold 5}_{{\bf e}_1,{\bf e}_2} , \; {\cal O}(1,1,-2,-2,0)$ & ${\cal O}(-1,-1,0,1,1)$\\ 
 \hline
  6770 & 14 & 1,2 & $\overline{\bold 5}_{{\bf e}_1,{\bf e}_2} , \; {\cal O}(1,1,-2,1,-2)$ & ${\cal O}(-1,-1,0,1,1)$ \\ 
\hline
 6777 & 17 & 1,2,3,4 & $\overline{\bold 5}_{{\bf e}_1,{\bf e}_3} , \; {\cal O}(1,1,-2,-3,1)$ & ${\cal O}(0,0,1,1,-1)$\\
 \hline
  6777 & 20 & 1,2,3,4 & $\overline{\bold 5}_{{\bf e}_2,{\bf e}_3} , \; {\cal O}(0,2,-1,2,-1)$ & ${\cal O}(1,-1,1,-2,0)$ \\ 
 \hline
   6890 & 1-2 16-17 & 1,2 & $\overline{\bold 5}_{{\bf e}_1,{\bf e}_4} , \; {\cal O}(0,2,2,-1,-1)$ & ${\cal O}(0,0,1,1,-1)$ \\ 
   \hline
 6890 & 5 & 1,2 & $\overline{\bold 5}_{{\bf e}_1,{\bf e}_3} , \; {\cal O}(1,1,-2,-3,1)$ & ${\cal O}(0,0,1,1,-1)$ \\ 
   \hline
 6890 & 18-19,22 & 1,2 & $\overline{\bold 5}_{{\bf e}_2,{\bf e}_5} , \; {\cal O}(0,-2,-2,1,1)$ & ${\cal O}(0,0,1,1,-1)$ \\ 
   \hline
 6890 & 20,21 & 1,2 & $\overline{\bold 5}_{{\bf e}_4,{\bf e}_5} , \; {\cal O}(-2,1,1,-3,1)$ & ${\cal O}(1,0,0,1,-1)$ \\ 
   \hline
 6890 & 24-27 & 1,2 & $\overline{\bold 5}_{{\bf e}_3,{\bf e}_4} , \; {\cal O}(0,-2,-2,2,1)$ & ${\cal O}(0,1,1,0,-1)$ \\ 
\hline
 \end{tabular}}
\end{center}
\caption{{\it Heterotic line bundle standard models whose spectrum can be forced to jump by a hidden sector bundle of the form (\ref{typeconsidered}). The first column specifies the CICY number of the manifold involved (according to the standard list \cite{Candelas:1987kf,database1}). The second and third columns specify the standard model numbers involved and the symmetries that are used in their construction according to the data sets provided in \cite{Anderson:2011ns,Anderson:2012yf,database2} and \cite{Braun:2010vc,database1} respectively. The fourth column gives an example of the component of the spectrum which can be forced to jump and the line bundle to which it is associated. Finally, the fifth column gives an example of an ${\cal L}$ which, when utilized in (\ref{typeconsidered}) would result in the change of spectrum being discussed.}} 
\label{resultstab}
\end{table}

Of the 2012 models in the data set, only 100 of them, approximately $5\%$, can be influenced by the moduli stabilization mechanism. This percentage is not very high but this figure is somewhat misleading. The issue is that in most cases in this list the standard model spectra is based upon line bundle cohomologies that do not jump on any locus in complex structure moduli space. If we focus on the 182 standard models which do have a spectrum that can jump at sub-loci of moduli space (which are listed in Appendix \ref{specstuff}), 100 is suddenly a large fraction. Perhaps a more useful figure then is that, within this data set, if the standard model spectrum can jump, then there is a $55\%$ chance that it will be forced to by the moduli stabilization mechanism. Clearly, in such a situation one should not consider moduli stabilization and model building separately. One should check if the cohomologies involved in model building can jump, and if they can it is important to check the effect of the hidden sector bundle on the spectrum.

We would like to emphasize that the above figure of $55\%$ can in some respects be regarded as a lower bound on the frequency at which this effect occurs in the line bundle standard model data set. As detailed above, we have not been able to perform a complete primary decomposition analysis of the jumping loci in all examples, and have had to restrict our attention to more crude analyses in many cases. These computations can easily miss loci associated with the hidden sector bundle that force the standard model spectrum jump. As such, interplay between moduli stabilization and model building structures could be even more pronounced than indicated here.

\section{Topological Vanishing of Yukawa Coupling in Heterotic Line Bundle Standard Models} \label{yuksec}

The tree level superpotential Yukawa couplings of Heterotic compactifications on Calabi-Yau threefolds are given by the following formula.
\begin{eqnarray} \label{suppot}
\lambda_{IJK} \propto \int_X \omega_I\wedge\omega_J\wedge\omega_K \wedge {\Omega}.
\end{eqnarray}
Here $I,J,K$ label the matter fields whose coupling is being computed and the $\omega$'s are the bundle valued one forms to which those matter fields are associated. The gauge structure of (\ref{suppot}) has been suppressed here: it is a gauge invariant combination of the three $\omega$'s that appears in the expression. We have left a `proportional to' sign explicitly in (\ref{suppot}) to emphasize that the absolute value of such a superpotential coupling is not physically meaningful in absence of knowledge of the K\"ahler potential. However, this formula can give us some information about the physical Yukawas, particularly concerning vanishings of such couplings.

There are several methods for computing quantities such as (\ref{suppot}) in the literature. These fall into two main approaches, using algebraic geometry \cite{Strominger:1985ks,Candelas:1987se,Candelas:1990pi,Greene:1986bm,Greene:1986jb,Distler:1987gg,Distler:1987ee,Greene:1987xh,Distler:1995bc,Braun:2006me,Bouchard:2006dn,Anderson:2009ge,Anderson:2010tc,Buchbinder:2014sya} and differential geometry \cite{Blesneag:2015pvz,Blesneag:2016yag} respectively. Here we will focus exclusively on the latter approach, which seems to be more powerful in the case of Line Bundle Standard Models. In particular, the approach of \cite{Blesneag:2015pvz,Blesneag:2016yag} makes it computationally easier to obtain moduli dependence of such couplings and leads to a powerful vanishing theorem. It is this latter result that we will make use of in what follows. We now discuss the statement of this theorem, leaving the details of its proof to the associated literature \cite{Blesneag:2015pvz,Blesneag:2016yag}.

\vspace{0.2cm}

Each cohomology group of which the $\omega_I$ are elements can be spanned by a basis, each element of which has a well defined ``type". Fortunately, in the Line Bundle Standard Model cases we will be interested in, this basis is compatible with the basis corresponding to standard model degrees of freedom. The type of a one form corresponding to a matter field is determined by how it descends from ambient space cohomologies in the Koszul sequence. In particular, if the form descends from a cohomology of the form $H^{\tau}(A,\wedge^{\tau-1}{\cal N}^{\vee} \otimes {\cal L})$, then it is said to be of type $\tau$.

The vanishing theorem proven in \cite{Blesneag:2015pvz,Blesneag:2016yag} then simply states that if the following condition is satisfied,
\begin{eqnarray} \label{van}
\tau_I+\tau_J+\tau_K<\text{dim}(A) \;,
\end{eqnarray}
where $\tau_I$ is the type of differential form and $\text{dim}(A)$ is the dimension of the ambient space, then the Yukawa coupling will vanish.

Using this result, it is possible to detect vanishings of Yukawa couplings in Heterotic Line Bundle Standard Models without heavy calculations. Such vanishings are, naively, topological in nature and need not be tied to any obvious symmetry property of the low energy effective theory (we will return to this issue at the end of this section). This is clearly of potential phenomenological interest as a mechanism of generating Yukawa textures of various types in such models.  Much like the `forced jumping' phenomena discussed in the previous section, such textures could be good or bad for the phenomenological viability of a given string theory standard model, depending upon their structure. For example, if all Yukawa couplings were found to vanish it might be difficult to achieve a sufficiently massive top quark in such a model. However, if the Yukawa matrix were forced to be rank one then this mechanism might provide a nice explanation as to why we observe one very heavy family in Nature. An example of another effect constraining couplings in such models are discussed in \cite{Buchbinder:2016jqr}.

In what follows we will investigate how common the vanishings of couplings we have discussed here are in the data set of Line Bundle Standard Models provided in \cite{Anderson:2011ns,Anderson:2012yf,database2}. Specifically we will examine those couplings which are consistent with all obvious symmetries of the models and compute which vanish due to (\ref{van}). We will begin with an example in the next sub-section and proceed to a general analysis in the following one.

\subsection{An example of topologically vanishing Yukawa couplings}
Let us illustrate the simple process of applying the vanishing theorem described above in an example. We will work on the manifold with CICY number 5256 according to the standard list \cite{Candelas:1987kf,database1}, which is defined by the following configuration matrix.
 \begin{eqnarray}
X=\left [ \begin{array}{c|cccc}
\mathbb{P}^{1} & 1 & 1 & 0 & 0  \\
\mathbb{P}^{1} & 2 & 0 & 0 & 0\\
\mathbb{P}^{1} & 0 & 0 & 1 & 1 \\
\mathbb{P}^{1} & 0 & 0 & 1 & 1\\
\mathbb{P}^{3} & 1 & 1 & 1 & 1
\end{array}
\right ],
\end{eqnarray}
We will quotient $X$ by the fourth discrete symmetry in the canonical list \cite{Braun:2010vc,database1}, which acts on the homogeneous coordinates in the following manner:
\begin{eqnarray}
\mathbb{Z}_2^{(1)}: \left\{ \begin{array}{c} x_{r,a}\rightarrow (-1)^a x_{r,a}  \\ x_{5,\alpha} \rightarrow (-1)^{\alpha}x_{5,\alpha} \end{array}\right. \\ \nonumber 
\mathbb{Z}_2^{(2)}: \left\{ \begin{array}{c} x_{r,a}\rightarrow x_{r,a+1} \\ x_{5,\alpha} \rightarrow  x_{5,\alpha+(-1)^{\alpha}} \end{array} \right. \;.
\end{eqnarray}
Here we make the identifications $x_{r,2}=x_{r,0}\;, \forall i$. In addition to this coordinate action there is a normal bundle action which descends to the following transformations on the defining polynomials.
\begin{eqnarray}
&&\mathbb{Z}_2^{(1)}: (p_1,p_2,p_3,p_4) \rightarrow(p_1,-p_2,p_3,-p_4)\\ \nonumber
&&\mathbb{Z}_2^{(2)}: (p_1,p_2,p_3,p_4) \rightarrow (p_1,-p_2,p_4,p_3)
\end{eqnarray}

On $X/\mathbb{Z}_2 \times \mathbb{Z}_2$ a Line Bundle Standard Model can be built of the form $V_{\textnormal{SM}} = \bigoplus_{i=1}^5 {\cal L}_i$ with the following line bundle content  \cite{Anderson:2011ns,Anderson:2012yf}.
\begin{eqnarray}
&&{\cal L}_1=\mathcal{O}_{X}(1, 0, -2, 1, 0)\;,\; {\cal L}_2 =\mathcal{O}_{X}(1, -2, 1, 0, 0)\;,\; {\cal L}_3=\mathcal{O}_{X}(0, 1, 1, -2, 0)\;, \\ \nonumber &&{\cal L}_4=\mathcal{O}_{X}(-1, 1, 0, 0, 0) \;,\; {\cal L}_5=\mathcal{O}_{X}(-1, 0, 0, 1, 0)
\end{eqnarray}
The non-trivial cohomology content of combinations of the line bundles ${\cal L}_i$ which are relevant for the standard model spectrum of this Line Bundle Standard Model are as follows.
\begin{eqnarray}
&&h^{*}(X, {\cal L}_1)=(0,4,0,0);\quad h^{*}(X,{\cal L}_2)=(0,4,0,0);\quad h^{*}(X, {\cal L}_3)=(0,4,0,0);
\\\nonumber
&&h^{*}(X, {\cal L}_1\otimes {\cal L}_4)=(0,4,0,0); \quad h^{*}(X, {\cal L}_1\otimes {\cal L}_5)=(0,3,3,0);\quad h^{*}(X, {\cal L}_2\otimes {\cal L}_5)=(0,4,0,0);
\\\nonumber
&&h^{*}(X, {\cal L}_4\otimes {\cal L}_5)=(0,4,0,0);\quad h^{*}(X, {\cal L}_1^{\vee}\otimes {\cal L}_5^{\vee})=(0,3,3,0);\quad 
h^{*}(X, {\cal L}_1\otimes {\cal L}_2^{\vee})=(0,12,0,0);\\\nonumber
&& h^{*}(X, {\cal L}_1\otimes {\cal L}_5^{\vee})=(0,3,3,0); \quad   h^{*}(X, {\cal L}_2\otimes {\cal L}_3^{\vee})=(0,12,0,0); \quad 
h^{*}(X, {\cal L}_2\otimes {\cal L}_4^{\vee})=(0,12,0,0);\\\nonumber
 &&   h^{*}(X, {\cal L}_3\otimes {\cal L}_4^{\vee})=(0,4,0,0); \quad   h^{*}(X, {\cal L}_3\otimes {\cal L}_5^{\vee})=(0,16,0,0);
\end{eqnarray}
  These cohomologies correspond respectively to the following multiplets on $X$ (before the quotient):
  \begin{eqnarray} \label{spectrumyeg}
  4\,\bold{10}_{{\bf e}_1};\; 4\,\bold{10}_{{\bf e}_2};\; 4\,\bold{10}_{{\bf e}_3};\;4\,\bar{\bold{5}}_{{{{\bf e}_1},{{\bf e}_4}}};\;3\,\bar{\bold{5}}_\bold{{{\bf e}_1},{{\bf e}_5}};\;4\,\bar{\bold{5}}_\bold{{{\bf e}_2},{{\bf e}_5}};\;4\,\bar{\bold{5}}_\bold{{{\bf e}_4},{{\bf e}_5}};\;3\,{\bold{5}}_\bold{{-{\bf e}_1},{-{\bf e}_5}};
  \\\nonumber  12\,\bold{1}_\bold{{{\bf e}_1},{-{\bf e}_2}};\; 3\,\bold{1}_\bold{{{\bf e}_1},{-{\bf e}_5}};\; 3\,\bold{1}_\bold{{{\bf e}_5},{-{\bf e}_1}};\; 12\,\bold{1}_\bold{{{\bf e}_2},{-{\bf e}_3}};\; 12\,\bold{1}_\bold{{{\bf e}_2},{-{\bf e}_4}};\; 4\,\bold{1}_\bold{{{\bf e}_3},{-{\bf e}_4}};\; 16\,\bold{1}_\bold{{{\bf e}_3},{-{\bf e}_5}};
  \end{eqnarray}
  As in earlier sections, we give spectra in this section in terms of GUT multiplets for conciseness, despite the fact that we use a Wilson line (whose exact form will not be needed here) to break the gauge group to that of the standard model.

  Given the spectrum of standard model representations and $U(1)$ charges given in (\ref{spectrumyeg}), one would naively expect the following Yukawa couplings to be present.
  \begin{eqnarray}
  \bar{\bold{5}}_{\bold{{{\bf e}_2},{{\bf e}_5}}}\bold{5}_{\bold{{-{\bf e}_1},{-{\bf e}_5}}}\bold{1}_{\bold{{{\bf e}_1},{-{\bf e}_2}}} \; , \; \bold{10}_{\bold{{{\bf e}_3}}}\bar{\bold{5}}_{\bold{{{\bf e}_1},{{\bf e}_4}}}\bar{\bold{5}}_{\bold{{{\bf e}_2},{{\bf e}_5}}}
 \end{eqnarray}
 Let us look at these two Yukawa couplings in more detail in the context of the vanishing theorem (\ref{van}). In order to use the theorem, we must first work out which ambient space cohomologies the relevant matter fields descend from in the Koszul  sequence. Beginning with the Yukawa coupling $\bar{\bold{5}}_{\bold{{{\bf e}_2},{{\bf e}_5}}}\bold{5}_{\bold{{-{\bf e}_1},-{{\bf e}_5}}}\bold{1}_{\bold{{{\bf e}_1},-{{\bf e}_2}}}$, the line bundles associated to the multiplets which appear are as follows.
 \begin{eqnarray}
 \bar{\bold{5}}_{\bold{{{\bf e}_2},{{\bf e}_5}}}:\mathcal{O}_{X}(0,-2,1,1,0)\;,\;\bold{5}_{\bold{-{{\bf e}_1},-{{\bf e}_5}}}:\mathcal{O}_{X}(0,0,2,-2,0)\;,\; \bold{1}_{\bold{{{\bf e}_1},-{{\bf e}_2}}}:\mathcal{O}_{X}(0,2,-3,1,0)
 \end{eqnarray}
 A short computation shows that $H^1(X,{\cal L})$ for all of these line bundles descends from the associated first cohomology on $A$, that is $H^1(A,{\cal L}_{2}\otimes {\cal L}_5)$, which is four dimensional, $H^1(A,{\cal L}_{1}^{\vee} \otimes{\cal L}_5^{\vee})$ which is three dimensional and $H^1(A,{\cal L}_{1}\otimes{\cal L}_2^{\vee})$, which is twelve dimensional, respectively.
 
From this analysis we can see that all three of the involved matter fields are of type one, and thus we have,
\begin{eqnarray}
\tau_{ \bar{\bold{5}}_{\bold{{{\bf e}_2},{{\bf e}_5}}}}+\tau_{\bold{5}_{\bold{{-{\bf e}_1},-{{\bf e}_5}}}}+\tau_{\bold{1}_{\bold{{{\bf e}_1},{-{\bf e}_2}}}}=3<\textnormal{dim}(A)=7\;.
\end{eqnarray}
Given this, the vanishing theorem tells us that this Yukawa coupling (or more precisely this set of 144 couplings) vanishes, despite the fact there is no obvious gauge theoretic restriction that would cause it to do so. Given that these upstairs couplings vanish, so do all of the associated downstairs couplings associated to the Line Bundle Standard Model itself.

For the second Yukawa coupling, $\bold{10}_{\bold{{{\bf e}_3}}}\bar{\bold{5}}_{\bold{{{\bf e}_1},{{\bf e}_4}}}\bar{\bold{5}}_{\bold{{{\bf e}_2},{{\bf e}_5}}}$ a similar procedure can be followed. We find that once again all three matter fields are of type 1, and thus the Yukawa coupling vanishes, naively due to topological restrictions with, once again, no obvious gauge theoretic restriction presenting itself.

\subsection{Scanning over the Line Bundle Standard Models}

We now proceed to apply an analysis of the form presented in the previous subsection to every Line Bundle Standard Model in the data set of \cite{Anderson:2011ns,Anderson:2012yf}. The procedure we apply is as follows. For each model, we first look at the multiplets which arise. That is, we examine the cohomology groups,
\begin{eqnarray}
H^1(X, {\cal L}_i)\;,\; H^1(X, {\cal L}_i\otimes {\cal L}_j)\;,\; H^1(X, {\cal L}_i^{\vee}\otimes {\cal L}_j^{\vee})\;,\; H^1(X, {\cal L}_i\otimes {\cal L}_j^{\vee}) \;,
\end{eqnarray}
which, as was detailed in Table \ref{spectab}, are the upstairs cohomologies that correspond to the following matter representations.
\begin{eqnarray} \label{poscoup}
\bold{10_{{{\bf e}_i}}}\;,\; \bar{\bold{5}}_{\bold{{{\bf e}_i}},\bold{{{\bf e}_j}}}\;,\; \bold{5_{-{{\bf e}_i},-{{\bf e}_j}}}\;,\; \bold{1_{{{\bf e}_i},-{{\bf e}_j}}}
\end{eqnarray}
Once we have extracted this list of multiplets from the Line Bundle Standard Model data set, we then extract all of the Yukawa couplings that are consistent with the constraints imposed by gauge symmetry. These are all of one of the following three forms.
\begin{eqnarray}
\bold{5}_{\bold{{-{\bf e}_i}},\bold{-{{\bf e}_j}}}\bold{10}_{\bold{{{\bf e}_i}}}\bold{10}_{\bold{{{\bf e}_j}}}, \quad \bar{\bold{5}}_{{{{\bf e}_i}},{{{\bf e}_j}}}\bold{\bar{{5}}_{{{{\bf e}_k}},{{{\bf e}_l}}}}\bold{{10}_{{{{\bf e}_m}}}},\quad \bold{1}_{\bold{{{\bf e}_i}},\bold{-{{\bf e}_j}}}\bold{5}_{\bold{-{{\bf e}_i}},\bold{-{{\bf e}_k}}}\bar{\bold{5}}_{\bold{{{\bf e}_j}},\bold{{{\bf e}_k}}}.
\end{eqnarray}
 Finally, for each Yukawa coupling that does not vanish due to gauge theoretic considerations, we examine the Koszul sequence associated to each of the line bundles giving rise to the matter multiplets involved and determine the types of the associated forms. We can then use the vanishing theorem (\ref{van}) to determine whether or not these couplings are actually present. We present the full results of this analysis in Appendix \ref{yukstuff} and will content ourselves here with some brief statistics on the results.

\vspace{0.2cm}

The number of Yukawa couplings which are non-zero after gauge theoretic considerations are taken into account is given in Table \ref{mrtab1}. Models are only listed in this table if they have at least one non-vanishing coupling at this level.
\begin{table}[!h]
\begin{center}
\begin{tabular}{|c|c|c|c|c|c|}
\hline
CICY No. & No. Sym.  & No. Models&$ \bold{5}_{\bold{{-{\bf e}_i}},\bold{-{{\bf e}_j}}}\bold{10}_{\bold{{{\bf e}_i}}}\bold{10}_{\bold{{{\bf e}_j}}}$& $\bar{\bold{5}}_{{{{\bf e}_i}},{{{\bf e}_j}}}\bold{\bar{{5}}_{{{{\bf e}_k}},{{{\bf e}_l}}}}\bold{{10}_{{{{\bf e}_m}}}}$ & $ \bold{1}_{\bold{{{\bf e}_i}},\bold{-{{\bf e}_j}}}\bold{5}_{\bold{-{{\bf e}_i}},\bold{-{{\bf e}_k}}}\bar{\bold{5}}_{\bold{{{\bf e}_j}},\bold{{{\bf e}_k}}}$ \\
\hline
{6784} & {4}&188 & 0 & 120 & 144  \\
\cline{1-6}
{6828} & {1}& 2& 0 & 5 & 0  \\
\cline{1-6}
{7862} & {1}& 14&18 & 53 & 19\\
\cline{1-6}
{5256} & {6}&84& 0 & 126 & 32  \\
\cline{1-6}
{5452} & {20}&800& 0 & 1376 & 208 \\
\cline{1-6}
{6947} & {1}&24& 0 & 24 & 12  \\
\cline{1-6}
{6732} & {2}&28& 24 & 68 & 12  \\
\cline{1-6}
{6770} & {2}&16& 16 & 32 & 0  \\
\cline{1-6}
{6777} & {4}&24& 48 & 64 & 80  \\
\cline{1-6}
{6890} & {2}&22& 24 & 50 & 12 \\
\cline{1-6}
{7447} & {1}&3& 0 & 5 & 4  \\
\cline{1-6}
{7487} & {4}&276& 164 & 580 & 444 \\
\cline{1-6}
\hline
\end{tabular}
\caption{\it The number of Yukawa couplings of various types that are permitted by the gauge symmetries of the set of Line Bundle Standard Models being considered \cite{Anderson:2011ns,Anderson:2012yf,database2}. The first column gives the CICY identification number of the manifold on which the models are based, according to the standard list \cite{Candelas:1987kf,database1}. The second column details how many symmetries are being considered, and thus the number of downstairs manifolds that each row corresponds to. `No. Models' gives the number of models with at least one Yukawa coupling that would be consistent with gauge invariance in the data set. The remaining three columns give the number of each type of such couplings that appear in this set of models}
\label{mrtab1}
\end{center}
\end{table} 

Given the data in Table \ref{mrtab1}, the question is now how many of these Yukawa couplings vanish due to the topological vanishing theorem (\ref{van}). The answer to this question is given in Table \ref{mrtab2}.
\begin{table}[!h]
\begin{center}
\begin{tabular}{|c|c|c|c|c|c|}
\hline
CICY No. & No. Sym. & No. Models &$ \bold{5}_{\bold{{-{\bf e}_i}},\bold{-{{\bf e}_j}}}\bold{10}_{\bold{{{\bf e}_i}}}\bold{10}_{\bold{{{\bf e}_j}}}$& $\bar{\bold{5}}_{{{{\bf e}_i}},{{{\bf e}_j}}}\bold{\bar{{5}}_{{{{\bf e}_k}},{{{\bf e}_l}}}}\bold{{10}_{{{{\bf e}_m}}}}$ & $ \bold{1}_{\bold{{{\bf e}_i}},\bold{-{{\bf e}_j}}}\bold{5}_{\bold{-{{\bf e}_i}},\bold{-{{\bf e}_k}}}\bar{\bold{5}}_{\bold{{{\bf e}_j}},\bold{{{\bf e}_k}}}$\\
\hline
{6784} & {4}& 0&0 & 0 & 0  \\
\cline{1-6}
{6828} & {1}& 0&0 & 0 & 0  \\
\cline{1-6}
{7862} & {1}&9& 8 & 5 & 3  \\
\cline{1-6}
{5256} & {6}&32& 0 & 32 & 24  \\
\cline{1-6}
{5452} & {20}&256& 0 & 240 & 192  \\
\cline{1-6}
{6947} & {1}&24& 0 & 24 & 12  \\
\cline{1-6}
{6732} & {2}&0& 0 & 0 & 0  \\
\cline{1-6}
{6770} & {2}&8& 16 & 0 & 0  \\
\cline{1-6}
{6777} & {4}&0& 0 & 0 & 0  \\
\cline{1-6}
{6890} & {2}&0& 0 & 0 & 0 \\
\cline{1-6}
{7447} & {1}&1& 0 & 1 & 3  \\
\cline{1-6}
{7487} & {4}&112& 80 & 32 & 0  \\
\cline{1-6}
\hline
\end{tabular}
\caption{\it The number of Yukawa couplings of various types that are permitted by the gauge symmetries but vanish due to the topological restriction (\ref{van}) for the set of Line Bundle Standard Models being considered \cite{Anderson:2011ns,Anderson:2012yf,database2}. The first column gives the CICY identification number of the manifold on which the models are based, according to the standard list \cite{Candelas:1987kf,database1}. The second column details how many symmetries are being considered, and thus the number of downstairs manifolds that each row corresponds to. `No. Models' gives the number of models with at least one Yukawa coupling that vanishes due to this topological consideration. The remaining three columns give the number of each type of such couplings that vanish due to (\ref{van}) in this set of models}
\label{mrtab2}
\end{center}
\end{table} 
Compiling this data into even more coarse overall figures, we obtain the percentage of the different types of coupling given in (\ref{poscoup}) which would be allowed by gauge invariance but which vanish due to these topological considerations. These figures are presented in Table \ref{mrtab3}.
\begin{table}[!h]
\begin{center}
\begin{tabular}{|c|c|c|c|}
\hline
Yukawa Type & Total Num. &  Top. Van. Num. & Percentage \\
\hline
$ \bold{5}_{\bold{{-{\bf e}_i}},\bold{-{{\bf e}_j}}}\bold{10}_{\bold{{{\bf e}_i}}}\bold{10}_{\bold{{{\bf e}_j}}}$ & 294 & 104 & 35.4\%  \\
\cline{1-4}
$\bar{\bold{5}}_{{{{\bf e}_i}},{{{\bf e}_j}}}\bold{\bar{{5}}_{{{{\bf e}_k}},{{{\bf e}_l}}}}\bold{{10}_{{{{\bf e}_m}}}}$  & 2503 & 334 & 13.3\%  \\
\cline{1-4}
$ \bold{1}_{\bold{{{\bf e}_i}},\bold{-{{\bf e}_j}}}\bold{5}_{\bold{-{{\bf e}_i}},\bold{-{{\bf e}_k}}}\bar{\bold{5}}_{\bold{{{\bf e}_j}},\bold{{{\bf e}_k}}}$& 967 & 234 & 24.2\%  \\
\cline{1-4}
In total & 3764 & 672 &17.9\% \\
\cline{1-4}
\hline
\end{tabular}
\caption{\it The total number of Yukawa couplings of each type in the Line Bundle Standard Model data set studied \cite{Anderson:2011ns,Anderson:2012yf,database2}. The column `Total Number' details the number of each type of coupling which are consistent with gauge invariance. The column `Top. Van. Num.' details the number of these couplings that are actually zero due to the vanishing theorem (\ref{van}).}
\label{mrtab3}
\end{center}
\end{table} 

In the final analysis there is a total of 1481 Standard Models in the data set which have at least one Yukawa coupling that would be expected to be non-zero based upon consideration of the obvious symmetries in the construction. Of these, 442 have have at least one such coupling which turns out to be zero due to the vanishing theorem (\ref{van}). This means that topological vanishing of Yukawa couplings plays a role in $29.8\%$ of these models.

\vspace{0.1cm}

A few comments are order about these results. Firstly, it is clear that this is not a rare phenomenon. A lot of couplings that would naively be allowed by gauge invariance in the theory actually vanish due to topological considerations. That this effect would be common was anticipated in \cite{Blesneag:2015pvz,Blesneag:2016yag}. It should also be mentioned that these results are reminiscent, for example, of long understood selection rules in orbifold compactifications \cite{Hamidi:1986vh,Dixon:1986qv,Font:1988tp,Font:1988nc,Kobayashi:2011cw}.

Obvious questions include whether or not such vanishings are also so ubiquitous in higher order couplings and whether there is some hidden reason, beyond quasi-topological restrictions, for the phenomenon. For the latter question, a potential hint is given by the results of \cite{Anderson:2010tc,Buchbinder:2014sya}. There it was shown that stability walls elsewhere in extended bundle and K\"ahler moduli space could have $U(1)$ symmetries that, while broken for the split bundle being studied, still restrict its couplings hugely. This effect can be very strong, essentially due to the holomorphic nature of the superpotential of the four dimensional theory. The bundles being studied in the Line Bundle Standard Model data set considered here are in larger K\"ahler cones than the simple examples considered in \cite{Anderson:2010tc}. In such cases, it is expected that the constraints on couplings will be even more restrictive (due to a larger number of stability walls being present). This could potentially explain the high percentages of topological vanishings found in Table \ref{mrtab3}

\section{Conclusions} \label{conc}

In this paper we have studied two effects which can arise in Line Bundle Standard Models \cite{Anderson:2011ns,Anderson:2012yf}. The first of these concerns the interaction of line bundle model building and the moduli stabilization mechanism of \cite{Anderson:2010mh,Anderson:2011ty}. In that work, the hidden sector gauge bundle is used to stabilize the complex structure to some higher co-dimensional sub-locus of moduli space. Here, we have investigated how often the system being forced to this special locus in complex structure moduli space causes the massless charged spectrum of the standard model to jump. The second effect we considered was concerned with the structure of Yukawa couplings. Couplings which are consistent with all obvious symmetries of the four dimensional effective theory can be zero due to seemingly topological restrictions. We have considered the form of topological vanishing presented in \cite{Blesneag:2015pvz,Blesneag:2016yag} and have determined how common such effects are in the known data set of Line Bundle Standard Models.

In our work on the first of these directions we have seen that, in the data set studied, if the standard model field content is capable of jumping, the hidden sector stabilization mechanism has a good chance of forcing it to do so. In particular, there is at least a $55\%$ chance that one (of the usually small number) of $SU(2)$ structure extension hidden sector bundles that can be consistently included in such a compactification will force the standard model bundle to a jumping locus. Such a strong interaction between the visible sector and hidden sector bundles may seem surprising at first. However, for the threefolds that are considered with non-vanishing first fundamental group, the second Chern characters are not that large. Given this, there are then not many choices of equivariant line bundles that can be used in the construction of a hidden sector bundle given any particular Line Bundle Standard Model. The restricted nature of the choices seems to lead to a relatively ubiquitous correlation between jumping loci of the cohomologies governing the hidden sector extension and the standard model spectrum. The basic message of Section \ref{specjump} of the paper is thus that such effects are something that should be considered in model building work, if the standard model bundles being considered have cohomologies which are capable of jumping. As we have emphasized in the main text, this effect could be either good or bad. It could force unwanted family/anti-family pairs to appear in the spectrum, but it could equally well force the generation of a Higgs-Higgs bar pair in a model that previously lacked such degrees of freedom.

In our investigation of topological vanishings of Yukawa couplings we have seen a similarly strong effect. We have seen that the vanishing theorem presented in \cite{Blesneag:2015pvz,Blesneag:2016yag} leads to an otherwise permitted Yukawa coupling being zero in $30\%$ of the models presented in \cite{Anderson:2011ns,Anderson:2012yf}. Indeed, almost $18\%$ of the couplings that are allowed by all of the obvious symmetries in these models actually vanish. As with the previous result, this effect can be either good or bad for the phenomenological viability of a model depending on the particular case at hand. It is clear, however, given the ubiquity of the effect, that such vanishings should be taken into account in phenomenological explorations of these constructions. Several natural questions follow from these results. For example, do similar, seemingly topological, vanishings of couplings happen for higher order interactions? We conjecture one possible explanation for the large number of vanishings that would answer this question in the affirmative. As was discussed in \cite{Anderson:2010tc}, stability walls elsewhere in combined bundle and K\"ahler moduli space, can have strong effects on superpotential couplings in backgrounds where those splittings are not manifest. In particular they can force such vanishings of couplings. Whether this really is the effect that is behind many of the vanishings that we have seen is a study that we leave for future work.

\section*{Acknowledgments}

The authors would like to thank Lara Anderson and Eric Sharpe for valuable discussions. The work of J.G. and J.W. is supported in part by NSF grant PHY-1720321. The authors would like to gratefully acknowledge the hospitality of the Simons Center for Geometry and Physics (and the semester long program, \emph{The Geometry and Physics of Hitchin Systems}) during the completion of this work.


\appendix

\section{Jumping Spectrum Results} \label{specstuff}

In this appendix we present data on the interplay between $\Sigma_{\textnormal{SM}}$ and $\Sigma_{\textnormal{H}}$, as defined in Section \ref{specjump}, for all of the Line Bundle Standard Models of \cite{Anderson:2011ns,Anderson:2012yf,database2} whose spectra are determined by cohomologies that could potentially jump in dimension. In particular, all cases where a non-trivial map is involved in the sequence chasing used to determine the spectrum are considered. In the tables below, `CICY No.', `Symmetry No.' and `Model No.' refer to the labels for the upstairs manifolds, symmetries and Line Bundle Standard Models that are being considered, relative to the relevant standard lists, \cite{Candelas:1987kf,database1}, \cite{Braun:2010vc,database1} and \cite{Anderson:2011ns,Anderson:2012yf,database2} respectively. The entries in the column entitled `Jump Line' specify the multiplet being considered in that row and the line bundle whose cohomology it is associated to. Finally, the columns `Jump Standard' and `Jump Extension' contain information about $\Sigma_{\textnormal{SM}}$ and $\Sigma_{\textnormal{H}}$ respectively.

For cases where no possible extension bundle of the form (\ref{typeconsidered}) exists, we place a `no extension' in the final column and perform no further computations. If the jumping locus for the standard model bundle only jumps on loci in complex structure moduli space where the associated Calabi-Yau manifold becomes singular we place a `singular' in the penultimate column (or $\textnormal{singular}^{*}$ if only a portion of this locus could be determined and that portion exhibited this property). In such cases, there is no need to perform any computations involving the hidden sector bundles and, as such, a `null' is placed in the final column. If, for a standard model which can indeed jump (indicated by a `y' in the `Jump Standard' column) there exists a hidden sector bundle for which we have been able to find an irreducible component to its jumping locus that lies entirely within $\Sigma_{\textnormal{SM}}$ then we put a `y' in the final column. If all such loci we have been able to find merely intersect the standard model bundle jumping locus we place a `g' in the final column. A `singular' in the last column indicates that all of the components of $\Sigma_{\textnormal{H}}$ that exist force the Calabi-Yau manifold to a singular locus in its moduli space. A $\textnormal{singular}^{*}$ in the final column means that all of the components of $\Sigma_{\textnormal{H}}$ that we were able to find have this property, but other loci may exist. An `unknown' in any column simply means that the system was so complicated that we were unable to extract any meaningful data in a reasonable amount of time.

\begin{center}
\resizebox{\textwidth}{!}{%
\begin{tabular}{|c|c|c|c|c|c|}
\hline
CICY No. & Symmetry No. & Model No. & Jump Line & Jump Standard & Jump Extension\\
\hline
\multirow{8}*{6784} & \multirow{8}*{3-6} & \multirow{3}*{1-5} & ${\bold{10}}_{{\bf e}_1} , \; {\cal O}(3,2,-2,-1)$ & singular & null\\
\cline{4-6}
&   &  &$\overline{\bold 5}_{{\bf e}_1,{\bf e}_3}, \; {\cal O}(2,2,-1,-1)$ & singular & null\\
\cline{4-6}
&   &  &$\overline{\bold 5}_{{\bf e}_1,{\bf e}_4}, \; {\cal O}(2,2,-1,-1)$ & singular & null\\
\cline{3-6}
&& \multirow{1}{*}{6} & $\overline{\bold 5}_{{\bf e}_1,{\bf e}_5}, \; {\cal O}(-1,2,2,-1)$ & y & y\\
\cline{3-6}
 &&  \multirow{1}{*}{7-10} & $\overline{\bold 5}_{{\bf e}_1,{\bf e}_2}, \; {\cal O} (2,2,-3,-1)$ & singular & null\\
\cline{3-6}
 &&  \multirow{1}{*}{11-50} &$\overline{\bold 5}_{{\bf e}_1,{\bf e}_2}, \; {\cal O}(2,2,-1,-1)$ & singular & null\\
\cline{3-6}
&&  \multirow{1}{*}{51} &$\overline{\bold 5}_{{\bf e}_4,{\bf e}_5}, \; {\cal O}(-3,2,2,-1)$ & y & y\\
\cline{3-6}
&&  \multirow{1}{*}{52} & $\overline{\bold 5}_{{\bf e}_1,{\bf e}_2}, \; {\cal O}(2,2,-1,-1)$ & singular & null\\
\cline{3-6}
&& \multirow{1}{*}{54} & $\overline{\bold 5}_{{\bf e}_1,{\bf e}_5}, \; \overline{\bold 5}_{{\bf e}_2,{\bf e}_5}, \; {\cal O}(-1,2,2,-1)$ & y & y\\
\cline{3-6}
\hline
\end{tabular}}
\end{center}

\begin{center}
\resizebox{\textwidth}{!}{%
\begin{tabular}{|c|c|c|c|c|c|}
\hline
CICY No. & Symmetry No. & Model No. & Jump Line & Jump Standard & Jump Extension\\
\hline
\multirow{4}*{6828} & \multirow{4}*{2} & \multirow{1}*{1} &$\overline{\bold 5}_{{\bf e}_1,{\bf e}_2}, \; {\cal O}(2,-3,2,-1)$ & y & y\\
\cline{3-6}
&& \multirow{1}{*}{2-5} & $\overline{\bold 5}_{{\bf e}_1,{\bf e}_2}, \; {\cal O}(2,-1,2,-1)$ & y & y\\
\cline{3-6}
 &&  \multirow{1}{*}{6} & ${\bold{10}}_{{\bf e}_1}, \; {\cal O}(2,-2,3,-1)$ & singular & null\\
\cline{3-6}
 &&  \multirow{1}{*}{7} & $\overline{\bold 5}_{{\bf e}_1,{\bf e}_2}, \; {\cal O} (2,2,-1,-1)$ & y & y\\
\cline{3-6}
\hline
\end{tabular}}
\end{center}

\begin{center}
\resizebox{\textwidth}{!}{%
\begin{tabular}{|c|c|c|c|c|c|}
\hline
CICY No. & Symmetry No. & Model No. & Jump Line & Jump Standard & Jump Extension\\
\hline
\multirow{22}*{7435} & \multirow{22}*{2} & \multirow{4}*{1} & ${\bold{10}}_{{\bf e}_5}, \; {\cal O}(-2,-2,0,1)$ & y & y\\
\cline{4-6}
&&& $\overline{\bold 5}_{{\bf e}_1,{\bf e}_2}, \; {\cal O} (4,2,-2,-1) $& y & y\\
\cline{4-6}
&&& $\overline{\bold 5}_{{\bf e}_3,{\bf e}_5}, \; {\cal O} (-3,-2,1,1) $& y & y\\
\cline{4-6}
&&& $\overline{\bold 5}_{{\bf e}_4,{\bf e}_5}, \; {\cal O} (-3,-2,1,1) $& y & y\\
\cline{3-6}
&& \multirow{4}{*}{2} & ${\bold{10}}_{{\bf e}_5}, \; {\cal O} (-2,0,-2,1)$ & y & y\\
\cline{4-6}
&&& $\overline{\bold 5}_{{\bf e}_1,{\bf e}_2}, \; {\cal O} (4,-2,2,-1) $& y & y\\
\cline{4-6}
&&& $\overline{\bold 5}_{{\bf e}_3,{\bf e}_5}, \; {\cal O} (-3,1,-2,1) $& y & y\\
\cline{4-6}
&&&$\overline{\bold 5}_{{\bf e}_4,{\bf e}_5}, \; {\cal O}(-3,1,-2,1) $& y & y\\
\cline{4-6}
\cline{3-6}
&& \multirow{3}{*}{3} & $\overline{\bold 5}_{{\bf e}_1,{\bf e}_5}, \; {\cal O} (-2,4,2,-1)$ & y & y\\
\cline{4-6}
&&& $\overline{\bold 5}_{{\bf e}_2,{\bf e}_4}, \; {\cal O} (1,-3,-2,1)$ & y & y\\
\cline{4-6}
&&& $\overline{\bold 5}_{{\bf e}_3,{\bf e}_4}, \; {\cal O}(1,-3,-2,1)$ & y & y\\
\cline{4-6}
\cline{3-6}
&& \multirow{3}{*}{4} & $\overline{\bold 5}_{{\bf e}_1,{\bf e}_5}, \; {\cal O}(-2,2,4,-1)$ & y & y\\
\cline{4-6}
&&& $\overline{\bold 5}_{{\bf e}_2,{\bf e}_4}, \; {\cal O}  (1,-2,-3,1)$ & y & y\\
\cline{4-6}
&&& $\overline{\bold 5}_{{\bf e}_3,{\bf e}_4}, \; {\cal O} (1,-2,-3,1)$ & y & y\\
\cline{4-6}
\cline{3-6}
\cline{3-6}
&& \multirow{4}{*}{5} &${\bold{10}}_{{\bf e}_5}, \; {\cal O}(-2,-2,0,1)$ & y & y\\
\cline{4-6}
&&& $\overline{\bold 5}_{{\bf e}_1,{\bf e}_2}, \; {\cal O}(2,4,-2,-1)$ & y & y\\
\cline{4-6}
&&& $\overline{\bold 5}_{{\bf e}_3,{\bf e}_5}, \; {\cal O}(-2,-3,1,1)$ & y & y\\
\cline{4-6}
&&& $\overline{\bold 5}_{{\bf e}_4,{\bf e}_5}, \; {\cal O} (-2,-3,1,1)$ & y & y\\
\cline{4-6}
\cline{3-6}
\cline{3-6}
&& \multirow{4}{*}{6} &${\bold{10}}_{{\bf e}_5}, \; {\cal O} (-2,0,-2,1)$ & y & y\\
\cline{4-6}
&&& $\overline{\bold 5}_{{\bf e}_1,{\bf e}_2}, \; {\cal O} (2,-2,4,-1) $& y & y\\
\cline{4-6}
&&& $\overline{\bold 5}_{{\bf e}_3,{\bf e}_5}, \; {\cal O}(-2,1,-3,1) $& y & y\\
\cline{4-6}
&&&  $\overline{\bold 5}_{{\bf e}_4,{\bf e}_5}, \; {\cal O}(-2,1,-3,1) $& y & y\\
\cline{4-6}
\cline{3-6}
\cline{3-6}
\hline
\end{tabular}}
\end{center}

\begin{center}
\resizebox{\textwidth}{!}{%
\begin{tabular}{|c|c|c|c|c|c|}
\hline
CICY No. & Symmetry No. & Model No. & Jump Line & Jump Standard & Jump Extension\\
\hline
\multirow{4}*{7862} & \multirow{4}*{3} & \multirow{1}*{2} & $\overline{\bold 5}_{{\bf e}_4,{\bf e}_5}, \; {\cal O} (-2,3,2,-3) $& y & g\\
\cline{3-6}
&& \multirow{1}{*}{3-6} & $\overline{\bold 5}_{{\bf e}_1,{\bf e}_2}, \; {\cal O}(2,-2,-2,2)$ & y & g\\
\cline{3-6}
 &&  \multirow{1}{*}{9-12} & $\overline{\bold 5}_{{\bf e}_1,{\bf e}_2}, \; {\cal O}(2,-2,-2,2)$ & y & g\\
\cline{3-6}
 &&  \multirow{1}{*}{15-18} & ${\bold{10}}_{{\bf e}_5}, \; {\cal O}  (-2,2,-2,2) $& y & g\\
\cline{3-6}
\hline
\end{tabular}}
\end{center}

\begin{center}
\resizebox{\textwidth}{!}{%
\begin{tabular}{|c|c|c|c|c|c|}
\hline
CICY No. & Symmetry No. & Model No. & Jump Line & Jump Standard & Jump Extension\\
\hline
\multirow{9}*{5256} & \multirow{9}*{3-6} & \multirow{1}*{1-4} &${\bold{10}}_{{\bf e}_2}, \; {\cal O}  (0,1,-2,-2,1)$ & singular & null\\
\cline{3-6}
&&\multirow{2}*{7-10} &${\bold{10}}_{{\bf e}_3}, \; {\cal O} (0,1,-2,-2,1)$ & singular & null\\
\cline{4-6}
&&& $\overline{\bold 5}_{{\bf e}_1,{\bf e}_2}, \; {\cal O}(2,-1,0,2,-1)$ & singular & null\\
\cline{3-6}
&& \multirow{2}{*}{20} & ${\bold{10}}_{{\bf e}_5}, \; {\cal O} (-2,-2,0,1,1)$ & singular & null\\
\cline{4-6}
&&& $\overline{\bold 5}_{{\bf e}_3,{\bf e}_4}, \; {\cal O} (0,2,2,-1,-1)$ & y & singular\\
\cline{3-6}
 &&  \multirow{1}{*}{21} & ${\bold{10}}_{{\bf e}_5}, \; {\cal O}  (-2,-2,1,0,1)$ & singular & null\\
\cline{3-6}
 &&  \multirow{2}{*}{22} &${\bold{10}}_{{\bf e}_5}, \; {\cal O} (-2,-2,1,0,1)$ & singular & null\\
 \cline{4-6}
 &&&$\overline{\bold 5}_{{\bf e}_1,{\bf e}_2}, \; {\cal O} (2,0,-1,2,-1)$ & y & g\\
 \cline{3-6}
 && \multirow{1}*{23} & ${\bold{10}}_{{\bf e}_5}, \; {\cal O}  (-2,-2,0,1,1)$ & singular & null\\
 \cline{3-6}
 && \multirow{1}*{24} & ${\bold{10}}_{{\bf e}_5}, \; {\cal O}  (-2,-2,1,0,1) $& singular & null\\
\cline{3-6}
&& \multirow{1}*{26} & ${\bold{10}}_{{\bf e}_5}, \; {\cal O}  (-2,-2,1,0,1)$ & singular & null\\
\cline{3-6}
\hline
\end{tabular}}
\end{center}

\begin{center}
\resizebox{\textwidth}{!}{%
\begin{tabular}{|c|c|c|c|c|c|}
\hline
CICY No. & Symmetry No. & Model No. & Jump Line & Jump Standard & Jump Extension\\
\hline
\multirow{17}*{5452} & \multirow{17}*{7-22} & \multirow{3}*{1} & $\overline{\bold 5}_{{\bf e}_1,{\bf e}_2}, \; {\cal O} (2,2,0,0,-2)$ & singular & null\\
\cline{4-6}
&&&$\overline{\bold 5}_{{\bf e}_4,{\bf e}_5}, \; {\cal O}(-3,0,0,0,1)$ & singular & null\\
\cline{4-6}
&&&${\bold{10}}_{{\bf e}_3}, \; {\cal O}  (1,-2,0,0,1)$ & singular & null\\
\cline{3-6}
&& \multirow{3}*{2} & $\overline{\bold 5}_{{\bf e}_1,{\bf e}_2}, \; {\cal O}(2,2,0,0,-2) $& singular & null\\
\cline{4-6}
&&&$\overline{\bold 5}_{{\bf e}_3,{\bf e}_5}, \; {\cal O} (-1,-2,-1,2,2)$ & singular & null\\
\cline{4-6}
&&& ${\bold{10}}_{{\bf e}_5}, \; {\cal O} (-2,0,0,1,1)$ & singular & null\\
\cline{3-6}
 &&  \multirow{1}{*}{3-6} &${\bold{10}}_{{\bf e}_5}, \; {\cal O}  (-2,0,-2,1,1)$ & singular & null\\
\cline{3-6}
 &&   \multirow{3}*{7} & $\overline{\bold 5}_{{\bf e}_1,{\bf e}_2}, \; {\cal O}(2,2,0,0,-2)$ & singular & null\\
\cline{4-6}
&&& $\overline{\bold 5}_{{\bf e}_4,{\bf e}_5}, \; {\cal O}(-2,-2,1,2,1)$ & singular & null\\
\cline{4-6}
&&& ${\bold{10}}_{{\bf e}_5}, \; {\cal O} (-2,0,0,1,1)$ & singular & null\\
\cline{3-6}
&&   \multirow{3}*{8} & $\overline{\bold 5}_{{\bf e}_1,{\bf e}_2}, \; {\cal O} (2,2,0,0,-2)$ & singular & null\\
\cline{4-6}
&&& $\overline{\bold 5}_{{\bf e}_4,{\bf e}_5}, \; {\cal O} (-2,-2,2,1,1)$ & singular & null\\
\cline{4-6}
&&&${\bold{10}}_{{\bf e}_4}, \; {\cal O}  (0,-2,1,0,1) $& singular & null\\
\cline{3-6}
&&   \multirow{2}*{9-12} &$\overline{\bold 5}_{{\bf e}_1,{\bf e}_2}, \; {\cal O} (2,2,0,-1,-1)$ & singular & null\\
\cline{4-6}
&&&${\bold{10}}_{{\bf e}_5}, \; {\cal O}  (-2,0,-2,1,1)$ & singular & null\\
\cline{3-6}
&&   \multirow{3}*{13} & $\overline{\bold 5}_{{\bf e}_1,{\bf e}_2}, \; {\cal O}(2,2,0,0,-2) $& singular & null\\
\cline{4-6}
&&& $\overline{\bold 5}_{{\bf e}_3,{\bf e}_4}, \; {\cal O} (0,-3,0,0,1) $& singular & null\\
\cline{4-6}
&&& ${\bold{10}}_{{\bf e}_5}, \; {\cal O}  (-2,1,0,0,1) $& singular & null\\
\cline{3-6}
&&   \multirow{2}*{14} & $\overline{\bold 5}_{{\bf e}_1,{\bf e}_2}, \; {\cal O}(2,2,0,0,-2)$ & singular & null\\
\cline{4-6}
&&& ${\bold{10}}_{{\bf e}_4}, \; {\cal O}  (0,-2,1,0,1) $& singular & null\\
\cline{3-6}
&&   \multirow{1}*{18-25} &${\bold{10}}_{{\bf e}_5}, \; {\cal O} (-2,0,-2,1,1)$ & singular & null\\
\cline{3-6}
&&   \multirow{1}*{30-33} & ${\bold{10}}_{{\bf e}_5}, \; {\cal O}  (-2,0,-2,1,1)$ & singular & null\\
\cline{3-6}
&&   \multirow{2}*{39-42} & $\overline{\bold 5}_{{\bf e}_3,{\bf e}_4}, \; {\cal O} (0,2,2,-1,-1)$ & singular & null\\
\cline{4-6}
&&&${\bold{10}}_{{\bf e}_5}, \; {\cal O}  (-2,0,-2,1,1)$ & singular & null\\
\cline{3-6}
&&   \multirow{2}*{43-46} & $\overline{\bold 5}_{{\bf e}_1,{\bf e}_2}, \; {\cal O}(2,2,-1,0,-1)$ & singular & null\\
\cline{4-6}
&&&${\bold{10}}_{{\bf e}_3}, \; {\cal O}  (0,-2,1,-2,1)$ & singular & null\\
\cline{3-6}
&&   \multirow{1}*{47-50} & ${\bold{10}}_{{\bf e}_3}, \; {\cal O}  (0,-2,1,-2,1)$ & singular & null\\
\cline{3-6}
&&   \multirow{1}*{52-55} & ${\bold{10}}_{{\bf e}_3}, \; {\cal O}  (0,-2,1,-2,1)$ & singular & null\\
\cline{3-6}
&&   \multirow{2}*{58-61} &${\bold{10}}_{{\bf e}_3}, \; {\cal O}  (0,-2,1,-2,1)$ & singular & null\\
\cline{4-6}
&&&$\overline{\bold 5}_{{\bf e}_1,{\bf e}_2}, \; {\cal O}(2,0,-1,2,-1)$ & singular & null\\
\cline{3-6}
&&   \multirow{1}*{63-66} & ${\bold{10}}_{{\bf e}_4}, \; {\cal O}  (0,-2,1,-2,1)$ & singular & null\\
\cline{3-6}
&&   \multirow{1}*{67-70} & ${\bold{10}}_{{\bf e}_4}, \; {\cal O}  (0,-2,1,-2,1)$ & singular & null\\
\cline{3-6}
\hline
\end{tabular}}
\end{center}
\begin{center}
\resizebox{\textwidth}{!}{%
\begin{tabular}{|c|c|c|c|c|c|}
\hline
CICY No. & Symmetry No. & Model No. & Jump Line & Jump Standard & Jump Extension\\
\hline
\multirow{21}*{6732} & \multirow{21}*{1-2} & \multirow{1}*{1} & $\overline{\bold 5}_{{\bf e}_1,{\bf e}_4}, \; {\cal O} (0,2,2,-1,-1)$ & y & y\\
\cline{3-6}
&&\multirow{1}*{2} &$\overline{\bold 5}_{{\bf e}_1,{\bf e}_4}, \; {\cal O} (0,2,2,-1,-1) $& y & y\\
\cline{3-6}
&& \multirow{2}{*}{3-4} & $\overline{\bold 5}_{{\bf e}_1,{\bf e}_2}, \; {\cal O}(2,0,2,-1,-1) $& y & y\\
\cline{4-6}
&&& $\overline{\bold 5}_{{\bf e}_4,{\bf e}_5}, \; {\cal O} (-3,1,-2,1,1)$ & $\textnormal{singular}^{*}$ & null\\
\cline{3-6}
 &&  \multirow{1}{*}{5-8} & $\overline{\bold 5}_{{\bf e}_4,{\bf e}_5}, \; {\cal O} (-2,-2,2,-2,2) $& unknown & unknown\\
\cline{3-6}
 &&  \multirow{1}{*}{9} &$\overline{\bold 5}_{{\bf e}_4,{\bf e}_5}, \; {\cal O}  (-2,-2,0,1,1) $& singular & null\\
 \cline{3-6}
 && \multirow{1}*{10-13} &$\overline{\bold 5}_{{\bf e}_4,{\bf e}_5}, \; {\cal O}  (-2,-2,1,0,1)$ & singular & null\\
 \cline{3-6}
 && \multirow{1}*{15-17} &$\overline{\bold 5}_{{\bf e}_3,{\bf e}_5}, \; {\cal O}  (-2,0,-2,1,1)$ & y & y\\
 \cline{3-6}
 && \multirow{2}*{19} & $\overline{\bold 5}_{{\bf e}_1,{\bf e}_5}, \; {\cal O}  (-1,2,0,2,-1)$ & y & y\\
\cline{4-6}
&&& $\overline{\bold 5}_{{\bf e}_2,{\bf e}_4}, \; {\cal O}  (1,-2,1,-3,1) $&  y & y\\
\cline{3-6}
&& \multirow{1}*{20} & $\overline{\bold 5}_{{\bf e}_4,{\bf e}_5}, \; {\cal O}  (-2,-2,0,1,1)$ & singular & null\\
\cline{3-6}
&& \multirow{1}*{21-24} &$\overline{\bold 5}_{{\bf e}_4,{\bf e}_5}, \; {\cal O} {(-2,-2,1,0,1)}$ & singular & null\\
\cline{3-6}
&& \multirow{1}*{26-28} &$\overline{\bold 5}_{{\bf e}_2,{\bf e}_5}, \; {\cal O}  {(0,-2,-2,1,1)} $& y & y\\
\cline{3-6}
&& \multirow{1}*{30-31} &$\overline{\bold 5}_{{\bf e}_1,{\bf e}_2}, \; {\cal O}  {(2,0,2,-1,-1)}$ & y & y\\
\cline{3-6}
&& \multirow{1}*{32} & $\overline{\bold 5}_{{\bf e}_4,{\bf e}_5}, \; {\cal O} {(-2,0,-2,2,1)}$ & y & y\\
\cline{3-6}
&& \multirow{2}*{33} &$\overline{\bold 5}_{{\bf e}_4,{\bf e}_5}, \; {\cal O} {(-2,1,1,-3,1)}$ &  y & y\\
\cline{4-6}
&&&$\overline{\bold 5}_{{\bf e}_1,{\bf e}_2}, \; {\cal O} (2,-1,0,2,-1)$ & y & y\\
\cline{3-6}
&& \multirow{2}*{34} &$\overline{\bold 5}_{{\bf e}_1,{\bf e}_4}, \; {\cal O}  (0,2,2,-1,-1)$ & y & y\\
\cline{4-6}
&&& $\overline{\bold 5}_{{\bf e}_2,{\bf e}_3}, \; {\cal O} (1,-3,-2,1,1)$ &  $\textnormal{singular}^{*}$ & null\\
\cline{3-6}
&& \multirow{2}*{35} & $\overline{\bold 5}_{{\bf e}_1,{\bf e}_4}, \; {\cal O}  (0,2,2,-1,-1)$ & y & y\\
\cline{4-6}
&&& $\overline{\bold 5}_{{\bf e}_2,{\bf e}_3}, \; {\cal O} (1,-3,-2,1,1)$ &  $\textnormal{singular}^{*}$ & null\\
\cline{3-6}
&& \multirow{1}*{36}& $\overline{\bold 5}_{{\bf e}_3,{\bf e}_4}, \; {\cal O}  (0,-2,-2,2,1) $& y & y\\
\cline{3-6}
\hline
\end{tabular}}
\end{center}

\begin{center}
\resizebox{\textwidth}{!}{%
\begin{tabular}{|c|c|c|c|c|c|}
\hline
CICY No. & Symmetry No. & Model No. & Jump Line & Jump Standard & Jump Extension\\
\hline
\multirow{2}*{6770} & \multirow{2}*{1-2} & \multirow{1}*{13} & $\overline{\bold 5}_{{\bf e}_1,{\bf e}_2}, \; {\cal O}(1,1,-2,-2,0)$ & y & y\\
\cline{3-6}
&& \multirow{1}{*}{14} & $\overline{\bold 5}_{{\bf e}_1,{\bf e}_2}, \; {\cal O} (1,1,-2,1,-2)$ & y & y\\
\cline{3-6}
\hline
\end{tabular}}
\end{center}

\begin{center}
\resizebox{\textwidth}{!}{%
\begin{tabular}{|c|c|c|c|c|c|}
\hline
CICY No. & Symmetry No. & Model No. & Jump Line & Jump Standard & Jump Extension\\
\hline
\multirow{10}*{6890} & \multirow{10}*{1-2} & \multirow{1}*{1-2} &$\overline{\bold 5}_{{\bf e}_1,{\bf e}_4}, \; {\cal O} (0,2,2,-1,-1)$ & y & y\\
\cline{3-6}
&& \multirow{1}{*}{4} &$\overline{\bold 5}_{{\bf e}_3,{\bf e}_5}, \; {\cal O} (-2,0,-2,1,1) $& singular & null\\
\cline{3-6}
&& \multirow{2}{*}{5} & $\overline{\bold 5}_{{\bf e}_1,{\bf e}_3}, \; {\cal O} (1,1,-2,-3,1)$ & y & y\\
\cline{4-6}
&&& $\overline{\bold 5}_{{\bf e}_2,{\bf e}_5}, \; {\cal O}(-1,0,2,2,-1)$ & singular & null\\
\cline{3-6}
&& \multirow{1}*{6-9} &$\overline{\bold 5}_{{\bf e}_3,{\bf e}_5}, \; {\cal O}(-2,2,-2,-2,2) $& unknown & unknown\\
\cline{3-6}
&& \multirow{1}*{10-13} &$\overline{\bold 5}_{{\bf e}_3,{\bf e}_5}, \; {\cal O}(-2,1,-2,0,1)$ & singular & null\\
\cline{3-6}
&& \multirow{2}{*}{16-17} & $\overline{\bold 5}_{{\bf e}_1,{\bf e}_4}, \; {\cal O} (0,2,2,-1,-1) $& y & y\\
\cline{4-6}
&&& $\overline{\bold 5}_{{\bf e}_2,{\bf e}_3}, \; {\cal O}(1,-2,-3,1,1)$ & $\textnormal{singular}^{*}$ & null\\
\cline{3-6}
&& \multirow{1}*{18-19} & $\overline{\bold 5}_{{\bf e}_2,{\bf e}_5}, \; {\cal O}(0,-2,-2,1,1)$ & y & y\\
\cline{3-6}
&& \multirow{1}*{20-21} & $\overline{\bold 5}_{{\bf e}_4,{\bf e}_5}, \; {\cal O} (-2,1,1,-3,1)$ &  y& y\\
\cline{3-6}
&& \multirow{1}*{22} &$\overline{\bold 5}_{{\bf e}_2,{\bf e}_5}, \; {\cal O} (0,-2,-2,1,1) $& y & y\\
\cline{3-6}
&& \multirow{2}{*}{24-27} & $\overline{\bold 5}_{{\bf e}_4,{\bf e}_5}, \; {\cal O} (-2,1,-2,0,1) $& singular & null\\
\cline{4-6}
&&&$\overline{\bold 5}_{{\bf e}_3,{\bf e}_4}, \; {\cal O} (0,-2,-2,2,1)$ & y & y\\
\cline{3-6}
\hline
\end{tabular}}
\end{center}

\begin{center}
\resizebox{\textwidth}{!}{%
\begin{tabular}{|c|c|c|c|c|c|}
\hline
CICY No. & Symmetry No. & Model No. & Jump Line & Jump Standard & Jump Extension\\
\hline
\multirow{7}*{6777} & \multirow{7}*{1-4} & \multirow{1}*{1-4} &$\overline{\bold 5}_{{\bf e}_2,{\bf e}_3}, \; {\cal O}(2,-2,-2,-2,2)$ & unknown & unknown\\
\cline{3-6}
&& \multirow{1}{*}{5-12} &$\overline{\bold 5}_{{\bf e}_2,{\bf e}_4}, \; {\cal O}(1,-2,-2,0,1)$ & singular & null\\
\cline{3-6}
&& \multirow{1}*{16} & $\overline{\bold 5}_{{\bf e}_3,{\bf e}_4}, \; {\cal O}(0,-2,-2,1,1)$ & singular & null\\
\cline{3-6}
&& \multirow{1}*{17} & $\overline{\bold 5}_{{\bf e}_1,{\bf e}_3}, \; {\cal O} (1,1,-2,-3,1)$ & y & y\\
\cline{3-6}
&& \multirow{1}*{19} &$\overline{\bold 5}_{{\bf e}_3,{\bf e}_4}, \; {\cal O}(0,-2,-2,1,1) $& singular & null\\
\cline{3-6}
&& \multirow{1}*{20} &$\overline{\bold 5}_{{\bf e}_2,{\bf e}_3}, \; {\cal O}(0,2,-1,2,-1)$ & y & y\\
\cline{3-6}
\hline
\end{tabular}}
\end{center}

\begin{center}
\resizebox{\textwidth}{!}{%
\begin{tabular}{|c|c|c|c|c|c|}
\hline
CICY No. & Symmetry No. & Model No. & Jump Line & Jump Standard & Jump Extension\\
\hline
\multirow{1}*{7447} & \multirow{1}*{2} & \multirow{1}*{3} & $\overline{\bold 5}_{{\bf e}_1,{\bf e}_3}, \; {\cal O} (1,-2,1,-2,2)$ & y & $\textnormal{singular}^{*}$\\
\cline{3-6}
\hline
\end{tabular}}
\end{center}

\begin{center}
\resizebox{\textwidth}{!}{%
\begin{tabular}{|c|c|c|c|c|c|}
\hline
CICY No. & Symmetry No. & Model No. & Jump Line & Jump Standard & Jump Extension\\
\hline
\multirow{10}*{7487} & \multirow{10}*{3-6} & \multirow{1}*{11-20} & $\overline{\bold 5}_{{\bf e}_1,{\bf e}_3}, \; {\cal O} (1,-2,1,-2,2)$ & singular & null\\
\cline{3-6}
&& \multirow{1}{*}{22} & $\overline{\bold 5}_{{\bf e}_3,{\bf e}_5}, \; {\cal O} (-2,2,1,1,-2)$ & y & $\textnormal{singular}^{*}$\\
\cline{3-6}
&& \multirow{3}{*}{23} &$\overline{\bold 5}_{{\bf e}_1,{\bf e}_2}, \; {\cal O}(2,-2,-2,1,1)$ &  & no extension\\
\cline{4-6}
&&&$\overline{\bold 5}_{{\bf e}_3,{\bf e}_5}, \; {\cal O}(-2,2,1,-2,1) $&  & no extension\\
\cline{4-6}
&&& $\overline{\bold 5}_{{\bf e}_4,{\bf e}_5}, \; {\cal O}(-2,1,2,1,-2) $&  & no extension\\
\cline{3-6}
&& \multirow{1}*{24} & $\overline{\bold 5}_{{\bf e}_4,{\bf e}_5}, \; {\cal O} (-2,1,1,2,-2)$ & y & $\textnormal{singular}^{*}$\\
\cline{3-6}
&& \multirow{1}*{26} &$\overline{\bold 5}_{{\bf e}_3,{\bf e}_5}, \; {\cal O} (-2,2,1,1,-2) $&  & no extension\\
\cline{3-6}
&& \multirow{1}*{27} &$\overline{\bold 5}_{{\bf e}_1,{\bf e}_2}, \; {\cal O} (2,-2,-2,1,1) $& y &$\textnormal{singular}^{*}$ \\
\cline{3-6}
&& \multirow{1}*{28} &$\overline{\bold 5}_{{\bf e}_1,{\bf e}_2}, \; {\cal O}(2,-2,-2,1,1) $& y &$\textnormal{singular}^{*}$ \\
\cline{3-6}
&& \multirow{1}*{36-39} &$\overline{\bold 5}_{{\bf e}_2,{\bf e}_4}, \; {\cal O} (1,-2,2,1,-2)$ & singular &null \\
\cline{3-6}
&& \multirow{1}*{61} & $\overline{\bold 5}_{{\bf e}_3,{\bf e}_5}, \; {\cal O}(-2,2,1,1,-2)$ & y &$\textnormal{singular}^{*}$ \\
\cline{3-6}
&& \multirow{1}*{72-81} & $\overline{\bold 5}_{{\bf e}_2,{\bf e}_3}, \; {\cal O} (2,-2,-2,1,1)$ & y &$\textnormal{singular}^{*}$\\
\cline{3-6}
\hline
\end{tabular}}
\end{center}

\section{Vanishing Coupling Results}\label{yukstuff}

In this appendix we give in detail, for every heterotic Line Bundle Standard Model in the data set of \cite{Anderson:2011ns,Anderson:2012yf,database2}, which couplings vanish due to the topological considerations discussed in Section \ref{yuksec}. In these tables, `CICY No.', `Sym. No.' and `Model No.' refer to the labels for the upstairs manifolds, symmetries and Line Bundle Standard Models that are being considered, relative to the relevant standard lists, \cite{Candelas:1987kf,database1}, \cite{Braun:2010vc,database1} and \cite{Anderson:2011ns,Anderson:2012yf,database2} respectively. The column `Yukawa Pattern'  lists the couplings that are consistent with the obvious symmetries of these models in each case. Finally the column `Top. Van.' details whether these couplings are affected by the topological vanishing condition (\ref{van}) of \cite{Blesneag:2015pvz,Blesneag:2016yag}.

\begin{center}
\begin{tabular}{|c|c|c|c|c|}
\hline
CICY No. & Model No. & Yukawa Pattern & Top. Van. & Sym. No.\\
\hline
\multirow{16}{*}{6784} & $1$ & $\textbf{10}_{\textbf{e}_{\textbf{1}}}\bar{\textbf{5}}_{\textbf{e}_{\textbf{2}},\textbf{e}_{\textbf{5}}}\bar{\textbf{5}}_{\textbf{e}_{\textbf{3}},\textbf{e}_{\textbf{4}}} $& n & 1,3\\
\cline{2-5}
& 2 & $\textbf{10}_{\textbf{e}_{\textbf{1}}}\bar{\textbf{5}}_{\textbf{e}_{\textbf{2}},\textbf{e}_{\textbf{5}}}\bar{\textbf{5}}_{\textbf{e}_{\textbf{3}},\textbf{e}_{\textbf{4}}}$ & n & 1-4\\
\cline{2-5}
& 3 & $\textbf{10}_{\textbf{e}_{\textbf{1}}}\bar{\textbf{5}}_{\textbf{e}_{\textbf{2}},\textbf{e}_{\textbf{5}}}\bar{\textbf{5}}_{\textbf{e}_{\textbf{3}},\textbf{e}_{\textbf{4}}}$ & n & 1-4\\
\cline{2-5}
& 4 & $\textbf{10}_{\textbf{e}_{\textbf{1}}}\bar{\textbf{5}}_{\textbf{e}_{\textbf{2}},\textbf{e}_{\textbf{5}}}\bar{\textbf{5}}_{\textbf{e}_{\textbf{3}},\textbf{e}_{\textbf{4}}}$ & n & 1-4\\
\cline{2-5}
& $5$ & $\textbf{10}_{\textbf{e}_{\textbf{1}}}\bar{\textbf{5}}_{\textbf{e}_{\textbf{2}},\textbf{e}_{\textbf{5}}}\bar{\textbf{5}}_{\textbf{e}_{\textbf{3}},\textbf{e}_{\textbf{4}}}$ & n & 2,4\\
\cline{2-5}
&\multirow{2}{*} 6 & $\textbf{10}_{\textbf{e}_{\textbf{1}}}\bar{\textbf{5}}_{\textbf{e}_{\textbf{2}},\textbf{e}_{\textbf{3}}}\bar{\textbf{5}}_{\textbf{e}_{\textbf{4}},\textbf{e}_{\textbf{5}}} $   & n & 1-4\\
\cline{3-5}
& & $\textbf{10}_{\textbf{e}_{\textbf{4}}}\bar{\textbf{5}}_{\textbf{e}_{\textbf{1}},\textbf{e}_{\textbf{5}}}\bar{\textbf{5}}_{\textbf{e}_{\textbf{2}},\textbf{e}_{\textbf{3}}}$ & n & 1-4\\
\cline{2-5}
& 7 & $\textbf{10}_{\textbf{e}_{\textbf{3}}}\bar{\textbf{5}}_{\textbf{e}_{\textbf{1}},\textbf{e}_{\textbf{2}}}\bar{\textbf{5}}_{\textbf{e}_{\textbf{4}},\textbf{e}_{\textbf{5}}} $ & n & 1-4\\
\cline{2-5}
& 8 & $\textbf{10}_{\textbf{e}_{\textbf{3}}}\bar{\textbf{5}}_{\textbf{e}_{\textbf{1}},\textbf{e}_{\textbf{2}}}\bar{\textbf{5}}_{\textbf{e}_{\textbf{4}},\textbf{e}_{\textbf{5}}} $ & n & 1-4\\
\cline{2-5}
& 9 & $\textbf{10}_{\textbf{e}_{\textbf{3}}}\bar{\textbf{5}}_{\textbf{e}_{\textbf{1}},\textbf{e}_{\textbf{2}}}\bar{\textbf{5}}_{\textbf{e}_{\textbf{4}},\textbf{e}_{\textbf{5}}}  $ & n & 1-4\\
\cline{2-5}
& 10 & $\textbf{10}_{\textbf{e}_{\textbf{3}}}\bar{\textbf{5}}_{\textbf{e}_{\textbf{1}},\textbf{e}_{\textbf{2}}}\bar{\textbf{5}}_{\textbf{e}_{\textbf{4}},\textbf{e}_{\textbf{5}}} $ & n & 1-4\\
\cline{2-5}
& 11 & $\textbf{1}_{\textbf{e}_{\textbf{4}},-\textbf{e}_{\textbf{3}}}\textbf{5}_{-\textbf{e}_{\textbf{4}},-\textbf{e}_{\textbf{5}}}\bar{\textbf{5}}_{\textbf{e}_{\textbf{3}}, \textbf{e}_{\textbf{5}}}$ & n & 1-4\\
\cline{2-5}
& 12 & $\textbf{1}_{\textbf{e}_{\textbf{4}},-\textbf{e}_{\textbf{3}}}\textbf{5}_{-\textbf{e}_{\textbf{4}},-\textbf{e}_{\textbf{5}}}\bar{\textbf{5}}_{\textbf{e}_{\textbf{3}}, \textbf{e}_{\textbf{5}}}$ & n & 1-4\\
\cline{2-5}
& 13 & $\textbf{1}_{\textbf{e}_{\textbf{4}},-\textbf{e}_{\textbf{3}}}\textbf{5}_{-\textbf{e}_{\textbf{4}},-\textbf{e}_{\textbf{5}}}\bar{\textbf{5}}_{\textbf{e}_{\textbf{3}}, \textbf{e}_{\textbf{5}}} $ & n & 1-4\\
\cline{2-5}
& 14 & $\textbf{1}_{\textbf{e}_{\textbf{4}},-\textbf{e}_{\textbf{3}}}\textbf{5}_{-\textbf{e}_{\textbf{4}},-\textbf{e}_{\textbf{5}}}\bar{\textbf{5}}_{\textbf{e}_{\textbf{3}}, \textbf{e}_{\textbf{5}}}$ & n & 1-4\\
\cline{2-5}
& 15 & $\textbf{1}_{\textbf{e}_{\textbf{4}},-\textbf{e}_{\textbf{3}}}\textbf{5}_{-\textbf{e}_{\textbf{4}},-\textbf{e}_{\textbf{5}}}\bar{\textbf{5}}_{\textbf{e}_{\textbf{3}}, \textbf{e}_{\textbf{5}}} $ & n & 1-4\\
\cline{2-5}
& 16 & $\textbf{1}_{\textbf{e}_{\textbf{4}},-\textbf{e}_{\textbf{3}}}\textbf{5}_{-\textbf{e}_{\textbf{4}},-\textbf{e}_{\textbf{5}}}\bar{\textbf{5}}_{\textbf{e}_{\textbf{3}}, \textbf{e}_{\textbf{5}}}$ & n & 1-4\\
\cline{2-5}
& 17 & $\textbf{1}_{\textbf{e}_{\textbf{4}},-\textbf{e}_{\textbf{3}}}\textbf{5}_{-\textbf{e}_{\textbf{4}},-\textbf{e}_{\textbf{5}}}\bar{\textbf{5}}_{\textbf{e}_{\textbf{3}}, \textbf{e}_{\textbf{5}}}$ & n & 1-4\\
\cline{2-5}
& 18 & $\textbf{1}_{\textbf{e}_{\textbf{4}},-\textbf{e}_{\textbf{3}}}\textbf{5}_{-\textbf{e}_{\textbf{4}},-\textbf{e}_{\textbf{5}}}\bar{\textbf{5}}_{\textbf{e}_{\textbf{3}}, \textbf{e}_{\textbf{5}}}$ & n & 1-4\\
\cline{2-5}
\hline
\end{tabular}
\end{center}
\begin{center}
\begin{tabular}{|c|c|c|c|c|}
\hline
CICY No. & Model No. & Yukawa Pattern & Top. Van. & Sym. No.\\
\hline
\multirow{36}{*}{6784} & 19 & $\textbf{1}_{\textbf{e}_{\textbf{4}},-\textbf{e}_{\textbf{3}}}\textbf{5}_{-\textbf{e}_{\textbf{4}},-\textbf{e}_{\textbf{5}}}\bar{\textbf{5}}_{\textbf{e}_{\textbf{3}}, \textbf{e}_{\textbf{5}}}$ & n & 1-4\\
\cline{2-5}
& 20 & $\textbf{1}_{\textbf{e}_{\textbf{4}},-\textbf{e}_{\textbf{3}}}\textbf{5}_{-\textbf{e}_{\textbf{4}},-\textbf{e}_{\textbf{5}}}\bar{\textbf{5}}_{\textbf{e}_{\textbf{3}}, \textbf{e}_{\textbf{5}}}$ & n & 2, 4\\
\cline{2-5}
& 21 & $\textbf{1}_{\textbf{e}_{\textbf{4}},-\textbf{e}_{\textbf{3}}}\textbf{5}_{-\textbf{e}_{\textbf{4}},-\textbf{e}_{\textbf{5}}}\bar{\textbf{5}}_{\textbf{e}_{\textbf{3}}, \textbf{e}_{\textbf{5}}} $ & n & 1, 3\\
\cline{2-5}
& 22 & $\textbf{1}_{\textbf{e}_{\textbf{4}},-\textbf{e}_{\textbf{3}}}\textbf{5}_{-\textbf{e}_{\textbf{4}},-\textbf{e}_{\textbf{5}}}\bar{\textbf{5}}_{\textbf{e}_{\textbf{3}}, \textbf{e}_{\textbf{5}}} $ & n & 1-4\\
\cline{2-5}
& 23 & $\textbf{1}_{\textbf{e}_{\textbf{4}},-\textbf{e}_{\textbf{3}}}\textbf{5}_{-\textbf{e}_{\textbf{4}},-\textbf{e}_{\textbf{5}}}\bar{\textbf{5}}_{\textbf{e}_{\textbf{3}}, \textbf{e}_{\textbf{5}}} $ & n & 1-4\\
\cline{2-5}
& 24 & $\textbf{1}_{\textbf{e}_{\textbf{4}},-\textbf{e}_{\textbf{3}}}\textbf{5}_{-\textbf{e}_{\textbf{4}},-\textbf{e}_{\textbf{5}}}\bar{\textbf{5}}_{\textbf{e}_{\textbf{3}}, \textbf{e}_{\textbf{5}}} $ & n & 1-4\\
\cline{2-5}
& 25 & $\textbf{1}_{\textbf{e}_{\textbf{4}},-\textbf{e}_{\textbf{3}}}\textbf{5}_{-\textbf{e}_{\textbf{4}},-\textbf{e}_{\textbf{5}}}\bar{\textbf{5}}_{\textbf{e}_{\textbf{3}}, \textbf{e}_{\textbf{5}}}$ & n & 1-4\\
\cline{2-5}
& 26 & $\textbf{1}_{\textbf{e}_{\textbf{4}},-\textbf{e}_{\textbf{3}}}\textbf{5}_{-\textbf{e}_{\textbf{4}},-\textbf{e}_{\textbf{5}}}\bar{\textbf{5}}_{\textbf{e}_{\textbf{3}}, \textbf{e}_{\textbf{5}}} $ & n & 1-4\\
\cline{2-5}
& 27 & $\textbf{1}_{\textbf{e}_{\textbf{4}},-\textbf{e}_{\textbf{3}}}\textbf{5}_{-\textbf{e}_{\textbf{4}},-\textbf{e}_{\textbf{5}}}\bar{\textbf{5}}_{\textbf{e}_{\textbf{3}}, \textbf{e}_{\textbf{5}}} $ & n & 1-4\\
\cline{2-5}
& 28 & $\textbf{1}_{\textbf{e}_{\textbf{4}},-\textbf{e}_{\textbf{3}}}\textbf{5}_{-\textbf{e}_{\textbf{4}},-\textbf{e}_{\textbf{5}}}\bar{\textbf{5}}_{\textbf{e}_{\textbf{3}}, \textbf{e}_{\textbf{5}}} $ & n & 1-4\\
\cline{2-5}
& 29 & $\textbf{1}_{\textbf{e}_{\textbf{4}},-\textbf{e}_{\textbf{3}}}\textbf{5}_{-\textbf{e}_{\textbf{4}},-\textbf{e}_{\textbf{5}}}\bar{\textbf{5}}_{\textbf{e}_{\textbf{3}}, \textbf{e}_{\textbf{5}}}$ & n & 1-4\\
\cline{2-5}
& 30 & $\textbf{1}_{\textbf{e}_{\textbf{4}},-\textbf{e}_{\textbf{3}}}\textbf{5}_{-\textbf{e}_{\textbf{4}},-\textbf{e}_{\textbf{5}}}\bar{\textbf{5}}_{\textbf{e}_{\textbf{3}}, \textbf{e}_{\textbf{5}}} $ & n & 2, 4\\
\cline{2-5}
& 31 & $\textbf{1}_{\textbf{e}_{\textbf{4}},-\textbf{e}_{\textbf{3}}}\textbf{5}_{-\textbf{e}_{\textbf{4}},-\textbf{e}_{\textbf{5}}}\bar{\textbf{5}}_{\textbf{e}_{\textbf{3}}, \textbf{e}_{\textbf{5}}} $ & n & 2, 4\\
\cline{2-5}
& 32 & $\textbf{1}_{\textbf{e}_{\textbf{4}},-\textbf{e}_{\textbf{3}}}\textbf{5}_{-\textbf{e}_{\textbf{4}},-\textbf{e}_{\textbf{5}}}\bar{\textbf{5}}_{\textbf{e}_{\textbf{3}}, \textbf{e}_{\textbf{5}}} $ & n & 1-4\\
\cline{2-5}
& 33 & $\textbf{1}_{\textbf{e}_{\textbf{4}},-\textbf{e}_{\textbf{3}}}\textbf{5}_{-\textbf{e}_{\textbf{4}},-\textbf{e}_{\textbf{5}}}\bar{\textbf{5}}_{\textbf{e}_{\textbf{3}}, \textbf{e}_{\textbf{5}}}$ & n & 1-4\\
\cline{2-5}
& 34 & $\textbf{1}_{\textbf{e}_{\textbf{4}},-\textbf{e}_{\textbf{3}}}\textbf{5}_{-\textbf{e}_{\textbf{4}},-\textbf{e}_{\textbf{5}}}\bar{\textbf{5}}_{\textbf{e}_{\textbf{3}}, \textbf{e}_{\textbf{5}}} $ & n & 1-4\\
\cline{2-5}
& 35 & $\textbf{1}_{\textbf{e}_{\textbf{4}},-\textbf{e}_{\textbf{3}}}\textbf{5}_{-\textbf{e}_{\textbf{4}},-\textbf{e}_{\textbf{5}}}\bar{\textbf{5}}_{\textbf{e}_{\textbf{3}}, \textbf{e}_{\textbf{5}}}$ & n & 1-4\\
\cline{2-5}
& 36 & $\textbf{1}_{\textbf{e}_{\textbf{4}},-\textbf{e}_{\textbf{3}}}\textbf{5}_{-\textbf{e}_{\textbf{4}},-\textbf{e}_{\textbf{5}}}\bar{\textbf{5}}_{\textbf{e}_{\textbf{3}}, \textbf{e}_{\textbf{5}}} $ & n & 1-4\\
\cline{2-5}
& 37 & $\textbf{1}_{\textbf{e}_{\textbf{4}},-\textbf{e}_{\textbf{3}}}\textbf{5}_{-\textbf{e}_{\textbf{4}},-\textbf{e}_{\textbf{5}}}\bar{\textbf{5}}_{\textbf{e}_{\textbf{3}}, \textbf{e}_{\textbf{5}}}$ & n & 1-4\\
\cline{2-5}
& 38 & $\textbf{1}_{\textbf{e}_{\textbf{4}},-\textbf{e}_{\textbf{3}}}\textbf{5}_{-\textbf{e}_{\textbf{4}},-\textbf{e}_{\textbf{5}}}\bar{\textbf{5}}_{\textbf{e}_{\textbf{3}}, \textbf{e}_{\textbf{5}}}$ & n & 1-4\\
\cline{2-5}
 & 39 & $\textbf{1}_{\textbf{e}_{\textbf{4}},-\textbf{e}_{\textbf{3}}}\textbf{5}_{-\textbf{e}_{\textbf{4}},-\textbf{e}_{\textbf{5}}}\bar{\textbf{5}}_{\textbf{e}_{\textbf{3}}, \textbf{e}_{\textbf{5}}}$ & n & 1-4\\
\cline{2-5}
& 40 & $\textbf{1}_{\textbf{e}_{\textbf{4}},-\textbf{e}_{\textbf{3}}}\textbf{5}_{-\textbf{e}_{\textbf{4}},-\textbf{e}_{\textbf{5}}}\bar{\textbf{5}}_{\textbf{e}_{\textbf{3}}, \textbf{e}_{\textbf{5}}} $ & n & 1,3\\
\cline{2-5}
& 41 & $\textbf{1}_{\textbf{e}_{\textbf{4}},-\textbf{e}_{\textbf{3}}}\textbf{5}_{-\textbf{e}_{\textbf{4}},-\textbf{e}_{\textbf{5}}}\bar{\textbf{5}}_{\textbf{e}_{\textbf{3}}, \textbf{e}_{\textbf{5}}}$ & n & 2,4\\
\cline{2-5}
& 42 & $\textbf{1}_{\textbf{e}_{\textbf{4}},-\textbf{e}_{\textbf{3}}}\textbf{5}_{-\textbf{e}_{\textbf{4}},-\textbf{e}_{\textbf{5}}}\bar{\textbf{5}}_{\textbf{e}_{\textbf{3}}, \textbf{e}_{\textbf{5}}} $ & n & 1-4\\
\cline{2-5}
& 43 & $\textbf{1}_{\textbf{e}_{\textbf{4}},-\textbf{e}_{\textbf{3}}}\textbf{5}_{-\textbf{e}_{\textbf{4}},-\textbf{e}_{\textbf{5}}}\bar{\textbf{5}}_{\textbf{e}_{\textbf{3}}, \textbf{e}_{\textbf{5}}} $ & n & 1-4\\
\cline{2-5}
& 44 & $\textbf{1}_{\textbf{e}_{\textbf{4}},-\textbf{e}_{\textbf{3}}}\textbf{5}_{-\textbf{e}_{\textbf{4}},-\textbf{e}_{\textbf{5}}}\bar{\textbf{5}}_{\textbf{e}_{\textbf{3}}, \textbf{e}_{\textbf{5}}}$ & n & 1-4\\
\cline{2-5}
& 45 & $\textbf{1}_{\textbf{e}_{\textbf{4}},-\textbf{e}_{\textbf{3}}}\textbf{5}_{-\textbf{e}_{\textbf{4}},-\textbf{e}_{\textbf{5}}}\bar{\textbf{5}}_{\textbf{e}_{\textbf{3}}, \textbf{e}_{\textbf{5}}}$ & n & 1-4\\
\cline{2-5}
& 46 & $\textbf{1}_{\textbf{e}_{\textbf{4}},-\textbf{e}_{\textbf{3}}}\textbf{5}_{-\textbf{e}_{\textbf{4}},-\textbf{e}_{\textbf{5}}}\bar{\textbf{5}}_{\textbf{e}_{\textbf{3}}, \textbf{e}_{\textbf{5}}}$ & n & 1-4\\
\cline{2-5}
& 47 & $\textbf{1}_{\textbf{e}_{\textbf{4}},-\textbf{e}_{\textbf{3}}}\textbf{5}_{-\textbf{e}_{\textbf{4}},-\textbf{e}_{\textbf{5}}}\bar{\textbf{5}}_{\textbf{e}_{\textbf{3}}, \textbf{e}_{\textbf{5}}}$ & n & 1-4\\
\cline{2-5}
& 48 & $\textbf{1}_{\textbf{e}_{\textbf{4}},-\textbf{e}_{\textbf{3}}}\textbf{5}_{-\textbf{e}_{\textbf{4}},-\textbf{e}_{\textbf{5}}}\bar{\textbf{5}}_{\textbf{e}_{\textbf{3}}, \textbf{e}_{\textbf{5}}}$ & n & 1-4\\
\cline{2-5}
& 49 & $\textbf{1}_{\textbf{e}_{\textbf{4}},-\textbf{e}_{\textbf{3}}}\textbf{5}_{-\textbf{e}_{\textbf{4}},-\textbf{e}_{\textbf{5}}}\bar{\textbf{5}}_{\textbf{e}_{\textbf{3}}, \textbf{e}_{\textbf{5}}}$ & n & 1-4\\
\cline{2-5}
& 50 & $\textbf{1}_{\textbf{e}_{\textbf{4}},-\textbf{e}_{\textbf{3}}}\textbf{5}_{-\textbf{e}_{\textbf{4}},-\textbf{e}_{\textbf{5}}}\bar{\textbf{5}}_{\textbf{e}_{\textbf{3}}, \textbf{e}_{\textbf{5}}}$ & n & 1,3\\
\cline{2-5}
& 51 & $\textbf{10}_{\textbf{e}_{\textbf{3}}}\bar{\textbf{5}}_{\textbf{e}_{\textbf{1}},\textbf{e}_{\textbf{2}}}\bar{\textbf{5}}_{\textbf{e}_{\textbf{4}},\textbf{e}_{\textbf{5}}} $ & n & 1-4\\
\cline{2-5}
& \multirow{2}{*}{52} & $\textbf{10}_{\textbf{e}_{\textbf{1}}}\bar{\textbf{5}}_{\textbf{e}_{\textbf{2}},\textbf{e}_{\textbf{5}}}\bar{\textbf{5}}_{\textbf{e}_{\textbf{3}},\textbf{e}_{\textbf{4}}}  $ & n & 1-4\\
\cline{3-5}
& & $\textbf{10}_{\textbf{e}_{\textbf{5}}}\bar{\textbf{5}}_{\textbf{e}_{\textbf{1}},\textbf{e}_{\textbf{2}}}\bar{\textbf{5}}_{\textbf{e}_{\textbf{3}},\textbf{e}_{\textbf{4}}}  $ & n & 1-4\\
\cline{2-5}
\cline{3-5}
\hline
\end{tabular}
\end{center}

\noindent On CICY 6784, a total of 188 models have Yukawa couplings consistent with the gauge symmetries of the models, with 264 Yukawa couplings being permitted in total. There are no allowed couplings of the form $ \bold{5}_{\bold{{-{\bf e}_i}},\bold{-{{\bf e}_j}}}\bold{10}_{\bold{{{\bf e}_i}}}\bold{10}_{\bold{{{\bf e}_j}}}$, 120 of the form $\textbf{10}_{\textbf{e}_{\textbf{i}}}\bar{\textbf{5}}_{\textbf{e}_{\textbf{j}},\textbf{e}_{\textbf{k}}}\bar{\textbf{5}}_{\textbf{e}_{\textbf{l}},\textbf{e}_{\textbf{m}}} $, and 144 of the form $\textbf{1}_{\textbf{e}_{\textbf{i}},-\textbf{e}_{\textbf{j}}}\textbf{5}_{-\textbf{e}_{\textbf{i}},-\textbf{e}_{\textbf{k}}}\bar{\textbf{5}}_{\textbf{e}_{\textbf{j}}, \textbf{e}_{\textbf{k}}}$. None of these couplings exhibit the topological vanishings we have studied here. 

\begin{center}
\begin{tabular}{|c|c|c|c|c|}
\hline
CICY No. & Model No. & Yukawa Pattern & Top. Van. & Sym. No.\\
\hline
\multirow{3}{*}{6828} & 1 & $\textbf{10}_{\textbf{e}_{\textbf{5}}}\bar{\textbf{5}}_{\textbf{e}_{\textbf{1}},\textbf{e}_{\textbf{2}}}\bar{\textbf{5}}_{\textbf{e}_{\textbf{3}},\textbf{e}_{\textbf{4}}} $& n & 2\\
\cline{2-5}
 & \multirow{2}{*}7 & $\textbf{10}_{\textbf{e}_{\textbf{1}}}\bar{\textbf{5}}_{\textbf{e}_{\textbf{2}},\textbf{e}_{\textbf{3}}}\bar{\textbf{5}}_{\textbf{e}_{\textbf{4}},\textbf{e}_{\textbf{5}}}  $ & n & 2\\
\cline{3-5}
&  & $\textbf{10}_{\textbf{e}_{\textbf{3}}}\bar{\textbf{5}}_{\textbf{e}_{\textbf{1}},\textbf{e}_{\textbf{2}}}\bar{\textbf{5}}_{\textbf{e}_{\textbf{4}},\textbf{e}_{\textbf{5}}} $ & n & 2\\
\cline{2-5}
\hline
\end{tabular}
\end{center}
On CICY 6828, a total of 2 models have Yukawa couplings consistent with the gauge symmetries of the models, with 5 Yukawa couplings being permitted in total. All of these couplings are of the form $\textbf{10}_{\textbf{e}_{\textbf{i}}}\bar{\textbf{5}}_{\textbf{e}_{\textbf{j}},\textbf{e}_{\textbf{k}}}\bar{\textbf{5}}_{\textbf{e}_{\textbf{l}},\textbf{e}_{\textbf{m}}} $. None of these couplings exhibit the topological vanishings we have studied here.


\begin{center}
\begin{tabular}{|c|c|c|c|c|}
\hline
CICY No. & Model No. & Yukawa Pattern & Top. Van. & Sym. No.\\
\hline
\multirow{14}{*}{7862} & 2 & $\textbf{5}_{-\textbf{e}_{\textbf{2}},-\textbf{e}_{\textbf{4}}}\textbf{10}_{\textbf{e}_{\textbf{2}}}\textbf{10}_{\textbf{e}_{\textbf{4}}} $& n & 3\\
\cline{2-5}
& \multirow{4}{*}9   & $\textbf{1}_{\textbf{e}_{\textbf{1}},-\textbf{e}_{\textbf{4}}}\textbf{5}_{-\textbf{e}_{\textbf{1}},-\textbf{e}_{\textbf{5}}}\bar{\textbf{5}}_{\textbf{e}_{\textbf{4}}, \textbf{e}_{\textbf{5}}}$& n & 3\\
\cline{3-5}
&  & $\textbf{1}_{\textbf{e}_{\textbf{2}},-\textbf{e}_{\textbf{4}}}\textbf{5}_{-\textbf{e}_{\textbf{2}},-\textbf{e}_{\textbf{5}}}\bar{\textbf{5}}_{\textbf{e}_{\textbf{4}}, \textbf{e}_{\textbf{5}}} $ & n & 3\\
\cline{3-5}
&  & $\textbf{10}_{\textbf{e}_{\textbf{1}}}\bar{\textbf{5}}_{\textbf{e}_{\textbf{2}},\textbf{e}_{\textbf{5}}}\bar{\textbf{5}}_{\textbf{e}_{\textbf{3}},\textbf{e}_{\textbf{4}}} $ & n & 3\\
\cline{3-5}
&  & $\textbf{10}_{\textbf{e}_{\textbf{2}}}\bar{\textbf{5}}_{\textbf{e}_{\textbf{1}},\textbf{e}_{\textbf{5}}}\bar{\textbf{5}}_{\textbf{e}_{\textbf{3}},\textbf{e}_{\textbf{4}}} $   & n & 3\\
\cline{2-5}
& \multirow{4}{*}{10}   & $\textbf{1}_{\textbf{e}_{\textbf{1}},-\textbf{e}_{\textbf{4}}}\textbf{5}_{-\textbf{e}_{\textbf{1}},-\textbf{e}_{\textbf{5}}}\bar{\textbf{5}}_{\textbf{e}_{\textbf{4}}, \textbf{e}_{\textbf{5}}}$& n & 3\\
\cline{3-5}
&  & $\textbf{1}_{\textbf{e}_{\textbf{2}},-\textbf{e}_{\textbf{4}}}\textbf{5}_{-\textbf{e}_{\textbf{2}},-\textbf{e}_{\textbf{5}}}\bar{\textbf{5}}_{\textbf{e}_{\textbf{4}}, \textbf{e}_{\textbf{5}}} $ & n & 3\\
\cline{3-5}
&  & $\textbf{10}_{\textbf{e}_{\textbf{1}}}\bar{\textbf{5}}_{\textbf{e}_{\textbf{2}},\textbf{e}_{\textbf{5}}}\bar{\textbf{5}}_{\textbf{e}_{\textbf{3}},\textbf{e}_{\textbf{4}}} $ & n & 3\\
\cline{3-5}
&  & $\textbf{10}_{\textbf{e}_{\textbf{2}}}\bar{\textbf{5}}_{\textbf{e}_{\textbf{1}},\textbf{e}_{\textbf{5}}}\bar{\textbf{5}}_{\textbf{e}_{\textbf{3}},\textbf{e}_{\textbf{4}}} $   & n & 3\\
\cline{2-5}
& \multirow{4}{*}{11}   & $\textbf{1}_{\textbf{e}_{\textbf{1}},-\textbf{e}_{\textbf{4}}}\textbf{5}_{-\textbf{e}_{\textbf{1}},-\textbf{e}_{\textbf{5}}}\bar{\textbf{5}}_{\textbf{e}_{\textbf{4}}, \textbf{e}_{\textbf{5}}}$& n & 3\\
\cline{3-5}
&  & $\textbf{1}_{\textbf{e}_{\textbf{2}},-\textbf{e}_{\textbf{4}}}\textbf{5}_{-\textbf{e}_{\textbf{2}},-\textbf{e}_{\textbf{5}}}\bar{\textbf{5}}_{\textbf{e}_{\textbf{4}}, \textbf{e}_{\textbf{5}}} $ & n & 3\\
\cline{3-5}
&  & $\textbf{10}_{\textbf{e}_{\textbf{1}}}\bar{\textbf{5}}_{\textbf{e}_{\textbf{2}},\textbf{e}_{\textbf{5}}}\bar{\textbf{5}}_{\textbf{e}_{\textbf{3}},\textbf{e}_{\textbf{4}}} $ & n & 3\\
\cline{3-5}
&  & $\textbf{10}_{\textbf{e}_{\textbf{2}}}\bar{\textbf{5}}_{\textbf{e}_{\textbf{1}},\textbf{e}_{\textbf{5}}}\bar{\textbf{5}}_{\textbf{e}_{\textbf{3}},\textbf{e}_{\textbf{4}}} $   & n & 3\\
\cline{2-5}
& \multirow{4}{*}{12}   & $\textbf{1}_{\textbf{e}_{\textbf{1}},-\textbf{e}_{\textbf{4}}}\textbf{5}_{-\textbf{e}_{\textbf{1}},-\textbf{e}_{\textbf{5}}}\bar{\textbf{5}}_{\textbf{e}_{\textbf{4}}, \textbf{e}_{\textbf{5}}}$& n & 3\\
\cline{3-5}
&  & $\textbf{1}_{\textbf{e}_{\textbf{2}},-\textbf{e}_{\textbf{4}}}\textbf{5}_{-\textbf{e}_{\textbf{2}},-\textbf{e}_{\textbf{5}}}\bar{\textbf{5}}_{\textbf{e}_{\textbf{4}}, \textbf{e}_{\textbf{5}}} $ & n & 3\\
\cline{3-5}
&  & $\textbf{10}_{\textbf{e}_{\textbf{1}}}\bar{\textbf{5}}_{\textbf{e}_{\textbf{2}},\textbf{e}_{\textbf{5}}}\bar{\textbf{5}}_{\textbf{e}_{\textbf{3}},\textbf{e}_{\textbf{4}}} $ & n & 3\\
\cline{3-5}
&  & $\textbf{10}_{\textbf{e}_{\textbf{2}}}\bar{\textbf{5}}_{\textbf{e}_{\textbf{1}},\textbf{e}_{\textbf{5}}}\bar{\textbf{5}}_{\textbf{e}_{\textbf{3}},\textbf{e}_{\textbf{4}}} $   & n & 3\\
\cline{2-5}
& \multirow{2}{*}{13} & $\textbf{1}_{\textbf{e}_{\textbf{1}},-\textbf{e}_{\textbf{2}}}\textbf{5}_{-\textbf{e}_{\textbf{1}},-\textbf{e}_{\textbf{5}}}\bar{\textbf{5}}_{\textbf{e}_{\textbf{2}}, \textbf{e}_{\textbf{5}}} $ & y & 3\\
\cline{3-5}
& & $\textbf{10}_{\textbf{e}_{\textbf{3}}}\bar{\textbf{5}}_{\textbf{e}_{\textbf{1}},\textbf{e}_{\textbf{4}}}\bar{\textbf{5}}_{\textbf{e}_{\textbf{2}},\textbf{e}_{\textbf{5}}}$ & y & 3\\
\cline{2-5}
& \multirow{2}{*}{15} & $\textbf{10}_{\textbf{e}_{\textbf{2}}}\bar{\textbf{5}}_{\textbf{e}_{\textbf{1}},\textbf{e}_{\textbf{5}}}\bar{\textbf{5}}_{\textbf{e}_{\textbf{3}},\textbf{e}_{\textbf{4}}} $ & n & 3\\
\cline{3-5}
& & $\textbf{5}_{-\textbf{e}_{\textbf{1}},-\textbf{e}_{\textbf{2}}}\textbf{10}_{\textbf{e}_{\textbf{1}}}\textbf{10}_{\textbf{e}_{\textbf{2}}} $ & y & 3\\
\cline{2-5}
& \multirow{2}{*}{16} & $\textbf{10}_{\textbf{e}_{\textbf{2}}}\bar{\textbf{5}}_{\textbf{e}_{\textbf{1}},\textbf{e}_{\textbf{5}}}\bar{\textbf{5}}_{\textbf{e}_{\textbf{3}},\textbf{e}_{\textbf{4}}} $ & n & 3\\
\cline{3-5}
& & $\textbf{5}_{-\textbf{e}_{\textbf{1}},-\textbf{e}_{\textbf{2}}}\textbf{10}_{\textbf{e}_{\textbf{1}}}\textbf{10}_{\textbf{e}_{\textbf{2}}} $ & y & 3\\
\cline{2-5}
& \multirow{2}{*}{17} & $\textbf{10}_{\textbf{e}_{\textbf{2}}}\bar{\textbf{5}}_{\textbf{e}_{\textbf{1}},\textbf{e}_{\textbf{5}}}\bar{\textbf{5}}_{\textbf{e}_{\textbf{3}},\textbf{e}_{\textbf{4}}} $ & n & 3\\
\cline{3-5}
& & $\textbf{5}_{-\textbf{e}_{\textbf{1}},-\textbf{e}_{\textbf{2}}}\textbf{10}_{\textbf{e}_{\textbf{1}}}\textbf{10}_{\textbf{e}_{\textbf{2}}} $ & y & 3\\
\cline{2-5}
& \multirow{2}{*}{18} & $\textbf{10}_{\textbf{e}_{\textbf{2}}}\bar{\textbf{5}}_{\textbf{e}_{\textbf{1}},\textbf{e}_{\textbf{5}}}\bar{\textbf{5}}_{\textbf{e}_{\textbf{3}},\textbf{e}_{\textbf{4}}} $ & n & 3\\
\cline{3-5}
& & $\textbf{5}_{-\textbf{e}_{\textbf{1}},-\textbf{e}_{\textbf{2}}}\textbf{10}_{\textbf{e}_{\textbf{1}}}\textbf{10}_{\textbf{e}_{\textbf{2}}} $ & y & 3\\
\cline{2-5}
& \multirow{3}{*}{19} & $\textbf{10}_{\textbf{e}_{\textbf{3}}}\bar{\textbf{5}}_{\textbf{e}_{\textbf{1}},\textbf{e}_{\textbf{4}}}\bar{\textbf{5}}_{\textbf{e}_{\textbf{2}},\textbf{e}_{\textbf{5}}}  $ & y & 3\\
\cline{3-5}
&   & $\textbf{10}_{\textbf{e}_{\textbf{5}}}\bar{\textbf{5}}_{\textbf{e}_{\textbf{1}},\textbf{e}_{\textbf{3}}}\bar{\textbf{5}}_{\textbf{e}_{\textbf{2}},\textbf{e}_{\textbf{4}}} $ & n & 3\\
\cline{3-5}
&   & $\textbf{5}_{-\textbf{e}_{\textbf{3}},-\textbf{e}_{\textbf{5}}}\textbf{10}_{\textbf{e}_{\textbf{3}}}\textbf{10}_{\textbf{e}_{\textbf{5}}} $ & n & 3\\
\cline{3-5}
\cline{2-5}
& \multirow{3}{*}{20} & $\textbf{10}_{\textbf{e}_{\textbf{3}}}\bar{\textbf{5}}_{\textbf{e}_{\textbf{1}},\textbf{e}_{\textbf{4}}}\bar{\textbf{5}}_{\textbf{e}_{\textbf{2}},\textbf{e}_{\textbf{5}}}  $ & y & 3\\
\cline{3-5}
&   & $\textbf{10}_{\textbf{e}_{\textbf{5}}}\bar{\textbf{5}}_{\textbf{e}_{\textbf{1}},\textbf{e}_{\textbf{3}}}\bar{\textbf{5}}_{\textbf{e}_{\textbf{2}},\textbf{e}_{\textbf{4}}} $ & n & 3\\
\cline{3-5}
&   & $\textbf{5}_{-\textbf{e}_{\textbf{3}},-\textbf{e}_{\textbf{5}}}\textbf{10}_{\textbf{e}_{\textbf{3}}}\textbf{10}_{\textbf{e}_{\textbf{5}}} $ & n & 3\\
\cline{3-5}
\cline{2-5}
& \multirow{3}{*}{21} & $\textbf{10}_{\textbf{e}_{\textbf{3}}}\bar{\textbf{5}}_{\textbf{e}_{\textbf{1}},\textbf{e}_{\textbf{4}}}\bar{\textbf{5}}_{\textbf{e}_{\textbf{2}},\textbf{e}_{\textbf{5}}}  $ & y & 3\\
\cline{3-5}
&   & $\textbf{10}_{\textbf{e}_{\textbf{5}}}\bar{\textbf{5}}_{\textbf{e}_{\textbf{1}},\textbf{e}_{\textbf{3}}}\bar{\textbf{5}}_{\textbf{e}_{\textbf{2}},\textbf{e}_{\textbf{4}}} $ & n & 3\\
\cline{3-5}
&   & $\textbf{5}_{-\textbf{e}_{\textbf{3}},-\textbf{e}_{\textbf{5}}}\textbf{10}_{\textbf{e}_{\textbf{3}}}\textbf{10}_{\textbf{e}_{\textbf{5}}} $ & n & 3\\
\cline{3-5}
\cline{2-5}
& \multirow{3}{*}{22} & $\textbf{10}_{\textbf{e}_{\textbf{3}}}\bar{\textbf{5}}_{\textbf{e}_{\textbf{1}},\textbf{e}_{\textbf{4}}}\bar{\textbf{5}}_{\textbf{e}_{\textbf{2}},\textbf{e}_{\textbf{5}}}  $ & y & 3\\
\cline{3-5}
&   & $\textbf{10}_{\textbf{e}_{\textbf{5}}}\bar{\textbf{5}}_{\textbf{e}_{\textbf{1}},\textbf{e}_{\textbf{3}}}\bar{\textbf{5}}_{\textbf{e}_{\textbf{2}},\textbf{e}_{\textbf{4}}} $ & n & 3\\
\cline{3-5}
&   & $\textbf{5}_{-\textbf{e}_{\textbf{3}},-\textbf{e}_{\textbf{5}}}\textbf{10}_{\textbf{e}_{\textbf{3}}}\textbf{10}_{\textbf{e}_{\textbf{5}}} $ & n & 3\\
\cline{3-5}
\hline
\end{tabular}
\end{center}
On CICY 7862, a total of 14 models have Yukawa couplings consistent with the gauge symmetries of the models, with 90 Yukawa couplings being permitted in total. Of these, 53 are of the form $\textbf{10}_{\textbf{e}_{\textbf{i}}}\bar{\textbf{5}}_{\textbf{e}_{\textbf{j}},\textbf{e}_{\textbf{k}}}\bar{\textbf{5}}_{\textbf{e}_{\textbf{l}},\textbf{e}_{\textbf{m}}} $, 19 are of the form $\textbf{1}_{\textbf{e}_{\textbf{i}},-\textbf{e}_{\textbf{j}}}\textbf{5}_{-\textbf{e}_{\textbf{i}},-\textbf{e}_{\textbf{k}}}\bar{\textbf{5}}_{\textbf{e}_{\textbf{j}}, \textbf{e}_{\textbf{k}}}$ and 18 are of the form $ \bold{5}_{\bold{{-{\bf e}_i}},\bold{-{{\bf e}_j}}}\bold{10}_{\bold{{{\bf e}_i}}}\bold{10}_{\bold{{{\bf e}_j}}}$. A total of 16 couplings exhibit the topological vanishing we have been studying in this paper, 3 of the form $\textbf{1}_{\textbf{e}_{\textbf{i}},-\textbf{e}_{\textbf{j}}}\textbf{5}_{-\textbf{e}_{\textbf{i}},-\textbf{e}_{\textbf{k}}}\bar{\textbf{5}}_{\textbf{e}_{\textbf{j}}, \textbf{e}_{\textbf{k}}}$, 5 of the form  $\textbf{10}_{\textbf{e}_{\textbf{i}}}\bar{\textbf{5}}_{\textbf{e}_{\textbf{j}},\textbf{e}_{\textbf{k}}}\bar{\textbf{5}}_{\textbf{e}_{\textbf{l}},\textbf{e}_{\textbf{m}}} $, and 8 of the form $ \bold{5}_{\bold{{-{\bf e}_i}},\bold{-{{\bf e}_j}}}\bold{10}_{\bold{{{\bf e}_i}}}\bold{10}_{\bold{{{\bf e}_j}}}$. A total of 9 out of the 14 models have at least one topologically vanishing coupling.

\begin{center}
\begin{tabular}{|c|c|c|c|c|}
\hline
CICY No.  & Model No. & Yukawa Pattern & Top. Van. & Sym. No.\\
\hline
\multirow{2}{*}{} & \multirow{2}{*}2 & $\textbf{1}_{\textbf{e}_{\textbf{3}},-\textbf{e}_{\textbf{4}}}\textbf{5}_{-\textbf{e}_{\textbf{3}},-\textbf{e}_{\textbf{5}}}\bar{\textbf{5}}_{\textbf{e}_{\textbf{4}}, \textbf{e}_{\textbf{5}}}$ & n & 1-2\\
\cline{3-5}
& & $\textbf{10}_{\textbf{e}_{\textbf{2}}}\bar{\textbf{5}}_{\textbf{e}_{\textbf{1}},\textbf{e}_{\textbf{4}}}\bar{\textbf{5}}_{\textbf{e}_{\textbf{3}},\textbf{e}_{\textbf{5}}} $ & n & 1-2\\
\cline{2-5}
& 3 & $\textbf{10}_{\textbf{e}_{\textbf{5}}}\bar{\textbf{5}}_{\textbf{e}_{\textbf{1}},\textbf{e}_{\textbf{3}}}\bar{\textbf{5}}_{\textbf{e}_{\textbf{2}},\textbf{e}_{\textbf{4}}} $ & n & 1-2\\
\cline{2-5}

\multirow{20}{*}{5256} & \multirow{2}{*} 5 & $\textbf{1}_{\textbf{e}_{\textbf{1}},-\textbf{e}_{\textbf{2}}}\textbf{5}_{-\textbf{e}_{\textbf{1}},-\textbf{e}_{\textbf{5}}}\bar{\textbf{5}}_{\textbf{e}_{\textbf{2}}, \textbf{e}_{\textbf{5}}}$& y & 3-6\\
\cline{3-5}
&  & $\textbf{10}_{\textbf{e}_{\textbf{3}}}\bar{\textbf{5}}_{\textbf{e}_{\textbf{1}},\textbf{e}_{\textbf{4}}}\bar{\textbf{5}}_{\textbf{e}_{\textbf{2}},\textbf{e}_{\textbf{5}}} $ & y & 3-6\\
\cline{2-5}
&\multirow{2}{*} 6 & $\textbf{1}_{\textbf{e}_{\textbf{1}},-\textbf{e}_{\textbf{2}}}\textbf{5}_{-\textbf{e}_{\textbf{1}},-\textbf{e}_{\textbf{5}}}\bar{\textbf{5}}_{\textbf{e}_{\textbf{2}}, \textbf{e}_{\textbf{5}}}$& y & 3-6\\
\cline{3-5}
&  & $\textbf{10}_{\textbf{e}_{\textbf{3}}}\bar{\textbf{5}}_{\textbf{e}_{\textbf{1}},\textbf{e}_{\textbf{4}}}\bar{\textbf{5}}_{\textbf{e}_{\textbf{2}},\textbf{e}_{\textbf{5}}} $ & y & 3-6\\
\cline{2-5}
& 7 & $\textbf{10}_{\textbf{e}_{\textbf{3}}}\bar{\textbf{5}}_{\textbf{e}_{\textbf{1}},\textbf{e}_{\textbf{2}}}\bar{\textbf{5}}_{\textbf{e}_{\textbf{4}},\textbf{e}_{\textbf{5}}} $ & n & 3-6\\
\cline{2-5}
& 8 & $\textbf{10}_{\textbf{e}_{\textbf{3}}}\bar{\textbf{5}}_{\textbf{e}_{\textbf{1}},\textbf{e}_{\textbf{2}}}\bar{\textbf{5}}_{\textbf{e}_{\textbf{4}},\textbf{e}_{\textbf{5}}}  $ & n & 3-6\\
\cline{2-5}
& 9 & $\textbf{10}_{\textbf{e}_{\textbf{3}}}\bar{\textbf{5}}_{\textbf{e}_{\textbf{1}},\textbf{e}_{\textbf{2}}}\bar{\textbf{5}}_{\textbf{e}_{\textbf{4}},\textbf{e}_{\textbf{5}}} $ & n & 3-6\\
\cline{2-5}
& 10 & $\textbf{10}_{\textbf{e}_{\textbf{3}}}\bar{\textbf{5}}_{\textbf{e}_{\textbf{1}},\textbf{e}_{\textbf{2}}}\bar{\textbf{5}}_{\textbf{e}_{\textbf{4}},\textbf{e}_{\textbf{5}}}  $ & n & 3-6\\
\cline{2-5}
& 11 & $\textbf{10}_{\textbf{e}_{\textbf{3}}}\bar{\textbf{5}}_{\textbf{e}_{\textbf{1}},\textbf{e}_{\textbf{2}}}\bar{\textbf{5}}_{\textbf{e}_{\textbf{4}},\textbf{e}_{\textbf{5}}}  $ & n & 3-6\\
\cline{2-5}
& 12 & $\textbf{10}_{\textbf{e}_{\textbf{3}}}\bar{\textbf{5}}_{\textbf{e}_{\textbf{1}},\textbf{e}_{\textbf{2}}}\bar{\textbf{5}}_{\textbf{e}_{\textbf{4}},\textbf{e}_{\textbf{5}}} $ & n & 3-6\\
\cline{2-5}
& 13 & $\textbf{10}_{\textbf{e}_{\textbf{3}}}\bar{\textbf{5}}_{\textbf{e}_{\textbf{1}},\textbf{e}_{\textbf{2}}}\bar{\textbf{5}}_{\textbf{e}_{\textbf{4}},\textbf{e}_{\textbf{5}}} $ & n & 3-6\\
\cline{2-5}
& 14 & $\textbf{10}_{\textbf{e}_{\textbf{3}}}\bar{\textbf{5}}_{\textbf{e}_{\textbf{1}},\textbf{e}_{\textbf{2}}}\bar{\textbf{5}}_{\textbf{e}_{\textbf{4}},\textbf{e}_{\textbf{5}}} $ & n & 3-6\\
\cline{2-5}
& 15 & $\textbf{10}_{\textbf{e}_{\textbf{2}}}\bar{\textbf{5}}_{\textbf{e}_{\textbf{1}},\textbf{e}_{\textbf{3}}}\bar{\textbf{5}}_{\textbf{e}_{\textbf{4}},\textbf{e}_{\textbf{5}}} $ & y & 3-6\\
\cline{2-5}
& 16 & $\textbf{10}_{\textbf{e}_{\textbf{5}}}\bar{\textbf{5}}_{\textbf{e}_{\textbf{1}},\textbf{e}_{\textbf{3}}}\bar{\textbf{5}}_{\textbf{e}_{\textbf{2}},\textbf{e}_{\textbf{4}}}  $ & y & 3-6\\
\cline{2-5}
& 17 & $\textbf{10}_{\textbf{e}_{\textbf{2}}}\bar{\textbf{5}}_{\textbf{e}_{\textbf{1}},\textbf{e}_{\textbf{3}}}\bar{\textbf{5}}_{\textbf{e}_{\textbf{4}},\textbf{e}_{\textbf{5}}}  $ & y & 3-6\\
\cline{2-5}
& 18 & $\textbf{10}_{\textbf{e}_{\textbf{1}}}\bar{\textbf{5}}_{\textbf{e}_{\textbf{2}},\textbf{e}_{\textbf{3}}}\bar{\textbf{5}}_{\textbf{e}_{\textbf{4}},\textbf{e}_{\textbf{5}}}  $ & y & 3-6\\
\cline{2-5}
& 19 & $\textbf{10}_{\textbf{e}_{\textbf{5}}}\bar{\textbf{5}}_{\textbf{e}_{\textbf{1}},\textbf{e}_{\textbf{4}}}\bar{\textbf{5}}_{\textbf{e}_{\textbf{2}},\textbf{e}_{\textbf{3}}} $ & y & 3-6\\
\cline{2-5}
& 20 & $\textbf{10}_{\textbf{e}_{\textbf{5}}}\bar{\textbf{5}}_{\textbf{e}_{\textbf{1}},\textbf{e}_{\textbf{2}}}\bar{\textbf{5}}_{\textbf{e}_{\textbf{3}},\textbf{e}_{\textbf{4}}}  $ & n & 3-6\\
\cline{2-5}
& 21 & $\textbf{10}_{\textbf{e}_{\textbf{5}}}\bar{\textbf{5}}_{\textbf{e}_{\textbf{1}},\textbf{e}_{\textbf{4}}}\bar{\textbf{5}}_{\textbf{e}_{\textbf{2}},\textbf{e}_{\textbf{3}}}  $ & n & 3-6\\
\cline{2-5}
& 22 & $\textbf{10}_{\textbf{e}_{\textbf{5}}}\bar{\textbf{5}}_{\textbf{e}_{\textbf{1}},\textbf{e}_{\textbf{2}}}\bar{\textbf{5}}_{\textbf{e}_{\textbf{3}},\textbf{e}_{\textbf{4}}}  $ & n & 3-6\\
\cline{2-5}
& 23 & $\textbf{10}_{\textbf{e}_{\textbf{5}}}\bar{\textbf{5}}_{\textbf{e}_{\textbf{1}},\textbf{e}_{\textbf{3}}}\bar{\textbf{5}}_{\textbf{e}_{\textbf{2}},\textbf{e}_{\textbf{4}}}  $ & n & 3-6\\
\cline{2-5}
& 25 & $\textbf{10}_{\textbf{e}_{\textbf{2}}}\bar{\textbf{5}}_{\textbf{e}_{\textbf{1}},\textbf{e}_{\textbf{4}}}\bar{\textbf{5}}_{\textbf{e}_{\textbf{3}},\textbf{e}_{\textbf{5}}} $ & y & 3-6\\
\cline{2-5}
\hline
\end{tabular}
\end{center}
On CICY 5256, a total of  84 models have Yukawa couplings consistent with the gauge symmetries of the models, with 158 Yukawa couplings being permitted in total. Of these, 126 are of the form $\textbf{10}_{\textbf{e}_{\textbf{i}}}\bar{\textbf{5}}_{\textbf{e}_{\textbf{j}},\textbf{e}_{\textbf{k}}}\bar{\textbf{5}}_{\textbf{e}_{\textbf{l}},\textbf{e}_{\textbf{m}}} $, 32 are of the form $\textbf{1}_{\textbf{e}_{\textbf{i}},-\textbf{e}_{\textbf{j}}}\textbf{5}_{-\textbf{e}_{\textbf{i}},-\textbf{e}_{\textbf{k}}}\bar{\textbf{5}}_{\textbf{e}_{\textbf{j}}, \textbf{e}_{\textbf{k}}}$. A total of 56 couplings exhibit the topological vanishing we have been studying in this paper, 24 of the form $\textbf{1}_{\textbf{e}_{\textbf{i}},-\textbf{e}_{\textbf{j}}}\textbf{5}_{-\textbf{e}_{\textbf{i}},-\textbf{e}_{\textbf{k}}}\bar{\textbf{5}}_{\textbf{e}_{\textbf{j}}, \textbf{e}_{\textbf{k}}}$, 32 of the form  $\textbf{10}_{\textbf{e}_{\textbf{i}}}\bar{\textbf{5}}_{\textbf{e}_{\textbf{j}},\textbf{e}_{\textbf{k}}}\bar{\textbf{5}}_{\textbf{e}_{\textbf{l}},\textbf{e}_{\textbf{m}}} $. A total of 32 out of the 84 models have at least one topologically vanishing coupling.

\begin{center}
\begin{tabular}{|c|c|c|c|c|}
\hline
CICY No.  & Model No. & Yukawa Pattern & Top. Van. & Sym. No.\\
\hline
\multirow{10}{*}{5452} & 1 & $\textbf{10}_{\textbf{e}_{\textbf{3}}}\bar{\textbf{5}}_{\textbf{e}_{\textbf{1}},\textbf{e}_{\textbf{2}}}\bar{\textbf{5}}_{\textbf{e}_{\textbf{4}},\textbf{e}_{\textbf{5}}}  $& n & 7-22\\
\cline{2-5}
& 2 & $\textbf{10}_{\textbf{e}_{\textbf{5}}}\bar{\textbf{5}}_{\textbf{e}_{\textbf{1}},\textbf{e}_{\textbf{2}}}\bar{\textbf{5}}_{\textbf{e}_{\textbf{3}},\textbf{e}_{\textbf{4}}} $ & n & 7-22\\
\cline{2-5}
 & 7 & $\textbf{10}_{\textbf{e}_{\textbf{5}}}\bar{\textbf{5}}_{\textbf{e}_{\textbf{1}},\textbf{e}_{\textbf{2}}}\bar{\textbf{5}}_{\textbf{e}_{\textbf{3}},\textbf{e}_{\textbf{4}}} $ & n & 7-22\\
\cline{2-5}
& 8 & $\textbf{10}_{\textbf{e}_{\textbf{4}}}\bar{\textbf{5}}_{\textbf{e}_{\textbf{1}},\textbf{e}_{\textbf{2}}}\bar{\textbf{5}}_{\textbf{e}_{\textbf{3}},\textbf{e}_{\textbf{5}}} $ & n & 7-22\\
\cline{2-5}
& 9 & $\textbf{10}_{\textbf{e}_{\textbf{5}}}\bar{\textbf{5}}_{\textbf{e}_{\textbf{1}},\textbf{e}_{\textbf{2}}}\bar{\textbf{5}}_{\textbf{e}_{\textbf{3}},\textbf{e}_{\textbf{4}}}$ & n & 7-22\\
\cline{2-5}
& 10 & $\textbf{10}_{\textbf{e}_{\textbf{5}}}\bar{\textbf{5}}_{\textbf{e}_{\textbf{1}},\textbf{e}_{\textbf{2}}}\bar{\textbf{5}}_{\textbf{e}_{\textbf{3}},\textbf{e}_{\textbf{4}}}$ & n & 7-22\\
\cline{2-5}
& 11 & $\textbf{10}_{\textbf{e}_{\textbf{5}}}\bar{\textbf{5}}_{\textbf{e}_{\textbf{1}},\textbf{e}_{\textbf{2}}}\bar{\textbf{5}}_{\textbf{e}_{\textbf{3}},\textbf{e}_{\textbf{4}}}$& n & 7-22\\
\cline{2-5}
& 12 & $\textbf{10}_{\textbf{e}_{\textbf{5}}}\bar{\textbf{5}}_{\textbf{e}_{\textbf{1}},\textbf{e}_{\textbf{2}}}\bar{\textbf{5}}_{\textbf{e}_{\textbf{3}},\textbf{e}_{\textbf{4}}} $ & n & 7-22\\
\cline{2-5}
& 13 &$\textbf{10}_{\textbf{e}_{\textbf{5}}}\bar{\textbf{5}}_{\textbf{e}_{\textbf{1}},\textbf{e}_{\textbf{2}}}\bar{\textbf{5}}_{\textbf{e}_{\textbf{3}},\textbf{e}_{\textbf{4}}} $ & n & 7-22\\
\cline{2-5}
& 14 & $\textbf{10}_{\textbf{e}_{\textbf{4}}}\bar{\textbf{5}}_{\textbf{e}_{\textbf{1}},\textbf{e}_{\textbf{2}}}\bar{\textbf{5}}_{\textbf{e}_{\textbf{3}},\textbf{e}_{\textbf{5}}}$ & n & 7-22\\
\cline{2-5}
& 15 & $\textbf{10}_{\textbf{e}_{\textbf{1}}}\bar{\textbf{5}}_{\textbf{e}_{\textbf{2}},\textbf{e}_{\textbf{4}}}\bar{\textbf{5}}_{\textbf{e}_{\textbf{3}},\textbf{e}_{\textbf{5}}}$ & y & 7-22\\
\cline{2-5}
 & 16 & $\textbf{10}_{\textbf{e}_{\textbf{1}}}\bar{\textbf{5}}_{\textbf{e}_{\textbf{2}},\textbf{e}_{\textbf{4}}}\bar{\textbf{5}}_{\textbf{e}_{\textbf{3}},\textbf{e}_{\textbf{5}}} $ & y & 7-22\\
\cline{2-5}
& 17 & $\textbf{10}_{\textbf{e}_{\textbf{2}}}\bar{\textbf{5}}_{\textbf{e}_{\textbf{1}},\textbf{e}_{\textbf{3}}}\bar{\textbf{5}}_{\textbf{e}_{\textbf{4}},\textbf{e}_{\textbf{5}}} $ & y & 7-22\\
\cline{2-5}
 & 18 & $\textbf{10}_{\textbf{e}_{\textbf{5}}}\bar{\textbf{5}}_{\textbf{e}_{\textbf{1}},\textbf{e}_{\textbf{3}}}\bar{\textbf{5}}_{\textbf{e}_{\textbf{2}},\textbf{e}_{\textbf{4}}} $ & n & 7-22\\
\cline{2-5}
\hline
\end{tabular}
\end{center}
\begin{center}
\begin{tabular}{|c|c|c|c|c|}
\hline
CICY No.  & Model No. & Yukawa Pattern & Top. Van. & Sym. No.\\
\hline
\multirow{40}{*}{5452} 
& 19 & $\textbf{10}_{\textbf{e}_{\textbf{5}}}\bar{\textbf{5}}_{\textbf{e}_{\textbf{1}},\textbf{e}_{\textbf{3}}}\bar{\textbf{5}}_{\textbf{e}_{\textbf{2}},\textbf{e}_{\textbf{4}}} $ & n & 7-22\\
\cline{2-5}
 & 20 & $\textbf{10}_{\textbf{e}_{\textbf{5}}}\bar{\textbf{5}}_{\textbf{e}_{\textbf{1}},\textbf{e}_{\textbf{3}}}\bar{\textbf{5}}_{\textbf{e}_{\textbf{2}},\textbf{e}_{\textbf{4}}} $ & n & 7-22\\
\cline{2-5}
& 21 & $\textbf{10}_{\textbf{e}_{\textbf{5}}}\bar{\textbf{5}}_{\textbf{e}_{\textbf{1}},\textbf{e}_{\textbf{3}}}\bar{\textbf{5}}_{\textbf{e}_{\textbf{2}},\textbf{e}_{\textbf{4}}}$ & n & 7-22\\
\cline{2-5}
& 22 & $\textbf{10}_{\textbf{e}_{\textbf{5}}}\bar{\textbf{5}}_{\textbf{e}_{\textbf{1}},\textbf{e}_{\textbf{3}}}\bar{\textbf{5}}_{\textbf{e}_{\textbf{2}},\textbf{e}_{\textbf{4}}} $& n & 7-22\\
\cline{2-5}
& 23 & $\textbf{10}_{\textbf{e}_{\textbf{5}}}\bar{\textbf{5}}_{\textbf{e}_{\textbf{1}},\textbf{e}_{\textbf{3}}}\bar{\textbf{5}}_{\textbf{e}_{\textbf{2}},\textbf{e}_{\textbf{4}}} $ & n & 7-22\\
\cline{2-5}
& 24 & $\textbf{10}_{\textbf{e}_{\textbf{5}}}\bar{\textbf{5}}_{\textbf{e}_{\textbf{1}},\textbf{e}_{\textbf{3}}}\bar{\textbf{5}}_{\textbf{e}_{\textbf{2}},\textbf{e}_{\textbf{4}}} $ & n & 7-22\\
\cline{2-5}
 & 25 & $\textbf{10}_{\textbf{e}_{\textbf{5}}}\bar{\textbf{5}}_{\textbf{e}_{\textbf{1}},\textbf{e}_{\textbf{3}}}\bar{\textbf{5}}_{\textbf{e}_{\textbf{2}},\textbf{e}_{\textbf{4}}}$ & n & 7-22\\
\cline{2-5}
& 26 & $\textbf{10}_{\textbf{e}_{\textbf{5}}}\bar{\textbf{5}}_{\textbf{e}_{\textbf{1}},\textbf{e}_{\textbf{4}}}\bar{\textbf{5}}_{\textbf{e}_{\textbf{2}},\textbf{e}_{\textbf{3}}} $ & y & 7-22\\
\cline{2-5}
& 27 & $\textbf{10}_{\textbf{e}_{\textbf{5}}}\bar{\textbf{5}}_{\textbf{e}_{\textbf{1}},\textbf{e}_{\textbf{4}}}\bar{\textbf{5}}_{\textbf{e}_{\textbf{2}},\textbf{e}_{\textbf{3}}}  $ & y & 7-22\\
\cline{2-5}
& 28 & $\textbf{10}_{\textbf{e}_{\textbf{1}}}\bar{\textbf{5}}_{\textbf{e}_{\textbf{2}},\textbf{e}_{\textbf{4}}}\bar{\textbf{5}}_{\textbf{e}_{\textbf{3}},\textbf{e}_{\textbf{5}}} $ & y & 7-22\\
\cline{2-5}
& 29 & $\textbf{10}_{\textbf{e}_{\textbf{1}}}\bar{\textbf{5}}_{\textbf{e}_{\textbf{2}},\textbf{e}_{\textbf{4}}}\bar{\textbf{5}}_{\textbf{e}_{\textbf{3}},\textbf{e}_{\textbf{5}}} $ & y & 7-22\\
\cline{2-5}
& 34 & $\textbf{10}_{\textbf{e}_{\textbf{5}}}\bar{\textbf{5}}_{\textbf{e}_{\textbf{1}},\textbf{e}_{\textbf{4}}}\bar{\textbf{5}}_{\textbf{e}_{\textbf{2}},\textbf{e}_{\textbf{3}}} $ & y & 7-22\\
\cline{2-5}
 & 35 & $\textbf{10}_{\textbf{e}_{\textbf{1}}}\bar{\textbf{5}}_{\textbf{e}_{\textbf{2}},\textbf{e}_{\textbf{3}}}\bar{\textbf{5}}_{\textbf{e}_{\textbf{4}},\textbf{e}_{\textbf{5}}}  $ & y & 7-22\\
\cline{2-5}
& 36 & $\textbf{10}_{\textbf{e}_{\textbf{2}}}\bar{\textbf{5}}_{\textbf{e}_{\textbf{1}},\textbf{e}_{\textbf{3}}}\bar{\textbf{5}}_{\textbf{e}_{\textbf{4}},\textbf{e}_{\textbf{5}}}  $ & y & 7-22\\
\cline{2-5}
& 37 & $\textbf{10}_{\textbf{e}_{\textbf{2}}}\bar{\textbf{5}}_{\textbf{e}_{\textbf{1}},\textbf{e}_{\textbf{3}}}\bar{\textbf{5}}_{\textbf{e}_{\textbf{4}},\textbf{e}_{\textbf{5}}} $ & y & 7-22\\
\cline{2-5}
& 38 & $\textbf{10}_{\textbf{e}_{\textbf{5}}}\bar{\textbf{5}}_{\textbf{e}_{\textbf{1}},\textbf{e}_{\textbf{3}}}\bar{\textbf{5}}_{\textbf{e}_{\textbf{2}},\textbf{e}_{\textbf{4}}} $ & y & 7-22\\
\cline{2-5}
& 39 & $\textbf{10}_{\textbf{e}_{\textbf{5}}}\bar{\textbf{5}}_{\textbf{e}_{\textbf{1}},\textbf{e}_{\textbf{2}}}\bar{\textbf{5}}_{\textbf{e}_{\textbf{3}},\textbf{e}_{\textbf{4}}} $ & n & 7-22\\
\cline{2-5}
& 40 & $\textbf{10}_{\textbf{e}_{\textbf{5}}}\bar{\textbf{5}}_{\textbf{e}_{\textbf{1}},\textbf{e}_{\textbf{2}}}\bar{\textbf{5}}_{\textbf{e}_{\textbf{3}},\textbf{e}_{\textbf{4}}} $ & n & 7-22\\
\cline{2-5}
& 41 & $\textbf{10}_{\textbf{e}_{\textbf{5}}}\bar{\textbf{5}}_{\textbf{e}_{\textbf{1}},\textbf{e}_{\textbf{2}}}\bar{\textbf{5}}_{\textbf{e}_{\textbf{3}},\textbf{e}_{\textbf{4}}}$ & n & 7-22\\
\cline{2-5}
& 42 & $\textbf{10}_{\textbf{e}_{\textbf{5}}}\bar{\textbf{5}}_{\textbf{e}_{\textbf{1}},\textbf{e}_{\textbf{2}}}\bar{\textbf{5}}_{\textbf{e}_{\textbf{3}},\textbf{e}_{\textbf{4}}} $ & n & 7-22\\
\cline{2-5}
& 43 & $\textbf{10}_{\textbf{e}_{\textbf{3}}}\bar{\textbf{5}}_{\textbf{e}_{\textbf{1}},\textbf{e}_{\textbf{2}}}\bar{\textbf{5}}_{\textbf{e}_{\textbf{4}},\textbf{e}_{\textbf{5}}}$ & n & 7-22\\
\cline{2-5}
& 44 & $\textbf{10}_{\textbf{e}_{\textbf{3}}}\bar{\textbf{5}}_{\textbf{e}_{\textbf{1}},\textbf{e}_{\textbf{2}}}\bar{\textbf{5}}_{\textbf{e}_{\textbf{4}},\textbf{e}_{\textbf{5}}} $ & n & 7-22\\
\cline{2-5}
& 45 & $\textbf{10}_{\textbf{e}_{\textbf{3}}}\bar{\textbf{5}}_{\textbf{e}_{\textbf{1}},\textbf{e}_{\textbf{2}}}\bar{\textbf{5}}_{\textbf{e}_{\textbf{4}},\textbf{e}_{\textbf{5}}} $ & n & 7-22\\
\cline{2-5}
& 46 & $\textbf{10}_{\textbf{e}_{\textbf{3}}}\bar{\textbf{5}}_{\textbf{e}_{\textbf{1}},\textbf{e}_{\textbf{2}}}\bar{\textbf{5}}_{\textbf{e}_{\textbf{4}},\textbf{e}_{\textbf{5}}} $ & n & 7-22\\
\cline{2-5}
& 47 & $\textbf{10}_{\textbf{e}_{\textbf{3}}}\bar{\textbf{5}}_{\textbf{e}_{\textbf{1}},\textbf{e}_{\textbf{2}}}\bar{\textbf{5}}_{\textbf{e}_{\textbf{4}},\textbf{e}_{\textbf{5}}} $ & n & 7-22\\
\cline{2-5}
& 48 & $\textbf{10}_{\textbf{e}_{\textbf{3}}}\bar{\textbf{5}}_{\textbf{e}_{\textbf{1}},\textbf{e}_{\textbf{2}}}\bar{\textbf{5}}_{\textbf{e}_{\textbf{4}},\textbf{e}_{\textbf{5}}}$ & n & 7-22\\
\cline{2-5}
 & 49 & $\textbf{10}_{\textbf{e}_{\textbf{3}}}\bar{\textbf{5}}_{\textbf{e}_{\textbf{1}},\textbf{e}_{\textbf{2}}}\bar{\textbf{5}}_{\textbf{e}_{\textbf{4}},\textbf{e}_{\textbf{5}}}$ & n & 7-22\\
\cline{2-5}
&50 & $\textbf{10}_{\textbf{e}_{\textbf{3}}}\bar{\textbf{5}}_{\textbf{e}_{\textbf{1}},\textbf{e}_{\textbf{2}}}\bar{\textbf{5}}_{\textbf{e}_{\textbf{4}},\textbf{e}_{\textbf{5}}} $ & n & 7-22\\
\cline{2-5}
& \multirow{2}{*}{51} & $\textbf{1}_{\textbf{e}_{\textbf{1}},-\textbf{e}_{\textbf{2}}}\textbf{5}_{-\textbf{e}_{\textbf{1}},-\textbf{e}_{\textbf{4}}}\bar{\textbf{5}}_{\textbf{e}_{\textbf{2}}, \textbf{e}_{\textbf{4}}} $ & y & 7-22\\
\cline{3-5}
& &$\textbf{10}_{\textbf{e}_{\textbf{3}}}\bar{\textbf{5}}_{\textbf{e}_{\textbf{1}},\textbf{e}_{\textbf{5}}}\bar{\textbf{5}}_{\textbf{e}_{\textbf{2}},\textbf{e}_{\textbf{4}}} $ & y & 7-22\\
\cline{2-5}
& 52 & $\textbf{10}_{\textbf{e}_{\textbf{3}}}\bar{\textbf{5}}_{\textbf{e}_{\textbf{1}},\textbf{e}_{\textbf{2}}}\bar{\textbf{5}}_{\textbf{e}_{\textbf{4}},\textbf{e}_{\textbf{5}}}$ & n & 7-22\\
\cline{2-5}
& 53 & $\textbf{10}_{\textbf{e}_{\textbf{3}}}\bar{\textbf{5}}_{\textbf{e}_{\textbf{1}},\textbf{e}_{\textbf{2}}}\bar{\textbf{5}}_{\textbf{e}_{\textbf{4}},\textbf{e}_{\textbf{5}}} $ & n & 7-22\\
\cline{2-5}
& 54 & $\textbf{10}_{\textbf{e}_{\textbf{3}}}\bar{\textbf{5}}_{\textbf{e}_{\textbf{1}},\textbf{e}_{\textbf{2}}}\bar{\textbf{5}}_{\textbf{e}_{\textbf{4}},\textbf{e}_{\textbf{5}}}$ & n & 7-22\\
\cline{2-5}
& 55 & $\textbf{10}_{\textbf{e}_{\textbf{3}}}\bar{\textbf{5}}_{\textbf{e}_{\textbf{1}},\textbf{e}_{\textbf{2}}}\bar{\textbf{5}}_{\textbf{e}_{\textbf{4}},\textbf{e}_{\textbf{5}}} $ & n & 7-22\\
\cline{2-5}
& \multirow{2}{*}{56} & $\textbf{1}_{\textbf{e}_{\textbf{1}},-\textbf{e}_{\textbf{2}}}\textbf{5}_{-\textbf{e}_{\textbf{1}},-\textbf{e}_{\textbf{4}}}\bar{\textbf{5}}_{\textbf{e}_{\textbf{2}}, \textbf{e}_{\textbf{4}}}  $ & y & 7-22\\
\cline{3-5}
&  & $\textbf{10}_{\textbf{e}_{\textbf{3}}}\bar{\textbf{5}}_{\textbf{e}_{\textbf{1}},\textbf{e}_{\textbf{5}}}\bar{\textbf{5}}_{\textbf{e}_{\textbf{2}},\textbf{e}_{\textbf{4}}}$ & y & 7-22\\
\cline{2-5}
& \multirow{2}{*}{57} & $\textbf{10}_{\textbf{e}_{\textbf{3}}}\bar{\textbf{5}}_{\textbf{e}_{\textbf{1}},\textbf{e}_{\textbf{5}}}\bar{\textbf{5}}_{\textbf{e}_{\textbf{2}},\textbf{e}_{\textbf{4}}} $ & n & 7-22\\
\cline{3-5}
& &$\textbf{1}_{\textbf{e}_{\textbf{2}},-\textbf{e}_{\textbf{1}}}\textbf{5}_{-\textbf{e}_{\textbf{2}},-\textbf{e}_{\textbf{5}}}\bar{\textbf{5}}_{\textbf{e}_{\textbf{1}}, \textbf{e}_{\textbf{5}}}  $ & y & 7-22\\
\cline{2-5}
& 58 & $\textbf{10}_{\textbf{e}_{\textbf{3}}}\bar{\textbf{5}}_{\textbf{e}_{\textbf{1}},\textbf{e}_{\textbf{2}}}\bar{\textbf{5}}_{\textbf{e}_{\textbf{4}},\textbf{e}_{\textbf{5}}} $ & n & 7-22\\
\cline{2-5}
 & 59 & $\textbf{10}_{\textbf{e}_{\textbf{3}}}\bar{\textbf{5}}_{\textbf{e}_{\textbf{1}},\textbf{e}_{\textbf{2}}}\bar{\textbf{5}}_{\textbf{e}_{\textbf{4}},\textbf{e}_{\textbf{5}}} $ & n & 7-22\\
\cline{2-5}
& 60 & $\textbf{10}_{\textbf{e}_{\textbf{3}}}\bar{\textbf{5}}_{\textbf{e}_{\textbf{1}},\textbf{e}_{\textbf{2}}}\bar{\textbf{5}}_{\textbf{e}_{\textbf{4}},\textbf{e}_{\textbf{5}}}$ & n & 7-22\\
\cline{2-5}
& 61 & $\textbf{10}_{\textbf{e}_{\textbf{3}}}\bar{\textbf{5}}_{\textbf{e}_{\textbf{1}},\textbf{e}_{\textbf{2}}}\bar{\textbf{5}}_{\textbf{e}_{\textbf{4}},\textbf{e}_{\textbf{5}}}$ & n & 7-22\\
\cline{2-5}
& \multirow{2}{*}{62} & $\textbf{1}_{\textbf{e}_{\textbf{1}},-\textbf{e}_{\textbf{2}}}\textbf{5}_{-\textbf{e}_{\textbf{1}},-\textbf{e}_{\textbf{5}}}\bar{\textbf{5}}_{\textbf{e}_{\textbf{2}}, \textbf{e}_{\textbf{5}}}$  & y & 7-22\\
\cline{3-5}
& & $\textbf{10}_{\textbf{e}_{\textbf{3}}}\bar{\textbf{5}}_{\textbf{e}_{\textbf{1}},\textbf{e}_{\textbf{4}}}\bar{\textbf{5}}_{\textbf{e}_{\textbf{2}},\textbf{e}_{\textbf{5}}}$ & y & 7-22\\
\cline{2-5}
\hline
\end{tabular}
\end{center}
\begin{center}
\begin{tabular}{|c|c|c|c|c|}
\hline
CICY No.  & Model No. & Yukawa Pattern & Top. Van. & Sym. No.\\
\hline
\multirow{30}{*}{5452} &50 & $\textbf{10}_{\textbf{e}_{\textbf{3}}}\bar{\textbf{5}}_{\textbf{e}_{\textbf{1}},\textbf{e}_{\textbf{2}}}\bar{\textbf{5}}_{\textbf{e}_{\textbf{4}},\textbf{e}_{\textbf{5}}} $ & n & 7-22\\
\cline{2-5}
& \multirow{2}{*}{51} & $\textbf{1}_{\textbf{e}_{\textbf{1}},-\textbf{e}_{\textbf{2}}}\textbf{5}_{-\textbf{e}_{\textbf{1}},-\textbf{e}_{\textbf{4}}}\bar{\textbf{5}}_{\textbf{e}_{\textbf{2}}, \textbf{e}_{\textbf{4}}} $ & y & 7-22\\
\cline{3-5}
& &$\textbf{10}_{\textbf{e}_{\textbf{3}}}\bar{\textbf{5}}_{\textbf{e}_{\textbf{1}},\textbf{e}_{\textbf{5}}}\bar{\textbf{5}}_{\textbf{e}_{\textbf{2}},\textbf{e}_{\textbf{4}}} $ & y & 7-22\\
\cline{2-5}
& 52 & $\textbf{10}_{\textbf{e}_{\textbf{3}}}\bar{\textbf{5}}_{\textbf{e}_{\textbf{1}},\textbf{e}_{\textbf{2}}}\bar{\textbf{5}}_{\textbf{e}_{\textbf{4}},\textbf{e}_{\textbf{5}}}$ & n & 7-22\\
\cline{2-5}
& 53 & $\textbf{10}_{\textbf{e}_{\textbf{3}}}\bar{\textbf{5}}_{\textbf{e}_{\textbf{1}},\textbf{e}_{\textbf{2}}}\bar{\textbf{5}}_{\textbf{e}_{\textbf{4}},\textbf{e}_{\textbf{5}}} $ & n & 7-22\\
\cline{2-5}
& 54 & $\textbf{10}_{\textbf{e}_{\textbf{3}}}\bar{\textbf{5}}_{\textbf{e}_{\textbf{1}},\textbf{e}_{\textbf{2}}}\bar{\textbf{5}}_{\textbf{e}_{\textbf{4}},\textbf{e}_{\textbf{5}}}$ & n & 7-22\\
\cline{2-5}
& 55 & $\textbf{10}_{\textbf{e}_{\textbf{3}}}\bar{\textbf{5}}_{\textbf{e}_{\textbf{1}},\textbf{e}_{\textbf{2}}}\bar{\textbf{5}}_{\textbf{e}_{\textbf{4}},\textbf{e}_{\textbf{5}}} $ & n & 7-22\\
\cline{2-5}
& \multirow{2}{*}{56} & $\textbf{1}_{\textbf{e}_{\textbf{1}},-\textbf{e}_{\textbf{2}}}\textbf{5}_{-\textbf{e}_{\textbf{1}},-\textbf{e}_{\textbf{4}}}\bar{\textbf{5}}_{\textbf{e}_{\textbf{2}}, \textbf{e}_{\textbf{4}}}  $ & y & 7-22\\
\cline{3-5}
&  & $\textbf{10}_{\textbf{e}_{\textbf{3}}}\bar{\textbf{5}}_{\textbf{e}_{\textbf{1}},\textbf{e}_{\textbf{5}}}\bar{\textbf{5}}_{\textbf{e}_{\textbf{2}},\textbf{e}_{\textbf{4}}}$ & y & 7-22\\
\cline{2-5}
& \multirow{2}{*}{57} & $\textbf{10}_{\textbf{e}_{\textbf{3}}}\bar{\textbf{5}}_{\textbf{e}_{\textbf{1}},\textbf{e}_{\textbf{5}}}\bar{\textbf{5}}_{\textbf{e}_{\textbf{2}},\textbf{e}_{\textbf{4}}} $ & n & 7-22\\
\cline{3-5}
& &$\textbf{1}_{\textbf{e}_{\textbf{2}},-\textbf{e}_{\textbf{1}}}\textbf{5}_{-\textbf{e}_{\textbf{2}},-\textbf{e}_{\textbf{5}}}\bar{\textbf{5}}_{\textbf{e}_{\textbf{1}}, \textbf{e}_{\textbf{5}}}  $ & y & 7-22\\
\cline{2-5}
& 58 & $\textbf{10}_{\textbf{e}_{\textbf{3}}}\bar{\textbf{5}}_{\textbf{e}_{\textbf{1}},\textbf{e}_{\textbf{2}}}\bar{\textbf{5}}_{\textbf{e}_{\textbf{4}},\textbf{e}_{\textbf{5}}} $ & n & 7-22\\
\cline{2-5}
 & 59 & $\textbf{10}_{\textbf{e}_{\textbf{3}}}\bar{\textbf{5}}_{\textbf{e}_{\textbf{1}},\textbf{e}_{\textbf{2}}}\bar{\textbf{5}}_{\textbf{e}_{\textbf{4}},\textbf{e}_{\textbf{5}}} $ & n & 7-22\\
\cline{2-5}
& 60 & $\textbf{10}_{\textbf{e}_{\textbf{3}}}\bar{\textbf{5}}_{\textbf{e}_{\textbf{1}},\textbf{e}_{\textbf{2}}}\bar{\textbf{5}}_{\textbf{e}_{\textbf{4}},\textbf{e}_{\textbf{5}}}$ & n & 7-22\\
\cline{2-5}
& 61 & $\textbf{10}_{\textbf{e}_{\textbf{3}}}\bar{\textbf{5}}_{\textbf{e}_{\textbf{1}},\textbf{e}_{\textbf{2}}}\bar{\textbf{5}}_{\textbf{e}_{\textbf{4}},\textbf{e}_{\textbf{5}}}$ & n & 7-22\\
\cline{2-5}
& \multirow{2}{*}{62} & $\textbf{1}_{\textbf{e}_{\textbf{1}},-\textbf{e}_{\textbf{2}}}\textbf{5}_{-\textbf{e}_{\textbf{1}},-\textbf{e}_{\textbf{5}}}\bar{\textbf{5}}_{\textbf{e}_{\textbf{2}}, \textbf{e}_{\textbf{5}}}$  & y & 7-22\\
\cline{3-5}
& & $\textbf{10}_{\textbf{e}_{\textbf{3}}}\bar{\textbf{5}}_{\textbf{e}_{\textbf{1}},\textbf{e}_{\textbf{4}}}\bar{\textbf{5}}_{\textbf{e}_{\textbf{2}},\textbf{e}_{\textbf{5}}}$ & y & 7-22\\
\cline{2-5}
 & \multirow{3}{*}{1} & $\textbf{10}_{\textbf{e}_{\textbf{1}}}\bar{\textbf{5}}_{\textbf{e}_{\textbf{2}},\textbf{e}_{\textbf{3}}}\bar{\textbf{5}}_{\textbf{e}_{\textbf{4}},\textbf{e}_{\textbf{5}}} $ & n & 1-4\\
\cline{3-5}
 &  & $\textbf{10}_{\textbf{e}_{\textbf{5}}}\bar{\textbf{5}}_{\textbf{e}_{\textbf{1}},\textbf{e}_{\textbf{4}}}\bar{\textbf{5}}_{\textbf{e}_{\textbf{2}},\textbf{e}_{\textbf{3}}}$ & n & 1-4\\
\cline{3-5}
   &  & $\textbf{1}_{\textbf{e}_{\textbf{4}},-\textbf{e}_{\textbf{3}}}\textbf{5}_{-\textbf{e}_{\textbf{4}},-\textbf{e}_{\textbf{5}}}\bar{\textbf{5}}_{\textbf{e}_{\textbf{3}}, \textbf{e}_{\textbf{5}}}$ & n & 1-4\\
\cline{2-5}
 & \multirow{3}{*}{2} & $\textbf{10}_{\textbf{e}_{\textbf{1}}}\bar{\textbf{5}}_{\textbf{e}_{\textbf{2}},\textbf{e}_{\textbf{3}}}\bar{\textbf{5}}_{\textbf{e}_{\textbf{4}},\textbf{e}_{\textbf{5}}} $ & n & 1-4\\
\cline{3-5}
 &  & $\textbf{10}_{\textbf{e}_{\textbf{5}}}\bar{\textbf{5}}_{\textbf{e}_{\textbf{1}},\textbf{e}_{\textbf{4}}}\bar{\textbf{5}}_{\textbf{e}_{\textbf{2}},\textbf{e}_{\textbf{3}}}$ & n & 1-4\\
\cline{3-5}
   &  & $\textbf{1}_{\textbf{e}_{\textbf{4}},-\textbf{e}_{\textbf{3}}}\textbf{5}_{-\textbf{e}_{\textbf{4}},-\textbf{e}_{\textbf{5}}}\bar{\textbf{5}}_{\textbf{e}_{\textbf{3}}, \textbf{e}_{\textbf{5}}}$ & n & 1-4\\
\cline{2-5}
 &  \multirow{3}{*}{3} & $\textbf{10}_{\textbf{e}_{\textbf{1}}}\bar{\textbf{5}}_{\textbf{e}_{\textbf{2}},\textbf{e}_{\textbf{3}}}\bar{\textbf{5}}_{\textbf{e}_{\textbf{4}},\textbf{e}_{\textbf{5}}} $ & n & 1-4\\
\cline{3-5}
 &  & $\textbf{10}_{\textbf{e}_{\textbf{5}}}\bar{\textbf{5}}_{\textbf{e}_{\textbf{1}},\textbf{e}_{\textbf{4}}}\bar{\textbf{5}}_{\textbf{e}_{\textbf{2}},\textbf{e}_{\textbf{3}}}$ & n & 1-4\\
\cline{3-5}
   &  & $\textbf{1}_{\textbf{e}_{\textbf{4}},-\textbf{e}_{\textbf{3}}}\textbf{5}_{-\textbf{e}_{\textbf{4}},-\textbf{e}_{\textbf{5}}}\bar{\textbf{5}}_{\textbf{e}_{\textbf{3}}, \textbf{e}_{\textbf{5}}}$ & n & 1-4\\
\cline{2-5}
 & \multirow{3}{*}{4} & $\textbf{10}_{\textbf{e}_{\textbf{1}}}\bar{\textbf{5}}_{\textbf{e}_{\textbf{2}},\textbf{e}_{\textbf{3}}}\bar{\textbf{5}}_{\textbf{e}_{\textbf{4}},\textbf{e}_{\textbf{5}}} $ & n & 1-4\\
\cline{3-5}
 &  & $\textbf{10}_{\textbf{e}_{\textbf{5}}}\bar{\textbf{5}}_{\textbf{e}_{\textbf{1}},\textbf{e}_{\textbf{4}}}\bar{\textbf{5}}_{\textbf{e}_{\textbf{2}},\textbf{e}_{\textbf{3}}}$ & n & 1-4\\
\cline{3-5}
   &  & $\textbf{1}_{\textbf{e}_{\textbf{4}},-\textbf{e}_{\textbf{3}}}\textbf{5}_{-\textbf{e}_{\textbf{4}},-\textbf{e}_{\textbf{5}}}\bar{\textbf{5}}_{\textbf{e}_{\textbf{3}}, \textbf{e}_{\textbf{5}}}$ & n & 1-4\\
\cline{2-5}
 & \multirow{2}{*}{6} & $\textbf{10}_{\textbf{e}_{\textbf{1}}}\bar{\textbf{5}}_{\textbf{e}_{\textbf{2}},\textbf{e}_{\textbf{3}}}\bar{\textbf{5}}_{\textbf{e}_{\textbf{4}},\textbf{e}_{\textbf{5}}} $ & n & 1-4\\
\cline{3-5}
   &  & $\textbf{10}_{\textbf{e}_{\textbf{4}}}\bar{\textbf{5}}_{\textbf{e}_{\textbf{1}},\textbf{e}_{\textbf{5}}}\bar{\textbf{5}}_{\textbf{e}_{\textbf{2}},\textbf{e}_{\textbf{3}}} $ & n & 1-4\\
\cline{2-5}
 & \multirow{2}{*}{7} & $\textbf{10}_{\textbf{e}_{\textbf{1}}}\bar{\textbf{5}}_{\textbf{e}_{\textbf{2}},\textbf{e}_{\textbf{3}}}\bar{\textbf{5}}_{\textbf{e}_{\textbf{4}},\textbf{e}_{\textbf{5}}} $ & n & 1-4\\
\cline{3-5}
   &  & $\textbf{10}_{\textbf{e}_{\textbf{4}}}\bar{\textbf{5}}_{\textbf{e}_{\textbf{1}},\textbf{e}_{\textbf{5}}}\bar{\textbf{5}}_{\textbf{e}_{\textbf{2}},\textbf{e}_{\textbf{3}}} $ & n & 1-4\\
\cline{2-5}
 & \multirow{2}{*}{8} & $\textbf{10}_{\textbf{e}_{\textbf{1}}}\bar{\textbf{5}}_{\textbf{e}_{\textbf{2}},\textbf{e}_{\textbf{3}}}\bar{\textbf{5}}_{\textbf{e}_{\textbf{4}},\textbf{e}_{\textbf{5}}} $ & n & 1-4\\
\cline{3-5}
   &  & $\textbf{10}_{\textbf{e}_{\textbf{4}}}\bar{\textbf{5}}_{\textbf{e}_{\textbf{1}},\textbf{e}_{\textbf{5}}}\bar{\textbf{5}}_{\textbf{e}_{\textbf{2}},\textbf{e}_{\textbf{3}}} $ & n & 1-4\\
\cline{2-5}
 & \multirow{2}{*}{9} & $\textbf{10}_{\textbf{e}_{\textbf{1}}}\bar{\textbf{5}}_{\textbf{e}_{\textbf{2}},\textbf{e}_{\textbf{3}}}\bar{\textbf{5}}_{\textbf{e}_{\textbf{4}},\textbf{e}_{\textbf{5}}} $ & n & 1-4\\
\cline{3-5}
   &  & $\textbf{10}_{\textbf{e}_{\textbf{4}}}\bar{\textbf{5}}_{\textbf{e}_{\textbf{1}},\textbf{e}_{\textbf{5}}}\bar{\textbf{5}}_{\textbf{e}_{\textbf{2}},\textbf{e}_{\textbf{3}}} $ & n & 1-4\\
\cline{2-5}
\hline
\end{tabular}
\end{center}
On CICY 5452, a total of 800 models have Yukawa couplings consistent with the gauge symmetries of the models, with 1584 Yukawa couplings being permitted in total. Of these, 1376 are of the form $\textbf{10}_{\textbf{e}_{\textbf{i}}}\bar{\textbf{5}}_{\textbf{e}_{\textbf{j}},\textbf{e}_{\textbf{k}}}\bar{\textbf{5}}_{\textbf{e}_{\textbf{l}},\textbf{e}_{\textbf{m}}} $ and 208 are of the form $\textbf{1}_{\textbf{e}_{\textbf{i}},-\textbf{e}_{\textbf{j}}}\textbf{5}_{-\textbf{e}_{\textbf{i}},-\textbf{e}_{\textbf{k}}}\bar{\textbf{5}}_{\textbf{e}_{\textbf{j}}, \textbf{e}_{\textbf{k}}}$ . A total of 432 couplings exhibit the topological vanishing we have been studying in this paper, 192 of the form $\textbf{1}_{\textbf{e}_{\textbf{i}},-\textbf{e}_{\textbf{j}}}\textbf{5}_{-\textbf{e}_{\textbf{i}},-\textbf{e}_{\textbf{k}}}\bar{\textbf{5}}_{\textbf{e}_{\textbf{j}}, \textbf{e}_{\textbf{k}}}$ and 240 of the form  $\textbf{10}_{\textbf{e}_{\textbf{i}}}\bar{\textbf{5}}_{\textbf{e}_{\textbf{j}},\textbf{e}_{\textbf{k}}}\bar{\textbf{5}}_{\textbf{e}_{\textbf{l}},\textbf{e}_{\textbf{m}}} $. A total of 256 out of the 800 models have at least one topologically vanishing coupling.

\begin{center}
\begin{tabular}{|c|c|c|c|c|c|}
\hline
CICY No.  & Model No. & Yukawa Pattern & Top. Van. & Sym. No.\\
\hline
\multirow{24}{*}{6947} &  1 &$\textbf{10}_{\textbf{e}_{\textbf{1}}}\bar{\textbf{5}}_{\textbf{e}_{\textbf{2}},\textbf{e}_{\textbf{4}}}\bar{\textbf{5}}_{\textbf{e}_{\textbf{3}},\textbf{e}_{\textbf{5}}}$ & y & 3\\
\cline{2-5}
   & 2 & $\textbf{10}_{\textbf{e}_{\textbf{1}}}\bar{\textbf{5}}_{\textbf{e}_{\textbf{2}},\textbf{e}_{\textbf{4}}}\bar{\textbf{5}}_{\textbf{e}_{\textbf{3}},\textbf{e}_{\textbf{5}}}$ & y & 3\\
\cline{2-5}
   & 3 & $\textbf{10}_{\textbf{e}_{\textbf{2}}}\bar{\textbf{5}}_{\textbf{e}_{\textbf{1}},\textbf{e}_{\textbf{3}}}\bar{\textbf{5}}_{\textbf{e}_{\textbf{4}},\textbf{e}_{\textbf{5}}}$ & y & 3\\
\cline{2-5}
   & 4 & $\textbf{10}_{\textbf{e}_{\textbf{2}}}\bar{\textbf{5}}_{\textbf{e}_{\textbf{1}},\textbf{e}_{\textbf{4}}}\bar{\textbf{5}}_{\textbf{e}_{\textbf{3}},\textbf{e}_{\textbf{5}}}$ & y & 3\\
\cline{2-5}
   & 5 & $\textbf{10}_{\textbf{e}_{\textbf{5}}}\bar{\textbf{5}}_{\textbf{e}_{\textbf{1}},\textbf{e}_{\textbf{4}}}\bar{\textbf{5}}_{\textbf{e}_{\textbf{2}},\textbf{e}_{\textbf{3}}}$ & y & 3\\
\cline{2-5}
   & 6 & $\textbf{10}_{\textbf{e}_{\textbf{5}}}\bar{\textbf{5}}_{\textbf{e}_{\textbf{1}},\textbf{e}_{\textbf{4}}}\bar{\textbf{5}}_{\textbf{e}_{\textbf{2}},\textbf{e}_{\textbf{3}}} $ & y & 3\\
\cline{2-5}
   & 7 & $\textbf{10}_{\textbf{e}_{\textbf{1}}}\bar{\textbf{5}}_{\textbf{e}_{\textbf{2}},\textbf{e}_{\textbf{4}}}\bar{\textbf{5}}_{\textbf{e}_{\textbf{3}},\textbf{e}_{\textbf{5}}}$ & y & 3\\
\cline{2-5}
   & 8 & $\textbf{10}_{\textbf{e}_{\textbf{1}}}\bar{\textbf{5}}_{\textbf{e}_{\textbf{2}},\textbf{e}_{\textbf{4}}}\bar{\textbf{5}}_{\textbf{e}_{\textbf{3}},\textbf{e}_{\textbf{5}}} $ & y & 3\\
\cline{2-5}
   & 9 & $\textbf{10}_{\textbf{e}_{\textbf{5}}}\bar{\textbf{5}}_{\textbf{e}_{\textbf{1}},\textbf{e}_{\textbf{3}}}\bar{\textbf{5}}_{\textbf{e}_{\textbf{2}},\textbf{e}_{\textbf{4}}}$ & y & 3\\
\cline{2-5}
   & 10 & $\textbf{10}_{\textbf{e}_{\textbf{5}}}\bar{\textbf{5}}_{\textbf{e}_{\textbf{1}},\textbf{e}_{\textbf{3}}}\bar{\textbf{5}}_{\textbf{e}_{\textbf{2}},\textbf{e}_{\textbf{4}}}$ & y & 3\\
\cline{2-5}
   & 11 & $\textbf{10}_{\textbf{e}_{\textbf{1}}}\bar{\textbf{5}}_{\textbf{e}_{\textbf{2}},\textbf{e}_{\textbf{3}}}\bar{\textbf{5}}_{\textbf{e}_{\textbf{4}},\textbf{e}_{\textbf{5}}}$ & y & 3\\
\cline{2-5}
   & 12 & $\textbf{10}_{\textbf{e}_{\textbf{1}}}\bar{\textbf{5}}_{\textbf{e}_{\textbf{2}},\textbf{e}_{\textbf{3}}}\bar{\textbf{5}}_{\textbf{e}_{\textbf{4}},\textbf{e}_{\textbf{5}}}$ & y & 3\\
\cline{2-5}
  & 13 & $\textbf{10}_{\textbf{e}_{\textbf{2}}}\bar{\textbf{5}}_{\textbf{e}_{\textbf{1}},\textbf{e}_{\textbf{4}}}\bar{\textbf{5}}_{\textbf{e}_{\textbf{3}},\textbf{e}_{\textbf{5}}} $ & y & 3\\
\cline{2-5}
   & 14 & $\textbf{10}_{\textbf{e}_{\textbf{2}}}\bar{\textbf{5}}_{\textbf{e}_{\textbf{1}},\textbf{e}_{\textbf{3}}}\bar{\textbf{5}}_{\textbf{e}_{\textbf{4}},\textbf{e}_{\textbf{5}}} $ & y & 3\\
\cline{2-5}
   & 15 & $\textbf{10}_{\textbf{e}_{\textbf{2}}}\bar{\textbf{5}}_{\textbf{e}_{\textbf{1}},\textbf{e}_{\textbf{3}}}\bar{\textbf{5}}_{\textbf{e}_{\textbf{4}},\textbf{e}_{\textbf{5}}} $ & y & 3\\
\cline{2-5}
   & 16 & $\textbf{10}_{\textbf{e}_{\textbf{5}}}\bar{\textbf{5}}_{\textbf{e}_{\textbf{1}},\textbf{e}_{\textbf{3}}}\bar{\textbf{5}}_{\textbf{e}_{\textbf{2}},\textbf{e}_{\textbf{4}}} $ & y & 3\\
\cline{2-5}
  & 17 & $\textbf{10}_{\textbf{e}_{\textbf{5}}}\bar{\textbf{5}}_{\textbf{e}_{\textbf{1}},\textbf{e}_{\textbf{3}}}\bar{\textbf{5}}_{\textbf{e}_{\textbf{2}},\textbf{e}_{\textbf{4}}} $ & y & 3\\
\cline{2-5}
   & 18 & $\textbf{10}_{\textbf{e}_{\textbf{2}}}\bar{\textbf{5}}_{\textbf{e}_{\textbf{1}},\textbf{e}_{\textbf{3}}}\bar{\textbf{5}}_{\textbf{e}_{\textbf{4}},\textbf{e}_{\textbf{5}}} $ & y & 3\\
\cline{2-5}
   & \multirow{2}{*}{19} & $\textbf{1}_{\textbf{e}_{\textbf{1}},-\textbf{e}_{\textbf{2}}}\textbf{5}_{-\textbf{e}_{\textbf{1}},-\textbf{e}_{\textbf{4}}}\bar{\textbf{5}}_{\textbf{e}_{\textbf{2}}, \textbf{e}_{\textbf{4}}}$& y & 3\\
\cline{3-5}
   &  & $\textbf{10}_{\textbf{e}_{\textbf{3}}}\bar{\textbf{5}}_{\textbf{e}_{\textbf{1}},\textbf{e}_{\textbf{5}}}\bar{\textbf{5}}_{\textbf{e}_{\textbf{2}},\textbf{e}_{\textbf{4}}}$ & y & 3\\
\cline{2-5}
   & \multirow{2}{*}{20} & $\textbf{1}_{\textbf{e}_{\textbf{1}},-\textbf{e}_{\textbf{2}}}\textbf{5}_{-\textbf{e}_{\textbf{1}},-\textbf{e}_{\textbf{4}}}\bar{\textbf{5}}_{\textbf{e}_{\textbf{2}}, \textbf{e}_{\textbf{4}}}$& y & 3\\
\cline{3-5}
   &  & $\textbf{10}_{\textbf{e}_{\textbf{3}}}\bar{\textbf{5}}_{\textbf{e}_{\textbf{1}},\textbf{e}_{\textbf{5}}}\bar{\textbf{5}}_{\textbf{e}_{\textbf{2}},\textbf{e}_{\textbf{4}}}$ & y & 3\\
\cline{2-5}
   & 21 & $\textbf{10}_{\textbf{e}_{\textbf{3}}}\bar{\textbf{5}}_{\textbf{e}_{\textbf{1}},\textbf{e}_{\textbf{4}}}\bar{\textbf{5}}_{\textbf{e}_{\textbf{2}},\textbf{e}_{\textbf{5}}}$ & y & 3\\
\cline{2-5}
   & 22 & $\textbf{10}_{\textbf{e}_{\textbf{3}}}\bar{\textbf{5}}_{\textbf{e}_{\textbf{1}},\textbf{e}_{\textbf{5}}}\bar{\textbf{5}}_{\textbf{e}_{\textbf{2}},\textbf{e}_{\textbf{4}}}$ & y & 3\\
\cline{2-5}
   & \multirow{2}{*}{23} & $\textbf{1}_{\textbf{e}_{\textbf{1}},-\textbf{e}_{\textbf{2}}}\textbf{5}_{-\textbf{e}_{\textbf{1}},-\textbf{e}_{\textbf{5}}}\bar{\textbf{5}}_{\textbf{e}_{\textbf{2}}, \textbf{e}_{\textbf{5}}}$& y & 3\\
\cline{3-5}
   &  & $\textbf{10}_{\textbf{e}_{\textbf{3}}}\bar{\textbf{5}}_{\textbf{e}_{\textbf{1}},\textbf{e}_{\textbf{4}}}\bar{\textbf{5}}_{\textbf{e}_{\textbf{2}},\textbf{e}_{\textbf{5}}} $ & y & 3\\
\cline{2-5}
   & \multirow{2}{*}{24} & $\textbf{1}_{\textbf{e}_{\textbf{1}},-\textbf{e}_{\textbf{2}}}\textbf{5}_{-\textbf{e}_{\textbf{1}},-\textbf{e}_{\textbf{5}}}\bar{\textbf{5}}_{\textbf{e}_{\textbf{2}}, \textbf{e}_{\textbf{5}}}$& y & 3\\
\cline{3-5}
   &  & $\textbf{10}_{\textbf{e}_{\textbf{3}}}\bar{\textbf{5}}_{\textbf{e}_{\textbf{1}},\textbf{e}_{\textbf{4}}}\bar{\textbf{5}}_{\textbf{e}_{\textbf{2}},\textbf{e}_{\textbf{5}}} $ & y & 3\\
\cline{2-5}
\hline
\end{tabular}
\end{center}
On CICY 6947, a total of 24 models have Yukawa couplings consistent with the gauge symmetries of the models, with 36 couplings being permitted in total. Of these, 24 are of the form $\textbf{10}_{\textbf{e}_{\textbf{i}}}\bar{\textbf{5}}_{\textbf{e}_{\textbf{j}},\textbf{e}_{\textbf{k}}}\bar{\textbf{5}}_{\textbf{e}_{\textbf{l}},\textbf{e}_{\textbf{m}}} $ and 12 are of the form $\textbf{1}_{\textbf{e}_{\textbf{i}},-\textbf{e}_{\textbf{j}}}\textbf{5}_{-\textbf{e}_{\textbf{i}},-\textbf{e}_{\textbf{k}}}\bar{\textbf{5}}_{\textbf{e}_{\textbf{j}}, \textbf{e}_{\textbf{k}}}$. All of these couplings exhibit topological vanishing.
\begin{center}
\begin{tabular}{|c|c|c|c|c|c|}
\hline
CICY No. & Model No. & Yukawa Pattern & Top. Van. & Sym. No.\\
\hline
\multirow{14}{*}{6732} & 1 &$\textbf{10}_{\textbf{e}_{\textbf{2}}}\bar{\textbf{5}}_{\textbf{e}_{\textbf{1}},\textbf{e}_{\textbf{3}}}\bar{\textbf{5}}_{\textbf{e}_{\textbf{4}},\textbf{e}_{\textbf{5}}} $ & n & 1-2\\
\cline{2-5}
   & 2 & $\textbf{10}_{\textbf{e}_{\textbf{2}}}\bar{\textbf{5}}_{\textbf{e}_{\textbf{1}},\textbf{e}_{\textbf{3}}}\bar{\textbf{5}}_{\textbf{e}_{\textbf{4}},\textbf{e}_{\textbf{5}}}$ & n & 1-2\\
\cline{2-5}
   & \multirow{5}{*}{5} & $\textbf{10}_{\textbf{e}_{\textbf{3}}}\bar{\textbf{5}}_{\textbf{e}_{\textbf{1}},\textbf{e}_{\textbf{2}}}\bar{\textbf{5}}_{\textbf{e}_{\textbf{4}},\textbf{e}_{\textbf{5}}} $ & n & 1-2\\
\cline{3-5}
   &  & $\textbf{10}_{\textbf{e}_{\textbf{4}}}\bar{\textbf{5}}_{\textbf{e}_{\textbf{1}},\textbf{e}_{\textbf{2}}}\bar{\textbf{5}}_{\textbf{e}_{\textbf{3}},\textbf{e}_{\textbf{5}}}$ & n & 1-2\\
\cline{3-5}
   &  & $\textbf{10}_{\textbf{e}_{\textbf{5}}}\bar{\textbf{5}}_{\textbf{e}_{\textbf{1}},\textbf{e}_{\textbf{2}}}\bar{\textbf{5}}_{\textbf{e}_{\textbf{3}},\textbf{e}_{\textbf{4}}} $ & n & 1-2\\
\cline{3-5}
   &  & $\textbf{5}_{-\textbf{e}_{\textbf{3}},-\textbf{e}_{\textbf{4}}}\textbf{10}_{\textbf{e}_{\textbf{3}}}\textbf{10}_{\textbf{e}_{\textbf{4}}}$ & n & 1-2\\
\cline{3-5}
   &  & $\textbf{5}_{-\textbf{e}_{\textbf{3}},-\textbf{e}_{\textbf{5}}}\textbf{10}_{\textbf{e}_{\textbf{3}}}\textbf{10}_{\textbf{e}_{\textbf{5}}}$ & n & 1-2\\
\cline{2-5}
   & \multirow{5}{*}{6} & $\textbf{10}_{\textbf{e}_{\textbf{3}}}\bar{\textbf{5}}_{\textbf{e}_{\textbf{1}},\textbf{e}_{\textbf{2}}}\bar{\textbf{5}}_{\textbf{e}_{\textbf{4}},\textbf{e}_{\textbf{5}}} $ & n & 1-2\\
\cline{3-5}
   &  & $\textbf{10}_{\textbf{e}_{\textbf{4}}}\bar{\textbf{5}}_{\textbf{e}_{\textbf{1}},\textbf{e}_{\textbf{2}}}\bar{\textbf{5}}_{\textbf{e}_{\textbf{3}},\textbf{e}_{\textbf{5}}}$ & n & 1-2\\
\cline{3-5}
   &  & $\textbf{10}_{\textbf{e}_{\textbf{5}}}\bar{\textbf{5}}_{\textbf{e}_{\textbf{1}},\textbf{e}_{\textbf{2}}}\bar{\textbf{5}}_{\textbf{e}_{\textbf{3}},\textbf{e}_{\textbf{4}}} $ & n & 1-2\\
\cline{3-5}
   &  & $\textbf{5}_{-\textbf{e}_{\textbf{3}},-\textbf{e}_{\textbf{4}}}\textbf{10}_{\textbf{e}_{\textbf{3}}}\textbf{10}_{\textbf{e}_{\textbf{4}}}$ & n & 1-2\\
\cline{3-5}
   &  & $\textbf{5}_{-\textbf{e}_{\textbf{3}},-\textbf{e}_{\textbf{5}}}\textbf{10}_{\textbf{e}_{\textbf{3}}}\textbf{10}_{\textbf{e}_{\textbf{5}}}$ & n & 1-2\\
\cline{2-5}
\hline
\end{tabular}
\end{center}
\begin{center}
\begin{tabular}{|c|c|c|c|c|c|}
\hline
CICY No. & Model No. & Yukawa Pattern & Top. Van. & Sym. No.\\
\hline
\multirow{14}{*}{6732} 
   &  \multirow{5}{*}{7} & $\textbf{10}_{\textbf{e}_{\textbf{3}}}\bar{\textbf{5}}_{\textbf{e}_{\textbf{1}},\textbf{e}_{\textbf{2}}}\bar{\textbf{5}}_{\textbf{e}_{\textbf{4}},\textbf{e}_{\textbf{5}}} $ & n & 1-2\\
\cline{3-5}
   &  & $\textbf{10}_{\textbf{e}_{\textbf{4}}}\bar{\textbf{5}}_{\textbf{e}_{\textbf{1}},\textbf{e}_{\textbf{2}}}\bar{\textbf{5}}_{\textbf{e}_{\textbf{3}},\textbf{e}_{\textbf{5}}}$ & n & 1-2\\
\cline{3-5}
   &  & $\textbf{10}_{\textbf{e}_{\textbf{5}}}\bar{\textbf{5}}_{\textbf{e}_{\textbf{1}},\textbf{e}_{\textbf{2}}}\bar{\textbf{5}}_{\textbf{e}_{\textbf{3}},\textbf{e}_{\textbf{4}}} $ & n & 1-2\\
\cline{3-5}
   &  & $\textbf{5}_{-\textbf{e}_{\textbf{3}},-\textbf{e}_{\textbf{4}}}\textbf{10}_{\textbf{e}_{\textbf{3}}}\textbf{10}_{\textbf{e}_{\textbf{4}}}$ & n & 1-2\\
\cline{3-5}
   &  & $\textbf{5}_{-\textbf{e}_{\textbf{3}},-\textbf{e}_{\textbf{5}}}\textbf{10}_{\textbf{e}_{\textbf{3}}}\textbf{10}_{\textbf{e}_{\textbf{5}}}$ & n & 1-2\\
\cline{2-5}
   & \multirow{5}{*}{8} & $\textbf{10}_{\textbf{e}_{\textbf{3}}}\bar{\textbf{5}}_{\textbf{e}_{\textbf{1}},\textbf{e}_{\textbf{2}}}\bar{\textbf{5}}_{\textbf{e}_{\textbf{4}},\textbf{e}_{\textbf{5}}} $ & n & 1-2\\
\cline{3-5}
   &  & $\textbf{10}_{\textbf{e}_{\textbf{4}}}\bar{\textbf{5}}_{\textbf{e}_{\textbf{1}},\textbf{e}_{\textbf{2}}}\bar{\textbf{5}}_{\textbf{e}_{\textbf{3}},\textbf{e}_{\textbf{5}}}$ & n & 1-2\\
\cline{3-5}
   &  & $\textbf{10}_{\textbf{e}_{\textbf{5}}}\bar{\textbf{5}}_{\textbf{e}_{\textbf{1}},\textbf{e}_{\textbf{2}}}\bar{\textbf{5}}_{\textbf{e}_{\textbf{3}},\textbf{e}_{\textbf{4}}} $ & n & 1-2\\
\cline{3-5}
   &  & $\textbf{5}_{-\textbf{e}_{\textbf{3}},-\textbf{e}_{\textbf{4}}}\textbf{10}_{\textbf{e}_{\textbf{3}}}\textbf{10}_{\textbf{e}_{\textbf{4}}}$ & n & 1-2\\
\cline{3-5}
   &  & $\textbf{5}_{-\textbf{e}_{\textbf{3}},-\textbf{e}_{\textbf{5}}}\textbf{10}_{\textbf{e}_{\textbf{3}}}\textbf{10}_{\textbf{e}_{\textbf{5}}}$ & n & 1-2\\
\cline{2-5}
   & 18 & $\textbf{5}_{-\textbf{e}_{\textbf{3}},-\textbf{e}_{\textbf{5}}}\textbf{10}_{\textbf{e}_{\textbf{3}}}\textbf{10}_{\textbf{e}_{\textbf{5}}}$ & n & 1-2\\
\cline{2-5}
  & {29} & $\textbf{5}_{-\textbf{e}_{\textbf{2}},-\textbf{e}_{\textbf{5}}}\textbf{10}_{\textbf{e}_{\textbf{2}}}\textbf{10}_{\textbf{e}_{\textbf{5}}} $& n & 1-2\\
\cline{2-5}
   & 30 & $\textbf{10}_{\textbf{e}_{\textbf{3}}}\bar{\textbf{5}}_{\textbf{e}_{\textbf{1}},\textbf{e}_{\textbf{4}}}\bar{\textbf{5}}_{\textbf{e}_{\textbf{2}},\textbf{e}_{\textbf{5}}}$ & n & 1-2\\
\cline{2-5}
   & 31 & $\textbf{10}_{\textbf{e}_{\textbf{3}}}\bar{\textbf{5}}_{\textbf{e}_{\textbf{1}},\textbf{e}_{\textbf{4}}}\bar{\textbf{5}}_{\textbf{e}_{\textbf{2}},\textbf{e}_{\textbf{5}}}$ & n & 1-2\\
\cline{2-5}
   & 32 & $\textbf{10}_{\textbf{e}_{\textbf{3}}}\bar{\textbf{5}}_{\textbf{e}_{\textbf{1}},\textbf{e}_{\textbf{2}}}\bar{\textbf{5}}_{\textbf{e}_{\textbf{4}},\textbf{e}_{\textbf{5}}} $ & n & 1-2\\
\cline{2-5}
   & 34 & $\textbf{1}_{\textbf{e}_{\textbf{1}},-\textbf{e}_{\textbf{3}}}\textbf{5}_{-\textbf{e}_{\textbf{1}},-\textbf{e}_{\textbf{5}}}\bar{\textbf{5}}_{\textbf{e}_{\textbf{3}}, \textbf{e}_{\textbf{5}}} $ & n & 1-2\\
\cline{2-5}
   & 35 & $\textbf{1}_{\textbf{e}_{\textbf{1}},-\textbf{e}_{\textbf{3}}}\textbf{5}_{-\textbf{e}_{\textbf{1}},-\textbf{e}_{\textbf{5}}}\bar{\textbf{5}}_{\textbf{e}_{\textbf{3}}, \textbf{e}_{\textbf{5}}} $ & n & 1-2\\
\cline{2-5}
   & 36 & $\textbf{10}_{\textbf{e}_{\textbf{2}}}\bar{\textbf{5}}_{\textbf{e}_{\textbf{1}},\textbf{e}_{\textbf{5}}}\bar{\textbf{5}}_{\textbf{e}_{\textbf{3}},\textbf{e}_{\textbf{4}}} $ & n & 1-2\\
\cline{2-5}
\hline
\end{tabular}
\end{center}
On CICY 6732, a total of 28 models have Yukawa couplings consistent with the gauge symmetries of the models, with 104 couplings being permitted in total. Of these, 68 are of the form $\textbf{10}_{\textbf{e}_{\textbf{i}}}\bar{\textbf{5}}_{\textbf{e}_{\textbf{j}},\textbf{e}_{\textbf{k}}}\bar{\textbf{5}}_{\textbf{e}_{\textbf{l}},\textbf{e}_{\textbf{m}}} $, 12 are of the form $\textbf{1}_{\textbf{e}_{\textbf{i}},-\textbf{e}_{\textbf{j}}}\textbf{5}_{-\textbf{e}_{\textbf{i}},-\textbf{e}_{\textbf{k}}}\bar{\textbf{5}}_{\textbf{e}_{\textbf{j}}, \textbf{e}_{\textbf{k}}}$ and 24 are of the form $ \bold{5}_{\bold{{-{\bf e}_i}},\bold{-{{\bf e}_j}}}\bold{10}_{\bold{{{\bf e}_i}}}\bold{10}_{\bold{{{\bf e}_j}}}$. None of these couplings exhibit the topological vanishings we have studied here.

\begin{center}
\begin{tabular}{|c|c|c|c|c|c|}
\hline
CICY No.  & Model No. & Yukawa Pattern & Top. Van. & Sym. No.\\
\hline
\multirow{8}{*}{6770} & {1} & $\textbf{5}_{-\textbf{e}_{\textbf{1}},-\textbf{e}_{\textbf{2}}}\textbf{10}_{\textbf{e}_{\textbf{1}}}\textbf{10}_{\textbf{e}_{\textbf{2}}}$ & y & 1-2\\
\cline{2-5}
   & 2 & $\textbf{5}_{-\textbf{e}_{\textbf{1}},-\textbf{e}_{\textbf{2}}}\textbf{10}_{\textbf{e}_{\textbf{1}}}\textbf{10}_{\textbf{e}_{\textbf{2}}}$ & y & 1-2\\
\cline{2-5}
 & {5} & $\textbf{5}_{-\textbf{e}_{\textbf{1}},-\textbf{e}_{\textbf{2}}}\textbf{10}_{\textbf{e}_{\textbf{1}}}\textbf{10}_{\textbf{e}_{\textbf{2}}} $ & y & 1-2\\
\cline{2-5}
   & 6 & $\textbf{5}_{-\textbf{e}_{\textbf{1}},-\textbf{e}_{\textbf{2}}}\textbf{10}_{\textbf{e}_{\textbf{1}}}\textbf{10}_{\textbf{e}_{\textbf{2}}} $& y & 1-2\\
\cline{2-5}
 & {7} & $\textbf{10}_{\textbf{e}_{\textbf{2}}}\bar{\textbf{5}}_{\textbf{e}_{\textbf{1}},\textbf{e}_{\textbf{4}}}\bar{\textbf{5}}_{\textbf{e}_{\textbf{3}},\textbf{e}_{\textbf{5}}} $ & n & 1-2\\
\cline{2-5}
   & 8 & $\textbf{10}_{\textbf{e}_{\textbf{2}}}\bar{\textbf{5}}_{\textbf{e}_{\textbf{1}},\textbf{e}_{\textbf{4}}}\bar{\textbf{5}}_{\textbf{e}_{\textbf{3}},\textbf{e}_{\textbf{5}}} $ & n & 1-2\\
\cline{2-5}
 & {11} & $\textbf{10}_{\textbf{e}_{\textbf{1}}}\bar{\textbf{5}}_{\textbf{e}_{\textbf{2}},\textbf{e}_{\textbf{5}}}\bar{\textbf{5}}_{\textbf{e}_{\textbf{3}},\textbf{e}_{\textbf{4}}}  $ & n & 1-2\\
\cline{2-5}
   & 12 & $\textbf{10}_{\textbf{e}_{\textbf{1}}}\bar{\textbf{5}}_{\textbf{e}_{\textbf{2}},\textbf{e}_{\textbf{5}}}\bar{\textbf{5}}_{\textbf{e}_{\textbf{3}},\textbf{e}_{\textbf{4}}}  $ & n & 1-2\\
\cline{2-5}
\hline
\end{tabular}
\end{center}
On CICY 6770, a total of 16 models have Yukawa couplings consistent with the gauge symmetries of the models, with 48 couplings being permitted in total. Of these, 32 are of the form $\textbf{10}_{\textbf{e}_{\textbf{i}}}\bar{\textbf{5}}_{\textbf{e}_{\textbf{j}},\textbf{e}_{\textbf{k}}}\bar{\textbf{5}}_{\textbf{e}_{\textbf{l}},\textbf{e}_{\textbf{m}}} $ and 16 are of the form $ \bold{5}_{\bold{{-{\bf e}_i}},\bold{-{{\bf e}_j}}}\bold{10}_{\bold{{{\bf e}_i}}}\bold{10}_{\bold{{{\bf e}_j}}}$. 
There are a total of 16 couplings that exhibit the topological vanishing we have been studying here. All of these are of the form $ \bold{5}_{\bold{{-{\bf e}_i}},\bold{-{{\bf e}_j}}}\bold{10}_{\bold{{{\bf e}_i}}}\bold{10}_{\bold{{{\bf e}_j}}}$. A total of 8 models have at least one coupling which exhibits this topological vanishing.


\begin{center}
\begin{tabular}{|c|c|c|c|c|c|}
\hline
CICY No.  & Model No. & Yukawa Pattern & Top. Van. & Sym. No.\\
\hline
\multirow{4}{*}{6777} &  \multirow{7}{*}{1} & $\textbf{1}_{\textbf{e}_{\textbf{2}},-\textbf{e}_{\textbf{3}}}\textbf{5}_{-\textbf{e}_{\textbf{2}},-\textbf{e}_{\textbf{5}}}\bar{\textbf{5}}_{\textbf{e}_{\textbf{3}}, \textbf{e}_{\textbf{5}}}  $ & n & 1-4\\
\cline{3-5}
   &  &$\textbf{1}_{\textbf{e}_{\textbf{3}},-\textbf{e}_{\textbf{2}}}\textbf{5}_{-\textbf{e}_{\textbf{3}},-\textbf{e}_{\textbf{5}}}\bar{\textbf{5}}_{\textbf{e}_{\textbf{2}}, \textbf{e}_{\textbf{5}}} $ & n & 1-4\\
\cline{3-5}
   &  & $\textbf{10}_{\textbf{e}_{\textbf{2}}}\bar{\textbf{5}}_{\textbf{e}_{\textbf{1}},\textbf{e}_{\textbf{4}}}\bar{\textbf{5}}_{\textbf{e}_{\textbf{3}},\textbf{e}_{\textbf{5}}} $ & n & 1-4\\
\cline{3-5}
 &  & $\textbf{10}_{\textbf{e}_{\textbf{3}}}\bar{\textbf{5}}_{\textbf{e}_{\textbf{1}},\textbf{e}_{\textbf{4}}}\bar{\textbf{5}}_{\textbf{e}_{\textbf{2}},\textbf{e}_{\textbf{5}}}  $ & n & 1-4\\
\cline{3-5}
   &  & $\textbf{10}_{\textbf{e}_{\textbf{5}}}\bar{\textbf{5}}_{\textbf{e}_{\textbf{1}},\textbf{e}_{\textbf{4}}}\bar{\textbf{5}}_{\textbf{e}_{\textbf{2}},\textbf{e}_{\textbf{3}}} $& n & 1-4\\
\cline{3-5}
 &  & $\textbf{5}_{-\textbf{e}_{\textbf{2}},-\textbf{e}_{\textbf{5}}}\textbf{10}_{\textbf{e}_{\textbf{2}}}\textbf{10}_{\textbf{e}_{\textbf{5}}}$  & n & 1-4\\
\cline{3-5}
   &  & $\textbf{5}_{-\textbf{e}_{\textbf{3}},-\textbf{e}_{\textbf{5}}}\textbf{10}_{\textbf{e}_{\textbf{3}}}\textbf{10}_{\textbf{e}_{\textbf{5}}}$ & n & 1-4\\
\cline{2-5}
\hline
\end{tabular}
\end{center}
\begin{center}
\begin{tabular}{|c|c|c|c|c|c|}
\hline
CICY No.  & Model No. & Yukawa Pattern & Top. Van. & Sym. No.\\
\hline
\multirow{16}{*}{6777} 
 &   \multirow{7}{*}{2} & $\textbf{1}_{\textbf{e}_{\textbf{2}},-\textbf{e}_{\textbf{3}}}\textbf{5}_{-\textbf{e}_{\textbf{2}},-\textbf{e}_{\textbf{5}}}\bar{\textbf{5}}_{\textbf{e}_{\textbf{3}}, \textbf{e}_{\textbf{5}}}  $ & n & 1-4\\
\cline{3-5}
   &  &$\textbf{1}_{\textbf{e}_{\textbf{3}},-\textbf{e}_{\textbf{2}}}\textbf{5}_{-\textbf{e}_{\textbf{3}},-\textbf{e}_{\textbf{5}}}\bar{\textbf{5}}_{\textbf{e}_{\textbf{2}}, \textbf{e}_{\textbf{5}}} $ & n & 1-4\\
\cline{3-5}
   &  & $\textbf{10}_{\textbf{e}_{\textbf{2}}}\bar{\textbf{5}}_{\textbf{e}_{\textbf{1}},\textbf{e}_{\textbf{4}}}\bar{\textbf{5}}_{\textbf{e}_{\textbf{3}},\textbf{e}_{\textbf{5}}} $ & n & 1-4\\
\cline{3-5}
 &  & $\textbf{10}_{\textbf{e}_{\textbf{3}}}\bar{\textbf{5}}_{\textbf{e}_{\textbf{1}},\textbf{e}_{\textbf{4}}}\bar{\textbf{5}}_{\textbf{e}_{\textbf{2}},\textbf{e}_{\textbf{5}}}  $ & n & 1-4\\
\cline{3-5}
   &  & $\textbf{10}_{\textbf{e}_{\textbf{5}}}\bar{\textbf{5}}_{\textbf{e}_{\textbf{1}},\textbf{e}_{\textbf{4}}}\bar{\textbf{5}}_{\textbf{e}_{\textbf{2}},\textbf{e}_{\textbf{3}}} $& n & 1-4\\
\cline{3-5}
 &  & $\textbf{5}_{-\textbf{e}_{\textbf{2}},-\textbf{e}_{\textbf{5}}}\textbf{10}_{\textbf{e}_{\textbf{2}}}\textbf{10}_{\textbf{e}_{\textbf{5}}}$  & n & 1-4\\
\cline{3-5}
   &  & $\textbf{5}_{-\textbf{e}_{\textbf{3}},-\textbf{e}_{\textbf{5}}}\textbf{10}_{\textbf{e}_{\textbf{3}}}\textbf{10}_{\textbf{e}_{\textbf{5}}}$ & n & 1-4\\
\cline{2-5}
 &  \multirow{7}{*}{3} & $\textbf{1}_{\textbf{e}_{\textbf{2}},-\textbf{e}_{\textbf{3}}}\textbf{5}_{-\textbf{e}_{\textbf{2}},-\textbf{e}_{\textbf{5}}}\bar{\textbf{5}}_{\textbf{e}_{\textbf{3}}, \textbf{e}_{\textbf{5}}}  $ & n & 1-4\\
\cline{3-5}
   &  &$\textbf{1}_{\textbf{e}_{\textbf{3}},-\textbf{e}_{\textbf{2}}}\textbf{5}_{-\textbf{e}_{\textbf{3}},-\textbf{e}_{\textbf{5}}}\bar{\textbf{5}}_{\textbf{e}_{\textbf{2}}, \textbf{e}_{\textbf{5}}} $ & n & 1-4\\
\cline{3-5}
   &  & $\textbf{10}_{\textbf{e}_{\textbf{2}}}\bar{\textbf{5}}_{\textbf{e}_{\textbf{1}},\textbf{e}_{\textbf{4}}}\bar{\textbf{5}}_{\textbf{e}_{\textbf{3}},\textbf{e}_{\textbf{5}}} $ & n & 1-4\\
\cline{3-5}
 &  & $\textbf{10}_{\textbf{e}_{\textbf{3}}}\bar{\textbf{5}}_{\textbf{e}_{\textbf{1}},\textbf{e}_{\textbf{4}}}\bar{\textbf{5}}_{\textbf{e}_{\textbf{2}},\textbf{e}_{\textbf{5}}}  $ & n & 1-4\\
\cline{3-5}
   &  & $\textbf{10}_{\textbf{e}_{\textbf{5}}}\bar{\textbf{5}}_{\textbf{e}_{\textbf{1}},\textbf{e}_{\textbf{4}}}\bar{\textbf{5}}_{\textbf{e}_{\textbf{2}},\textbf{e}_{\textbf{3}}} $& n & 1-4\\
\cline{3-5}
 &  & $\textbf{5}_{-\textbf{e}_{\textbf{2}},-\textbf{e}_{\textbf{5}}}\textbf{10}_{\textbf{e}_{\textbf{2}}}\textbf{10}_{\textbf{e}_{\textbf{5}}}$  & n & 1-4\\
\cline{3-5}
   &  & $\textbf{5}_{-\textbf{e}_{\textbf{3}},-\textbf{e}_{\textbf{5}}}\textbf{10}_{\textbf{e}_{\textbf{3}}}\textbf{10}_{\textbf{e}_{\textbf{5}}}$ & n & 1-4\\
\cline{2-5}
 &  \multirow{7}{*}{4} & $\textbf{1}_{\textbf{e}_{\textbf{2}},-\textbf{e}_{\textbf{3}}}\textbf{5}_{-\textbf{e}_{\textbf{2}},-\textbf{e}_{\textbf{5}}}\bar{\textbf{5}}_{\textbf{e}_{\textbf{3}}, \textbf{e}_{\textbf{5}}}  $ & n & 1-4\\
\cline{3-5}
   &  &$\textbf{1}_{\textbf{e}_{\textbf{3}},-\textbf{e}_{\textbf{2}}}\textbf{5}_{-\textbf{e}_{\textbf{3}},-\textbf{e}_{\textbf{5}}}\bar{\textbf{5}}_{\textbf{e}_{\textbf{2}}, \textbf{e}_{\textbf{5}}} $ & n & 1-4\\
\cline{3-5}
   &  & $\textbf{10}_{\textbf{e}_{\textbf{2}}}\bar{\textbf{5}}_{\textbf{e}_{\textbf{1}},\textbf{e}_{\textbf{4}}}\bar{\textbf{5}}_{\textbf{e}_{\textbf{3}},\textbf{e}_{\textbf{5}}} $ & n & 1-4\\
\cline{3-5}
 &  & $\textbf{10}_{\textbf{e}_{\textbf{3}}}\bar{\textbf{5}}_{\textbf{e}_{\textbf{1}},\textbf{e}_{\textbf{4}}}\bar{\textbf{5}}_{\textbf{e}_{\textbf{2}},\textbf{e}_{\textbf{5}}}  $ & n & 1-4\\
\cline{3-5}
   &  & $\textbf{10}_{\textbf{e}_{\textbf{5}}}\bar{\textbf{5}}_{\textbf{e}_{\textbf{1}},\textbf{e}_{\textbf{4}}}\bar{\textbf{5}}_{\textbf{e}_{\textbf{2}},\textbf{e}_{\textbf{3}}} $& n & 1-4\\
\cline{3-5}
 &  & $\textbf{5}_{-\textbf{e}_{\textbf{2}},-\textbf{e}_{\textbf{5}}}\textbf{10}_{\textbf{e}_{\textbf{2}}}\textbf{10}_{\textbf{e}_{\textbf{5}}}$  & n & 1-4\\
\cline{3-5}
   &  & $\textbf{5}_{-\textbf{e}_{\textbf{3}},-\textbf{e}_{\textbf{5}}}\textbf{10}_{\textbf{e}_{\textbf{3}}}\textbf{10}_{\textbf{e}_{\textbf{5}}}$ & n & 1-4\\
\cline{2-5}
 & \multirow{2}{*}{13} & $\textbf{1}_{\textbf{e}_{\textbf{2}},-\textbf{e}_{\textbf{4}}}\textbf{5}_{-\textbf{e}_{\textbf{2}},-\textbf{e}_{\textbf{5}}}\bar{\textbf{5}}_{\textbf{e}_{\textbf{4}}, \textbf{e}_{\textbf{5}}}$ & n & 1-4\\
\cline{3-5}
 &  & $\textbf{5}_{-\textbf{e}_{\textbf{2}},-\textbf{e}_{\textbf{5}}}\textbf{10}_{\textbf{e}_{\textbf{2}}}\textbf{10}_{\textbf{e}_{\textbf{5}}}$ & n & 1-4\\
\cline{2-5}
 & \multirow{2}{*}{14} & $\textbf{1}_{\textbf{e}_{\textbf{2}},-\textbf{e}_{\textbf{4}}}\textbf{5}_{-\textbf{e}_{\textbf{2}},-\textbf{e}_{\textbf{5}}}\bar{\textbf{5}}_{\textbf{e}_{\textbf{4}}, \textbf{e}_{\textbf{5}}}$ & n & 1-4\\
\cline{3-5}
 &  & $\textbf{5}_{-\textbf{e}_{\textbf{2}},-\textbf{e}_{\textbf{5}}}\textbf{10}_{\textbf{e}_{\textbf{2}}}\textbf{10}_{\textbf{e}_{\textbf{5}}}$ & n & 1-4\\
\cline{2-5}\hline
\end{tabular}
\end{center}
On CICY 6777, a total of 24 models have Yukawa couplings consistent with the gauge symmetries of the models, with 192 Yukawa couplings being permitted in total. Of these, 64 are of the form $\textbf{10}_{\textbf{e}_{\textbf{i}}}\bar{\textbf{5}}_{\textbf{e}_{\textbf{j}},\textbf{e}_{\textbf{k}}}\bar{\textbf{5}}_{\textbf{e}_{\textbf{l}},\textbf{e}_{\textbf{m}}} $, 80 are of the form $\textbf{1}_{\textbf{e}_{\textbf{i}},-\textbf{e}_{\textbf{j}}}\textbf{5}_{-\textbf{e}_{\textbf{i}},-\textbf{e}_{\textbf{k}}}\bar{\textbf{5}}_{\textbf{e}_{\textbf{j}}, \textbf{e}_{\textbf{k}}}$ and 48 are of the form $ \bold{5}_{\bold{{-{\bf e}_i}},\bold{-{{\bf e}_j}}}\bold{10}_{\bold{{{\bf e}_i}}}\bold{10}_{\bold{{{\bf e}_j}}}$. None of these couplings exhibit the topological vanishings we have studied here.


\begin{center}
\begin{tabular}{|c|c|c|c|c|}
\hline
CICY No.  & Model No. & Yukawa Pattern & Top. Van. & Sym. No.\\
\hline
\multirow{12}{*}{6890}  & 1& $\textbf{10}_{\textbf{e}_{\textbf{3}}}\bar{\textbf{5}}_{\textbf{e}_{\textbf{1}},\textbf{e}_{\textbf{2}}}\bar{\textbf{5}}_{\textbf{e}_{\textbf{4}},\textbf{e}_{\textbf{5}}} $ & n & 1-2\\
\cline{2-5}
   & 2 & $\textbf{10}_{\textbf{e}_{\textbf{3}}}\bar{\textbf{5}}_{\textbf{e}_{\textbf{1}},\textbf{e}_{\textbf{2}}}\bar{\textbf{5}}_{\textbf{e}_{\textbf{4}},\textbf{e}_{\textbf{5}}}  $ & n & 1-2\\
\cline{2-5}
 & \multirow{5}{*} 6 & $\textbf{10}_{\textbf{e}_{\textbf{3}}}\bar{\textbf{5}}_{\textbf{e}_{\textbf{1}},\textbf{e}_{\textbf{2}}}\bar{\textbf{5}}_{\textbf{e}_{\textbf{4}},\textbf{e}_{\textbf{5}}}  $ & n & 1-2\\
\cline{3-5}
   &  & $\textbf{10}_{\textbf{e}_{\textbf{4}}}\bar{\textbf{5}}_{\textbf{e}_{\textbf{1}},\textbf{e}_{\textbf{2}}}\bar{\textbf{5}}_{\textbf{e}_{\textbf{3}},\textbf{e}_{\textbf{5}}} $& n & 1-2\\
\cline{3-5}
 &  & $\textbf{10}_{\textbf{e}_{\textbf{5}}}\bar{\textbf{5}}_{\textbf{e}_{\textbf{1}},\textbf{e}_{\textbf{2}}}\bar{\textbf{5}}_{\textbf{e}_{\textbf{3}},\textbf{e}_{\textbf{4}}}  $ & n & 1-2\\
\cline{3-5}
   &  & $\textbf{5}_{-\textbf{e}_{\textbf{3}},-\textbf{e}_{\textbf{4}}}\textbf{10}_{\textbf{e}_{\textbf{3}}}\textbf{10}_{\textbf{e}_{\textbf{4}}}$ & n & 1-2\\
\cline{3-5}
 &  & $\textbf{5}_{-\textbf{e}_{\textbf{4}},-\textbf{e}_{\textbf{5}}}\textbf{10}_{\textbf{e}_{\textbf{4}}}\textbf{10}_{\textbf{e}_{\textbf{5}}} $ & n & 1-2\\
\cline{2-5}
 &\multirow{5}{*} 7 & $\textbf{10}_{\textbf{e}_{\textbf{3}}}\bar{\textbf{5}}_{\textbf{e}_{\textbf{1}},\textbf{e}_{\textbf{2}}}\bar{\textbf{5}}_{\textbf{e}_{\textbf{4}},\textbf{e}_{\textbf{5}}}  $ & n & 1-2\\
\cline{3-5}
   &  & $\textbf{10}_{\textbf{e}_{\textbf{4}}}\bar{\textbf{5}}_{\textbf{e}_{\textbf{1}},\textbf{e}_{\textbf{2}}}\bar{\textbf{5}}_{\textbf{e}_{\textbf{3}},\textbf{e}_{\textbf{5}}} $& n & 1-2\\
\cline{3-5}
 &  & $\textbf{10}_{\textbf{e}_{\textbf{5}}}\bar{\textbf{5}}_{\textbf{e}_{\textbf{1}},\textbf{e}_{\textbf{2}}}\bar{\textbf{5}}_{\textbf{e}_{\textbf{3}},\textbf{e}_{\textbf{4}}}  $ & n & 1-2\\
\cline{3-5}
   &  & $\textbf{5}_{-\textbf{e}_{\textbf{3}},-\textbf{e}_{\textbf{4}}}\textbf{10}_{\textbf{e}_{\textbf{3}}}\textbf{10}_{\textbf{e}_{\textbf{4}}}$ & n & 1-2\\
\cline{3-5}
 &  & $\textbf{5}_{-\textbf{e}_{\textbf{4}},-\textbf{e}_{\textbf{5}}}\textbf{10}_{\textbf{e}_{\textbf{4}}}\textbf{10}_{\textbf{e}_{\textbf{5}}} $ & n & 1-2\\
\cline{2-5}
\hline
\end{tabular}
\end{center}
\begin{center}
\begin{tabular}{|c|c|c|c|c|}
\hline
CICY No.  & Model No. & Yukawa Pattern & Top. Van. & Sym. No.\\
\hline
\multirow{12}{*}{6890} 
 & \multirow{5}{*} 8 & $\textbf{10}_{\textbf{e}_{\textbf{3}}}\bar{\textbf{5}}_{\textbf{e}_{\textbf{1}},\textbf{e}_{\textbf{2}}}\bar{\textbf{5}}_{\textbf{e}_{\textbf{4}},\textbf{e}_{\textbf{5}}}  $ & n & 1-2\\
\cline{3-5}
   &  & $\textbf{10}_{\textbf{e}_{\textbf{4}}}\bar{\textbf{5}}_{\textbf{e}_{\textbf{1}},\textbf{e}_{\textbf{2}}}\bar{\textbf{5}}_{\textbf{e}_{\textbf{3}},\textbf{e}_{\textbf{5}}} $& n & 1-2\\
\cline{3-5}
 &  & $\textbf{10}_{\textbf{e}_{\textbf{5}}}\bar{\textbf{5}}_{\textbf{e}_{\textbf{1}},\textbf{e}_{\textbf{2}}}\bar{\textbf{5}}_{\textbf{e}_{\textbf{3}},\textbf{e}_{\textbf{4}}}  $ & n & 1-2\\
\cline{3-5}
   &  & $\textbf{5}_{-\textbf{e}_{\textbf{3}},-\textbf{e}_{\textbf{4}}}\textbf{10}_{\textbf{e}_{\textbf{3}}}\textbf{10}_{\textbf{e}_{\textbf{4}}}$ & n & 1-2\\
\cline{3-5}
 &  & $\textbf{5}_{-\textbf{e}_{\textbf{4}},-\textbf{e}_{\textbf{5}}}\textbf{10}_{\textbf{e}_{\textbf{4}}}\textbf{10}_{\textbf{e}_{\textbf{5}}} $ & n & 1-2\\
\cline{2-5}
 & \multirow{5}{*} 9 & $\textbf{10}_{\textbf{e}_{\textbf{3}}}\bar{\textbf{5}}_{\textbf{e}_{\textbf{1}},\textbf{e}_{\textbf{2}}}\bar{\textbf{5}}_{\textbf{e}_{\textbf{4}},\textbf{e}_{\textbf{5}}}  $ & n & 1-2\\
\cline{3-5}
   &  & $\textbf{10}_{\textbf{e}_{\textbf{4}}}\bar{\textbf{5}}_{\textbf{e}_{\textbf{1}},\textbf{e}_{\textbf{2}}}\bar{\textbf{5}}_{\textbf{e}_{\textbf{3}},\textbf{e}_{\textbf{5}}} $& n & 1-2\\
\cline{3-5}
 &  & $\textbf{10}_{\textbf{e}_{\textbf{5}}}\bar{\textbf{5}}_{\textbf{e}_{\textbf{1}},\textbf{e}_{\textbf{2}}}\bar{\textbf{5}}_{\textbf{e}_{\textbf{3}},\textbf{e}_{\textbf{4}}}  $ & n & 1-2\\
\cline{3-5}
   &  & $\textbf{5}_{-\textbf{e}_{\textbf{3}},-\textbf{e}_{\textbf{4}}}\textbf{10}_{\textbf{e}_{\textbf{3}}}\textbf{10}_{\textbf{e}_{\textbf{4}}}$ & n & 1-2\\
\cline{3-5}
 &  & $\textbf{5}_{-\textbf{e}_{\textbf{4}},-\textbf{e}_{\textbf{5}}}\textbf{10}_{\textbf{e}_{\textbf{4}}}\textbf{10}_{\textbf{e}_{\textbf{5}}} $ & n & 1-2\\
\cline{2-5}
   & 14 & $\textbf{5}_{-\textbf{e}_{\textbf{4}},-\textbf{e}_{\textbf{5}}}\textbf{10}_{\textbf{e}_{\textbf{4}}}\textbf{10}_{\textbf{e}_{\textbf{5}}} $& n & 1-2\\
\cline{2-5}
   & 16 & $\textbf{1}_{\textbf{e}_{\textbf{1}},-\textbf{e}_{\textbf{3}}}\textbf{5}_{-\textbf{e}_{\textbf{1}},-\textbf{e}_{\textbf{5}}}\bar{\textbf{5}}_{\textbf{e}_{\textbf{3}}, \textbf{e}_{\textbf{5}}} $& n & 1-2\\
\cline{2-5}
   & 17 & $\textbf{1}_{\textbf{e}_{\textbf{1}},-\textbf{e}_{\textbf{3}}}\textbf{5}_{-\textbf{e}_{\textbf{1}},-\textbf{e}_{\textbf{5}}}\bar{\textbf{5}}_{\textbf{e}_{\textbf{3}}, \textbf{e}_{\textbf{5}}}$& n & 1-2\\
\cline{2-5}
   & 23 & $\textbf{5}_{-\textbf{e}_{\textbf{2}},-\textbf{e}_{\textbf{4}}}\textbf{10}_{\textbf{e}_{\textbf{2}}}\textbf{10}_{\textbf{e}_{\textbf{4}}} $& n & 1-2\\
\cline{2-5}
   & 28 & $\textbf{10}_{\textbf{e}_{\textbf{2}}}\bar{\textbf{5}}_{\textbf{e}_{\textbf{1}},\textbf{e}_{\textbf{5}}}\bar{\textbf{5}}_{\textbf{e}_{\textbf{3}},\textbf{e}_{\textbf{4}}}$& n & 1-2\\
\cline{2-5}
\hline
\end{tabular}
\end{center}
On CICY 6777, a total of 22 models have Yukawa couplings consistent with the gauge symmetries of the models, with 86 Yukawa couplings being permitted in total. Of these, 50 are of the form $\textbf{10}_{\textbf{e}_{\textbf{i}}}\bar{\textbf{5}}_{\textbf{e}_{\textbf{j}},\textbf{e}_{\textbf{k}}}\bar{\textbf{5}}_{\textbf{e}_{\textbf{l}},\textbf{e}_{\textbf{m}}} $, 12 are of the form $\textbf{1}_{\textbf{e}_{\textbf{i}},-\textbf{e}_{\textbf{j}}}\textbf{5}_{-\textbf{e}_{\textbf{i}},-\textbf{e}_{\textbf{k}}}\bar{\textbf{5}}_{\textbf{e}_{\textbf{j}}, \textbf{e}_{\textbf{k}}}$ and 24 are of the form $ \bold{5}_{\bold{{-{\bf e}_i}},\bold{-{{\bf e}_j}}}\bold{10}_{\bold{{{\bf e}_i}}}\bold{10}_{\bold{{{\bf e}_j}}}$. None of these couplings exhibit the topological vanishings we have studied here.
\begin{center}
\begin{tabular}{|c|c|c|c|c|c|}
\hline
CICY No.  & Model No. & Yukawa Pattern & Top. Van. & Sym. No.\\
\hline
\multirow{4}{*}{7447} & \multirow{2}{*}{1} & $\textbf{1}_{\textbf{e}_{\textbf{1}},-\textbf{e}_{\textbf{2}}}\textbf{5}_{-\textbf{e}_{\textbf{1}},-\textbf{e}_{\textbf{5}}}\bar{\textbf{5}}_{\textbf{e}_{\textbf{2}}, \textbf{e}_{\textbf{5}}}$ & y & 2\\
\cline{3-5}
  &  & $\textbf{10}_{\textbf{e}_{\textbf{3}}}\bar{\textbf{5}}_{\textbf{e}_{\textbf{1}},\textbf{e}_{\textbf{4}}}\bar{\textbf{5}}_{\textbf{e}_{\textbf{2}},\textbf{e}_{\textbf{5}}} $ & y & 2\\
\cline{2-5}
   & \multirow{2}{*}{2 }& $\textbf{10}_{\textbf{e}_{\textbf{2}}}\bar{\textbf{5}}_{\textbf{e}_{\textbf{1}},\textbf{e}_{\textbf{4}}}\bar{\textbf{5}}_{\textbf{e}_{\textbf{3}},\textbf{e}_{\textbf{5}}} $ & n & 2\\
\cline{3-5}
 & &$\textbf{1}_{\textbf{e}_{\textbf{3}},-\textbf{e}_{\textbf{1}}}\textbf{5}_{-\textbf{e}_{\textbf{3}},-\textbf{e}_{\textbf{4}}}\bar{\textbf{5}}_{\textbf{e}_{\textbf{1}}, \textbf{e}_{\textbf{4}}}$ & n & 2\\
\cline{2-5}
   & 3 & $\textbf{10}_{\textbf{e}_{\textbf{1}}}\bar{\textbf{5}}_{\textbf{e}_{\textbf{2}},\textbf{e}_{\textbf{5}}}\bar{\textbf{5}}_{\textbf{e}_{\textbf{3}},\textbf{e}_{\textbf{4}}}  $& n & 2\\
\cline{2-5}
   & \multirow{2}{*}{4} & $\textbf{10}_{\textbf{e}_{\textbf{1}}}\bar{\textbf{5}}_{\textbf{e}_{\textbf{2}},\textbf{e}_{\textbf{5}}}\bar{\textbf{5}}_{\textbf{e}_{\textbf{3}},\textbf{e}_{\textbf{4}}} $ & n & 2\\
\cline{3-5}
   &  & $\textbf{10}_{\textbf{e}_{\textbf{4}}}\bar{\textbf{5}}_{\textbf{e}_{\textbf{1}},\textbf{e}_{\textbf{3}}}\bar{\textbf{5}}_{\textbf{e}_{\textbf{2}},\textbf{e}_{\textbf{5}}} $ & n & 2\\
\cline{2-5}
\hline
\end{tabular}
\end{center}
On CICY 7447, a total of 4 models have Yukawa couplings consistent with the gauge symmetries of the models, with 9 couplings being permitted in total. Of these, 5 are of the form $\textbf{10}_{\textbf{e}_{\textbf{i}}}\bar{\textbf{5}}_{\textbf{e}_{\textbf{j}},\textbf{e}_{\textbf{k}}}\bar{\textbf{5}}_{\textbf{e}_{\textbf{l}},\textbf{e}_{\textbf{m}}} $ and 4 are of the form $\textbf{1}_{\textbf{e}_{\textbf{i}},-\textbf{e}_{\textbf{j}}}\textbf{5}_{-\textbf{e}_{\textbf{i}},-\textbf{e}_{\textbf{k}}}\bar{\textbf{5}}_{\textbf{e}_{\textbf{j}}, \textbf{e}_{\textbf{k}}}$. There are a total of 4 couplings that exhibit the topological vanishing we have been studying here,  3 of the form $\textbf{1}_{\textbf{e}_{\textbf{i}},-\textbf{e}_{\textbf{j}}}\textbf{5}_{-\textbf{e}_{\textbf{i}},-\textbf{e}_{\textbf{k}}}\bar{\textbf{5}}_{\textbf{e}_{\textbf{j}}, \textbf{e}_{\textbf{k}}}$ and 1 of the form $\textbf{10}_{\textbf{e}_{\textbf{i}}}\bar{\textbf{5}}_{\textbf{e}_{\textbf{j}},\textbf{e}_{\textbf{k}}}\bar{\textbf{5}}_{\textbf{e}_{\textbf{l}},\textbf{e}_{\textbf{m}}} $. All of the topologically vanishing couplings occur in 1 model.
\begin{center}
\begin{tabular}{|c|c|c|c|c|}
\hline
CICY No.  & Model No. & Yukawa Pattern & Top. Van. & Sym. No.\\
\hline
\multirow{12}{*}{7487} &  \multirow{6}{*}{1} & $\textbf{1}_{\textbf{e}_{\textbf{2}},-\textbf{e}_{\textbf{3}}}\textbf{5}_{-\textbf{e}_{\textbf{2}},-\textbf{e}_{\textbf{4}}}\bar{\textbf{5}}_{\textbf{e}_{\textbf{3}}, \textbf{e}_{\textbf{4}}} $ & n & 1,3\\
\cline{3-5}
   & & $\textbf{1}_{\textbf{e}_{\textbf{3}},-\textbf{e}_{\textbf{1}}}\textbf{5}_{-\textbf{e}_{\textbf{3}},-\textbf{e}_{\textbf{4}}}\bar{\textbf{5}}_{\textbf{e}_{\textbf{1}}, \textbf{e}_{\textbf{4}}}$ & n & 1-4\\
\cline{3-5}
   & & $\textbf{1}_{\textbf{e}_{\textbf{3}},-\textbf{e}_{\textbf{2}}}\textbf{5}_{-\textbf{e}_{\textbf{3}},-\textbf{e}_{\textbf{4}}}\bar{\textbf{5}}_{\textbf{e}_{\textbf{2}}, \textbf{e}_{\textbf{4}}}$ & n & 1,3\\
\cline{3-5}
   &  & $\textbf{10}_{\textbf{e}_{\textbf{1}}}\bar{\textbf{5}}_{\textbf{e}_{\textbf{2}},\textbf{e}_{\textbf{4}}}\bar{\textbf{5}}_{\textbf{e}_{\textbf{3}},\textbf{e}_{\textbf{5}}} $ & n & 1,3\\
\cline{3-5}
  &  & $\textbf{10}_{\textbf{e}_{\textbf{2}}}\bar{\textbf{5}}_{\textbf{e}_{\textbf{1}},\textbf{e}_{\textbf{4}}}\bar{\textbf{5}}_{\textbf{e}_{\textbf{3}},\textbf{e}_{\textbf{5}}} $ & n & 1-4\\
\cline{3-5}
   &  & $\textbf{5}_{-\textbf{e}_{\textbf{2}},-\textbf{e}_{\textbf{3}}}\textbf{10}_{\textbf{e}_{\textbf{2}}}\textbf{10}_{\textbf{e}_{\textbf{3}}} $& y & 1,3\\
\cline{2-5}
 & \multirow{6}{*}{2} & $\textbf{1}_{\textbf{e}_{\textbf{2}},-\textbf{e}_{\textbf{3}}}\textbf{5}_{-\textbf{e}_{\textbf{2}},-\textbf{e}_{\textbf{4}}}\bar{\textbf{5}}_{\textbf{e}_{\textbf{3}}, \textbf{e}_{\textbf{4}}} $ & n & 1-4\\
\cline{3-5}
   & & $\textbf{1}_{\textbf{e}_{\textbf{3}},-\textbf{e}_{\textbf{1}}}\textbf{5}_{-\textbf{e}_{\textbf{3}},-\textbf{e}_{\textbf{4}}}\bar{\textbf{5}}_{\textbf{e}_{\textbf{1}}, \textbf{e}_{\textbf{4}}}$ & n & 1-4\\
\cline{3-5}
   & & $\textbf{1}_{\textbf{e}_{\textbf{3}},-\textbf{e}_{\textbf{2}}}\textbf{5}_{-\textbf{e}_{\textbf{3}},-\textbf{e}_{\textbf{4}}}\bar{\textbf{5}}_{\textbf{e}_{\textbf{2}}, \textbf{e}_{\textbf{4}}}$ & n & 1-4\\
\cline{3-5}
   &  & $\textbf{10}_{\textbf{e}_{\textbf{1}}}\bar{\textbf{5}}_{\textbf{e}_{\textbf{2}},\textbf{e}_{\textbf{4}}}\bar{\textbf{5}}_{\textbf{e}_{\textbf{3}},\textbf{e}_{\textbf{5}}} $ & n & 1-4\\
\cline{3-5}
  &  & $\textbf{10}_{\textbf{e}_{\textbf{2}}}\bar{\textbf{5}}_{\textbf{e}_{\textbf{1}},\textbf{e}_{\textbf{4}}}\bar{\textbf{5}}_{\textbf{e}_{\textbf{3}},\textbf{e}_{\textbf{5}}} $ & n & 1-4\\
\cline{3-5}
   &  & $\textbf{5}_{-\textbf{e}_{\textbf{2}},-\textbf{e}_{\textbf{3}}}\textbf{10}_{\textbf{e}_{\textbf{2}}}\textbf{10}_{\textbf{e}_{\textbf{3}}} $& y & 1-4\\
\cline{2-5}
\hline
\end{tabular}
\end{center}

\begin{center}
\begin{tabular}{|c|c|c|c|c|}
\hline
CICY No.  & Model No. & Yukawa Pattern & Top. Van. & Sym. No.\\
\hline
\multirow{40}{*}{7487}
 &\multirow{6}{*}{3} & $\textbf{1}_{\textbf{e}_{\textbf{2}},-\textbf{e}_{\textbf{3}}}\textbf{5}_{-\textbf{e}_{\textbf{2}},-\textbf{e}_{\textbf{4}}}\bar{\textbf{5}}_{\textbf{e}_{\textbf{3}}, \textbf{e}_{\textbf{4}}} $ & n & 1-4\\
\cline{3-5}
   & & $\textbf{1}_{\textbf{e}_{\textbf{3}},-\textbf{e}_{\textbf{1}}}\textbf{5}_{-\textbf{e}_{\textbf{3}},-\textbf{e}_{\textbf{4}}}\bar{\textbf{5}}_{\textbf{e}_{\textbf{1}}, \textbf{e}_{\textbf{4}}}$ & n & 1-4\\
\cline{3-5}
   & & $\textbf{1}_{\textbf{e}_{\textbf{3}},-\textbf{e}_{\textbf{2}}}\textbf{5}_{-\textbf{e}_{\textbf{3}},-\textbf{e}_{\textbf{4}}}\bar{\textbf{5}}_{\textbf{e}_{\textbf{2}}, \textbf{e}_{\textbf{4}}}$ & n & 1-4\\
\cline{3-5}
   &  & $\textbf{10}_{\textbf{e}_{\textbf{1}}}\bar{\textbf{5}}_{\textbf{e}_{\textbf{2}},\textbf{e}_{\textbf{4}}}\bar{\textbf{5}}_{\textbf{e}_{\textbf{3}},\textbf{e}_{\textbf{5}}} $ & n & 1-4\\
\cline{3-5}
  &  & $\textbf{10}_{\textbf{e}_{\textbf{2}}}\bar{\textbf{5}}_{\textbf{e}_{\textbf{1}},\textbf{e}_{\textbf{4}}}\bar{\textbf{5}}_{\textbf{e}_{\textbf{3}},\textbf{e}_{\textbf{5}}} $ & n & 1-4\\
\cline{3-5}
   &  & $\textbf{5}_{-\textbf{e}_{\textbf{2}},-\textbf{e}_{\textbf{3}}}\textbf{10}_{\textbf{e}_{\textbf{2}}}\textbf{10}_{\textbf{e}_{\textbf{3}}} $& y & 1-4\\
\cline{2-5}

 &\multirow{6}{*}{4} & $\textbf{1}_{\textbf{e}_{\textbf{2}},-\textbf{e}_{\textbf{3}}}\textbf{5}_{-\textbf{e}_{\textbf{2}},-\textbf{e}_{\textbf{4}}}\bar{\textbf{5}}_{\textbf{e}_{\textbf{3}}, \textbf{e}_{\textbf{4}}} $ & n & 1-4\\
\cline{3-5}
   & & $\textbf{1}_{\textbf{e}_{\textbf{3}},-\textbf{e}_{\textbf{1}}}\textbf{5}_{-\textbf{e}_{\textbf{3}},-\textbf{e}_{\textbf{4}}}\bar{\textbf{5}}_{\textbf{e}_{\textbf{1}}, \textbf{e}_{\textbf{4}}}$ & n & 1-4\\
\cline{3-5}
   & & $\textbf{1}_{\textbf{e}_{\textbf{3}},-\textbf{e}_{\textbf{2}}}\textbf{5}_{-\textbf{e}_{\textbf{3}},-\textbf{e}_{\textbf{4}}}\bar{\textbf{5}}_{\textbf{e}_{\textbf{2}}, \textbf{e}_{\textbf{4}}}$ & n & 1-4\\
\cline{3-5}
   &  & $\textbf{10}_{\textbf{e}_{\textbf{1}}}\bar{\textbf{5}}_{\textbf{e}_{\textbf{2}},\textbf{e}_{\textbf{4}}}\bar{\textbf{5}}_{\textbf{e}_{\textbf{3}},\textbf{e}_{\textbf{5}}} $ & n & 1-4\\
\cline{3-5}
  &  & $\textbf{10}_{\textbf{e}_{\textbf{2}}}\bar{\textbf{5}}_{\textbf{e}_{\textbf{1}},\textbf{e}_{\textbf{4}}}\bar{\textbf{5}}_{\textbf{e}_{\textbf{3}},\textbf{e}_{\textbf{5}}} $ & n & 1-4\\
\cline{3-5}
   &  & $\textbf{5}_{-\textbf{e}_{\textbf{2}},-\textbf{e}_{\textbf{3}}}\textbf{10}_{\textbf{e}_{\textbf{2}}}\textbf{10}_{\textbf{e}_{\textbf{3}}} $& y & 1-4\\
\cline{2-5}

 & \multirow{6}{*}{5} & $\textbf{1}_{\textbf{e}_{\textbf{2}},-\textbf{e}_{\textbf{3}}}\textbf{5}_{-\textbf{e}_{\textbf{2}},-\textbf{e}_{\textbf{4}}}\bar{\textbf{5}}_{\textbf{e}_{\textbf{3}}, \textbf{e}_{\textbf{4}}} $ & n & 1-4\\
\cline{3-5}
   & & $\textbf{1}_{\textbf{e}_{\textbf{3}},-\textbf{e}_{\textbf{1}}}\textbf{5}_{-\textbf{e}_{\textbf{3}},-\textbf{e}_{\textbf{4}}}\bar{\textbf{5}}_{\textbf{e}_{\textbf{1}}, \textbf{e}_{\textbf{4}}}$ & n & 1-4\\
\cline{3-5}
   & & $\textbf{1}_{\textbf{e}_{\textbf{3}},-\textbf{e}_{\textbf{2}}}\textbf{5}_{-\textbf{e}_{\textbf{3}},-\textbf{e}_{\textbf{4}}}\bar{\textbf{5}}_{\textbf{e}_{\textbf{2}}, \textbf{e}_{\textbf{4}}}$ & n & 1-4\\
\cline{3-5}
   &  & $\textbf{10}_{\textbf{e}_{\textbf{1}}}\bar{\textbf{5}}_{\textbf{e}_{\textbf{2}},\textbf{e}_{\textbf{4}}}\bar{\textbf{5}}_{\textbf{e}_{\textbf{3}},\textbf{e}_{\textbf{5}}} $ & n & 1-4\\
\cline{3-5}
  &  & $\textbf{10}_{\textbf{e}_{\textbf{2}}}\bar{\textbf{5}}_{\textbf{e}_{\textbf{1}},\textbf{e}_{\textbf{4}}}\bar{\textbf{5}}_{\textbf{e}_{\textbf{3}},\textbf{e}_{\textbf{5}}} $ & n & 1-4\\
\cline{3-5}
   &  & $\textbf{5}_{-\textbf{e}_{\textbf{2}},-\textbf{e}_{\textbf{3}}}\textbf{10}_{\textbf{e}_{\textbf{2}}}\textbf{10}_{\textbf{e}_{\textbf{3}}} $& y & 1-4\\
\cline{2-5}

 & \multirow{6}{*}{6} & $\textbf{1}_{\textbf{e}_{\textbf{2}},-\textbf{e}_{\textbf{3}}}\textbf{5}_{-\textbf{e}_{\textbf{2}},-\textbf{e}_{\textbf{4}}}\bar{\textbf{5}}_{\textbf{e}_{\textbf{3}}, \textbf{e}_{\textbf{4}}} $ & n & 1-4\\
\cline{3-5}
   & & $\textbf{1}_{\textbf{e}_{\textbf{3}},-\textbf{e}_{\textbf{1}}}\textbf{5}_{-\textbf{e}_{\textbf{3}},-\textbf{e}_{\textbf{4}}}\bar{\textbf{5}}_{\textbf{e}_{\textbf{1}}, \textbf{e}_{\textbf{4}}}$ & n & 1-4\\
\cline{3-5}
   & & $\textbf{1}_{\textbf{e}_{\textbf{3}},-\textbf{e}_{\textbf{2}}}\textbf{5}_{-\textbf{e}_{\textbf{3}},-\textbf{e}_{\textbf{4}}}\bar{\textbf{5}}_{\textbf{e}_{\textbf{2}}, \textbf{e}_{\textbf{4}}}$ & n & 1-4\\
\cline{3-5}
   &  & $\textbf{10}_{\textbf{e}_{\textbf{1}}}\bar{\textbf{5}}_{\textbf{e}_{\textbf{2}},\textbf{e}_{\textbf{4}}}\bar{\textbf{5}}_{\textbf{e}_{\textbf{3}},\textbf{e}_{\textbf{5}}} $ & n & 1-4\\
\cline{3-5}
  &  & $\textbf{10}_{\textbf{e}_{\textbf{2}}}\bar{\textbf{5}}_{\textbf{e}_{\textbf{1}},\textbf{e}_{\textbf{4}}}\bar{\textbf{5}}_{\textbf{e}_{\textbf{3}},\textbf{e}_{\textbf{5}}} $ & n & 1-4\\
\cline{3-5}
   &  & $\textbf{5}_{-\textbf{e}_{\textbf{2}},-\textbf{e}_{\textbf{3}}}\textbf{10}_{\textbf{e}_{\textbf{2}}}\textbf{10}_{\textbf{e}_{\textbf{3}}} $& y & 1-4\\
\cline{2-5}
& \multirow{6}{*}{7} & $\textbf{1}_{\textbf{e}_{\textbf{2}},-\textbf{e}_{\textbf{3}}}\textbf{5}_{-\textbf{e}_{\textbf{2}},-\textbf{e}_{\textbf{4}}}\bar{\textbf{5}}_{\textbf{e}_{\textbf{3}}, \textbf{e}_{\textbf{4}}} $ & n & 1-4\\
\cline{3-5}
   & & $\textbf{1}_{\textbf{e}_{\textbf{3}},-\textbf{e}_{\textbf{1}}}\textbf{5}_{-\textbf{e}_{\textbf{3}},-\textbf{e}_{\textbf{4}}}\bar{\textbf{5}}_{\textbf{e}_{\textbf{1}}, \textbf{e}_{\textbf{4}}}$ & n & 1-4\\
\cline{3-5}
   & & $\textbf{1}_{\textbf{e}_{\textbf{3}},-\textbf{e}_{\textbf{2}}}\textbf{5}_{-\textbf{e}_{\textbf{3}},-\textbf{e}_{\textbf{4}}}\bar{\textbf{5}}_{\textbf{e}_{\textbf{2}}, \textbf{e}_{\textbf{4}}}$ & n & 1-4\\
\cline{3-5}
   &  & $\textbf{10}_{\textbf{e}_{\textbf{1}}}\bar{\textbf{5}}_{\textbf{e}_{\textbf{2}},\textbf{e}_{\textbf{4}}}\bar{\textbf{5}}_{\textbf{e}_{\textbf{3}},\textbf{e}_{\textbf{5}}} $ & n & 1-4\\
\cline{3-5}
  &  & $\textbf{10}_{\textbf{e}_{\textbf{2}}}\bar{\textbf{5}}_{\textbf{e}_{\textbf{1}},\textbf{e}_{\textbf{4}}}\bar{\textbf{5}}_{\textbf{e}_{\textbf{3}},\textbf{e}_{\textbf{5}}} $ & n & 1-4\\
\cline{3-5}
   &  & $\textbf{5}_{-\textbf{e}_{\textbf{2}},-\textbf{e}_{\textbf{3}}}\textbf{10}_{\textbf{e}_{\textbf{2}}}\textbf{10}_{\textbf{e}_{\textbf{3}}} $& y & 1-4\\
\cline{2-5}
 &  \multirow{6}{*}{8} & $\textbf{1}_{\textbf{e}_{\textbf{2}},-\textbf{e}_{\textbf{3}}}\textbf{5}_{-\textbf{e}_{\textbf{2}},-\textbf{e}_{\textbf{4}}}\bar{\textbf{5}}_{\textbf{e}_{\textbf{3}}, \textbf{e}_{\textbf{4}}} $ & n & 1-4\\
\cline{3-5}
   & & $\textbf{1}_{\textbf{e}_{\textbf{3}},-\textbf{e}_{\textbf{1}}}\textbf{5}_{-\textbf{e}_{\textbf{3}},-\textbf{e}_{\textbf{4}}}\bar{\textbf{5}}_{\textbf{e}_{\textbf{1}}, \textbf{e}_{\textbf{4}}}$ & n & 1-4\\
\cline{3-5}
   & & $\textbf{1}_{\textbf{e}_{\textbf{3}},-\textbf{e}_{\textbf{2}}}\textbf{5}_{-\textbf{e}_{\textbf{3}},-\textbf{e}_{\textbf{4}}}\bar{\textbf{5}}_{\textbf{e}_{\textbf{2}}, \textbf{e}_{\textbf{4}}}$ & n & 1-4\\
\cline{3-5}
   &  & $\textbf{10}_{\textbf{e}_{\textbf{1}}}\bar{\textbf{5}}_{\textbf{e}_{\textbf{2}},\textbf{e}_{\textbf{4}}}\bar{\textbf{5}}_{\textbf{e}_{\textbf{3}},\textbf{e}_{\textbf{5}}} $ & n & 1-4\\
\cline{3-5}
  &  & $\textbf{10}_{\textbf{e}_{\textbf{2}}}\bar{\textbf{5}}_{\textbf{e}_{\textbf{1}},\textbf{e}_{\textbf{4}}}\bar{\textbf{5}}_{\textbf{e}_{\textbf{3}},\textbf{e}_{\textbf{5}}} $ & n & 1-4\\
\cline{3-5}
   &  & $\textbf{5}_{-\textbf{e}_{\textbf{2}},-\textbf{e}_{\textbf{3}}}\textbf{10}_{\textbf{e}_{\textbf{2}}}\textbf{10}_{\textbf{e}_{\textbf{3}}} $& y & 1-4\\
\cline{2-5}
 & \multirow{6}{*}{9} & $\textbf{1}_{\textbf{e}_{\textbf{2}},-\textbf{e}_{\textbf{3}}}\textbf{5}_{-\textbf{e}_{\textbf{2}},-\textbf{e}_{\textbf{4}}}\bar{\textbf{5}}_{\textbf{e}_{\textbf{3}}, \textbf{e}_{\textbf{4}}} $ & n & 1-4\\
\cline{3-5}
   & & $\textbf{1}_{\textbf{e}_{\textbf{3}},-\textbf{e}_{\textbf{1}}}\textbf{5}_{-\textbf{e}_{\textbf{3}},-\textbf{e}_{\textbf{4}}}\bar{\textbf{5}}_{\textbf{e}_{\textbf{1}}, \textbf{e}_{\textbf{4}}}$ & n & 1-4\\
\cline{3-5}
   & & $\textbf{1}_{\textbf{e}_{\textbf{3}},-\textbf{e}_{\textbf{2}}}\textbf{5}_{-\textbf{e}_{\textbf{3}},-\textbf{e}_{\textbf{4}}}\bar{\textbf{5}}_{\textbf{e}_{\textbf{2}}, \textbf{e}_{\textbf{4}}}$ & n & 1-4\\
\cline{3-5}
   &  & $\textbf{10}_{\textbf{e}_{\textbf{1}}}\bar{\textbf{5}}_{\textbf{e}_{\textbf{2}},\textbf{e}_{\textbf{4}}}\bar{\textbf{5}}_{\textbf{e}_{\textbf{3}},\textbf{e}_{\textbf{5}}} $ & n & 1-4\\
\cline{3-5}
  &  & $\textbf{10}_{\textbf{e}_{\textbf{2}}}\bar{\textbf{5}}_{\textbf{e}_{\textbf{1}},\textbf{e}_{\textbf{4}}}\bar{\textbf{5}}_{\textbf{e}_{\textbf{3}},\textbf{e}_{\textbf{5}}} $ & n & 1-4\\
\cline{3-5}
   &  & $\textbf{5}_{-\textbf{e}_{\textbf{2}},-\textbf{e}_{\textbf{3}}}\textbf{10}_{\textbf{e}_{\textbf{2}}}\textbf{10}_{\textbf{e}_{\textbf{3}}} $& y & 1-4\\
\cline{2-5}
 & \multirow{6}{*}{10} & $\textbf{1}_{\textbf{e}_{\textbf{2}},-\textbf{e}_{\textbf{3}}}\textbf{5}_{-\textbf{e}_{\textbf{2}},-\textbf{e}_{\textbf{4}}}\bar{\textbf{5}}_{\textbf{e}_{\textbf{3}}, \textbf{e}_{\textbf{4}}} $ & n & 2,4\\
\cline{3-5}
   & & $\textbf{1}_{\textbf{e}_{\textbf{3}},-\textbf{e}_{\textbf{1}}}\textbf{5}_{-\textbf{e}_{\textbf{3}},-\textbf{e}_{\textbf{4}}}\bar{\textbf{5}}_{\textbf{e}_{\textbf{1}}, \textbf{e}_{\textbf{4}}}$ & n & 1-4\\
\cline{3-5}
   & & $\textbf{1}_{\textbf{e}_{\textbf{3}},-\textbf{e}_{\textbf{2}}}\textbf{5}_{-\textbf{e}_{\textbf{3}},-\textbf{e}_{\textbf{4}}}\bar{\textbf{5}}_{\textbf{e}_{\textbf{2}}, \textbf{e}_{\textbf{4}}}$ & n & 2,4\\
\cline{3-5}
   &  & $\textbf{10}_{\textbf{e}_{\textbf{1}}}\bar{\textbf{5}}_{\textbf{e}_{\textbf{2}},\textbf{e}_{\textbf{4}}}\bar{\textbf{5}}_{\textbf{e}_{\textbf{3}},\textbf{e}_{\textbf{5}}} $ & n & 2,4\\
\cline{3-5}
  &  & $\textbf{10}_{\textbf{e}_{\textbf{2}}}\bar{\textbf{5}}_{\textbf{e}_{\textbf{1}},\textbf{e}_{\textbf{4}}}\bar{\textbf{5}}_{\textbf{e}_{\textbf{3}},\textbf{e}_{\textbf{5}}} $ & n & 1-4\\
\cline{3-5}
   &  & $\textbf{5}_{-\textbf{e}_{\textbf{2}},-\textbf{e}_{\textbf{3}}}\textbf{10}_{\textbf{e}_{\textbf{2}}}\textbf{10}_{\textbf{e}_{\textbf{3}}} $& y & 2,4\\
\cline{2-5}

\hline
\end{tabular}
\end{center}

\begin{center}
\begin{tabular}{|c|c|c|c|c|}
\hline
CICY No. & Model No. & Yukawa Pattern & Top. Van. & Sym. No.\\
\hline
\multirow{30}{*}{7487} & \multirow{4}{*}{11} & $\textbf{1}_{\textbf{e}_{\textbf{2}},-\textbf{e}_{\textbf{3}}}\textbf{5}_{-\textbf{e}_{\textbf{2}},-\textbf{e}_{\textbf{4}}}\bar{\textbf{5}}_{\textbf{e}_{\textbf{3}}, \textbf{e}_{\textbf{4}}}$ & n & 1,3\\
\cline{3-5}
    & & $\textbf{1}_{\textbf{e}_{\textbf{3}},-\textbf{e}_{\textbf{2}}}\textbf{5}_{-\textbf{e}_{\textbf{3}},-\textbf{e}_{\textbf{4}}}\bar{\textbf{5}}_{\textbf{e}_{\textbf{2}}, \textbf{e}_{\textbf{4}}}$ & n & 1,3\\
\cline{3-5}
   &  & $\textbf{10}_{\textbf{e}_{\textbf{1}}}\bar{\textbf{5}}_{\textbf{e}_{\textbf{2}},\textbf{e}_{\textbf{5}}}\bar{\textbf{5}}_{\textbf{e}_{\textbf{3}},\textbf{e}_{\textbf{4}}} $ & n & 1-4\\
\cline{3-5}
   &  & $\textbf{5}_{-\textbf{e}_{\textbf{2}},-\textbf{e}_{\textbf{3}}}\textbf{10}_{\textbf{e}_{\textbf{2}}}\textbf{10}_{\textbf{e}_{\textbf{3}}}  $& y & 1,3\\
\cline{2-5}
 & \multirow{4}{*}{12} & $\textbf{1}_{\textbf{e}_{\textbf{2}},-\textbf{e}_{\textbf{3}}}\textbf{5}_{-\textbf{e}_{\textbf{2}},-\textbf{e}_{\textbf{4}}}\bar{\textbf{5}}_{\textbf{e}_{\textbf{3}}, \textbf{e}_{\textbf{4}}}$ & n & 1-4\\
\cline{3-5}
    & & $\textbf{1}_{\textbf{e}_{\textbf{3}},-\textbf{e}_{\textbf{2}}}\textbf{5}_{-\textbf{e}_{\textbf{3}},-\textbf{e}_{\textbf{4}}}\bar{\textbf{5}}_{\textbf{e}_{\textbf{2}}, \textbf{e}_{\textbf{4}}}$ & n & 1-4\\
\cline{3-5}
   &  & $\textbf{10}_{\textbf{e}_{\textbf{1}}}\bar{\textbf{5}}_{\textbf{e}_{\textbf{2}},\textbf{e}_{\textbf{5}}}\bar{\textbf{5}}_{\textbf{e}_{\textbf{3}},\textbf{e}_{\textbf{4}}} $ & n & 1-4\\
\cline{3-5}
   &  & $\textbf{5}_{-\textbf{e}_{\textbf{2}},-\textbf{e}_{\textbf{3}}}\textbf{10}_{\textbf{e}_{\textbf{2}}}\textbf{10}_{\textbf{e}_{\textbf{3}}}  $& y & 1-4\\
\cline{2-5}
 &  \multirow{4}{*}{13} & $\textbf{1}_{\textbf{e}_{\textbf{2}},-\textbf{e}_{\textbf{3}}}\textbf{5}_{-\textbf{e}_{\textbf{2}},-\textbf{e}_{\textbf{4}}}\bar{\textbf{5}}_{\textbf{e}_{\textbf{3}}, \textbf{e}_{\textbf{4}}}$ & n & 1-4\\
\cline{3-5}
    & & $\textbf{1}_{\textbf{e}_{\textbf{3}},-\textbf{e}_{\textbf{2}}}\textbf{5}_{-\textbf{e}_{\textbf{3}},-\textbf{e}_{\textbf{4}}}\bar{\textbf{5}}_{\textbf{e}_{\textbf{2}}, \textbf{e}_{\textbf{4}}}$ & n & 1-4\\
\cline{3-5}
   &  & $\textbf{10}_{\textbf{e}_{\textbf{1}}}\bar{\textbf{5}}_{\textbf{e}_{\textbf{2}},\textbf{e}_{\textbf{5}}}\bar{\textbf{5}}_{\textbf{e}_{\textbf{3}},\textbf{e}_{\textbf{4}}} $ & n & 1-4\\
\cline{3-5}
   &  & $\textbf{5}_{-\textbf{e}_{\textbf{2}},-\textbf{e}_{\textbf{3}}}\textbf{10}_{\textbf{e}_{\textbf{2}}}\textbf{10}_{\textbf{e}_{\textbf{3}}}  $& y & 1-4\\
\cline{2-5}
 & \multirow{4}{*}{14} & $\textbf{1}_{\textbf{e}_{\textbf{2}},-\textbf{e}_{\textbf{3}}}\textbf{5}_{-\textbf{e}_{\textbf{2}},-\textbf{e}_{\textbf{4}}}\bar{\textbf{5}}_{\textbf{e}_{\textbf{3}}, \textbf{e}_{\textbf{4}}}$ & n & 1-4\\
\cline{3-5}
    & & $\textbf{1}_{\textbf{e}_{\textbf{3}},-\textbf{e}_{\textbf{2}}}\textbf{5}_{-\textbf{e}_{\textbf{3}},-\textbf{e}_{\textbf{4}}}\bar{\textbf{5}}_{\textbf{e}_{\textbf{2}}, \textbf{e}_{\textbf{4}}}$ & n & 1-4\\
\cline{3-5}
   &  & $\textbf{10}_{\textbf{e}_{\textbf{1}}}\bar{\textbf{5}}_{\textbf{e}_{\textbf{2}},\textbf{e}_{\textbf{5}}}\bar{\textbf{5}}_{\textbf{e}_{\textbf{3}},\textbf{e}_{\textbf{4}}} $ & n & 1-4\\
\cline{3-5}
   &  & $\textbf{5}_{-\textbf{e}_{\textbf{2}},-\textbf{e}_{\textbf{3}}}\textbf{10}_{\textbf{e}_{\textbf{2}}}\textbf{10}_{\textbf{e}_{\textbf{3}}}  $& y & 1-4\\
\cline{2-5}
 &\multirow{4}{*}{15} & $\textbf{1}_{\textbf{e}_{\textbf{2}},-\textbf{e}_{\textbf{3}}}\textbf{5}_{-\textbf{e}_{\textbf{2}},-\textbf{e}_{\textbf{4}}}\bar{\textbf{5}}_{\textbf{e}_{\textbf{3}}, \textbf{e}_{\textbf{4}}}$ & n & 1-4\\
\cline{3-5}
    & & $\textbf{1}_{\textbf{e}_{\textbf{3}},-\textbf{e}_{\textbf{2}}}\textbf{5}_{-\textbf{e}_{\textbf{3}},-\textbf{e}_{\textbf{4}}}\bar{\textbf{5}}_{\textbf{e}_{\textbf{2}}, \textbf{e}_{\textbf{4}}}$ & n & 1-4\\
\cline{3-5}
   &  & $\textbf{10}_{\textbf{e}_{\textbf{1}}}\bar{\textbf{5}}_{\textbf{e}_{\textbf{2}},\textbf{e}_{\textbf{5}}}\bar{\textbf{5}}_{\textbf{e}_{\textbf{3}},\textbf{e}_{\textbf{4}}} $ & n & 1-4\\
\cline{3-5}
   &  & $\textbf{5}_{-\textbf{e}_{\textbf{2}},-\textbf{e}_{\textbf{3}}}\textbf{10}_{\textbf{e}_{\textbf{2}}}\textbf{10}_{\textbf{e}_{\textbf{3}}}  $& y & 1-4\\
\cline{2-5}
 & \multirow{4}{*}{16} & $\textbf{1}_{\textbf{e}_{\textbf{2}},-\textbf{e}_{\textbf{3}}}\textbf{5}_{-\textbf{e}_{\textbf{2}},-\textbf{e}_{\textbf{4}}}\bar{\textbf{5}}_{\textbf{e}_{\textbf{3}}, \textbf{e}_{\textbf{4}}}$ & n & 1-4\\
\cline{3-5}
    & & $\textbf{1}_{\textbf{e}_{\textbf{3}},-\textbf{e}_{\textbf{2}}}\textbf{5}_{-\textbf{e}_{\textbf{3}},-\textbf{e}_{\textbf{4}}}\bar{\textbf{5}}_{\textbf{e}_{\textbf{2}}, \textbf{e}_{\textbf{4}}}$ & n & 1-4\\
\cline{3-5}
   &  & $\textbf{10}_{\textbf{e}_{\textbf{1}}}\bar{\textbf{5}}_{\textbf{e}_{\textbf{2}},\textbf{e}_{\textbf{5}}}\bar{\textbf{5}}_{\textbf{e}_{\textbf{3}},\textbf{e}_{\textbf{4}}} $ & n & 1-4\\
\cline{3-5}
   &  & $\textbf{5}_{-\textbf{e}_{\textbf{2}},-\textbf{e}_{\textbf{3}}}\textbf{10}_{\textbf{e}_{\textbf{2}}}\textbf{10}_{\textbf{e}_{\textbf{3}}}  $& y & 1-4\\
\cline{2-5}
 & \multirow{4}{*}{17} & $\textbf{1}_{\textbf{e}_{\textbf{2}},-\textbf{e}_{\textbf{3}}}\textbf{5}_{-\textbf{e}_{\textbf{2}},-\textbf{e}_{\textbf{4}}}\bar{\textbf{5}}_{\textbf{e}_{\textbf{3}}, \textbf{e}_{\textbf{4}}}$ & n & 1-4\\
\cline{3-5}
    & & $\textbf{1}_{\textbf{e}_{\textbf{3}},-\textbf{e}_{\textbf{2}}}\textbf{5}_{-\textbf{e}_{\textbf{3}},-\textbf{e}_{\textbf{4}}}\bar{\textbf{5}}_{\textbf{e}_{\textbf{2}}, \textbf{e}_{\textbf{4}}}$ & n & 1-4\\
\cline{3-5}
   &  & $\textbf{10}_{\textbf{e}_{\textbf{1}}}\bar{\textbf{5}}_{\textbf{e}_{\textbf{2}},\textbf{e}_{\textbf{5}}}\bar{\textbf{5}}_{\textbf{e}_{\textbf{3}},\textbf{e}_{\textbf{4}}} $ & n & 1-4\\
\cline{3-5}
   &  & $\textbf{5}_{-\textbf{e}_{\textbf{2}},-\textbf{e}_{\textbf{3}}}\textbf{10}_{\textbf{e}_{\textbf{2}}}\textbf{10}_{\textbf{e}_{\textbf{3}}}  $& y & 1-4\\
\cline{2-5}
 & \multirow{4}{*}{18} & $\textbf{1}_{\textbf{e}_{\textbf{2}},-\textbf{e}_{\textbf{3}}}\textbf{5}_{-\textbf{e}_{\textbf{2}},-\textbf{e}_{\textbf{4}}}\bar{\textbf{5}}_{\textbf{e}_{\textbf{3}}, \textbf{e}_{\textbf{4}}}$ & n & 1-4\\
\cline{3-5}
    & & $\textbf{1}_{\textbf{e}_{\textbf{3}},-\textbf{e}_{\textbf{2}}}\textbf{5}_{-\textbf{e}_{\textbf{3}},-\textbf{e}_{\textbf{4}}}\bar{\textbf{5}}_{\textbf{e}_{\textbf{2}}, \textbf{e}_{\textbf{4}}}$ & n & 1-4\\
\cline{3-5}
   &  & $\textbf{10}_{\textbf{e}_{\textbf{1}}}\bar{\textbf{5}}_{\textbf{e}_{\textbf{2}},\textbf{e}_{\textbf{5}}}\bar{\textbf{5}}_{\textbf{e}_{\textbf{3}},\textbf{e}_{\textbf{4}}} $ & n & 1-4\\
\cline{3-5}
   &  & $\textbf{5}_{-\textbf{e}_{\textbf{2}},-\textbf{e}_{\textbf{3}}}\textbf{10}_{\textbf{e}_{\textbf{2}}}\textbf{10}_{\textbf{e}_{\textbf{3}}}  $& y & 1-4\\
\cline{2-5}

 & \multirow{4}{*}{19} & $\textbf{1}_{\textbf{e}_{\textbf{2}},-\textbf{e}_{\textbf{3}}}\textbf{5}_{-\textbf{e}_{\textbf{2}},-\textbf{e}_{\textbf{4}}}\bar{\textbf{5}}_{\textbf{e}_{\textbf{3}}, \textbf{e}_{\textbf{4}}}$ & n & 1-4\\
\cline{3-5}
    & & $\textbf{1}_{\textbf{e}_{\textbf{3}},-\textbf{e}_{\textbf{2}}}\textbf{5}_{-\textbf{e}_{\textbf{3}},-\textbf{e}_{\textbf{4}}}\bar{\textbf{5}}_{\textbf{e}_{\textbf{2}}, \textbf{e}_{\textbf{4}}}$ & n & 1-4\\
\cline{3-5}
   &  & $\textbf{10}_{\textbf{e}_{\textbf{1}}}\bar{\textbf{5}}_{\textbf{e}_{\textbf{2}},\textbf{e}_{\textbf{5}}}\bar{\textbf{5}}_{\textbf{e}_{\textbf{3}},\textbf{e}_{\textbf{4}}} $ & n & 1-4\\
\cline{3-5}
   &  & $\textbf{5}_{-\textbf{e}_{\textbf{2}},-\textbf{e}_{\textbf{3}}}\textbf{10}_{\textbf{e}_{\textbf{2}}}\textbf{10}_{\textbf{e}_{\textbf{3}}}  $& y & 1-4\\
\cline{2-5}

 & \multirow{4}{*}{20} & $\textbf{1}_{\textbf{e}_{\textbf{2}},-\textbf{e}_{\textbf{3}}}\textbf{5}_{-\textbf{e}_{\textbf{2}},-\textbf{e}_{\textbf{4}}}\bar{\textbf{5}}_{\textbf{e}_{\textbf{3}}, \textbf{e}_{\textbf{4}}}$ & n & 2,4\\
\cline{3-5}
    & & $\textbf{1}_{\textbf{e}_{\textbf{3}},-\textbf{e}_{\textbf{2}}}\textbf{5}_{-\textbf{e}_{\textbf{3}},-\textbf{e}_{\textbf{4}}}\bar{\textbf{5}}_{\textbf{e}_{\textbf{2}}, \textbf{e}_{\textbf{4}}}$ & n & 2,4\\
\cline{3-5}
   &  & $\textbf{10}_{\textbf{e}_{\textbf{1}}}\bar{\textbf{5}}_{\textbf{e}_{\textbf{2}},\textbf{e}_{\textbf{5}}}\bar{\textbf{5}}_{\textbf{e}_{\textbf{3}},\textbf{e}_{\textbf{4}}} $ & n & 1-4\\
\cline{3-5}
   &  & $\textbf{5}_{-\textbf{e}_{\textbf{2}},-\textbf{e}_{\textbf{3}}}\textbf{10}_{\textbf{e}_{\textbf{2}}}\textbf{10}_{\textbf{e}_{\textbf{3}}}  $& y & 2,4\\
\cline{2-5}
 & \multirow{2}{*}{21} & $\textbf{1}_{\textbf{e}_{\textbf{1}},-\textbf{e}_{\textbf{2}}}\textbf{5}_{-\textbf{e}_{\textbf{1}},-\textbf{e}_{\textbf{5}}}\bar{\textbf{5}}_{\textbf{e}_{\textbf{2}}, \textbf{e}_{\textbf{5}}} $ & n & 1-4\\
\cline{3-5}
   &  & $\textbf{10}_{\textbf{e}_{\textbf{3}}}\bar{\textbf{5}}_{\textbf{e}_{\textbf{1}},\textbf{e}_{\textbf{4}}}\bar{\textbf{5}}_{\textbf{e}_{\textbf{2}},\textbf{e}_{\textbf{5}}} $ & y & 1-4\\
\cline{2-5}
 &  {22} & $\textbf{10}_{\textbf{e}_{\textbf{5}}}\bar{\textbf{5}}_{\textbf{e}_{\textbf{1}},\textbf{e}_{\textbf{2}}}\bar{\textbf{5}}_{\textbf{e}_{\textbf{3}},\textbf{e}_{\textbf{4}}} $ & n & 1-4\\
\cline{2-5}
 &  \multirow{4}{*}{23} &$\textbf{10}_{\textbf{e}_{\textbf{3}}}\bar{\textbf{5}}_{\textbf{e}_{\textbf{1}},\textbf{e}_{\textbf{2}}}\bar{\textbf{5}}_{\textbf{e}_{\textbf{4}},\textbf{e}_{\textbf{5}}}  $& n & 1-4\\
\cline{3-5}
   &  & $\textbf{10}_{\textbf{e}_{\textbf{4}}}\bar{\textbf{5}}_{\textbf{e}_{\textbf{1}},\textbf{e}_{\textbf{2}}}\bar{\textbf{5}}_{\textbf{e}_{\textbf{3}},\textbf{e}_{\textbf{5}}} $ & n & 1-4\\
\cline{3-5}
   &  & $\textbf{10}_{\textbf{e}_{\textbf{5}}}\bar{\textbf{5}}_{\textbf{e}_{\textbf{1}},\textbf{e}_{\textbf{2}}}\bar{\textbf{5}}_{\textbf{e}_{\textbf{3}},\textbf{e}_{\textbf{4}}}  $& n & 1-4\\
\cline{3-5}
   &  & $\textbf{5}_{-\textbf{e}_{\textbf{3}},-\textbf{e}_{\textbf{4}}}\textbf{10}_{\textbf{e}_{\textbf{3}}}\textbf{10}_{\textbf{e}_{\textbf{4}}}  $& y & 1-4\\
\cline{2-5}
\hline
\end{tabular}
\end{center}

\begin{center}
\begin{tabular}{|c|c|c|c|c|}
\hline
CICY No.  & Model No. & Yukawa Pattern & Top. Van. & Sym. No.\\
\hline
\multirow{30}{*}{7487} 
 &  {24} &$\textbf{5}_{-\textbf{e}_{\textbf{3}},-\textbf{e}_{\textbf{4}}}\textbf{10}_{\textbf{e}_{\textbf{3}}}\textbf{10}_{\textbf{e}_{\textbf{4}}}  $ & y & 1-4\\
\cline{2-5}
 &  {26} &$\textbf{10}_{\textbf{e}_{\textbf{2}}}\bar{\textbf{5}}_{\textbf{e}_{\textbf{1}},\textbf{e}_{\textbf{4}}}\bar{\textbf{5}}_{\textbf{e}_{\textbf{3}},\textbf{e}_{\textbf{5}}}  $& n & 1-4\\
\cline{2-5}
 &  \multirow{2}{*}{27} & $\textbf{10}_{\textbf{e}_{\textbf{5}}}\bar{\textbf{5}}_{\textbf{e}_{\textbf{1}},\textbf{e}_{\textbf{2}}}\bar{\textbf{5}}_{\textbf{e}_{\textbf{3}},\textbf{e}_{\textbf{4}}} $ & n & 1-4\\
\cline{3-5}
   &  & $\textbf{10}_{\textbf{e}_{\textbf{5}}}\bar{\textbf{5}}_{\textbf{e}_{\textbf{1}},\textbf{e}_{\textbf{3}}}\bar{\textbf{5}}_{\textbf{e}_{\textbf{2}},\textbf{e}_{\textbf{4}}} $ & n & 1-4\\
\cline{2-5}
 &  \multirow{2}{*}{28} & $\textbf{10}_{\textbf{e}_{\textbf{5}}}\bar{\textbf{5}}_{\textbf{e}_{\textbf{1}},\textbf{e}_{\textbf{2}}}\bar{\textbf{5}}_{\textbf{e}_{\textbf{3}},\textbf{e}_{\textbf{4}}} $ & n & 1-4\\
\cline{3-5}
   &  & $\textbf{10}_{\textbf{e}_{\textbf{5}}}\bar{\textbf{5}}_{\textbf{e}_{\textbf{1}},\textbf{e}_{\textbf{3}}}\bar{\textbf{5}}_{\textbf{e}_{\textbf{2}},\textbf{e}_{\textbf{4}}} $ & y & 1-4\\
\cline{2-5}
 &  \multirow{2}{*}{29}  & $\textbf{10}_{\textbf{e}_{\textbf{2}}}\bar{\textbf{5}}_{\textbf{e}_{\textbf{1}},\textbf{e}_{\textbf{3}}}\bar{\textbf{5}}_{\textbf{e}_{\textbf{4}},\textbf{e}_{\textbf{5}}} $ & n & 1-4\\
\cline{3-5}
  &  & $\textbf{10}_{\textbf{e}_{\textbf{2}}}\bar{\textbf{5}}_{\textbf{e}_{\textbf{1}},\textbf{e}_{\textbf{5}}}\bar{\textbf{5}}_{\textbf{e}_{\textbf{3}},\textbf{e}_{\textbf{4}}} $& n & 1,3\\
\cline{2-5}
 &  \multirow{2}{*}{30}  & $\textbf{10}_{\textbf{e}_{\textbf{2}}}\bar{\textbf{5}}_{\textbf{e}_{\textbf{1}},\textbf{e}_{\textbf{3}}}\bar{\textbf{5}}_{\textbf{e}_{\textbf{4}},\textbf{e}_{\textbf{5}}} $ & n & 1-4\\
\cline{3-5}
   &  & $\textbf{10}_{\textbf{e}_{\textbf{2}}}\bar{\textbf{5}}_{\textbf{e}_{\textbf{1}},\textbf{e}_{\textbf{5}}}\bar{\textbf{5}}_{\textbf{e}_{\textbf{3}},\textbf{e}_{\textbf{4}}} $& n & 1-4\\
\cline{2-5}
 &  \multirow{2}{*}{31}  & $\textbf{10}_{\textbf{e}_{\textbf{2}}}\bar{\textbf{5}}_{\textbf{e}_{\textbf{1}},\textbf{e}_{\textbf{3}}}\bar{\textbf{5}}_{\textbf{e}_{\textbf{4}},\textbf{e}_{\textbf{5}}} $ & n & 1-4\\
\cline{3-5}
   &  & $\textbf{10}_{\textbf{e}_{\textbf{2}}}\bar{\textbf{5}}_{\textbf{e}_{\textbf{1}},\textbf{e}_{\textbf{5}}}\bar{\textbf{5}}_{\textbf{e}_{\textbf{3}},\textbf{e}_{\textbf{4}}} $& n & 1-4\\
\cline{2-5}
 &   \multirow{2}{*}{32}  & $\textbf{10}_{\textbf{e}_{\textbf{2}}}\bar{\textbf{5}}_{\textbf{e}_{\textbf{1}},\textbf{e}_{\textbf{3}}}\bar{\textbf{5}}_{\textbf{e}_{\textbf{4}},\textbf{e}_{\textbf{5}}} $ & n & 1-4\\
\cline{3-5}
   &  & $\textbf{10}_{\textbf{e}_{\textbf{2}}}\bar{\textbf{5}}_{\textbf{e}_{\textbf{1}},\textbf{e}_{\textbf{5}}}\bar{\textbf{5}}_{\textbf{e}_{\textbf{3}},\textbf{e}_{\textbf{4}}} $& n & 1-4\\
\cline{2-5}
 &  \multirow{2}{*}{33}  & $\textbf{10}_{\textbf{e}_{\textbf{2}}}\bar{\textbf{5}}_{\textbf{e}_{\textbf{1}},\textbf{e}_{\textbf{3}}}\bar{\textbf{5}}_{\textbf{e}_{\textbf{4}},\textbf{e}_{\textbf{5}}} $ & n & 1-4\\
\cline{3-5}
   &  & $\textbf{10}_{\textbf{e}_{\textbf{2}}}\bar{\textbf{5}}_{\textbf{e}_{\textbf{1}},\textbf{e}_{\textbf{5}}}\bar{\textbf{5}}_{\textbf{e}_{\textbf{3}},\textbf{e}_{\textbf{4}}} $& n & 2,4\\
\cline{2-5}
 & \multirow{2}{*}{34}  & $\textbf{10}_{\textbf{e}_{\textbf{2}}}\bar{\textbf{5}}_{\textbf{e}_{\textbf{1}},\textbf{e}_{\textbf{3}}}\bar{\textbf{5}}_{\textbf{e}_{\textbf{4}},\textbf{e}_{\textbf{5}}} $ & y & 1-4\\
\cline{3-5}
   &  & $\textbf{10}_{\textbf{e}_{\textbf{2}}}\bar{\textbf{5}}_{\textbf{e}_{\textbf{1}},\textbf{e}_{\textbf{5}}}\bar{\textbf{5}}_{\textbf{e}_{\textbf{3}},\textbf{e}_{\textbf{4}}} $& n & 1-4\\
\cline{2-5}
 &  {35} & $\textbf{10}_{\textbf{e}_{\textbf{2}}}\bar{\textbf{5}}_{\textbf{e}_{\textbf{1}},\textbf{e}_{\textbf{3}}}\bar{\textbf{5}}_{\textbf{e}_{\textbf{4}},\textbf{e}_{\textbf{5}}} $ & y & 1-4\\
\cline{2-5}
 &  \multirow{2}{*}{36} &$\textbf{10}_{\textbf{e}_{\textbf{2}}}\bar{\textbf{5}}_{\textbf{e}_{\textbf{1}},\textbf{e}_{\textbf{3}}}\bar{\textbf{5}}_{\textbf{e}_{\textbf{4}},\textbf{e}_{\textbf{5}}} $& n & 1-4\\
\cline{3-5}
   &  & $\textbf{10}_{\textbf{e}_{\textbf{5}}}\bar{\textbf{5}}_{\textbf{e}_{\textbf{1}},\textbf{e}_{\textbf{3}}}\bar{\textbf{5}}_{\textbf{e}_{\textbf{2}},\textbf{e}_{\textbf{4}}} $ & n & 1-4\\
\cline{2-5}
 &  \multirow{2}{*}{37} &$\textbf{10}_{\textbf{e}_{\textbf{2}}}\bar{\textbf{5}}_{\textbf{e}_{\textbf{1}},\textbf{e}_{\textbf{3}}}\bar{\textbf{5}}_{\textbf{e}_{\textbf{4}},\textbf{e}_{\textbf{5}}} $& n & 1-4\\
\cline{3-5}
   &  & $\textbf{10}_{\textbf{e}_{\textbf{5}}}\bar{\textbf{5}}_{\textbf{e}_{\textbf{1}},\textbf{e}_{\textbf{3}}}\bar{\textbf{5}}_{\textbf{e}_{\textbf{2}},\textbf{e}_{\textbf{4}}} $ & n & 1-4\\
\cline{2-5}
 &  \multirow{2}{*}{38} &$\textbf{10}_{\textbf{e}_{\textbf{2}}}\bar{\textbf{5}}_{\textbf{e}_{\textbf{1}},\textbf{e}_{\textbf{3}}}\bar{\textbf{5}}_{\textbf{e}_{\textbf{4}},\textbf{e}_{\textbf{5}}} $& n & 1-4\\
\cline{3-5}
   &  & $\textbf{10}_{\textbf{e}_{\textbf{5}}}\bar{\textbf{5}}_{\textbf{e}_{\textbf{1}},\textbf{e}_{\textbf{3}}}\bar{\textbf{5}}_{\textbf{e}_{\textbf{2}},\textbf{e}_{\textbf{4}}} $ & n & 1-4\\
\cline{2-5}
 & \multirow{2}{*}{39} &$\textbf{10}_{\textbf{e}_{\textbf{2}}}\bar{\textbf{5}}_{\textbf{e}_{\textbf{1}},\textbf{e}_{\textbf{3}}}\bar{\textbf{5}}_{\textbf{e}_{\textbf{4}},\textbf{e}_{\textbf{5}}} $& n & 1-4\\
\cline{3-5}
   &  & $\textbf{10}_{\textbf{e}_{\textbf{5}}}\bar{\textbf{5}}_{\textbf{e}_{\textbf{1}},\textbf{e}_{\textbf{3}}}\bar{\textbf{5}}_{\textbf{e}_{\textbf{2}},\textbf{e}_{\textbf{4}}} $ & n & 1-4\\
\cline{2-5}
 &  \multirow{2}{*}{40}& $\textbf{10}_{\textbf{e}_{\textbf{1}}}\bar{\textbf{5}}_{\textbf{e}_{\textbf{2}},\textbf{e}_{\textbf{3}}}\bar{\textbf{5}}_{\textbf{e}_{\textbf{4}},\textbf{e}_{\textbf{5}}} $ & n & 1-4\\
\cline{3-5}
   &  & $\textbf{10}_{\textbf{e}_{\textbf{2}}}\bar{\textbf{5}}_{\textbf{e}_{\textbf{1}},\textbf{e}_{\textbf{3}}}\bar{\textbf{5}}_{\textbf{e}_{\textbf{4}},\textbf{e}_{\textbf{5}}}  $& n & 1-4\\
\cline{2-5}
 &   \multirow{2}{*}{41}& $\textbf{10}_{\textbf{e}_{\textbf{1}}}\bar{\textbf{5}}_{\textbf{e}_{\textbf{2}},\textbf{e}_{\textbf{3}}}\bar{\textbf{5}}_{\textbf{e}_{\textbf{4}},\textbf{e}_{\textbf{5}}} $ & n & 1-4\\
\cline{3-5}
   &  & $\textbf{10}_{\textbf{e}_{\textbf{2}}}\bar{\textbf{5}}_{\textbf{e}_{\textbf{1}},\textbf{e}_{\textbf{3}}}\bar{\textbf{5}}_{\textbf{e}_{\textbf{4}},\textbf{e}_{\textbf{5}}}  $& n & 1-4\\
\cline{2-5}
 &   \multirow{2}{*}{42}& $\textbf{10}_{\textbf{e}_{\textbf{1}}}\bar{\textbf{5}}_{\textbf{e}_{\textbf{2}},\textbf{e}_{\textbf{3}}}\bar{\textbf{5}}_{\textbf{e}_{\textbf{4}},\textbf{e}_{\textbf{5}}} $ & n & 1-4\\
\cline{3-5}
   &  & $\textbf{10}_{\textbf{e}_{\textbf{2}}}\bar{\textbf{5}}_{\textbf{e}_{\textbf{1}},\textbf{e}_{\textbf{3}}}\bar{\textbf{5}}_{\textbf{e}_{\textbf{4}},\textbf{e}_{\textbf{5}}}  $& n & 1-4\\
\cline{2-5}
 &   \multirow{2}{*}{43}& $\textbf{10}_{\textbf{e}_{\textbf{1}}}\bar{\textbf{5}}_{\textbf{e}_{\textbf{2}},\textbf{e}_{\textbf{3}}}\bar{\textbf{5}}_{\textbf{e}_{\textbf{4}},\textbf{e}_{\textbf{5}}} $ & n & 1-4\\
\cline{3-5}
   &  & $\textbf{10}_{\textbf{e}_{\textbf{2}}}\bar{\textbf{5}}_{\textbf{e}_{\textbf{1}},\textbf{e}_{\textbf{3}}}\bar{\textbf{5}}_{\textbf{e}_{\textbf{4}},\textbf{e}_{\textbf{5}}}  $& n & 1-4\\
\cline{2-5}
 &  \multirow{2}{*}{44}& $\textbf{10}_{\textbf{e}_{\textbf{1}}}\bar{\textbf{5}}_{\textbf{e}_{\textbf{2}},\textbf{e}_{\textbf{3}}}\bar{\textbf{5}}_{\textbf{e}_{\textbf{4}},\textbf{e}_{\textbf{5}}} $ & n & 1-4\\
\cline{3-5}
   &  & $\textbf{10}_{\textbf{e}_{\textbf{2}}}\bar{\textbf{5}}_{\textbf{e}_{\textbf{1}},\textbf{e}_{\textbf{3}}}\bar{\textbf{5}}_{\textbf{e}_{\textbf{4}},\textbf{e}_{\textbf{5}}}  $& n & 1-4\\
\cline{2-5}
 &   \multirow{2}{*}{45}& $\textbf{10}_{\textbf{e}_{\textbf{1}}}\bar{\textbf{5}}_{\textbf{e}_{\textbf{2}},\textbf{e}_{\textbf{3}}}\bar{\textbf{5}}_{\textbf{e}_{\textbf{4}},\textbf{e}_{\textbf{5}}} $ & n & 1-4\\
\cline{3-5}
   &  & $\textbf{10}_{\textbf{e}_{\textbf{2}}}\bar{\textbf{5}}_{\textbf{e}_{\textbf{1}},\textbf{e}_{\textbf{3}}}\bar{\textbf{5}}_{\textbf{e}_{\textbf{4}},\textbf{e}_{\textbf{5}}}  $& n & 1-4\\
\cline{2-5}
 &   \multirow{2}{*}{46}& $\textbf{10}_{\textbf{e}_{\textbf{1}}}\bar{\textbf{5}}_{\textbf{e}_{\textbf{2}},\textbf{e}_{\textbf{3}}}\bar{\textbf{5}}_{\textbf{e}_{\textbf{4}},\textbf{e}_{\textbf{5}}} $ & n & 1-4\\
\cline{3-5}
   &  & $\textbf{10}_{\textbf{e}_{\textbf{2}}}\bar{\textbf{5}}_{\textbf{e}_{\textbf{1}},\textbf{e}_{\textbf{3}}}\bar{\textbf{5}}_{\textbf{e}_{\textbf{4}},\textbf{e}_{\textbf{5}}}  $& n & 1-4\\
\cline{2-5}
 &   \multirow{2}{*}{47}& $\textbf{10}_{\textbf{e}_{\textbf{1}}}\bar{\textbf{5}}_{\textbf{e}_{\textbf{2}},\textbf{e}_{\textbf{3}}}\bar{\textbf{5}}_{\textbf{e}_{\textbf{4}},\textbf{e}_{\textbf{5}}} $ & n & 1-4\\
\cline{3-5}
   &  & $\textbf{10}_{\textbf{e}_{\textbf{2}}}\bar{\textbf{5}}_{\textbf{e}_{\textbf{1}},\textbf{e}_{\textbf{3}}}\bar{\textbf{5}}_{\textbf{e}_{\textbf{4}},\textbf{e}_{\textbf{5}}}  $& n & 1-4\\
\cline{2-5}
 &   \multirow{2}{*}{48}& $\textbf{10}_{\textbf{e}_{\textbf{1}}}\bar{\textbf{5}}_{\textbf{e}_{\textbf{2}},\textbf{e}_{\textbf{3}}}\bar{\textbf{5}}_{\textbf{e}_{\textbf{4}},\textbf{e}_{\textbf{5}}} $ & n & 1-4\\
\cline{3-5}
   &  & $\textbf{10}_{\textbf{e}_{\textbf{2}}}\bar{\textbf{5}}_{\textbf{e}_{\textbf{1}},\textbf{e}_{\textbf{3}}}\bar{\textbf{5}}_{\textbf{e}_{\textbf{4}},\textbf{e}_{\textbf{5}}}  $& n & 1-4\\
\cline{2-5}
\hline
\end{tabular}
\end{center}

\begin{center}
\begin{tabular}{|c|c|c|c|c|}
\hline
CICY No.  & Model No. & Yukawa Pattern & Top. Van. & Sym. No.\\
\hline
\multirow{30}{*}{7487}  &  \multirow{2}{*}{49}& $\textbf{10}_{\textbf{e}_{\textbf{1}}}\bar{\textbf{5}}_{\textbf{e}_{\textbf{2}},\textbf{e}_{\textbf{3}}}\bar{\textbf{5}}_{\textbf{e}_{\textbf{4}},\textbf{e}_{\textbf{5}}} $ & n & 1-4\\
\cline{3-5}
   &  & $\textbf{10}_{\textbf{e}_{\textbf{2}}}\bar{\textbf{5}}_{\textbf{e}_{\textbf{1}},\textbf{e}_{\textbf{3}}}\bar{\textbf{5}}_{\textbf{e}_{\textbf{4}},\textbf{e}_{\textbf{5}}}  $& n & 1-4\\
\cline{2-5}
&  \multirow{2}{*}{50}& $\textbf{10}_{\textbf{e}_{\textbf{1}}}\bar{\textbf{5}}_{\textbf{e}_{\textbf{2}},\textbf{e}_{\textbf{3}}}\bar{\textbf{5}}_{\textbf{e}_{\textbf{4}},\textbf{e}_{\textbf{5}}} $ & n & 1-4\\
\cline{3-5}
   &  & $\textbf{10}_{\textbf{e}_{\textbf{2}}}\bar{\textbf{5}}_{\textbf{e}_{\textbf{1}},\textbf{e}_{\textbf{3}}}\bar{\textbf{5}}_{\textbf{e}_{\textbf{4}},\textbf{e}_{\textbf{5}}}  $& n & 1-4\\
\cline{2-5}
 & \multirow{2}{*}{51}& $\textbf{10}_{\textbf{e}_{\textbf{1}}}\bar{\textbf{5}}_{\textbf{e}_{\textbf{2}},\textbf{e}_{\textbf{3}}}\bar{\textbf{5}}_{\textbf{e}_{\textbf{4}},\textbf{e}_{\textbf{5}}} $ & n & 1-4\\
\cline{3-5}
   &  & $\textbf{10}_{\textbf{e}_{\textbf{2}}}\bar{\textbf{5}}_{\textbf{e}_{\textbf{1}},\textbf{e}_{\textbf{3}}}\bar{\textbf{5}}_{\textbf{e}_{\textbf{4}},\textbf{e}_{\textbf{5}}}  $& n & 1-4\\
\cline{2-5}
 & \multirow{2}{*}{52}& $\textbf{10}_{\textbf{e}_{\textbf{1}}}\bar{\textbf{5}}_{\textbf{e}_{\textbf{2}},\textbf{e}_{\textbf{3}}}\bar{\textbf{5}}_{\textbf{e}_{\textbf{4}},\textbf{e}_{\textbf{5}}} $ & n & 1-4\\
\cline{3-5}
   &  & $\textbf{10}_{\textbf{e}_{\textbf{2}}}\bar{\textbf{5}}_{\textbf{e}_{\textbf{1}},\textbf{e}_{\textbf{3}}}\bar{\textbf{5}}_{\textbf{e}_{\textbf{4}},\textbf{e}_{\textbf{5}}}  $& n & 1-4\\
\cline{2-5}
& \multirow{2}{*}{53}& $\textbf{10}_{\textbf{e}_{\textbf{1}}}\bar{\textbf{5}}_{\textbf{e}_{\textbf{2}},\textbf{e}_{\textbf{3}}}\bar{\textbf{5}}_{\textbf{e}_{\textbf{4}},\textbf{e}_{\textbf{5}}} $ & n & 1-4\\
\cline{3-5}
   &  & $\textbf{10}_{\textbf{e}_{\textbf{2}}}\bar{\textbf{5}}_{\textbf{e}_{\textbf{1}},\textbf{e}_{\textbf{3}}}\bar{\textbf{5}}_{\textbf{e}_{\textbf{4}},\textbf{e}_{\textbf{5}}}  $& n & 1-4\\
\cline{2-5}
&  \multirow{2}{*}{54}& $\textbf{10}_{\textbf{e}_{\textbf{1}}}\bar{\textbf{5}}_{\textbf{e}_{\textbf{2}},\textbf{e}_{\textbf{3}}}\bar{\textbf{5}}_{\textbf{e}_{\textbf{4}},\textbf{e}_{\textbf{5}}} $ & n & 1-4\\
\cline{3-5}
   &  & $\textbf{10}_{\textbf{e}_{\textbf{2}}}\bar{\textbf{5}}_{\textbf{e}_{\textbf{1}},\textbf{e}_{\textbf{3}}}\bar{\textbf{5}}_{\textbf{e}_{\textbf{4}},\textbf{e}_{\textbf{5}}}  $& n & 1-4\\
\cline{2-5}
 & \multirow{2}{*}{55}& $\textbf{10}_{\textbf{e}_{\textbf{1}}}\bar{\textbf{5}}_{\textbf{e}_{\textbf{2}},\textbf{e}_{\textbf{3}}}\bar{\textbf{5}}_{\textbf{e}_{\textbf{4}},\textbf{e}_{\textbf{5}}} $ & n & 1-4\\
\cline{3-5}
   &  & $\textbf{10}_{\textbf{e}_{\textbf{2}}}\bar{\textbf{5}}_{\textbf{e}_{\textbf{1}},\textbf{e}_{\textbf{3}}}\bar{\textbf{5}}_{\textbf{e}_{\textbf{4}},\textbf{e}_{\textbf{5}}}  $& n & 1-4\\
\cline{2-5}
 & \multirow{2}{*}{56}& $\textbf{10}_{\textbf{e}_{\textbf{1}}}\bar{\textbf{5}}_{\textbf{e}_{\textbf{2}},\textbf{e}_{\textbf{3}}}\bar{\textbf{5}}_{\textbf{e}_{\textbf{4}},\textbf{e}_{\textbf{5}}} $ & n & 1-4\\
\cline{3-5}
   &  & $\textbf{10}_{\textbf{e}_{\textbf{2}}}\bar{\textbf{5}}_{\textbf{e}_{\textbf{1}},\textbf{e}_{\textbf{3}}}\bar{\textbf{5}}_{\textbf{e}_{\textbf{4}},\textbf{e}_{\textbf{5}}}  $& n & 1-4\\
\cline{2-5}
&  \multirow{2}{*}{57}& $\textbf{10}_{\textbf{e}_{\textbf{1}}}\bar{\textbf{5}}_{\textbf{e}_{\textbf{2}},\textbf{e}_{\textbf{3}}}\bar{\textbf{5}}_{\textbf{e}_{\textbf{4}},\textbf{e}_{\textbf{5}}} $ & n & 1-4\\
\cline{3-5}
   &  & $\textbf{10}_{\textbf{e}_{\textbf{2}}}\bar{\textbf{5}}_{\textbf{e}_{\textbf{1}},\textbf{e}_{\textbf{3}}}\bar{\textbf{5}}_{\textbf{e}_{\textbf{4}},\textbf{e}_{\textbf{5}}}  $& n & 1-4\\
\cline{2-5}
& \multirow{2}{*}{58}& $\textbf{10}_{\textbf{e}_{\textbf{1}}}\bar{\textbf{5}}_{\textbf{e}_{\textbf{2}},\textbf{e}_{\textbf{3}}}\bar{\textbf{5}}_{\textbf{e}_{\textbf{4}},\textbf{e}_{\textbf{5}}} $ & n & 1-4\\
\cline{3-5}
   &  & $\textbf{10}_{\textbf{e}_{\textbf{2}}}\bar{\textbf{5}}_{\textbf{e}_{\textbf{1}},\textbf{e}_{\textbf{3}}}\bar{\textbf{5}}_{\textbf{e}_{\textbf{4}},\textbf{e}_{\textbf{5}}}  $& n & 1-4\\
\cline{2-5}
&  \multirow{2}{*}{59}& $\textbf{10}_{\textbf{e}_{\textbf{1}}}\bar{\textbf{5}}_{\textbf{e}_{\textbf{2}},\textbf{e}_{\textbf{3}}}\bar{\textbf{5}}_{\textbf{e}_{\textbf{4}},\textbf{e}_{\textbf{5}}} $ & n & 1-4\\
\cline{3-5}
   &  & $\textbf{10}_{\textbf{e}_{\textbf{2}}}\bar{\textbf{5}}_{\textbf{e}_{\textbf{1}},\textbf{e}_{\textbf{3}}}\bar{\textbf{5}}_{\textbf{e}_{\textbf{4}},\textbf{e}_{\textbf{5}}}  $& n & 1-4\\
\cline{2-5}
   & 60 & $\textbf{10}_{\textbf{e}_{\textbf{5}}}\bar{\textbf{5}}_{\textbf{e}_{\textbf{1}},\textbf{e}_{\textbf{3}}}\bar{\textbf{5}}_{\textbf{e}_{\textbf{2}},\textbf{e}_{\textbf{4}}} $& y & 1-4\\
\cline{2-5}
   & 61 & $\textbf{10}_{\textbf{e}_{\textbf{2}}}\bar{\textbf{5}}_{\textbf{e}_{\textbf{1}},\textbf{e}_{\textbf{3}}}\bar{\textbf{5}}_{\textbf{e}_{\textbf{4}},\textbf{e}_{\textbf{5}}} $& y & 1-4\\
\cline{2-5}
   & 62 & $\textbf{10}_{\textbf{e}_{\textbf{5}}}\bar{\textbf{5}}_{\textbf{e}_{\textbf{1}},\textbf{e}_{\textbf{3}}}\bar{\textbf{5}}_{\textbf{e}_{\textbf{2}},\textbf{e}_{\textbf{4}}} $& n & 1-4\\
\cline{2-5}
   & 63 & $\textbf{10}_{\textbf{e}_{\textbf{5}}}\bar{\textbf{5}}_{\textbf{e}_{\textbf{1}},\textbf{e}_{\textbf{3}}}\bar{\textbf{5}}_{\textbf{e}_{\textbf{2}},\textbf{e}_{\textbf{4}}} $& y & 1-4\\
   & \multirow{4}{*}{64} & $\textbf{10}_{\textbf{e}_{\textbf{2}}}\bar{\textbf{5}}_{\textbf{e}_{\textbf{1}},\textbf{e}_{\textbf{5}}}\bar{\textbf{5}}_{\textbf{e}_{\textbf{3}},\textbf{e}_{\textbf{4}}} $ & n & 1,3\\
\cline{3-5}
   &  & $\textbf{10}_{\textbf{e}_{\textbf{3}}}\bar{\textbf{5}}_{\textbf{e}_{\textbf{1}},\textbf{e}_{\textbf{5}}}\bar{\textbf{5}}_{\textbf{e}_{\textbf{2}},\textbf{e}_{\textbf{4}}}  $ & n & 1-4\\
\cline{3-5}
   &  & $\textbf{10}_{\textbf{e}_{\textbf{5}}}\bar{\textbf{5}}_{\textbf{e}_{\textbf{1}},\textbf{e}_{\textbf{2}}}\bar{\textbf{5}}_{\textbf{e}_{\textbf{3}},\textbf{e}_{\textbf{4}}} $& n & 1,3\\
\cline{3-5}
   &  & $\textbf{10}_{\textbf{e}_{\textbf{5}}}\bar{\textbf{5}}_{\textbf{e}_{\textbf{1}},\textbf{e}_{\textbf{3}}}\bar{\textbf{5}}_{\textbf{e}_{\textbf{2}},\textbf{e}_{\textbf{4}}} $& n & 1-4\\
\cline{2-5}
&  \multirow{4}{*}{65} & $\textbf{10}_{\textbf{e}_{\textbf{2}}}\bar{\textbf{5}}_{\textbf{e}_{\textbf{1}},\textbf{e}_{\textbf{5}}}\bar{\textbf{5}}_{\textbf{e}_{\textbf{3}},\textbf{e}_{\textbf{4}}} $ & n & 1-4\\
\cline{3-5}
   &  & $\textbf{10}_{\textbf{e}_{\textbf{3}}}\bar{\textbf{5}}_{\textbf{e}_{\textbf{1}},\textbf{e}_{\textbf{5}}}\bar{\textbf{5}}_{\textbf{e}_{\textbf{2}},\textbf{e}_{\textbf{4}}}  $ & n & 1-4\\
\cline{3-5}
   &  & $\textbf{10}_{\textbf{e}_{\textbf{5}}}\bar{\textbf{5}}_{\textbf{e}_{\textbf{1}},\textbf{e}_{\textbf{2}}}\bar{\textbf{5}}_{\textbf{e}_{\textbf{3}},\textbf{e}_{\textbf{4}}} $& n & 1-4\\
\cline{3-5}
   &  & $\textbf{10}_{\textbf{e}_{\textbf{5}}}\bar{\textbf{5}}_{\textbf{e}_{\textbf{1}},\textbf{e}_{\textbf{3}}}\bar{\textbf{5}}_{\textbf{e}_{\textbf{2}},\textbf{e}_{\textbf{4}}} $& n & 1-4\\
\cline{2-5}
 & \multirow{4}{*}{66} & $\textbf{10}_{\textbf{e}_{\textbf{2}}}\bar{\textbf{5}}_{\textbf{e}_{\textbf{1}},\textbf{e}_{\textbf{5}}}\bar{\textbf{5}}_{\textbf{e}_{\textbf{3}},\textbf{e}_{\textbf{4}}} $ & n & 1-4\\
\cline{3-5}
   &  & $\textbf{10}_{\textbf{e}_{\textbf{3}}}\bar{\textbf{5}}_{\textbf{e}_{\textbf{1}},\textbf{e}_{\textbf{5}}}\bar{\textbf{5}}_{\textbf{e}_{\textbf{2}},\textbf{e}_{\textbf{4}}}  $ & n & 1-4\\
\cline{3-5}
   &  & $\textbf{10}_{\textbf{e}_{\textbf{5}}}\bar{\textbf{5}}_{\textbf{e}_{\textbf{1}},\textbf{e}_{\textbf{2}}}\bar{\textbf{5}}_{\textbf{e}_{\textbf{3}},\textbf{e}_{\textbf{4}}} $& n & 1-4\\
\cline{3-5}
   &  & $\textbf{10}_{\textbf{e}_{\textbf{5}}}\bar{\textbf{5}}_{\textbf{e}_{\textbf{1}},\textbf{e}_{\textbf{3}}}\bar{\textbf{5}}_{\textbf{e}_{\textbf{2}},\textbf{e}_{\textbf{4}}} $& n & 1-4\\
\cline{2-5}
 &  \multirow{4}{*}{67} & $\textbf{10}_{\textbf{e}_{\textbf{2}}}\bar{\textbf{5}}_{\textbf{e}_{\textbf{1}},\textbf{e}_{\textbf{5}}}\bar{\textbf{5}}_{\textbf{e}_{\textbf{3}},\textbf{e}_{\textbf{4}}} $ & n & 1-4\\
\cline{3-5}
   &  & $\textbf{10}_{\textbf{e}_{\textbf{3}}}\bar{\textbf{5}}_{\textbf{e}_{\textbf{1}},\textbf{e}_{\textbf{5}}}\bar{\textbf{5}}_{\textbf{e}_{\textbf{2}},\textbf{e}_{\textbf{4}}}  $ & n & 1-4\\
\cline{3-5}
   &  & $\textbf{10}_{\textbf{e}_{\textbf{5}}}\bar{\textbf{5}}_{\textbf{e}_{\textbf{1}},\textbf{e}_{\textbf{2}}}\bar{\textbf{5}}_{\textbf{e}_{\textbf{3}},\textbf{e}_{\textbf{4}}} $& n & 1-4\\
\cline{3-5}
   &  & $\textbf{10}_{\textbf{e}_{\textbf{5}}}\bar{\textbf{5}}_{\textbf{e}_{\textbf{1}},\textbf{e}_{\textbf{3}}}\bar{\textbf{5}}_{\textbf{e}_{\textbf{2}},\textbf{e}_{\textbf{4}}} $& n & 1-4\\
   \cline{2-5}
   & \multirow{4}{*}{68} & $\textbf{10}_{\textbf{e}_{\textbf{2}}}\bar{\textbf{5}}_{\textbf{e}_{\textbf{1}},\textbf{e}_{\textbf{5}}}\bar{\textbf{5}}_{\textbf{e}_{\textbf{3}},\textbf{e}_{\textbf{4}}} $ & n & 2,4\\
\cline{3-5}
   &  & $\textbf{10}_{\textbf{e}_{\textbf{3}}}\bar{\textbf{5}}_{\textbf{e}_{\textbf{1}},\textbf{e}_{\textbf{5}}}\bar{\textbf{5}}_{\textbf{e}_{\textbf{2}},\textbf{e}_{\textbf{4}}}  $ & n & 1-4\\
\cline{3-5}
   &  & $\textbf{10}_{\textbf{e}_{\textbf{5}}}\bar{\textbf{5}}_{\textbf{e}_{\textbf{1}},\textbf{e}_{\textbf{2}}}\bar{\textbf{5}}_{\textbf{e}_{\textbf{3}},\textbf{e}_{\textbf{4}}} $& n & 2,4\\
\cline{3-5}
   &  & $\textbf{10}_{\textbf{e}_{\textbf{5}}}\bar{\textbf{5}}_{\textbf{e}_{\textbf{1}},\textbf{e}_{\textbf{3}}}\bar{\textbf{5}}_{\textbf{e}_{\textbf{2}},\textbf{e}_{\textbf{4}}} $& n & 1-4\\
\cline{2-5}
\hline
\end{tabular}
\end{center}

\begin{center}
\begin{tabular}{|c|c|c|c|c|c|}
\hline
CICY No.  & Model No. & Yukawa Pattern & Top. Van. & Sym. No.\\
\hline
\multirow{16}{*}{7487}  

& {69} &   $\textbf{10}_{\textbf{e}_{\textbf{2}}}\bar{\textbf{5}}_{\textbf{e}_{\textbf{1}},\textbf{e}_{\textbf{3}}}\bar{\textbf{5}}_{\textbf{e}_{\textbf{4}},\textbf{e}_{\textbf{5}}} $& n & 1-4\\
\cline{2-5}
 & {70} &   $\textbf{10}_{\textbf{e}_{\textbf{2}}}\bar{\textbf{5}}_{\textbf{e}_{\textbf{1}},\textbf{e}_{\textbf{3}}}\bar{\textbf{5}}_{\textbf{e}_{\textbf{4}},\textbf{e}_{\textbf{5}}}$& y & 1-4\\
\cline{2-5}
 &\multirow{2}{*}{71} &  $\textbf{10}_{\textbf{e}_{\textbf{2}}}\bar{\textbf{5}}_{\textbf{e}_{\textbf{1}},\textbf{e}_{\textbf{3}}}\bar{\textbf{5}}_{\textbf{e}_{\textbf{4}},\textbf{e}_{\textbf{5}}} $ & n & 1-4\\
\cline{3-5}
  &  & $\textbf{10}_{\textbf{e}_{\textbf{5}}}\bar{\textbf{5}}_{\textbf{e}_{\textbf{1}},\textbf{e}_{\textbf{3}}}\bar{\textbf{5}}_{\textbf{e}_{\textbf{2}},\textbf{e}_{\textbf{4}}} $& n & 1-4\\
\cline{2-5}
  & \multirow{2}{*}{72} & $\textbf{10}_{\textbf{e}_{\textbf{2}}}\bar{\textbf{5}}_{\textbf{e}_{\textbf{1}},\textbf{e}_{\textbf{3}}}\bar{\textbf{5}}_{\textbf{e}_{\textbf{4}},\textbf{e}_{\textbf{5}}}$ & n & 1-4\\
\cline{3-5}
   &  & $\textbf{10}_{\textbf{e}_{\textbf{5}}}\bar{\textbf{5}}_{\textbf{e}_{\textbf{1}},\textbf{e}_{\textbf{4}}}\bar{\textbf{5}}_{\textbf{e}_{\textbf{2}},\textbf{e}_{\textbf{3}}}$& n & 1,3\\
\cline{2-5}
&  \multirow{2}{*}{73} & $\textbf{10}_{\textbf{e}_{\textbf{2}}}\bar{\textbf{5}}_{\textbf{e}_{\textbf{1}},\textbf{e}_{\textbf{3}}}\bar{\textbf{5}}_{\textbf{e}_{\textbf{4}},\textbf{e}_{\textbf{5}}}$ & n & 1-4\\
\cline{3-5}
   &  & $\textbf{10}_{\textbf{e}_{\textbf{5}}}\bar{\textbf{5}}_{\textbf{e}_{\textbf{1}},\textbf{e}_{\textbf{4}}}\bar{\textbf{5}}_{\textbf{e}_{\textbf{2}},\textbf{e}_{\textbf{3}}}$& n & 1-4\\
\cline{2-5}
 &  \multirow{2}{*}{74} & $\textbf{10}_{\textbf{e}_{\textbf{2}}}\bar{\textbf{5}}_{\textbf{e}_{\textbf{1}},\textbf{e}_{\textbf{3}}}\bar{\textbf{5}}_{\textbf{e}_{\textbf{4}},\textbf{e}_{\textbf{5}}}$ & n & 1-4\\
\cline{3-5}
   &  & $\textbf{10}_{\textbf{e}_{\textbf{5}}}\bar{\textbf{5}}_{\textbf{e}_{\textbf{1}},\textbf{e}_{\textbf{4}}}\bar{\textbf{5}}_{\textbf{e}_{\textbf{2}},\textbf{e}_{\textbf{3}}}$& n & 1-4\\
\cline{2-5}
 &  \multirow{2}{*}{75} & $\textbf{10}_{\textbf{e}_{\textbf{2}}}\bar{\textbf{5}}_{\textbf{e}_{\textbf{1}},\textbf{e}_{\textbf{3}}}\bar{\textbf{5}}_{\textbf{e}_{\textbf{4}},\textbf{e}_{\textbf{5}}}$ & n & 1-4\\
\cline{3-5}
   &  & $\textbf{10}_{\textbf{e}_{\textbf{5}}}\bar{\textbf{5}}_{\textbf{e}_{\textbf{1}},\textbf{e}_{\textbf{4}}}\bar{\textbf{5}}_{\textbf{e}_{\textbf{2}},\textbf{e}_{\textbf{3}}}$& n & 1-4\\
\cline{2-5}
&  \multirow{2}{*}{76} & $\textbf{10}_{\textbf{e}_{\textbf{2}}}\bar{\textbf{5}}_{\textbf{e}_{\textbf{1}},\textbf{e}_{\textbf{3}}}\bar{\textbf{5}}_{\textbf{e}_{\textbf{4}},\textbf{e}_{\textbf{5}}}$ & n & 1-4\\
\cline{3-5}
   &  & $\textbf{10}_{\textbf{e}_{\textbf{5}}}\bar{\textbf{5}}_{\textbf{e}_{\textbf{1}},\textbf{e}_{\textbf{4}}}\bar{\textbf{5}}_{\textbf{e}_{\textbf{2}},\textbf{e}_{\textbf{3}}}$& n & 1-4\\
\cline{2-5}
&  \multirow{2}{*}{77} & $\textbf{10}_{\textbf{e}_{\textbf{2}}}\bar{\textbf{5}}_{\textbf{e}_{\textbf{1}},\textbf{e}_{\textbf{3}}}\bar{\textbf{5}}_{\textbf{e}_{\textbf{4}},\textbf{e}_{\textbf{5}}}$ & n & 1-4\\
\cline{3-5}
   &  & $\textbf{10}_{\textbf{e}_{\textbf{5}}}\bar{\textbf{5}}_{\textbf{e}_{\textbf{1}},\textbf{e}_{\textbf{4}}}\bar{\textbf{5}}_{\textbf{e}_{\textbf{2}},\textbf{e}_{\textbf{3}}}$& n & 1-4\\
\cline{2-5}
& \multirow{2}{*}{78} & $\textbf{10}_{\textbf{e}_{\textbf{2}}}\bar{\textbf{5}}_{\textbf{e}_{\textbf{1}},\textbf{e}_{\textbf{3}}}\bar{\textbf{5}}_{\textbf{e}_{\textbf{4}},\textbf{e}_{\textbf{5}}}$ & n & 1-4\\
\cline{3-5}
   &  & $\textbf{10}_{\textbf{e}_{\textbf{5}}}\bar{\textbf{5}}_{\textbf{e}_{\textbf{1}},\textbf{e}_{\textbf{4}}}\bar{\textbf{5}}_{\textbf{e}_{\textbf{2}},\textbf{e}_{\textbf{3}}}$& n & 1-4\\
\cline{2-5}
 & \multirow{2}{*}{79} & $\textbf{10}_{\textbf{e}_{\textbf{2}}}\bar{\textbf{5}}_{\textbf{e}_{\textbf{1}},\textbf{e}_{\textbf{3}}}\bar{\textbf{5}}_{\textbf{e}_{\textbf{4}},\textbf{e}_{\textbf{5}}}$ & n & 1-4\\
\cline{3-5}
   &  & $\textbf{10}_{\textbf{e}_{\textbf{5}}}\bar{\textbf{5}}_{\textbf{e}_{\textbf{1}},\textbf{e}_{\textbf{4}}}\bar{\textbf{5}}_{\textbf{e}_{\textbf{2}},\textbf{e}_{\textbf{3}}}$& n & 1-4\\
\cline{2-5}
& \multirow{2}{*}{80} & $\textbf{10}_{\textbf{e}_{\textbf{2}}}\bar{\textbf{5}}_{\textbf{e}_{\textbf{1}},\textbf{e}_{\textbf{3}}}\bar{\textbf{5}}_{\textbf{e}_{\textbf{4}},\textbf{e}_{\textbf{5}}}$ & n & 1-4\\
\cline{3-5}
   &  & $\textbf{10}_{\textbf{e}_{\textbf{5}}}\bar{\textbf{5}}_{\textbf{e}_{\textbf{1}},\textbf{e}_{\textbf{4}}}\bar{\textbf{5}}_{\textbf{e}_{\textbf{2}},\textbf{e}_{\textbf{3}}}$& n & 1-4\\
\cline{2-5}
& \multirow{2}{*}{81} & $\textbf{10}_{\textbf{e}_{\textbf{2}}}\bar{\textbf{5}}_{\textbf{e}_{\textbf{1}},\textbf{e}_{\textbf{3}}}\bar{\textbf{5}}_{\textbf{e}_{\textbf{4}},\textbf{e}_{\textbf{5}}}$ & n & 1-4\\
\cline{3-5}
   &  & $\textbf{10}_{\textbf{e}_{\textbf{5}}}\bar{\textbf{5}}_{\textbf{e}_{\textbf{1}},\textbf{e}_{\textbf{4}}}\bar{\textbf{5}}_{\textbf{e}_{\textbf{2}},\textbf{e}_{\textbf{3}}}$& n & 2,4\\
\cline{2-5}
\hline
\end{tabular}
\end{center}
On CICY 7487, a total of 276 models have Yukawa couplings consistent with the gauge symmetries of the models, with 1188 Yukawa couplings being permitted in total. Of these, 580 are of the form $\textbf{10}_{\textbf{e}_{\textbf{i}}}\bar{\textbf{5}}_{\textbf{e}_{\textbf{j}},\textbf{e}_{\textbf{k}}}\bar{\textbf{5}}_{\textbf{e}_{\textbf{l}},\textbf{e}_{\textbf{m}}} $, 444 are of the form $\textbf{1}_{\textbf{e}_{\textbf{i}},-\textbf{e}_{\textbf{j}}}\textbf{5}_{-\textbf{e}_{\textbf{i}},-\textbf{e}_{\textbf{k}}}\bar{\textbf{5}}_{\textbf{e}_{\textbf{j}}, \textbf{e}_{\textbf{k}}}$ and 164 are of the form $ \bold{5}_{\bold{{-{\bf e}_i}},\bold{-{{\bf e}_j}}}\bold{10}_{\bold{{{\bf e}_i}}}\bold{10}_{\bold{{{\bf e}_j}}}$. A total of 112 couplings exhibit the topological vanishing we have been studying in this paper, 32 of the form  $\textbf{10}_{\textbf{e}_{\textbf{i}}}\bar{\textbf{5}}_{\textbf{e}_{\textbf{j}},\textbf{e}_{\textbf{k}}}\bar{\textbf{5}}_{\textbf{e}_{\textbf{l}},\textbf{e}_{\textbf{m}}} $ and 80 of the form $ \bold{5}_{\bold{{-{\bf e}_i}},\bold{-{{\bf e}_j}}}\bold{10}_{\bold{{{\bf e}_i}}}\bold{10}_{\bold{{{\bf e}_j}}}$. A total of 112 out of the 276 models have at least one topologically vanishing coupling.



\begin{thebibliography}{99}
\ifx\doiref\asklfhas\newcommand{\doiref}[2]{\href{http://dx.doi.org/#1}{#2}}\fi
\raggedright 
\ifx\arxivref\asklfhas\newcommand{\arxivref}[2]{\href{http://arxiv.org/abs/#1}{arXiv:#1}}\fi
\raggedright


\bibitem{Candelas:1985en}
  P.~Candelas, G.~T.~Horowitz, A.~Strominger, E.~Witten,
  ``Vacuum Configurations for Superstrings,''
  Nucl.\ Phys.\  {\bf B258 } (1985)  46-74.

\bibitem{gsw}  
  M.~B.~Green, J.~H.~Schwarz and E.~Witten,
  ``Superstring theory. Vol. 2: Loop amplitudes, anomalies and phenomenology,''
{\it  Cambridge, Uk: Univ. Pr. ( 1987) 596 P. ( Cambridge Monographs On Mathematical Physics)}

\bibitem{Greene:1986bm}
  B.~R.~Greene, K.~H.~Kirklin, P.~J.~Miron and G.~G.~Ross,
  ``A Three Generation Superstring Model. 1. Compactification and Discrete Symmetries,''
  Nucl.\ Phys.\ B {\bf 278} (1986) 667.

\bibitem{Greene:1986jb}
  B.~R.~Greene, K.~H.~Kirklin, P.~J.~Miron and G.~G.~Ross,
  ``A Three Generation Superstring Model. 2. Symmetry Breaking and the Low-Energy Theory,''
  Nucl.\ Phys.\ B {\bf 292} (1987) 606.

\bibitem{Braun:2011ni}
  V.~Braun, P.~Candelas, R.~Davies and R.~Donagi,
  ``The MSSM Spectrum from (0,2)-Deformations of the Heterotic Standard Embedding,''
  arXiv:1112.1097 [hep-th].

\bibitem{Braun:2005ux}
  V.~Braun, Y.~H.~He, B.~A.~Ovrut and T.~Pantev,
  ``A Heterotic standard model,''
  Phys.\ Lett.\  B {\bf 618}, 252 (2005)
  [arXiv:hep-th/0501070].

\bibitem{Braun:2005bw}
  V.~Braun, Y.~H.~He, B.~A.~Ovrut and T.~Pantev,
  ``A Standard model from the E(8) x E(8) heterotic superstring,''
  JHEP {\bf 0506}, 039 (2005)
  [arXiv:hep-th/0502155].

\bibitem{Braun:2005zv}
  V.~Braun, Y.~H.~He, B.~A.~Ovrut and T.~Pantev,
  ``Vector bundle extensions, sheaf cohomology, and the heterotic standard
  model,''
  Adv.\ Theor.\ Math.\ Phys.\  {\bf 10}, 4 (2006)
  [arXiv:hep-th/0505041].
  
\bibitem{Distler:1987ee}
  J.~Distler and B.~R.~Greene,
  ``Aspects of (2,0) String Compactifications,''
  Nucl.\ Phys.\ B {\bf 304} (1988) 1.
  
\bibitem{Kachru:1995em}
  S.~Kachru,
  ``Some three generation (0,2) Calabi-Yau models,''
  Phys.\ Lett.\ B {\bf 349} (1995) 76
  [hep-th/9501131].

\bibitem{Bouchard:2005ag}
  V.~Bouchard and R.~Donagi,
  ``An SU(5) heterotic standard model,''
  Phys.\ Lett.\  B {\bf 633}, 783 (2006)
  [arXiv:hep-th/0512149].

\bibitem{Braun:2005nv}
  V.~Braun, Y.~H.~He, B.~A.~Ovrut and T.~Pantev,
  ``The Exact MSSM spectrum from string theory,''
  JHEP {\bf 0605}, 043 (2006)
  [arXiv:hep-th/0512177].
  
\bibitem{Bouchard:2006dn}
  V.~Bouchard, M.~Cvetic and R.~Donagi,
  ``Tri-linear couplings in an heterotic minimal supersymmetric standard model,''
  Nucl.\ Phys.\  B {\bf 745}, 62 (2006)
  [arXiv:hep-th/0602096].

\bibitem{Blumenhagen:2006ux}
  R.~Blumenhagen, S.~Moster and T.~Weigand,
  ``Heterotic GUT and standard model vacua from simply connected Calabi-Yau
  manifolds,''
  Nucl.\ Phys.\  B {\bf 751}, 186 (2006)
  [arXiv:hep-th/0603015].

\bibitem{Blumenhagen:2006wj}
  R.~Blumenhagen, S.~Moster, R.~Reinbacher and T.~Weigand,
  ``Massless Spectra of Three Generation U(N) Heterotic String Vacua,''
  JHEP {\bf 0705}, 041 (2007)
  [arXiv:hep-th/0612039].
  
\bibitem{Anderson:2007nc}
  L.~B.~Anderson, Y.~H.~He and A.~Lukas,
  ``Heterotic Compactification, An Algorithmic Approach,''
  JHEP {\bf 0707}, 049 (2007)
  [arXiv:hep-th/0702210].

\bibitem{Anderson:2008uw}
  L.~B.~Anderson, Y.~H.~He and A.~Lukas,
  ``Monad Bundles in Heterotic String Compactifications,''
  JHEP {\bf 0807}, 104 (2008)
  [arXiv:0805.2875 [hep-th]].

\bibitem{Anderson:2009mh} 
  L.~B.~Anderson, J.~Gray, Y.~H.~He and A.~Lukas,
  ``Exploring Positive Monad Bundles And A New Heterotic Standard Model,''
  JHEP {\bf 1002}, 054 (2010)
  doi:10.1007/JHEP02(2010)054
  [arXiv:0911.1569 [hep-th]].

\bibitem{Anderson:2011ns} 
  L.~B.~Anderson, J.~Gray, A.~Lukas and E.~Palti,
  ``Two Hundred Heterotic Standard Models on Smooth Calabi-Yau Threefolds,''
  Phys.\ Rev.\ D {\bf 84}, 106005 (2011)
  doi:10.1103/PhysRevD.84.106005
  [arXiv:1106.4804 [hep-th]].
  
\bibitem{Anderson:2012yf} 
  L.~B.~Anderson, J.~Gray, A.~Lukas and E.~Palti,
  ``Heterotic Line Bundle Standard Models,''
  JHEP {\bf 1206}, 113 (2012)
  doi:10.1007/JHEP06(2012)113
  [arXiv:1202.1757 [hep-th]].

\bibitem{Anderson:2013xka} 
  L.~B.~Anderson, A.~Constantin, J.~Gray, A.~Lukas and E.~Palti,
  ``A Comprehensive Scan for Heterotic SU(5) GUT models,''
  JHEP {\bf 1401}, 047 (2014)
  doi:10.1007/JHEP01(2014)047
  [arXiv:1307.4787 [hep-th]].
  
  \bibitem{He:2013ofa} 
  Y.~H.~He, S.~J.~Lee, A.~Lukas and C.~Sun,
  ``Heterotic Model Building: 16 Special Manifolds,''
  JHEP {\bf 1406}, 077 (2014)
  doi:10.1007/JHEP06(2014)077
  [arXiv:1309.0223 [hep-th]].
  
  \bibitem{Constantin:2015bea} 
  A.~Constantin, A.~Lukas and C.~Mishra,
  ``The Family Problem: Hints from Heterotic Line Bundle Models,''
  JHEP {\bf 1603}, 173 (2016)
  doi:10.1007/JHEP03(2016)173
  [arXiv:1509.02729 [hep-th]].

\bibitem{Braun:2017feb} 
  A.~P.~Braun, C.~R.~Brodie and A.~Lukas,
  ``Heterotic Line Bundle Models on Elliptically Fibered Calabi-Yau Three-folds,''
  JHEP {\bf 1804}, 087 (2018)
  doi:10.1007/JHEP04(2018)087
  [arXiv:1706.07688 [hep-th]].
  
  \bibitem{Constantin:2018xkj} 
  A.~Constantin, Y.~H.~He and A.~Lukas,
  ``Counting String Theory Standard Models,''
  Phys.\ Lett.\ B {\bf 792}, 258 (2019)
  doi:10.1016/j.physletb.2019.03.048
  [arXiv:1810.00444 [hep-th]].







\bibitem{Buchmuller:2005jr}
  W.~Buchmuller, K.~Hamaguchi, O.~Lebedev, M.~Ratz,
  ``Supersymmetric standard model from the heterotic string,''
  Phys.\ Rev.\ Lett.\  {\bf 96}, 121602 (2006).
  [hep-ph/0511035].

\bibitem{Buchmuller:2006ik}
  W.~Buchmuller, K.~Hamaguchi, O.~Lebedev, M.~Ratz,
  ``Supersymmetric Standard Model from the Heterotic String (II),''
  Nucl.\ Phys.\  {\bf B785}, 149-209 (2007).
  [hep-th/0606187].

\bibitem{Lebedev:2006kn}
  O.~Lebedev, H.~P.~Nilles, S.~Raby, S.~Ramos-Sanchez, M.~Ratz, P.~K.~S.~Vaudrevange, A.~Wingerter,
  ``A Mini-landscape of exact MSSM spectra in heterotic orbifolds,''
  Phys.\ Lett.\  {\bf B645}, 88-94 (2007).
  [hep-th/0611095].

\bibitem{Kim:2007mt}
  J.~E.~Kim, J.~-H.~Kim, B.~Kyae,
  ``Superstring standard model from Z(12-I) orbifold compactification with and without exotics, and effective R-parity,''
  JHEP {\bf 0706 } (2007)  034.
  [hep-ph/0702278 [HEP-PH]].

\bibitem{Lebedev:2007hv}
  O.~Lebedev, H.~P.~Nilles, S.~Raby, S.~Ramos-Sanchez, M.~Ratz, P.~K.~S.~Vaudrevange, A.~Wingerter,
  ``The Heterotic Road to the MSSM with R parity,''
  Phys.\ Rev.\  {\bf D77 } (2008)  046013.
  [arXiv:0708.2691 [hep-th]].

\bibitem{Lebedev:2008un}
  O.~Lebedev, H.~P.~Nilles, S.~Ramos-Sanchez, M.~Ratz, P.~K.~S.~Vaudrevange,
  ``Heterotic mini-landscape. (II). Completing the search for MSSM vacua in a Z(6) orbifold,''
  Phys.\ Lett.\  {\bf B668}, 331-335 (2008).
  [arXiv:0807.4384 [hep-th]].

\bibitem{Nibbelink:2009sp}
  S.~G.~Nibbelink, J.~Held, F.~Ruehle, M.~Trapletti, P.~K.~S.~Vaudrevange,
  ``Heterotic Z(6-II) MSSM Orbifolds in Blowup,''
  JHEP {\bf 0903}, 005 (2009).
  [arXiv:0901.3059 [hep-th]].

\bibitem{Blaszczyk:2009in}
  M.~Blaszczyk, S.~G.~Nibbelink, M.~Ratz, F.~Ruehle, M.~Trapletti, P.~K.~S.~Vaudrevange,
  ``A Z2xZ2 standard model,''
  Phys.\ Lett.\  {\bf B683}, 340-348 (2010).
  [arXiv:0911.4905 [hep-th]].

\bibitem{Blaszczyk:2010db}
  M.~Blaszczyk, S.~G.~Nibbelink, F.~Ruehle, M.~Trapletti, P.~K.~S.~Vaudrevange,
  ``Heterotic MSSM on a Resolved Orbifold,''
  JHEP {\bf 1009}, 065 (2010).
  [arXiv:1007.0203 [hep-th]].

\bibitem{Kappl:2010yu}
  R.~Kappl, B.~Petersen, S.~Raby, M.~Ratz, R.~Schieren, P.~K.~S.~Vaudrevange,
  ``String-derived MSSM vacua with residual R symmetries,''
  Nucl.\ Phys.\  {\bf B847}, 325-349 (2011).
  [arXiv:1012.4574 [hep-th]].


\bibitem{Assel:2009xa}
  B.~Assel, K.~Christodoulides, A.~E.~Faraggi, C.~Kounnas and J.~Rizos,
  ``Exophobic Quasi-Realistic Heterotic String Vacua,''
  Phys.\ Lett.\  B {\bf 683} (2010) 306
  [arXiv:0910.3697 [hep-th]].

\bibitem{Christodoulides:2011zs}
  K.~Christodoulides, A.~E.~Faraggi, J.~Rizos,
  ``Top Quark Mass in Exophobic Pati-Salam Heterotic String Model,''  
  [arXiv:1104.2264 [hep-ph]].

\bibitem{Cleaver:2011ir}
  G.~Cleaver, A.~E.~Faraggi, J.~Greenwald, D.~Moore, K.~Pechan, E.~Remkus, T.~Renner,
  ``Investigation of Quasi--Realistic Heterotic String Models with Reduced Higgs Spectrum,''
  [arXiv:1105.0447 [hep-ph]].


\bibitem{Mutter:2018sra} 
  A.~Mütter, E.~Parr and P.~K.~S.~Vaudrevange,
  ``Deep learning in the heterotic orbifold landscape,''
  Nucl.\ Phys.\ B {\bf 940}, 113 (2019)
  doi:10.1016/j.nuclphysb.2019.01.013
  [arXiv:1811.05993 [hep-th]].
  
  \bibitem{Faraggi:2017cnh} 
  A.~E.~Faraggi, J.~Rizos and H.~Sonmez,
  ``Classification of Standard-like Heterotic-String Vacua,''
  Nucl.\ Phys.\ B {\bf 927}, 1 (2018)
  doi:10.1016/j.nuclphysb.2017.12.006
  [arXiv:1709.08229 [hep-th]].
  



\bibitem{Maio:2011qn}
  M.~Maio and A.~N.~Schellekens,
  ``Permutation orbifolds of heterotic Gepner models,''
  Nucl.\ Phys.\ B {\bf 848} (2011) 594
  [arXiv:1102.5293 [hep-th]].
  
\bibitem{GatoRivera:2009yt}
  B.~Gato-Rivera and A.~N.~Schellekens,
  ``Heterotic Weight Lifting,''
  Nucl.\ Phys.\ B {\bf 828} (2010) 375
  [arXiv:0910.1526 [hep-th]].
  
\bibitem{GatoRivera:2010xn}
  B.~Gato-Rivera and A.~N.~Schellekens,
  ``Asymmetric Gepner Models II. Heterotic Weight Lifting,''
  Nucl.\ Phys.\ B {\bf 846} (2011) 429
  [arXiv:1009.1320 [hep-th]].
  
  
  \bibitem{Strominger:1985ks} 
  A.~Strominger,
  ``Yukawa Couplings in Superstring Compactification,''
  Phys.\ Rev.\ Lett.\  {\bf 55}, 2547 (1985).
  doi:10.1103/PhysRevLett.55.2547
  
  \bibitem{Candelas:1987se} 
  P.~Candelas,
  ``Yukawa Couplings Between (2,1) Forms,''
  Nucl.\ Phys.\ B {\bf 298}, 458 (1988).
  doi:10.1016/0550-3213(88)90351-3
  
  \bibitem{Candelas:1990pi} 
  P.~Candelas and X.~de la Ossa,
  ``Moduli Space of {Calabi-Yau} Manifolds,''
  Nucl.\ Phys.\ B {\bf 355}, 455 (1991).
  doi:10.1016/0550-3213(91)90122-E
      
      \bibitem{Distler:1987gg} 
  J.~Distler, B.~R.~Greene, K.~Kirklin and P.~Miron,
  ``Evaluation of 27-bar**3 Yukawa Couplings in a Three Generation Superstring Model,''
  Phys.\ Lett.\ B {\bf 195}, 41 (1987).
  doi:10.1016/0370-2693(87)90883-5
      
  \bibitem{Greene:1987xh} 
  B.~R.~Greene, K.~H.~Kirklin, P.~J.~Miron and G.~G.~Ross,
  ``27**3 Yukawa Couplings for a Three Generation Superstring Model,''
  Phys.\ Lett.\ B {\bf 192}, 111 (1987).
  doi:10.1016/0370-2693(87)91151-8

\bibitem{Distler:1995bc} 
  J.~Distler and S.~Kachru,
  ``Duality of (0,2) string vacua,''
  Nucl.\ Phys.\ B {\bf 442}, 64 (1995)
  doi:10.1016/S0550-3213(95)00130-1
  [hep-th/9501111].

  \bibitem{Braun:2006me} 
  V.~Braun, Y.~H.~He and B.~A.~Ovrut,
  ``Yukawa couplings in heterotic standard models,''
  JHEP {\bf 0604}, 019 (2006)
  doi:10.1088/1126-6708/2006/04/019
  [hep-th/0601204].
    
  \bibitem{Anderson:2009ge} 
  L.~B.~Anderson, J.~Gray, D.~Grayson, Y.~H.~He and A.~Lukas,
  ``Yukawa Couplings in Heterotic Compactification,''
  Commun.\ Math.\ Phys.\  {\bf 297}, 95 (2010)
  doi:10.1007/s00220-010-1033-8
  [arXiv:0904.2186 [hep-th]].

  \bibitem{Anderson:2010tc} 
  L.~B.~Anderson, J.~Gray and B.~Ovrut,
  ``Yukawa Textures From Heterotic Stability Walls,''
  JHEP {\bf 1005}, 086 (2010)
  doi:10.1007/JHEP05(2010)086
  [arXiv:1001.2317 [hep-th]].
 
\bibitem{Buchbinder:2014sya} 
  E.~I.~Buchbinder, A.~Constantin and A.~Lukas,
  ``Non-generic Couplings in Supersymmetric Standard Models,''
  Phys.\ Lett.\ B {\bf 748}, 251 (2015)
  doi:10.1016/j.physletb.2015.07.012
  [arXiv:1409.2412 [hep-th]].

  \bibitem{Blesneag:2015pvz} 
  S.~Blesneag, E.~I.~Buchbinder, P.~Candelas and A.~Lukas,
  ``Holomorphic Yukawa Couplings in Heterotic String Theory,''
  JHEP {\bf 1601}, 152 (2016)
  doi:10.1007/JHEP01(2016)152
  [arXiv:1512.05322 [hep-th]].
  
\bibitem{Blesneag:2016yag} 
  S.~Blesneag, E.~I.~Buchbinder and A.~Lukas,
  ``Holomorphic Yukawa Couplings for Complete Intersection Calabi-Yau Manifolds,''
  JHEP {\bf 1701}, 119 (2017)
  doi:10.1007/JHEP01(2017)119
  [arXiv:1607.03461 [hep-th]].

 
 \bibitem{Donagi:2009ra} 
  R.~Donagi and M.~Wijnholt,
  ``Higgs Bundles and UV Completion in F-Theory,''
  Commun.\ Math.\ Phys.\  {\bf 326}, 287 (2014)
  doi:10.1007/s00220-013-1878-8
  [arXiv:0904.1218 [hep-th]].

 \bibitem{Anderson:2010mh} 
  L.~B.~Anderson, J.~Gray, A.~Lukas and B.~Ovrut,
  ``Stabilizing the Complex Structure in Heterotic Calabi-Yau Vacua,''
  JHEP {\bf 1102}, 088 (2011)
  doi:10.1007/JHEP02(2011)088
  [arXiv:1010.0255 [hep-th]].

\bibitem{Anderson:2011ty} 
  L.~B.~Anderson, J.~Gray, A.~Lukas and B.~Ovrut,
  ``The Atiyah Class and Complex Structure Stabilization in Heterotic Calabi-Yau Compactifications,''
  JHEP {\bf 1110}, 032 (2011)
  doi:10.1007/JHEP10(2011)032
  [arXiv:1107.5076 [hep-th]].

\bibitem{Anderson:2011cza} 
  L.~B.~Anderson, J.~Gray, A.~Lukas and B.~Ovrut,
  ``Stabilizing All Geometric Moduli in Heterotic Calabi-Yau Vacua,''
  Phys.\ Rev.\ D {\bf 83}, 106011 (2011)
  doi:10.1103/PhysRevD.83.106011
  [arXiv:1102.0011 [hep-th]].

\bibitem{Anderson:2013qca} 
  L.~B.~Anderson, J.~Gray, A.~Lukas and B.~Ovrut,
  ``Vacuum Varieties, Holomorphic Bundles and Complex Structure Stabilization in Heterotic Theories,''
  JHEP {\bf 1307}, 017 (2013)
  doi:10.1007/JHEP07(2013)017
  [arXiv:1304.2704 [hep-th]].


\bibitem{Donagi:2004qk} 
  R.~Donagi, Y.~H.~He, B.~A.~Ovrut and R.~Reinbacher,
  ``Moduli dependent spectra of heterotic compactifications,''
  Phys.\ Lett.\ B {\bf 598}, 279 (2004)
  doi:10.1016/j.physletb.2004.08.010
  [hep-th/0403291].

\bibitem{Donagi:2004ia} 
  R.~Donagi, Y.~H.~He, B.~A.~Ovrut and R.~Reinbacher,
  ``The Particle spectrum of heterotic compactifications,''
  JHEP {\bf 0412}, 054 (2004)
  doi:10.1088/1126-6708/2004/12/054
  [hep-th/0405014].


\bibitem{Brandle:2002fa}
  M.~Brandle and A.~Lukas,
  ``Flop transitions in M theory cosmology,''
  Phys.\ Rev.\ D {\bf 68} (2003) 024030
  doi:10.1103/PhysRevD.68.024030
  [hep-th/0212263].

\bibitem{Jarv:2003qy} 
  L.~Jarv, T.~Mohaupt and F.~Saueressig,
  ``M theory cosmologies from singular Calabi-Yau compactifications,''
  JCAP {\bf 0402}, 012 (2004)
  doi:10.1088/1475-7516/2004/02/012
  [hep-th/0310174].

\bibitem{Kofman:2004yc} 
  L.~Kofman, A.~D.~Linde, X.~Liu, A.~Maloney, L.~McAllister and E.~Silverstein,
  ``Beauty is attractive: Moduli trapping at enhanced symmetry points,''
  JHEP {\bf 0405}, 030 (2004)
  doi:10.1088/1126-6708/2004/05/030
  [hep-th/0403001].

\bibitem{Mohaupt:2004pr} 
  T.~Mohaupt and F.~Saueressig,
  ``Dynamical conifold transitions and moduli trapping in M-theory cosmology,''
  JCAP {\bf 0501}, 006 (2005)
  doi:10.1088/1475-7516/2005/01/006
  [hep-th/0410273].

\bibitem{Lukas:2004du} 
  A.~Lukas, E.~Palti and P.~M.~Saffin,
  ``Type IIB conifold transitions in cosmology,''
  Phys.\ Rev.\ D {\bf 71}, 066001 (2005)
  doi:10.1103/PhysRevD.71.066001
  [hep-th/0411033].

\bibitem{Abel:2005jx} 
  S.~A.~Abel and J.~Gray,
  ``On the chaos of D-brane phase transitions,''
  JHEP {\bf 0511}, 018 (2005)
  doi:10.1088/1126-6708/2005/11/018
  [hep-th/0504170].

\bibitem{Greene:2007sa} 
  B.~Greene, S.~Judes, J.~Levin, S.~Watson and A.~Weltman,
  ``Cosmological moduli dynamics,''
  JHEP {\bf 0707}, 060 (2007)
  doi:10.1088/1126-6708/2007/07/060
  [hep-th/0702220].


\bibitem{Yau:1986gu} 
  S.~T.~Yau,
  ``Compact Three-dimensional Kahler Manifolds With Zero Ricci Curvature,''
  In *Argonne/Chicago 1985, Proceedings, Anomalies, Geometry, Topology*, 395-406

\bibitem{Hubsch:1986ny} 
  T.~Hubsch,
  ``Calabi-yau Manifolds: Motivations and Constructions,''
  Commun.\ Math.\ Phys.\  {\bf 108}, 291 (1987).
  
\bibitem{Candelas:1987kf} 
  P.~Candelas, A.~M.~Dale, C.~A.~Lutken and R.~Schimmrigk,
  ``Complete Intersection Calabi-Yau Manifolds,''
  Nucl.\ Phys.\ B {\bf 298}, 493 (1988).
  
\bibitem{Candelas:1987du} 
  P.~Candelas, C.~A.~Lutken and R.~Schimmrigk,
  ``Complete Intersection Calabi-yau Manifolds. 2. Three Generation Manifolds,''
  Nucl.\ Phys.\ B {\bf 306}, 113 (1988).
  
\bibitem{Green:1986ck} 
  P.~Green and T.~Hubsch,
  ``Calabi-yau Manifolds as Complete Intersections in Products of Complex Projective Spaces,''
  Commun.\ Math.\ Phys.\  {\bf 109}, 99 (1987).
  
\bibitem{Gray:2013mja} 
  J.~Gray, A.~S.~Haupt and A.~Lukas,
  ``All Complete Intersection Calabi-Yau Four-Folds,''
  JHEP {\bf 1307}, 070 (2013)
  [arXiv:1303.1832 [hep-th]].
  
  \bibitem{Gray:2014fla} 
  J.~Gray, A.~S.~Haupt and A.~Lukas,
  ``Topological Invariants and Fibration Structure of Complete Intersection Calabi-Yau Four-Folds,''
  JHEP {\bf 1409}, 093 (2014)
  [arXiv:1405.2073 [hep-th]].
 
 \bibitem{Anderson:2015iia} 
  L.~B.~Anderson, F.~Apruzzi, X.~Gao, J.~Gray and S.~J.~Lee,
  ``A new construction of Calabi?Yau manifolds: Generalized CICYs,''
  Nucl.\ Phys.\ B {\bf 906}, 441 (2016)
  doi:10.1016/j.nuclphysb.2016.03.016
  [arXiv:1507.03235 [hep-th]].

 
 \bibitem{Batyrev}
V.~V.~Batyrev,
``Dual Polyhedra and Mirror Symmetry for Calabi-Yau Hypersurfaces in Toric Varieties,"
[alg-geom/9310003].

\bibitem{Berglund:1991pp} 
  P.~Berglund and T.~Hubsch,
  ``A Generalized construction of mirror manifolds,''
  Nucl.\ Phys.\ B {\bf 393}, 377 (1993)
  [AMS/IP Stud.\ Adv.\ Math.\  {\bf 9}, 327 (1998)]
  doi:10.1016/0550-3213(93)90250-S
  [hep-th/9201014].

\bibitem{Kreuzer:2000qv} 
  M.~Kreuzer and H.~Skarke,
  ``Reflexive polyhedra, weights and toric Calabi-Yau fibrations,''
  Rev.\ Math.\ Phys.\  {\bf 14}, 343 (2002)
  doi:10.1142/S0129055X0200120X
  [math/0001106 [math-ag]].

\bibitem{Kreuzer:2000xy} 
  M.~Kreuzer and H.~Skarke,
  ``Complete classification of reflexive polyhedra in four-dimensions,''
  Adv.\ Theor.\ Math.\ Phys.\  {\bf 4}, 1209 (2002)
  doi:10.4310/ATMP.2000.v4.n6.a2
  [hep-th/0002240].

\bibitem{Altman:2014bfa} 
  R.~Altman, J.~Gray, Y.~H.~He, V.~Jejjala and B.~D.~Nelson,
  ``A Calabi-Yau Database: Threefolds Constructed from the Kreuzer-Skarke List,''
  JHEP {\bf 1502}, 158 (2015)
  doi:10.1007/JHEP02(2015)158
  [arXiv:1411.1418 [hep-th]].
  

\bibitem{Braun:2010vc} 
  V.~Braun,
  ``On Free Quotients of Complete Intersection Calabi-Yau Manifolds,''
  JHEP {\bf 1104}, 005 (2011)
  doi:10.1007/JHEP04(2011)005
  [arXiv:1003.3235 [hep-th]].
  

\bibitem{Hubsch:1992nu} 
  T.~Hubsch,
  ``Calabi-Yau manifolds: A Bestiary for physicists,''

\bibitem{Anderson:2008ex} 
  L.~B.~Anderson,
  ``Heterotic and M-theory Compactifications for String Phenomenology,''
  arXiv:0808.3621 [hep-th].
  

\bibitem{Gray:2018kss} 
  J.~Gray and H.~Parsian,
  ``Moduli identification methods in Type II compactifications,''
  JHEP {\bf 1807}, 158 (2018)
  doi:10.1007/JHEP07(2018)158
  [arXiv:1803.08176 [hep-th]].

\bibitem{atiyah}
M.~F.~Atiyah,
``Complex Analytic Connections in Fibre Bundles,''
Trans. AMS, Vol. 85, No. 1 (May, 1957), pp. 181-207.
  
  \bibitem{Gray:2006gn} 
  J.~Gray, Y.~H.~He and A.~Lukas,
  ``Algorithmic Algebraic Geometry and Flux Vacua,''
  JHEP {\bf 0609}, 031 (2006)
  doi:10.1088/1126-6708/2006/09/031
  [hep-th/0606122].
  
  \bibitem{Gray:2009fy} 
  J.~Gray,
  ``A Simple Introduction to Grobner Basis Methods in String Phenomenology,''
  Adv.\ High Energy Phys.\  {\bf 2011}, 217035 (2011)
  doi:10.1155/2011/217035
  [arXiv:0901.1662 [hep-th]].
  
\bibitem{Gray:2008zs} 
  J.~Gray, Y.~H.~He, A.~Ilderton and A.~Lukas,
  ``STRINGVACUA: A Mathematica Package for Studying Vacuum Configurations in String Phenomenology,''
  Comput.\ Phys.\ Commun.\  {\bf 180}, 107 (2009)
  doi:10.1016/j.cpc.2008.08.009
  [arXiv:0801.1508 [hep-th]].
  
\bibitem{cicypackage}
L.~B.~Anderson, J.~Gray, Y.-H.~He, S.~J. Lee, and A.~Lukas, ``CICY package", based on methods described in arXiv:0911.1569, arXiv:0911.0865, arXiv:0805.2875, hep-th/0703249, hep-th/0702210.
  
  \bibitem{Anderson:2014hia} 
  L.~B.~Anderson, A.~Constantin, S.~J.~Lee and A.~Lukas,
  ``Hypercharge Flux in Heterotic Compactifications,''
  Phys.\ Rev.\ D {\bf 91}, no. 4, 046008 (2015)
  doi:10.1103/PhysRevD.91.046008
  [arXiv:1411.0034 [hep-th]].


\bibitem{database1}
A machine readable version of the CICY list and the symmetries used in this paper can be \\ found here: \href{http://www-thphys.physics.ox.ac.uk/projects/CalabiYau/CicyQuotients/Cicy_Quotients/Cicy_Quotients.html}{``http://www-thphys.physics.ox.ac.uk/projects/CalabiYau/CicyQuotients/Cicy\_Quotients/Cicy\_Quotients.html"}.

\bibitem{database2}
A machine readable version of the list of Line Bundle Standard Models being studied in this paper can be found here: \href{http://www-thphys.physics.ox.ac.uk/projects/CalabiYau/linebundlemodels/index.html}{``http://www-thphys.physics.ox.ac.uk/projects/CalabiYau/linebundlemodels/index.html"}.


\bibitem{Buchbinder:2016jqr} 
  E.~I.~Buchbinder, A.~Constantin, J.~Gray and A.~Lukas,
  ``Yukawa Unification in Heterotic String Theory,''
  Phys.\ Rev.\ D {\bf 94}, no. 4, 046005 (2016)
  doi:10.1103/PhysRevD.94.046005
  [arXiv:1606.04032 [hep-th]].


\bibitem{Hamidi:1986vh} 
  S.~Hamidi and C.~Vafa,
  ``Interactions on Orbifolds,''
  Nucl.\ Phys.\ B {\bf 279}, 465 (1987).
  doi:10.1016/0550-3213(87)90006-X

\bibitem{Dixon:1986qv} 
  L.~J.~Dixon, D.~Friedan, E.~J.~Martinec and S.~H.~Shenker,
  ``The Conformal Field Theory of Orbifolds,''
  Nucl.\ Phys.\ B {\bf 282}, 13 (1987).
  doi:10.1016/0550-3213(87)90676-6

\bibitem{Font:1988tp} 
  A.~Font, L.~E.~Ibanez, H.~P.~Nilles and F.~Quevedo,
  ``Degenerate Orbifolds,''
  Nucl.\ Phys.\ B {\bf 307}, 109 (1988)
  Erratum: [Nucl.\ Phys.\ B {\bf 310}, 764 (1988)].
  doi:10.1016/0550-3213(88)90102-2, 10.1016/0550-3213(88)90524-X

\bibitem{Font:1988nc} 
  A.~Font, L.~E.~Ibanez, H.~P.~Nilles and F.~Quevedo,
  ``On the Concept of Naturalness in String Theories,''
  Phys.\ Lett.\ B {\bf 213}, 274 (1988).
  doi:10.1016/0370-2693(88)91760-1

\bibitem{Kobayashi:2011cw} 
  T.~Kobayashi, S.~L.~Parameswaran, S.~Ramos-Sanchez and I.~Zavala,
  ``Revisiting Coupling Selection Rules in Heterotic Orbifold Models,''
  JHEP {\bf 1205}, 008 (2012)
  Erratum: [JHEP {\bf 1212}, 049 (2012)]
  doi:10.1007/JHEP12(2012)049, 10.1007/JHEP05(2012)008
  [arXiv:1107.2137 [hep-th]].



 \end{thebibliography}
\end{document}